\documentclass[12pt]{elsarticle}

\usepackage{graphicx, import}
\usepackage{amsfonts,textgreek}
\usepackage{mathtools}
\usepackage{dsfont}
\usepackage{multicol}
\usepackage{bbm}
\usepackage{url}
\usepackage[dvipsnames]{xcolor}
\usepackage[colorlinks=true,allcolors=blue]{hyperref}
\usepackage[nameinlink,noabbrev,capitalize]{cleveref}
\usepackage{xfrac}
\usepackage{caption}
\usepackage{subcaption}
\usepackage{xspace}
\usepackage{stmaryrd}
\usepackage{enumerate}
\usepackage{soul}

\usepackage[utf8]{inputenc}
\usepackage{wrapfig,bm}
\usepackage{lineno}
\usepackage{siunitx} 
\usepackage{caption}
\usepackage{psfrag}
\usepackage{upgreek}
\usepackage{isomath}
\usepackage{latexsym}
\usepackage{array}
\usepackage{multirow}
\usepackage{booktabs}
\usepackage{colortbl}
\usepackage{verbatim}
\usepackage{graphicx}
\usepackage{epstopdf,overpic}
\usepackage{float}
\usepackage{soul}
\usepackage{color}
\usepackage{dirtytalk}
\usepackage{todonotes}
\usepackage{textcomp}
\usepackage{algorithm, algpseudocode, lmodern, mathtools}
\usepackage{tikz}
\usetikzlibrary{shapes.geometric, arrows}
\usepackage{amssymb}
\usepackage{geometry}
\usepackage{amsmath}
\usepackage{mathrsfs}  
\usepackage{graphicx}
\usepackage{titlesec}
\usepackage{indentfirst}
\usepackage{bigints}
\usepackage{breqn}
\usepackage[export]{adjustbox}
\usepackage{pgfplots}
\usetikzlibrary{plotmarks,spy,calc}
\newlength{\figureheight}
\newlength{\figurewidth}

\usepackage{multirow}
\usepackage[normalem]{ulem}
\newcommand{\eref}[1]{Equation~(\ref{#1})}
\newcommand{\erefs}[1]{Equations~(\ref{#1})}
\newcommand{\fref}[1]{Figure~\ref{#1}}
\newcommand{\frefs}[1]{Figures~\ref{#1}}

\newcommand{\sref}[1]{Section~\ref{#1}}

\usetikzlibrary{calc,fit,intersections,shapes}

\pgfdeclareradialshading{ballshading}{\pgfpoint{-10bp}{10bp}}
 {color(0bp)=(gray!40!white); 
 color(9bp)=(gray!75!white);
 color(18bp)=(gray!70!black); 
 color(25bp)=(gray!50!black); 
 color(50bp)=(black)}

\begin{document}

\begin{frontmatter}

\title{Level set topology optimization of metamaterial-based heat manipulators using isogeometric analysis}

\author[label1]{Chintan Jansari}
\author[label1,label3]{St\'ephane P.A. Bordas\corref{cor1}}\ead{stephane.bordas@alum.northwestern.edu}
\author[label2]{Elena Atroshchenko}

\cortext[cor1]{Corresponding author}
\address[label1]{Institute of Computational Engineering, Faculty of Sciences, Technology and Medicine, University of Luxembourg, Luxembourg City, Luxembourg.}
\address[label3]{Clyde Visiting Fellow, Department of Mechanical Engineering, The University of Utah, Salt Lake City, Utah, United States.}
\address[label2]{School of Civil and Environmental Engineering, University of New South Wales, Sydney, Australia.}


\begin{abstract}
We exploit level set topology optimization to find the optimal material distribution for metamaterial-based heat manipulators. The level set function, geometry, and solution field are parameterized using the non-uniform rational B-spline (NURBS) basis functions in order to take advantage of easy control of smoothness and continuity. In addition, NURBS approximations can produce conic geometries exactly and provide higher efficiency for higher-order elements. The values of the level set function at the control points (called expansion coefficients) are utilized as design variables. For optimization, we use an advanced mathematical programming technique, Sequential Quadratic Programming (SQP). Taking into account a large number of design variables and the small number of constraints associated with our optimization problem, the adjoint method is utilized to calculate the required sensitivities with respect to the design variables. The efficiency and robustness of the proposed method are demonstrated by solving three numerical examples. We have also shown that the current method can handle different geometries and types of objective functions. In addition, regularization techniques such as Tikhonov regularization and volume regularization have been explored to reduce unnecessary complexity and increase the manufacturability of optimized topologies.  

\end{abstract}


\end{frontmatter}

\section{Introduction}
\label{sec:Introduction}
\par Due to the special arrangement of the constituent materials, thermal metamaterials can have heat transfer capabilities superior to those of materials available in nature. Therefore, researchers have proposed the use of artificially created thermal metamaterials to control heat fluxes. The concept opened research opportunities in heat transfer applications. Creating devices that are thermal equivalent to resistors, capacitors, inductors, diodes, transistors, etc. is one such opportunity. These devices are called heat manipulators, and several of them have already been proposed, such as thermal cloak~\cite{Narayana2012,Guenneau2012,schittnyExperimentsTransformationThermodynamics2013b,Han2014,Han2014_2,Sklan2016,Li2019,FUJII2019}, thermal concentrator~\cite{Narayana2012,Guenneau2012,schittnyExperimentsTransformationThermodynamics2013b,Li2019,Shen2016}, thermal camouflage~\cite{Han2014,Peng2020}, heat flux inverter~\cite{Narayana2012} etc. The spatial layout of the member materials of a metamaterial-based heat manipulator has a significant impact on its performance. Because of this, structural optimization can be a useful technique for understanding the impact of the spatial arrangement of member materials and for developing superior designs. To the knowledge of the authors, work on optimization of the metamaterial-based heat manipulator is limited~\cite{fujiiExploringOptimalTopology2018,fujiiCloakingConcentratorThermal2020,fujiiOptimizingStructuralTopology2019}. The authors already published an article on isogeometric shape optimization in combination with the gradient-free Particle Swarm Optimization (PSO) algorithm~\cite{jansari2022design}. The present work can be considered as an extension of the previous work, as in this work, we are using the topology optimization to avoid the limitations imposed by shape optimization and explore even larger design space.
\par The goal of structural optimization is to determine the material distribution in the design domain that gives the best-desired performance (of a device or structure). In the early days of its inception, structural optimization primarily focused on the optimization of mechanical structures. Over a period of time, optimization transpired as a more general numerical technique that can cover a wider range of physical problems including fluids, optics, acoustics, thermodynamics, etc. Lately, topology optimization has become increasingly popular compared to other structural optimization techniques such as shape and size optimization. The reason for its popularity is its ability to allow changes in topology during optimization, hence avoiding the need for a close-to-optimal initial design. Several topology optimization approaches have been developed to date. The most common ones are density-based, level set-based, phase field-based, and evolutionary algorithm-based. Despite the fact that these approaches define diverse directions for topology optimizations, there are few conceptual differences between them. It is difficult to determine which technique is best for a given problem due to the lack of direct-relevant comparisons~\cite{sigmundTopologyOptimizationApproaches2013}. In addition, not simply the choice of a method but also factors like optimizers, filters, constraints, etc., define the overall performance. However, particularly in the level set based approach, the interface is clearly defined through out the optimization process, and therefore, we focus on level set topology optimization considering their possible advantage in applying objective function or constraints on the interface in future work~\cite{sethian2000structural,wang2003level,allaire2004structural,vandijkLevelsetMethodsStructural2013}. The remaining approaches are covered in detail in review articles~\cite{sigmundTopologyOptimizationApproaches2013,munkTopologyShapeOptimization2015,rozvanyCriticalReviewEstablished2009,deatonSurveyStructuralMultidisciplinary2014}.
\par The level set method (LSM), proposed by Osher and Sethian~\cite{osher1988fronts}, captures moving interfaces in multi-phase flow. The idea to incorporate LSM with topology optimization was suggested by Haber and Bendsoe~\cite{haber1998problem}. Following this, several research groups began working and subsequently published the level set topology optimization method~\cite{sethian2000structural,de2000topology}. In Osher and Santosa~\cite{osher2001level}, Allaire et al.~\cite{allaire2004structural,allaire2002level}, and Wang et al.~\cite{wang2003level}, the shape-sensitivity-based level set topology optimization framework was introduced. Specifically for heat manipulators, Fujii and Akimoto have explored the level set topology optimization method. They used the method to optimize a thermal cloak~\cite{fujiiExploringOptimalTopology2018}, a thermal cloak-concentrator~\cite{fujiiCloakingConcentratorThermal2020}, as well as a combined thermal-electric cloak~\cite{fujiiOptimizingStructuralTopology2019}. In their work, the finite element method, for the boundary value problem, and a stochastic evolution strategy, for optimization, are utilized. Conversely, we use the isogeometric analysis for the boundary value problem and a gradient-based Sequential Quadratic Programming (SQP) method for optimization. In the next few paragraphs, we discuss the level set topology optimization method in addition to the other aspects of the optimization process.
\par  In level set topology optimization, the isocontours of a level set function (LSF) implicitly define the interfaces between the material phases. Accordingly, the topology changes with the motion of these isocontours. Additionally, the mesh remains unchanged, eliminating the cost associated with creating a new mesh at each iteration. 
\begin{figure}[!htbp]
\centering
\includegraphics[width=5.8 in]{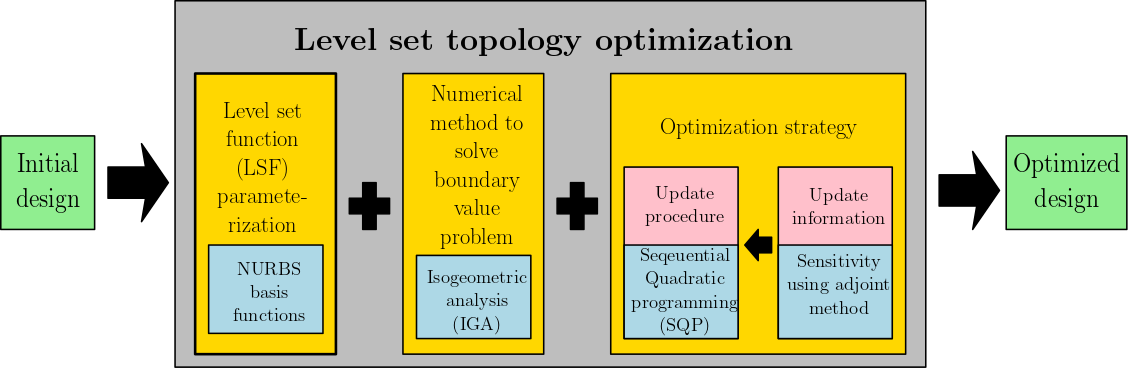}
\caption {The components of the level set topology optimization.} 
\label{fig:method overview}
\end{figure}
\par There are three main components of the level set topology optimization: the level set function (LSF) parameterization (design space), an efficient numerical method to solve the boundary value problem, and the optimization strategy. In this work, NURBS basis functions are employed to parameterize the LSF, geometry, and solution field (i.e., the temperature distribution). Isogeometric analysis (IGA)~\cite{hughes2005isogeometric} is utilized to solve the thermal boundary value problem. Here, we decouple LSF parameterization from geometry and solution field by taking two different NURBS bases. However, it will be advantageous to decouple all three parameterizations using tools such as Geometry Independent Field approximaTion (GIFT)~\cite{atroshchenko_weakening_2018,jansari2022adaptive}. For optimization, we use a mathematical programming approach-Sequential Quadratic Programming (SQP)~\cite{nocedalNumericalOptimization2006}. The motivation behind the above-mentioned choices and their respective alternatives are discussed in the following paragraphs. \fref{fig:method overview} shows the overview of the method with its components and particular choices for each component.
\par At first, we discuss LSF parameterization. The LSF is approximated over the whole design domain from a few point values using LSF parameterization, and these values at nodes/control points work as design variables. The LSF parameterization decides the design freedom, level of detail, and nature of the optimization problem. Hence, the effort required for optimization strongly depends on the LSF parameterization. In the present work, the parameterization for the LSF is decoupled from the geometry and solution field parameterizations. Decoupling allows for securing the required accuracy of the solution without increasing the complexity of the optimization. This idea is standard in LSM for free boundary problems, as discussed in~\cite{Duddu2008}.
\par The use of NURBS parameterizations and IGA provides several advantages over Lagrange parameterizations and the conventional finite element method (FEM)~\cite{wangStructuralDesignOptimization2018,gaoComprehensiveReviewIsogeometric2020}: (i) easy control of smoothness and inter-element continuity, (ii) an exact representation of conic geometries, and (iii) higher efficiency for higher-order elements. In \cite{hughes2005isogeometric}, Hughes \textit{et al.} proposed Isogeometric Analysis (IGA) based on the discretization of a Galerkin formulation using NURBS basis functions.
Later, several other variants of IGA such as isogeomteric collocation method (IGA-C)~\cite{auricchio2010isogeometric}, isogeometric boundary element method (IGABEM)~\cite{simpson2012two,simpson2013isogeometric}, geometry independent field approximation (GIFT)~\cite{atroshchenko_weakening_2018,jansari2022adaptive} have been proposed. IGA and its variants have been successfully implemented in shape and topology optimization framework~\cite{wangStructuralDesignOptimization2018,gaoComprehensiveReviewIsogeometric2020,Lian2016,lian2017}. In our work, we exploit the parameterized level set topology optimization incorporating IGA proposed by Wang \textit{et al.}~\cite{wang2016isogeometric}. However, instead of the immersed boundary technique to map the geometry to a numerical model as in~\cite{wang2016isogeometric}, we use density-based point geometry mapping due to its relatively easy implementation. Also, density-based mapping avoids ill-conditioning issues which are common in immersed boundary techniques. In density-based mapping, no special treatment is performed for the integration. Thermal conductivity at an integration point for numerical analysis is defined directly from the LSF value. However, the material definition by this simplistic approach is not very accurate near the interface. Therefore, by using a fine enough solution mesh, it ensured that the density-based mapping does not deteriorate the solution accuracy.
\par Lastly, the optimization procedure is discussed. It includes two parts: (i) update information and (ii) update procedure.  Update information often consists of the objective function and constraint sensitivities with respect to design variables. The most common methods for sensitivity analysis are (a) the direct method, (b) the finite difference method, (c) the semi-analytical method, and (d) the adjoint method~\cite{wangStructuralDesignOptimization2018}. The direct method requires to solve an extra system for each design variable; therefore, the direct method is significantly costly for complex problems. Similarly, the finite difference method requires to solve ${n+1}$ boundary value problems for $n$ design variables. Furthermore, the accuracy of the finite difference method depends on the perturbation size. The semi-analytical approach is computationally efficient; however, its accuracy can be relatively unsatisfactory for special cases due to the incompatibility of the design sensitivity field with the structure~\cite{Bruno1990Accuracy}. The adjoint method requires to solve ${n+1}$ system for $n$ objectives and constraints that depend on the field solution. The adjoint system is efficient for a problem with a large number of design variables and a small number of constraints, which aligns with our case. Therefore, in this work, the adjoint method~\cite{allaireStructuralOptimizationUsing}~is used to calculate the sensitivity.
\par The update procedure decides how to use the update information to advance the level set interfaces. For the update procedure, two classes of methods are available in the literature (a) Hamilton-Jacobi (HJ) equation-based procedures and (b) mathematical programming. In the first class, the problem is considered a quasi-temporal problem. The interface motion is calculated based on the solution of the Hamilton-Jacobi (HJ) equation in pseudo-time~\cite{burger2005survey,sethian1999level,sethian2001evolution}. The second class, mathematical programming~\cite{haber2004multilevel,luo2007shape,maute2011topology,norato2004geometry} often equipped with sophisticated step selection, constraint handling strategies, as well as optimized speed and efficiency. In this work, Sequential Quadratic Programming (SQP)~\cite{nocedalNumericalOptimization2006} is chosen due to its ability to accurately solve nonlinear constrained problems. In addition, MATLAB has an inbuilt subroutine on SQP in its `fmincon' optimization tool, which provides an advantage from an implementation point of view. 
\par The remainder of the paper is organized as follows: \sref{sec:Level set topology optimization} provides the numerical formulation of the level set topology optimization that includes boundary value problem formulation, implicit interface representation, and numerical approximations. The optimization problem, the sensitivity analysis, the SQP algorithm, and the regularization techniques are explained in \sref{sec:Optimization problem}. In~\sref{sec:Numerical Examples}, three numerical examples are demonstrated: a toy problem of an annular ring with the known solutions for the state and adjoint BVPs (\sref{sec:BenchAnncase}), the thermal cloak problem (\sref{sec:Chen2015case_cloak}), and the thermal camouflage problem (\sref{sec:Yang2016case_cmflg}), which corroborates the efficiency and robustness of the proposed method. \sref{sec:Conclusions} presents the main conclusions of the current work.

\section{Level set topology optimization with isogeometric analysis}
\label{sec:Level set topology optimization}
\subsection{Boundary value problem description}
\label{sec:Boundary value problem}
\begin{figure}[!htbp]
\centering
\includegraphics[width=5.2 in]{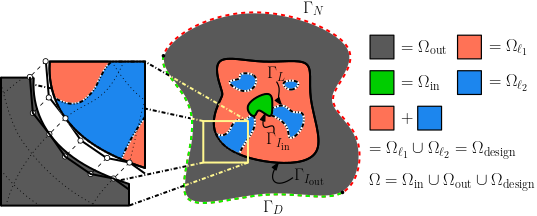}
\caption {Domain description of the boundary value problem. $\Omega_{\rm design}$ represents the region of the heat manipulator, $\Omega_{\rm in}~\&~\Omega_{\rm out}$ are, respectively, the inside and outside regions with respect to $\Omega_{\rm design}$. $\Omega$=$\Omega_{\rm in}\cup\Omega_{\rm out}\cup\Omega_{\rm design}$ . The solid black line shows an explicitly defined interface $\Gamma_I$, while the broken black \& white line shows an implicitly defined interface $\Gamma_L$ (that separates $\Omega_{\rm design}$ in two parts $\Omega_{\ell_1}$ and $\Omega_{\ell_2}$). The detailed view highlights the matching of control points of connecting patches at the interface $\Gamma_I$.} 
\label{fig:BVP domain}
\end{figure}

As shown in \fref{fig:BVP domain}, a heat manipulator embedded in the domain $\mathrm{\Omega} \in\mathbb{R}^{2}$ is considered. Note that all formulations given here stand for three dimensional physical space. The domain is externally bounded by $\mathrm{\Gamma}=\partial\mathrm{\Omega}=\mathrm{\Gamma}_D \cup \mathrm{\Gamma}_N$, where $\mathrm{\Gamma}_D$ and $\mathrm{\Gamma}_N$  are two parts of the boundary, where the Dirichlet and Neumann boundary conditions are applied, respectively. $\mathrm{\Gamma}_D \cap \mathrm{\Gamma}_N =\emptyset$. The embedded heat manipulator uniquely divides the domain into 3 different parts; the inside region $\mathrm{\Omega}_{\rm in}$, the heat manipulator region $\mathrm{\Omega}_{\rm design}$, and the outside region $\mathrm{\Omega}_{\rm out}$. $\mathrm{\Omega}=\mathrm{\Omega}_{\rm in} \cup \mathrm{\Omega}_{\rm design} \cup \mathrm{\Omega}_{\rm out}$. The internal boundaries between these parts are collectively denoted by $\mathrm{\Gamma}_I=\mathrm{\Gamma}_{I_{\rm in}} \cup \mathrm{\Gamma}_{I_{\rm out}}$. 
\par In this work, we restrict our scope to metamaterials made of two isotropic member materials. Therefore, in addition to above-mentioned partition, $\mathrm{\Omega}_{\rm design}$ is divided into two parts $\mathrm{\Omega}_{l_1}$ and $\mathrm{\Omega}_{l_2}$, representing two member materials ($\mathrm{\Omega}_{\rm design}=\mathrm{\Omega}_{\ell_1}\cup\mathrm{\Omega}_{\ell_2}$). An interface $\mathrm{\Gamma}_L$ between the member materials is implicitly defined by the level set function. The description of the level set function and corresponding interface $\mathrm{\Gamma}_L$ will be explained in detail in \sref{sec:Implicit bound with LSF}. We simplify the problem with the following assumptions: the temperature and normal flux are continuous along $\mathrm{\Gamma}_I$, the heat conduction is the only present form of heat transfer, and there is no internal heat generation. The steady-state thermal boundary value problem for the given temperature field $T$ can be written as, 
\begin{subequations}
\begin{align} 
\nabla \cdot \left( \boldsymbol{\kappa}\nabla T\right) &= 0 \quad &&\textrm{in} \quad \mathrm{\Omega}, \label{eq:Laplace equation}\\
 T &= T_D \quad &&\textrm{on} \quad \mathrm{\Gamma}_D, \label{eq:Boundary conditions a}\\
  \nabla T\cdot \boldsymbol{n} &= Q_N \quad &&\textrm{on} \quad \mathrm{\Gamma}_N, \label{eq:Boundary conditions b}\\
  \left\llbracket T\right\rrbracket &=0 \quad &&\textrm{on} \quad \mathrm{\Gamma}_I, \label{eq:Interface conditions a}\\
  \boldsymbol{n}\cdot \left\llbracket \boldsymbol{\kappa}\nabla T\right\rrbracket &= 0 \quad &&\textrm{on} \quad \mathrm{\Gamma}_I, \label{eq:Interface conditions b}
\end{align}
\label{eq:Heat conduction BVP}%
\end{subequations}
where $\boldsymbol{\kappa}$ is the thermal conductivity matrix (for isotropic material, $\boldsymbol{\kappa}=\kappa \mathrm{\mathbf{I}_2}$ with $\mathrm{\mathbf{I}_2}$ be an identity matrix of $\mathbb{R}^{2}$), $Q_N$ is the flux applied on $\mathrm{\Gamma}_N$, $T_D$ is the prescribed temperature on $\mathrm{\Gamma}_D$,  $\boldsymbol{n}$ is the unit normal on the boundary, $\llbracket\cdot\rrbracket$ is the jump operator, and $\nabla= \small{\left({ \dfrac{\textstyle\partial}{\textstyle\partial x},\dfrac{\textstyle\partial}{\textstyle\partial y}}\right)}$. On the internal boundary $\mathrm{\Gamma}_I$, $\boldsymbol{n}=\boldsymbol{n}^1= -\boldsymbol{n}^2$  where the connected patches at $\mathrm{\Gamma}_I$ are denoted by 1 and 2. 
\par To solve the boundary value problem, the strong form described in \eref{eq:Heat conduction BVP} is transformed into the weak form using the standard Bubnov-Galerkin formulation. The weak formulation is given as follows: Find $
T^h \in \mathscr{T}^h \subseteq \mathscr{T} = \big\lbrace T \in \mathbb{H}^1 ({\mathrm{\Omega}}),   T= T_D \hspace{0.15cm} \textrm{on} \ {\mathrm{\Gamma}}_D \big\rbrace
$ such that $\forall S^h \in \mathscr{S}^h_0 \subseteq \mathscr{S}_0 = \left\lbrace S \in \mathbb{H}^1 ({\mathrm{\Omega}}), S=0 \hspace{0.15cm} \textrm{on} \ {\mathrm{\Gamma}}_D \right\rbrace$,  
\begin{equation}
a(T^h,S^h) = \ell(S^h),
\label{eq:weak_form}
\end{equation}
with 
\begin{equation}
a(T^h,S^h) = \int _{\mathrm{\Omega}} (\nabla  S^h)^{\rm T} \boldsymbol{\kappa}\nabla T^h d\mathrm{\Omega},
\label{eq:weak_form_a}
\end{equation}
\begin{equation}
\ell{(S^h)} = \int _{\mathrm{\Gamma}_{N}} (S^h)^{\rm T} Q_N  d\mathrm{\Gamma}.
\label{eq:weak_form_l}
\end{equation} 
\par The interface continuity conditions described in \erefs{eq:Interface conditions a}-(\ref{eq:Interface conditions b}) are applied using Nitsche's method~\cite{Nguyen2014}. Nitsche's method applies the interface conditions weakly while preserving the coercivity and consistency of the bilinear form. Nitsche's method lies between the Lagrange multiplier method and the penalty method, designed to overcome some of the limitations of these conventional methods such as over-sensitivity to the penalty parameter, inconsistency of variational form, limitations imposed by stability conditions etc. Nitsche's method modifies the bilinear form by substituting the Lagrange multipliers of the Lagrange method by their actual physical representation, i.e. normal flux.  As shown in~\cite{Nguyen2014,HU2018}, the bilinear form after modification appears as follows, 
\begin{multline}
a(T^h,S^h) = \int _{\mathrm{\Omega}} (\nabla  S^h)^{\rm T} \boldsymbol{\kappa}\nabla T^h d\mathrm{\Omega} - \int _{\mathrm{\Gamma}_I} \left(\boldsymbol{n}\cdot \{\boldsymbol{\kappa}\nabla S^h\}\right)^{\rm T}\llbracket T^h \rrbracket~d\mathrm{\Gamma} \\- \int _{\mathrm{\Gamma}_I} \llbracket S^h  \rrbracket^{\rm T} \left(\boldsymbol{n}\cdot\{\boldsymbol{\kappa}\nabla T^h\}\right)  d\mathrm{\Gamma} + \int _{\mathrm{\Gamma}_I} \beta~\llbracket S^h  \rrbracket^{\rm T} \llbracket T^h  \rrbracket~d\mathrm{\Gamma}, 
\label{eq:modified_weak_form_a}%
\end{multline}
where $\beta$ is the stabilization parameter and $\{\cdot\}$ is the averaging operator defined as $\{\theta\}=\gamma\theta^1 + (1-\gamma)\theta^2$ with $\gamma$ being the averaging parameter ($0<\gamma<1$)~\cite{Nguyen2014,HU2018}. For the current work, $\beta=1\times 10^{12}$ and $\gamma=0.5$. In the literature~\cite{Nguyen2014,HU2018}, it is also reported that the large stabilization parameter might cause ill-conditioning of the system, but we did not face any conditioning issue for our boundary value problem.

\subsection{Implicit boundary representation with level set function}
\label{sec:Implicit bound with LSF}
As described in \sref{sec:Boundary value problem}, the  interface $\Gamma_L$, inside $\Omega_{\rm design}$ that separates the member materials, is not explicitly defined in \eref{eq:Heat conduction BVP}. It is defined by a level set function, and its movement is the main act of level set topology optimization.
\par
\begin{figure}
    \centering
    \begin{subfigure}[b]{0.55\textwidth}{\centering\includegraphics[width=1\textwidth]{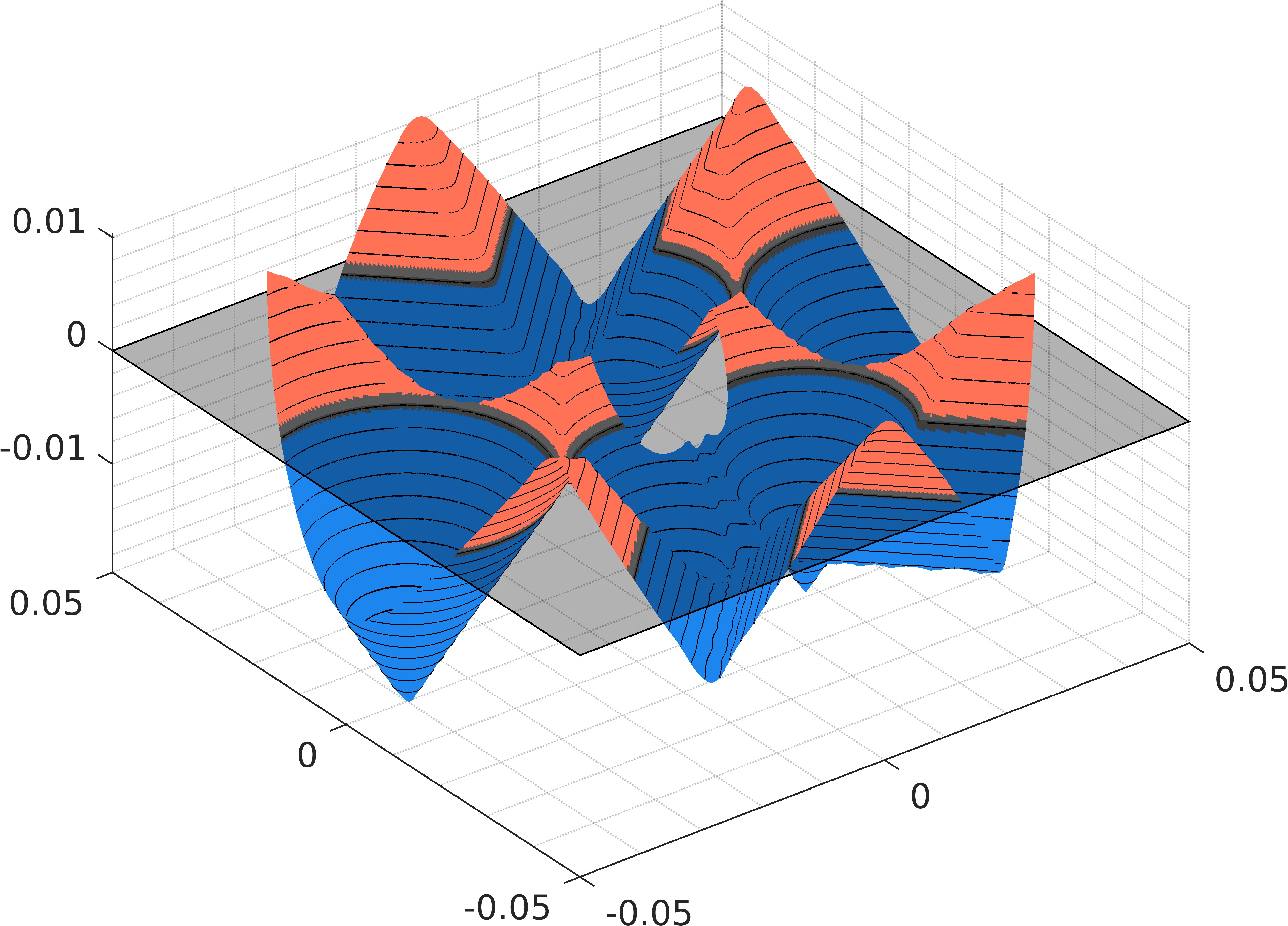}}
         \caption{\centering Level set function $\varPhi$ in 3D}
         \label{fig:levelset representation a}%
    \end{subfigure}
    \begin{subfigure}[b]{0.42\textwidth}{\centering\includegraphics[width=1\textwidth]{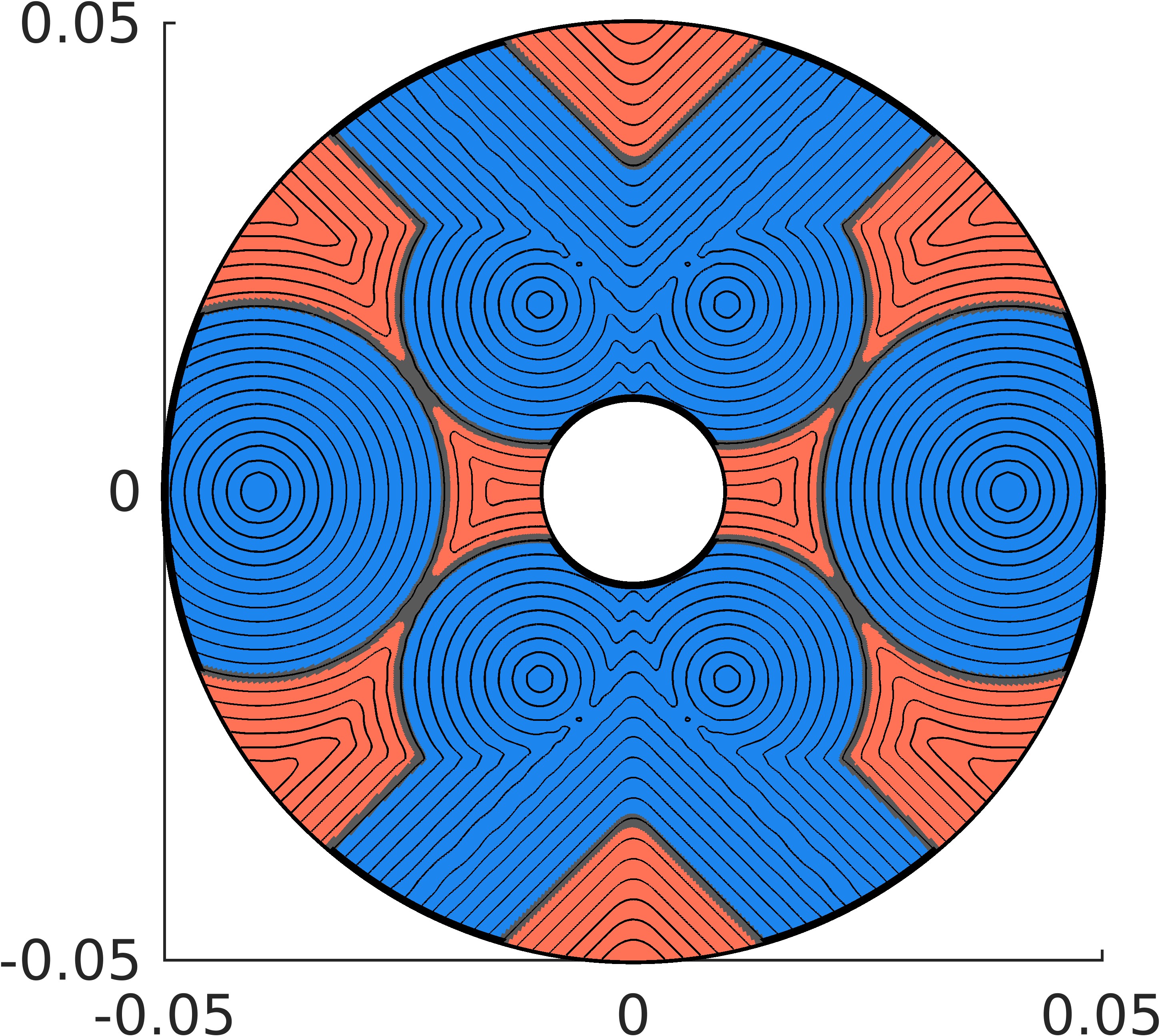}}
         \caption{\centering Level set function $\varPhi$ in 2D}
         \label{fig:levelset representation b}%
    \end{subfigure}
    \caption{Level set representation to define the material distribution. (a) The level set function $\varPhi$ in 3D. Height represents the value of $\varPhi$. $\varPhi$ above and below the plane, $\varPhi=0$, represent two member materials (shown by two different colors). (b) The level set function $\varPhi$ in 2D. The material interface is calculated by the intersection of function $\varPhi$ with the plane, $\varPhi=0$.}
    \label{fig:levelset representation}%
\end{figure}

 In the level set method in $\mathbb{R}^d$, the interface between two materials is implicitly represented by an isosurface of a scalar function  $\varPhi:\mathbb{R}^{d} \rightarrow \mathbb{R}$, called level set function (LSF). Accordingly, in our work, an isosurface of LSF $\varPhi:\mathbb{R}^{2} \rightarrow \mathbb{R}$, $\varPhi=0$, defines the interface $\Gamma_{L}$ separating $\Omega_{\ell_1}$ and $\Omega_{\ell_2}$ in $\Omega_{\rm design}$ as shown in \fref{fig:levelset representation}. The overall level set representation can be given as,
\begin{subequations}
\begin{align}
    \varPhi(\boldsymbol{x})&>0\quad \forall\boldsymbol{x} \in \Omega_{\ell_1}\setminus \Gamma_L, \\
    \varPhi(\boldsymbol{x})&=0 \quad\forall\boldsymbol{x} \in \Gamma_L,\\
    \varPhi(\boldsymbol{x})&<0 \quad\forall\boldsymbol{x} \in \Omega_{\ell_2}\setminus \Gamma_L,
\end{align}
\label{eq:LSF}%
\end{subequations}
During optimization, the movement of the interface is defined by the evolution of the isosurface $\varPhi=0$, while the background Eulerian mesh remains fixed.

\par  In the level set method, the geometric mapping defines how the LSF information is utilized in the numerical solution of the boundary value problem. Via accuracy of the numerical solution, the geometric mapping affects the optimization results. There are three common geometric mapping approaches~\cite{vandijkLevelsetMethodsStructural2013}: (a) conformal discretization, (b) immersed boundary techniques~\cite{fries2010extended,Duprez2020}, and (c) density-based mapping. In conformal discretization, the mesh conforms to the interface defined by the LSF. The method is distinct from shape optimization, as the LSF governs the changes in shape. The approach provides a crisp interface representation and the most accurate solution. However, it becomes expensive due to remeshing at each iteration. On the other hand, the immersed boundary techniques allow the nonconforming mesh. In these methods, the mesh remains fixed and the interface is captured in the numerical model using special treatment. Immersed boundary techniques also have a crisp interface representation and allow the enforcement of interface conditions directly. A specialized code for numerical integration and field approximation is needed for the elements cut by level set interfaces. Sometimes immersed boundary techniques face issues of noise and ill-conditioning due to small intersections. The last approach, density-based mapping, is the most common because of its easy implementation.
For the same advantage, we utilize density-based geometric mapping. We explore point-wise density mapping. In other words, the LSF value at an integration point defines the material density at that particular point. Hence, the thermal conductivity $\boldsymbol{\kappa}$ becomes directly a function of the LSF value. It takes the following form,
\begin{equation}
    \boldsymbol{\kappa} = \begin{cases}
                 \boldsymbol{\kappa}^1 = \kappa^1 \mathrm{\mathbf{I}_2}  \quad\text{if } \boldsymbol{x}\in\Omega_{l_1} \quad \textit{i.e.}~\varPhi \geqslant 0,\\
                 \boldsymbol{\kappa}^2 = \kappa^2 \mathrm{\mathbf{I}_2} \quad\text{if } \boldsymbol{x}\in\Omega_{l_2}\setminus \Gamma_L \quad \textit{i.e.}~\varPhi < 0,
               \end{cases}
\end{equation}
which can be represented using the Heaviside function $H$ as,
\begin{equation}
    \boldsymbol{\kappa}(\varPhi) = \boldsymbol{\kappa}^1 H(\varPhi) + \boldsymbol{\kappa}^2 (1 - H(\varPhi)),
    \label{eq:kappa based on LSF}
\end{equation}
with 
\begin{equation}
 H(\varPhi)=\begin{cases}
                 1 \quad\text{if } \varPhi \geqslant 0,\\
                 0 \quad\text{if } \varPhi < 0.
               \end{cases}
\label{eq:Heaviside function}
\end{equation}
\par For the sensitivity analysis (see \sref{sec:Sensitivity analysis}), we need the derivative of $\boldsymbol{\kappa}$ with respect to $\varPhi$, and hence the derivative of the Heaviside function with respect to $\varPhi$, i.e. the Dirac delta function $\delta(\varPhi)$. The singularity of the Dirac delta function brings numerical issues while calculating the derivatives; therefore, both functions are often replaced by their approximations. There are several forms of an approximate Heaviside function available in the literature~\cite{vandijkLevelsetMethodsStructural2013}. Here, we use a polynomial form given as,  
\begin{equation}\label{eq:Heavised-approx}
     H(\varPhi)=\begin{cases}
                 \alpha \quad &\text{if} \quad \varPhi < -\Delta,\\
                 \dfrac{3(1-\alpha)}{4} \left(\dfrac{\varPhi}{\Delta}-\dfrac{\varPhi^{3}}{3 \Delta^{3}}\right) +\dfrac{1+\alpha}{2} \quad&\text{if} -\Delta \leqslant \varPhi < \Delta,\\
                 1 \quad &\text{if} \quad\varPhi  \geqslant \Delta,\\
               \end{cases}
 \end{equation}
where $\alpha$ is a small positive value (here we take $\alpha=0$) and $\Delta$ is the support bandwidth. 
\par
\begin{figure}[!htbp]
\centering
\includegraphics[width=3.8in]{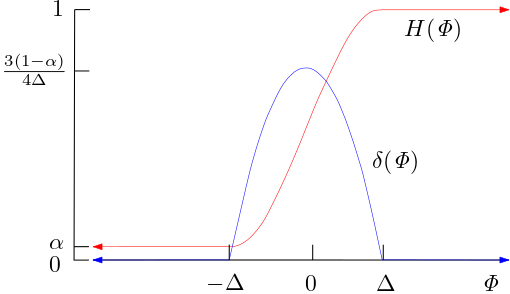}
\caption {Smoothed Heaviside function and smoothed Dirac delta function.} 
\label{fig:smoothed H and delta fn}
\end{figure}

 Consequently, the derivative of $\boldsymbol{\kappa}$ can be written as, 
\begin{equation}
    \dfrac{d \boldsymbol{\kappa} (\varPhi)}{d \varPhi} = (\boldsymbol{\kappa}^1 - \boldsymbol{\kappa}^2)\delta(\varPhi),
    \label{eq:kappa der based ontakes LSF}
\end{equation}
where one-dimensional Dirac delta function $\delta(\varPhi)$ is approximated by,
\begin{equation}\label{delta-approx}
 \delta(\varPhi)=
 \begin{cases}
                 \dfrac{3(1-\alpha)}{4\Delta} \left(1-\dfrac{\varPhi^{2}}{ \Delta^{2}}\right)  \quad&\text{if} \quad |\varPhi|\leqslant \Delta,\\
                 0 \quad &\text{if} \quad|\varPhi| > \Delta.\\
               \end{cases}
 \end{equation}
The smoothed Heaviside and smoothed Dirac delta functions in the given polynomial form are shown in \fref{fig:smoothed H and delta fn}.

\subsection{Solution of the boundary value problem using IGA}
\par Following the implicit representation of the interface by the level-set function, the weak form (\eref{eq:modified_weak_form_a}) is modified again to accommodate the conductivity matrix functional $\boldsymbol{\kappa}(\varPhi)$. Since the LSF is defined in domain $\Omega_{\text{design}}$, which does not have a common boundary with $\partial\Omega$, the applied boundary condition is not a function of the LSF. Therefore, the linear form (\eref{eq:weak_form_l}) remains unchanged. The modified weak form transforms to,
\begin{equation}\label{eq:mmodified_weak_form}
a(T^h,S^h,\varPhi) = \ell(S^h),
\end{equation}
with
\begin{multline}\label{eq:mmodified_weak_form_a}
a(T^h,S^h,\varPhi) = \int _{\Omega} (\nabla  S^h)^{\rm T} \boldsymbol{\kappa}(\varPhi)\nabla T^h d\Omega - \int _{\Gamma_I} \left(\boldsymbol{n}\cdot \{\boldsymbol{\kappa}(\varPhi)\nabla S^h\}\right)^{\rm T}\llbracket T^h \rrbracket~d\Gamma \\- \int _{\Gamma_I} \llbracket S^h  \rrbracket^{\rm T} \left(\boldsymbol{n}\cdot\{\boldsymbol{\kappa}(\varPhi)\nabla T^h\}\right)  d\Gamma + \int _{\Gamma_I} \beta~\llbracket S^h  \rrbracket^{\rm T} \llbracket T^h  \rrbracket~d\Gamma, 
 \end{multline}
\begin{figure}[!htbp]
\centering
\includegraphics[width=3.8in]{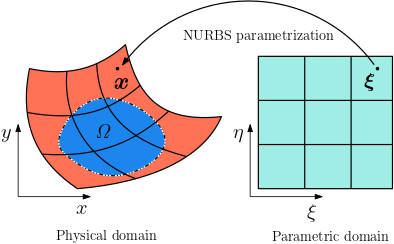}
\caption {Parametrization of a point from the parametric domain to a point in the physical domain using NURBS basis functions.} 
\label{fig:NURBS parameterization}
\end{figure} 
\par The next step is to define the geometry and solution approximations to discretize the weak form. As mentioned in \sref{sec:Introduction}, we use NURBS basis functions for geometry and solution field approximations. Here we use standard IGA, where both geometry and solution fields are approximated with same NURBS basis functions. Let $\boldsymbol{x}\in \Omega$, and $\boldsymbol{\xi}$ be the corresponding point in parametric domain as shown in \fref{fig:NURBS parameterization}, then the domain using $n$ NURBS functions $N_{i}$ and $n$ control points $\mathbf{X}_i$ is approximated as,
\begin{equation}
    \boldsymbol{x} = \sum _{i=1}^{n}\mathbf{X}_{i}N_{i}(\boldsymbol{\xi}),
    \label{eq:NURBs_approx}
\end{equation}
\noindent and accordingly, the test and trial functions are also approximated with the same NURBS shape functions as,
\begin{equation}
  T^h(\boldsymbol{\xi}) = \sum_{i=1}^{n}T_{i}N_{i}(\boldsymbol{\xi}), \quad \textrm{and} \quad 
 S^h(\boldsymbol{\xi}) = \sum_{i=1}^{n}S_{i}N_{i}(\boldsymbol{\xi}),
\label{eq:Trail test approx}     
\end{equation}
\noindent Here, $T_i$ and $S_{i}$ are the temperature and the arbitrary temperature at the $i^{th}$ control point.
\par By substituting \eref{eq:Trail test approx} in \eref{eq:mmodified_weak_form}, a linear system is obtained,
\begin{equation}
\label{eq:Linear matrix system}
\mathbf{K} \mathbf{T} = \mathbf{F},
\end{equation}
where $\mathbf{T}$ is the vector of unknown temperatures at the control points. The global stiffness matrix $\mathbf{K}$ and the global flux vector $\mathbf{F}$ are written as,
\begin{equation} \label{eq:K}
\mathbf{K}= \mathbf{K}^b + \mathbf{K}^n+ (\mathbf{K}^n)^{\rm T}+\mathbf{K}^s,
\end{equation}
\begin{equation}
\mathbf{F}= \int _{\Gamma_{N}} \mathbf{N}^{\textrm{T}} Q_N~d\Gamma,
\end{equation} 
where $\mathbf{K}^b$ is the bulk stiffness matrix. As $\Omega_{\rm in}$, $\Omega_{\rm out}$ and $\Omega_{\rm design}$ are considered separate patches, $\mathbf{K}^b$ is defined as follows,
\begin{equation}
\mathbf{K}^b =\sum_{k \in {\rm \{in,design,out \}}} \int _{\Omega_k} (\mathbf{B}^k)^{\textrm{T}}\boldsymbol{\kappa}^k(\varPhi)\mathbf{B}^k~d\Omega. 
\label{eq:Kb}
\end{equation}
where $\mathbf{B}$ is the shape function derivative matrix (with superscript $k$ representing the patch index). 
\par $\mathbf{K}^n$ and $\mathbf{K}^s$ are the interfacial stiffness matrices, since these matrices are used to couple adjacent patches with the conditions given in \erefs{eq:Interface conditions a}-(\ref{eq:Interface conditions b}). A point to note, before defining $\mathbf{K}^n$ and $\mathbf{K}^s$, is that the connecting patches have matching control points at the interface as shown in \fref{fig:BVP domain}. Following the notation 
 used in \sref{sec:Boundary value problem}, the connecting patches at interface $\Gamma_I$ are denoted as 1 and 2 (and by superscripts 1 \& 2 in the following equations). Consequently, $\mathbf{K}^n$ and $\mathbf{K}^s$ are given by the following equations,
\renewcommand\arraystretch{2}
\begin{equation}
\mathbf{K}^n =
\begin{bmatrix} 
-\gamma\displaystyle\int_{\Gamma_I} (\mathbf{N}^1)^{\textrm{T}}\boldsymbol{n}\boldsymbol{\kappa}^1(\varPhi)\mathbf{B}^1~d\Gamma 
& -(1-\gamma)\displaystyle\int_{\Gamma_I} (\mathbf{N}^1)^{\textrm{T}}\boldsymbol{n}\boldsymbol{\kappa}^2(\varPhi)\mathbf{B}^2~d\Gamma    \\[0.1em]
\gamma\displaystyle\int_{\Gamma_I} (\mathbf{N}^2)^{\textrm{T}}\boldsymbol{n}\boldsymbol{\kappa}^1(\varPhi)\mathbf{B}^1~d\Gamma  & (1-\gamma)\displaystyle\int_{\Gamma_I} (\mathbf{N}^2)^{\textrm{T}}\boldsymbol{n}\boldsymbol{\kappa}^2(\varPhi)\mathbf{B}^2~d\Gamma 
\end{bmatrix},
\label{eq:Kn}
\end{equation}
\begin{equation}
\mathbf{K}^s =
\begin{bmatrix} 
\beta\displaystyle\int_{\Gamma_I} (\mathbf{N}^1)^{\textrm{T}}\mathbf{N}^1~d\Gamma 
& -\beta\displaystyle\int_{\Gamma_I} (\mathbf{N}^1)^{\textrm{T}}\mathbf{N}^2~d\Gamma  \\[0.1em]
-\beta\displaystyle\int_{\Gamma_I} (\mathbf{N}^2)^{\textrm{T}}\mathbf{N}^1~d\Gamma  & \beta\displaystyle\int_{\Gamma_I} (\mathbf{N}^2)^{\textrm{T}}\mathbf{N}^2~d\Gamma 
\end{bmatrix},
\label{eq:Ks}%
\end{equation}
where $\mathbf{N}$ is the vector of shape functions. For a given patch $k$, the shape function derivative matrix $\mathbf{B}^k$ and the vector of shape functions $\mathbf{N}^k$ are given as follows,
\begin{equation}
\mathbf{B}^k =
\begin{bmatrix} 
{N\strut_{1,x}^k} & {N\strut_{2,x}^k}&... & {N\strut_{I,x}^k} &...   \\[0.1em]
{N\strut_{1,y}^k} & {N\strut_{2,y}^k}&... & {N\strut_{I,y}^k} &... 
\end{bmatrix}, \quad
\mathbf{N}^k =
\begin{bmatrix} 
{N\strut_{1}^k} & {N\strut_{2}^k} & ... & {N\strut_{I}^k} & ... 
\end{bmatrix}.
 \end{equation}
\par In \sref{sec:Sensitivity analysis}, the derivative of global stiffness matrix with respect to LSF $\varPhi$ will be needed in the sensitivity calculation. It is defined by differentiating  \eref{eq:K} as follows, 
\begin{equation} \label{eq:dKdphi}
\dfrac{d\mathbf{K}}{d\varPhi}= \dfrac{d\mathbf{K}^b}{d\varPhi} + \dfrac{d\mathbf{K}^n}{d\varPhi}+ \left(\dfrac{d\mathbf{K}^n}{d\varPhi}\right)^{\rm T}+\dfrac{d\mathbf{K}^s}{d\varPhi},
\end{equation}
where
\begin{equation}\label{eq:dKbdphi}  
\dfrac{d\mathbf{K}^b}{d\varPhi}=\sum_{k \in {\rm \{in,design,out \}}}\int _{\Omega_k} (\mathbf{B}^k)^{\textrm{T}}\dfrac{d\boldsymbol{\kappa}^k(\varPhi)}{d\varPhi}\mathbf{B}^k~d\Omega,
\end{equation}
and $\dfrac{d\mathbf{K}^n}{d\varPhi}$ \& $\dfrac{d\mathbf{K}^s}{d\varPhi}$ are defined similarly by differentiating \eref{eq:Kn} \& \eref{eq:Ks}, respectively.

\subsection{Level set function parameterization}
\label{sec:LSF Parameterization}
\begin{figure}[htbp!]
\centering
\includegraphics[width=6in]{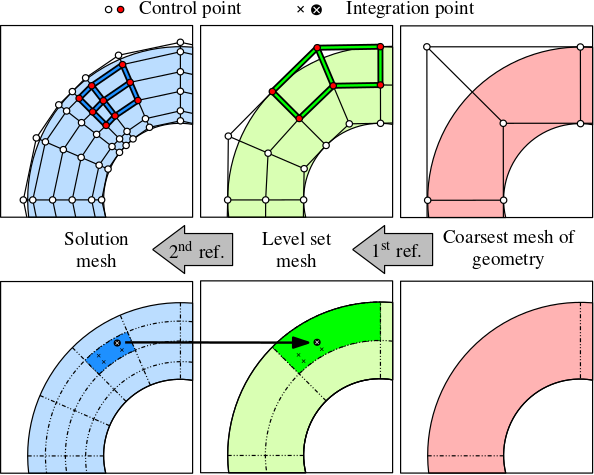}
\caption {The discretization strategy to decouple the LSF parametrization from the geometry and  solution field parameterizations. Two stages of refinement are provided. The first is to create the design/LSF mesh, and the second is to create the solution mesh to ensure the solution accuracy.} \label{fig:mesh independancy}
\end{figure}
\par The objective of LSF parameterization is to use the basis functions (in our case, the NURBS basis functions) to approximate LSF on the entire domain via values at nodes/control points. Values at nodes/control points are called expansion coefficients and are utilized as design variables for optimization. The choice of LSF parameterization also decides the design freedom as well as the detailedness of the level set interface, and eventually influences the optimization results. A finer mesh for LSF parameterization often means a more complex optimization problem with higher design variables, which naturally requires more computational effort. On the other hand, the numerical analysis method requires a finer mesh to ensure adequate solution accuracy. Taking this into account, it is advantageous to decouple the LSF parameterization from the structural mesh. This way, the structural mesh can be refined without changing the design variables. 
\par In this paper, we utilize two stages of refinement to achieve the above-mentioned decoupling. After creating a geometry on the coarsest possible NURBS mesh, we provide the first stage of refinement to get the required LSF parameterization (or the required number of design variables). After the first stage, the mesh is refined again for the second stage to ensure the required structural accuracy. The detailed refinement procedure is shown in \fref{fig:mesh independancy}. The degree elevation and knot insertion algorithms are used for both stages of refinement. All refinements are provided uniformly. For future work, it will be advantageous to define a refinement criterion~\cite{bordas2008simple,bordas2007derivative,duflot2008posteriori,jansari2019adaptive} based on the LSF, and employ local refinement using advanced splines~\cite{jansari2022adaptive}. The LSF $\varPhi$ is parameterized using $m$ NURBS basis functions $R_i$ as,
\begin{equation}\label{eq:LSF parameterization}
    \varPhi(\boldsymbol{\xi})=\sum_{i=1}^m R_i(\boldsymbol{\xi})\varPhi_i
\end{equation}
where $\varPhi_i$ is the expansion coefficient corresponding to the $i^{th}$ control point. 
\par We often need the expansion coefficients corresponding to a predefined LSF, for example, to initialize the optimization. To obtain these expansion coefficients $\varPhi_i$, a simple linear mass system is solved. The mass system can be written as,
\begin{equation}
    \label{eq:Mass matrix system}
    \mathbf{M}\boldsymbol{\Phi}= \mathbf{\Psi},
\end{equation}
where $\boldsymbol{\Phi}=[\varPhi_1\hspace{0.7em}\varPhi_2\hspace{0.7em}...\hspace{0.7em}\varPhi_m]^{\rm T}$, mass matrix $\mathbf{M}$ and the right-side vector $\mathbf{\Psi}$ are defined as,
\begin{equation}
\label{eq:Mass matrix}
\mathbf{M}=\int_{\Omega_{\rm design}} (\mathbf{R})^{\textrm{T}}\mathbf{R}~d\Omega,
\end{equation}
\begin{equation}
\label{eq:Rhs vector}
\mathbf{\Psi}= \int_{\Omega_{\rm design}}(\mathbf{R})^{\textrm{T}} \varPhi~d\Omega.
\end{equation}

\section{Optimization problem}
\label{sec:Optimization problem}
\subsection{Optimization problem description}
\label{sec:Optimization problem description}
\par In the level set topology optimization method, the expansion coefficients, as mentioned in \eref{eq:LSF parameterization}, are used as design variables. The goal is to find the values of these design variables, such that the corresponding topology yields the optimal value of the function of interest (called objective function). In our case, $\boldsymbol{\Phi}=[\varPhi_1\hspace{0.7em}\varPhi_2\hspace{0.7em}...\hspace{0.7em}\varPhi_{N_{\rm var}}]^{\rm T}$ is the vector of the $N_{\mathrm{var}}$ design variables,  and $J$ is the objective function. In a general case, $N_{\text{var}}$ can be different from the number of basis functions $m$ in \eref{eq:LSF parameterization}. For the most of numerical examples in the next section, $N_{\text{var}}$ are less than $m$ considering imposed $x$ and $y$ axial symmetry. The topology optimization problem for a heat manipulator can be defined in a mathematical form as,
\begin{equation}
\label{eq:optimization problem}
\min_{\boldsymbol{\Phi} \in  \mathbb{R}^{N_{\mathrm{var}}} }~J(T^h,\varPhi),
\end{equation}
with
\begin{subequations}
\begin{alignat}{1}
&J : \mathbb{R}^{N_{\mathrm{var}}}  \rightarrow \mathbb{R},\\
&J : \boldsymbol{\Phi} \mapsto J(T^h(\boldsymbol{\Phi}),\varPhi(\boldsymbol{\Phi})),
\end{alignat}
\label{eq:obj fun}
\end{subequations}
\noindent\textrm{such that the following constraints are satisfied,}
\begin{align}
&\text{Equality constraint:}\quad &a(T^h,S^h,\varPhi)= \ell(S^h), \forall S^h \in \mathscr{S}^h_0 \quad &\textrm{in} \quad \Omega, \label{eq:equality constraint_1}\\
&\text{Equality constraint:}\quad &T= T_D \quad &\textrm{on} \quad \Gamma_D,   \label{eq:equality constraint_2}\\
&\text{Box constraints:}\quad &\varPhi_{i,\rm{min}}\leq~\varPhi_{i}\leq~\varPhi_{i,\rm{max}} \quad &i=1,2,...,N_{\mathrm{var}} ,  \label{eq:Box constraints}
\end{align} 
where $\varPhi_{i,\rm{min}}$ and $\varPhi_{i,\rm{max}}$ are the lower and upper bounds of the design variable $\varPhi_{i}$.
\par As mentioned in \sref{sec:Introduction}, the given optimization problem is solved using a mathematical programming technique - Sequential Quadratic programming (SQP). In SQP, the boundary value problem described in \sref{sec:Boundary value problem} is solved in each iteration. Since SQP is a gradient-based algorithm, the sensitivity of the objective functions $J$ with respect to each design variable $\varPhi_i$ needs to be calculated (using the solution of the boundary value problem) at the end of each iteration. Later, the sensitivities are fed into the algorithm to generate the new values of the design variables. The following two sections will, respectively, explain the adjoint method to calculate the sensitivity at each iteration and SQP in detail.

\subsection{Sensitivity analysis}
\label{sec:Sensitivity analysis}
In this section, we outline the sensitivity analysis method called the adjoint method~\cite{allaireStructuralOptimizationUsing,WANG2003227,LUO2007680}. In the most general case, the performance objective function $J(T,\varPhi)$ can be written as a sum of two terms, corresponding to the volume and the surface integrals, respectively, i.e.
\begin{equation}
J(T,\varPhi)= \int _{\Omega_b} J_b(T,\varPhi)~d\Omega + \int _{\Gamma_s} J_s(T,\varPhi)~d\Gamma,
\label{eq:Objective fun definition}
\end{equation}
where ${\Omega_b}$ is the domain where the volume term is calculated, and ${\Gamma_s}$ is part of the boundary where the surface term is calculated. For our numerical examples, only the volume term of the objective function is considered. 
\par Next, the Lagrangian can be obtained by augmenting the objective functional $J(T,\varPhi)$ with the weak form constraint as well as Dirichlet boundary constraint (with the help of the Lagrange multipliers $P$ and $\lambda$, respectively) that $T$ should satisfy. It is defined as,
\begin{equation}\label{eq:Langrangian 1}
    \mathcal{L}:\mathbb{H}^1(\mathbb{R}^2) \times \mathbb{H}^1(\mathbb{R}^2) \times \mathbb{H}^1(\mathbb{R}^2) \times \mathbb{H}^1(\mathbb{R}^2) \rightarrow \mathbb{R},
\end{equation} with
\begin{equation}\label{eq:Langrangian 2}
    \mathcal{L}(T,P,\varPhi,\lambda)=J(T,\varPhi)+a(T,P,\varPhi)-\ell(P)+\int_{\mathrm{\Gamma}_D}\lambda (T-T_D)~d\mathrm{\Gamma}.
\end{equation}
\par The optimality conditions of the minimization problem are derived as the stationary conditions of the Lagrangian. The first stationary condition is obtained by equating the partial Fr\'echet derivative of $\mathcal{L}$ with respect to $P$ (in any arbitrary direction $\delta P \in \mathbb{H}^1$) to zero, i.e.
\begin{equation} \label{eq:Langrangian der wrt P 1}
    \left<\dfrac{\partial \mathcal{L}(T,P,\varPhi,\lambda)}{\partial P}, \delta P \right> = \left<\dfrac{\partial a(T, P, \varPhi)}{\partial P}, \delta P \right> -  \left<\dfrac{\partial \ell(P)}{\partial P}, \delta P \right>=0.
\end{equation}
Using the bilinear form and the linear form from \eref{eq:mmodified_weak_form_a} and \eref{eq:weak_form_l}, it can be derived that,
\begin{multline} \label{eq:bilinear-linear form properties}
    \left<\dfrac{\partial a(T, P, \varPhi)}{\partial T}, \delta T \right>= a(\delta T, P, \varPhi); \quad \left<\dfrac{\partial a(T, P, \varPhi)}{\partial P}, \delta P \right>= a(T,\delta P, \varPhi);\\ \text{and} \quad \left<\dfrac{\partial \ell(P)}{\partial P}, \delta P \right>=\ell(\delta P).
\end{multline}
Substituting the second and third equations of \eref{eq:bilinear-linear form properties} in \eref{eq:Langrangian der wrt P 1},
\begin{equation} \label{eq:Langrangian der wrt P 2}
    a(T, \delta P, \varPhi)= \ell(\delta P),
\end{equation}
which shows that $\forall \delta P \in \mathbb{H}^1$, given state variable $T \in \mathbb{H}^1$ satisfies the weak form.
\par The second stationary condition is obtained by taking the partial Fr\'echet derivative of $\mathcal{L}$ with respect to $\lambda$ (in any arbitrary direction $\delta \lambda \in \mathbb{H}^1$) and equating it to zero as follows,
\begin{equation} \label{eq:Langrangian der wrt lambda 1}
    \left<\dfrac{\partial \mathcal{L}(T,P,\varPhi,\lambda)}{\partial \lambda}, \delta \lambda \right> = \int_{\mathrm{\Gamma}_D}\delta \lambda (T-T_D)~d\mathrm{\Gamma} =0.
\end{equation}
which readily implies that $T=T_D$ on $\mathrm{\Gamma}_D$. Together \eref{eq:Langrangian der wrt P 2} and \eref{eq:Langrangian der wrt lambda 1} solve the boundary value problem in the state variable $T$ (as described in \eref{eq:Heat conduction BVP}). 
\par The third stationary condition is obtained by taking the partial Fr\'echet derivative of $\mathcal{L}$ with respect to $T$ (in any arbitrary direction $\delta T \in \mathbb{H}^1$) and equating it to zero as follows,
\begin{equation} \label{eq:Langrangian der wrt T 1}
    \left<\dfrac{\partial \mathcal{L}(T,P,\varPhi,\lambda)}{\partial T}, \delta T \right> =  \left<\dfrac{\partial J(T,\varPhi)}{\partial T}, \delta T \right> + \left<\dfrac{\partial a(T, P, \varPhi)}{\partial T}, \delta T \right> +\int_{\mathrm{\Gamma}_D}\lambda \delta T~d\mathrm{\Gamma}=0.
\end{equation}
Using \eref{eq:bilinear-linear form properties} \& the fact that $T$ is constant, i.e.$\delta T=0$, on $\mathrm{\Gamma}_D$,
\begin{equation} \label{eq:Langrangian der wrt T 2}
    a(\delta T, P, \varPhi)=-\left<\dfrac{\partial J(T,\varPhi)}{\partial T}, \delta T \right>. 
\end{equation}
\par \eref{eq:Langrangian der wrt T 2} is valid for $\forall\delta T \in \mathbb{H}^1$, therefore, with an arbitrary normal derivative ${\small \dfrac{\partial \delta T}{d\boldsymbol{n}}}$ on $\mathrm{\Gamma}_D$ and vanishing trace $\delta T=0$, it gives, $P=0$ on $\mathrm{\Gamma}_D$. Combining this fact with \eref{eq:Langrangian der wrt T 2}, we can generate a well-posed adjoint problem as,
\begin{subequations}  \label{eq:Adjoint BVP}
\begin{align} 
\nabla \cdot \left( \boldsymbol{\kappa}\nabla P\right) &= -\dfrac{\partial J(T,\varPhi)}{\partial T} \quad &&\textrm{in} \quad \mathrm{\Omega}, \label{eq:Adjoint Laplace equation}\\
 P &= 0 \quad &&\textrm{on} \quad \mathrm{\Gamma}_D, \label{eq:Adjoint Boundary conditions a}\\
  \nabla P\cdot \boldsymbol{n} &= 0 \quad &&\textrm{on} \quad \mathrm{\Gamma}_N, \label{eq:Adjoint Boundary conditions b}\\
  \left\llbracket P\right\rrbracket &=0 \quad &&\textrm{on} \quad \mathrm{\Gamma}_I, \label{eq:Adjoint Interface conditions a}\\
  \boldsymbol{n}\cdot \left\llbracket \boldsymbol{\kappa}\nabla P\right\rrbracket &= 0 \quad &&\textrm{on} \quad \mathrm{\Gamma}_I, \label{eq:Adjoint Interface conditions b}
\end{align}
\end{subequations}
where $T$ is the primary temperature field, and $P$ is the adjoint temperature field. By employing the trial and test  function approximations, the adjoint problem is discretized into the following linear system,
\begin{equation} \label{eq:adjoint eq. matrix form}
\mathbf{K}^{\rm T} \mathbf{P}= \mathbf{F}_{\rm adj},
\end{equation}
where $\mathbf{P}$ is the vector of adjoint temperature at control points, $\mathbf{F}_{\rm adj}$ is the global adjoint flux vector defined as,
\begin{equation}
\mathbf{F}_{\rm adj}=-\int _{\Omega_b} \mathbf{N}^{\textrm{T}} \dfrac{d J_b}{d T}~d\Omega - \int _{\Gamma_s} \mathbf{N}^{\textrm{T}} \dfrac{d J_s}{d T}~d\Gamma,
\end{equation} 
\par At last, with the fulfillment of all three stationary conditions, the sensitivity of the objective functional $\left(\dfrac{dJ}{d\varPhi}\right)$ becomes equal to the total derivative of Lagrangian $\mathcal{L}$ with respect to the LSF $\varPhi$. Therefore, it can be written as follows, 
\begin{multline}\label{eq:Langrangian der wrt phi 1}
      \left<\dfrac{d J(T,\varPhi)}{d \varPhi},\delta \varPhi \right>:=\left<\dfrac{d \mathcal{L}(T,P,\varPhi,\lambda)}{d \varPhi}, \delta \varPhi \right> =  \left<\dfrac{\partial J(T,\varPhi)}{\partial \varPhi}, \delta \varPhi \right>+ \left<\dfrac{\partial J(T,\varPhi)}{\partial T}, \delta T \right> \\+ \left<\dfrac{\partial a(T, P, \varPhi)}{\partial \varPhi}, \delta \varPhi \right> + \left<\dfrac{\partial a(T, P, \varPhi)}{\partial T}, \delta T \right>.   
\end{multline}
By substituting \eref{eq:Langrangian der wrt T 2},
\begin{equation}\label{eq:Langrangian der wrt phi 2}
    \left<\dfrac{d J(T,\varPhi)}{d \varPhi},\delta \varPhi  \right>=\left<\dfrac{\partial J(T,\varPhi)}{\partial \varPhi}, \delta \varPhi \right> + \left<\dfrac{\partial a(T, P, \varPhi)}{\partial \varPhi}, \delta \varPhi \right>, 
\end{equation}
\begin{equation}\label{eq:Langrangian der wrt phi 3}
    \therefore \dfrac{d J(T,\varPhi)}{d \varPhi}=\dfrac{\partial J(T,\varPhi)}{\partial \varPhi} +\dfrac{\partial a(T, P, \varPhi)}{\partial \varPhi}.  
\end{equation}
By employing trial and test function approximations, \eref{eq:Langrangian der wrt phi 3} can be written as,
\begin{equation} \label{eq:dJ_dPhi matrix form}
\dfrac{d J}{d\varPhi} = \int _{\Omega_b} \dfrac{d J_b}{d\varPhi}~d\Omega + \int _{\Gamma_s} \dfrac{d J_s}{d\varPhi}~d\Gamma +(\mathbf{P})^{\rm T}~\dfrac{d \mathbf{K}}{d\varPhi}~\mathbf{T},
\end{equation}
where $\dfrac{d \mathbf{K}}{d\varPhi}$ is the derivative of the global stiffness matrix with respect to $\varPhi$ (as given in \eref{eq:dKdphi}). Using the LSF parameterization (\eref{eq:LSF parameterization}), the sensitivity with respect to a particular expansion coefficient $\varPhi_i$ is given as, 
\begin{equation} \label{eq:dJ_dPhi matrix form_2}
\dfrac{d J}{d\varPhi_i} = \int _{\Omega_b} \dfrac{d J_b}{d\varPhi}~\dfrac{d \varPhi}{d\varPhi_i}~d\Omega + \int _{\Gamma_s} \dfrac{d J_s}{d\varPhi}~\dfrac{d \varPhi}{d\varPhi_i}~d\Gamma +(\mathbf{P})^{\rm T}~\dfrac{d \mathbf{K}}{d\varPhi}~\dfrac{d \varPhi}{d\varPhi_i}~\mathbf{T}.
\end{equation}

\subsection{Update scheme - Sequential Quadratic Programming (SQP)}
\label{sec:Update Scheme-SQP}

\par In the Sequential Quadratic Programming (SQP) approach, the optimization problem is approximated as a quadratic sub-problem at each iteration~\cite{nocedalNumericalOptimization2006}.  For the SQP method, the main challenge is to form a good quadratic sub-problem that generates a suitable step for optimization. For our optimization problem, the quadratic subproblem in the $m^{th}$ iteration is modeled as,
 \begin{equation}
\label{eq:Quadratic problem}
\min_{\boldsymbol{p} \in  \mathbb{R}^{N_{\mathrm{var}}} }~J_m+ \nabla J_m^{\rm T}\boldsymbol{p}+\dfrac{1}{2}\boldsymbol{p}^{\rm T}\boldsymbol{\mathcal{H}}_m\boldsymbol{p},
\end{equation} 
where $J_m$, $\nabla J_m$ and $\boldsymbol{\mathcal{H}}_m$ are the objective function value, the gradient of the objective function, and the Hessian approximation of the Lagrangian $\mathcal{L}$ in the $m^{th}$ iteration, respectively.  Later, the minimizer of the subproblem  $\boldsymbol{p}=\boldsymbol{p}_m$ is used to get the next approximation $\boldsymbol{\Phi}_{m+1}$ based on a line search algorithm or a trust region algorithm. One point to note here is that equality constraints are excluded. As the temperature $T$ is evaluated from the linear system \eref{eq:Linear matrix system}, the equality constraints will be satisfied explicitly at each iteration.
\par In this work, we use `fmincon' optimization toolbox from MATLAB with the `sqp' option. The `sqp' option solves the above-mentioned quadratic subproblem with an active set strategy. At each iteration, the user provides the objective function $J_m$ and the gradient $\nabla J_m$ (using the sensitivity analysis from the last section). For the calculation of Hessian approximation $\boldsymbol{\mathcal{H}}_m$, it uses the BFGS-update scheme. Once the solution to the subproblem $\boldsymbol{p}_m$ is found, the size of step $\alpha_m$ is found by using an inbuilt line search algorithm. Then, the design variables are updated as, 
\begin{equation}
    \boldsymbol{\Phi}_{m+1} = \boldsymbol{\Phi}_{m} + \alpha_m   \boldsymbol{p}_m.
\end{equation}
\subsection{Regularizations}
\label{sec:Regularizations}
\par More often, the optimization problem is found to suffer from a lack of well-posedness, numerical artifacts, slower convergence, and entrapment into local minima with poor performance. To overcome these issues, regularization techniques can be used. Sometimes, regularization is utilized to control geometrical features in optimization results. Generally, a single regularization technique addresses more than one of the above-mentioned problems. For our numerical examples, we explored three regularization techniques: LSF reinitialization, Tikhonov regularization, and volume regularization. 
\subsubsection{Level set function reinitialization}
\label{sec:LSF reinitialization}
\par As the interfaces are defined by the zero-level contours of LSF, only local region near these contours are uniquely described in the optimal solution. Therefore, the LSF is not unique especially in the region far from interfaces. As a consequence of this non-uniqueness property, the LSF sometimes becomes too flat or too steep during optimization. This phenomenon can deteriorate the convergence rate. To alleviate the problem, the LSF is initialized as well as maintained as a sign distance function ($||\nabla \varPhi||_{2} = 1$) during optimization. One way to maintain the LSF as a sign distance function is to reinitialize it after several iterations while maintaining the locations of zero-level contours.
\begin{figure}
    \centering
    \begin{subfigure}[b]{0.3\textwidth}{\centering\includegraphics[width=1\textwidth]{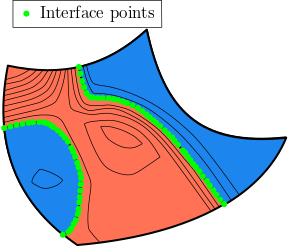}}
         \caption{\centering $\varPhi$ before reinitialization. Interface points are evaluated using LSF parameterization.}
         \label{fig:Geometry-based constrained LSF reinitialization a}
    \end{subfigure}
    \begin{subfigure}[b]{0.3\textwidth}{\centering\includegraphics[width=1\textwidth]{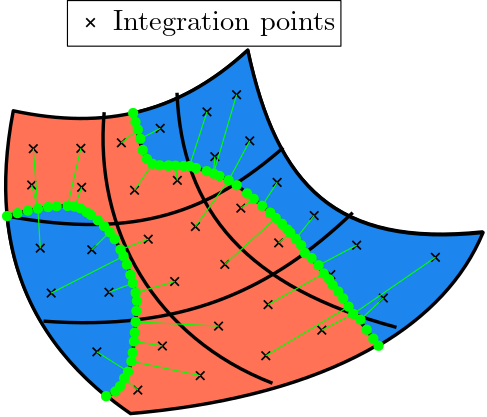}}
         \caption{\centering Evaluation of sign distance from integration points to closest interface points.}
         \label{fig:Geometry-based constrained LSF reinitialization b}
    \end{subfigure}
    \begin{subfigure}[b]{0.3\textwidth}{\centering\includegraphics[width=1\textwidth]{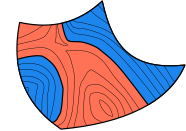}}
         \caption{\centering Reinitialized $\varPhi$ after solving the mass matrix system using the sign distance at integration points.}
        \label{fig:Geometry-based constrained LSF reinitialization c}
    \end{subfigure}
    \caption{Geometry-based constrained LSF reinitialization technique}
    \label{fig:Geometry-based constrained LSF reinitialization}
\end{figure}
\par
\begin{figure}
    \centering
    \begin{subfigure}[b]{0.35\textwidth}{\centering\includegraphics[width=1\textwidth]{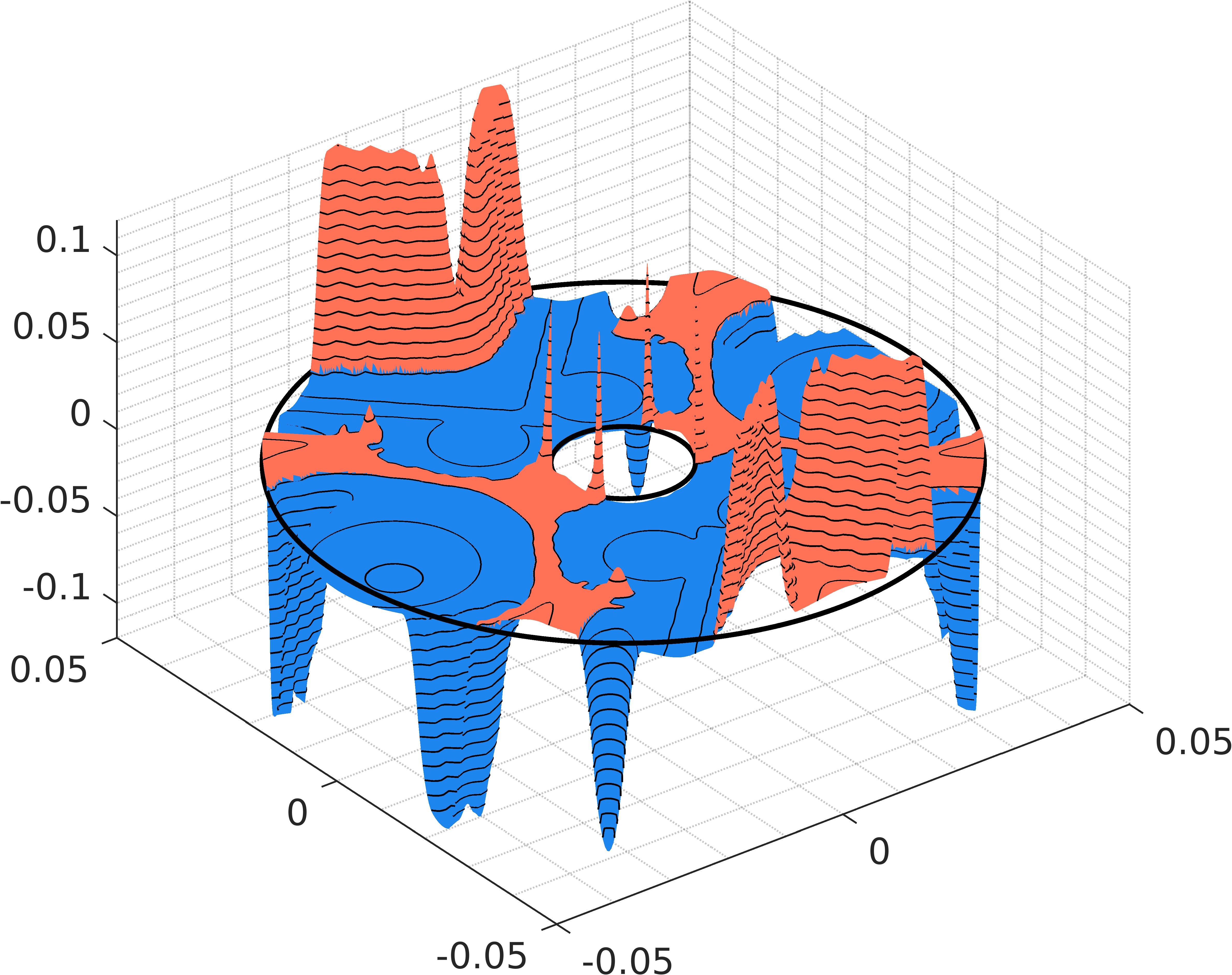}}
         \caption{\centering $\varPhi$ before reinitialization}
    \end{subfigure}
    \begin{subfigure}[b]{0.35\textwidth}{\centering\includegraphics[width=1\textwidth]{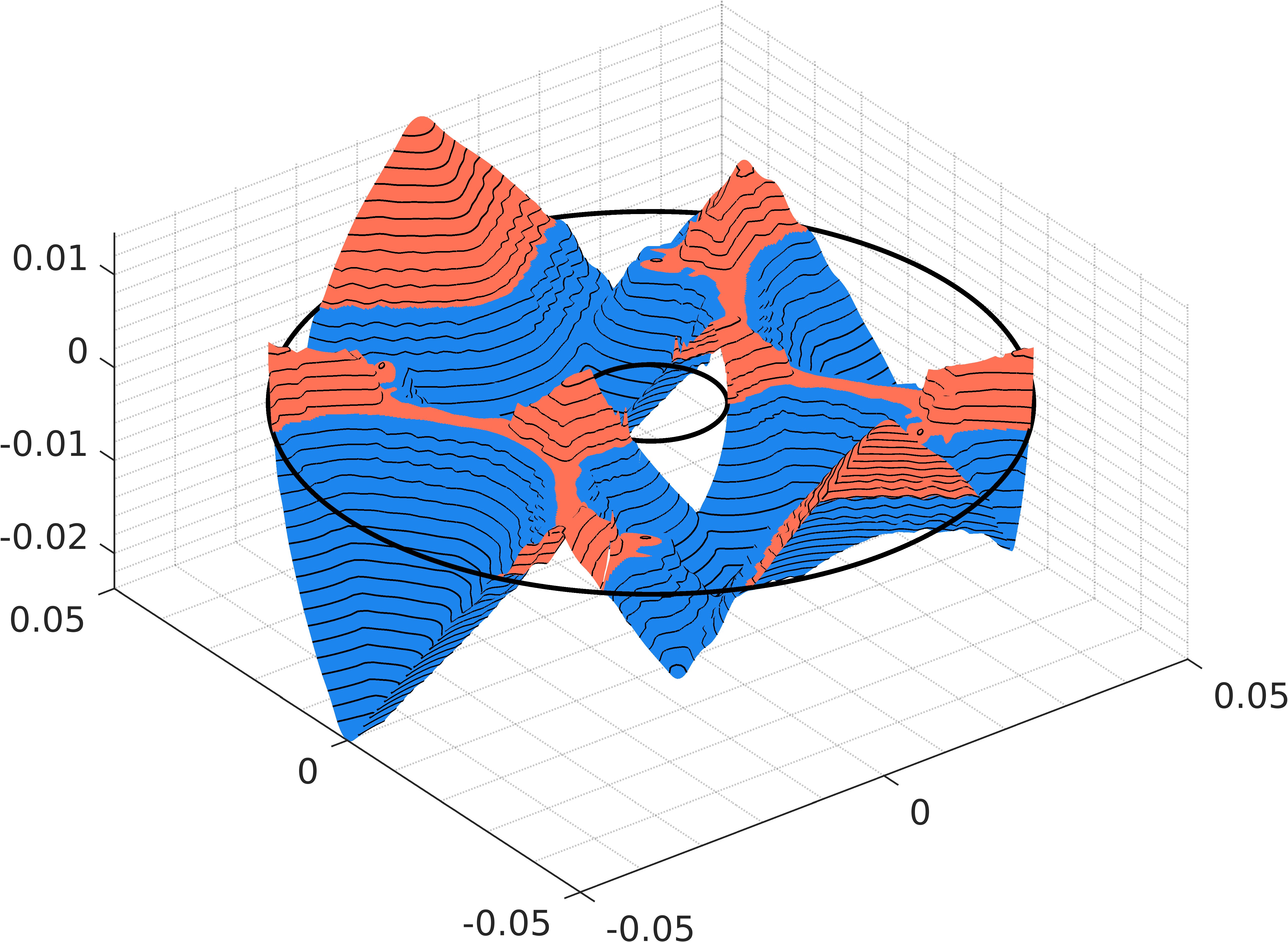}}
         \caption{\centering $\varPhi$ after reinitialization}
    \end{subfigure}
    \begin{subfigure}[b]{0.22\textwidth}{\centering\includegraphics[width=1\textwidth]{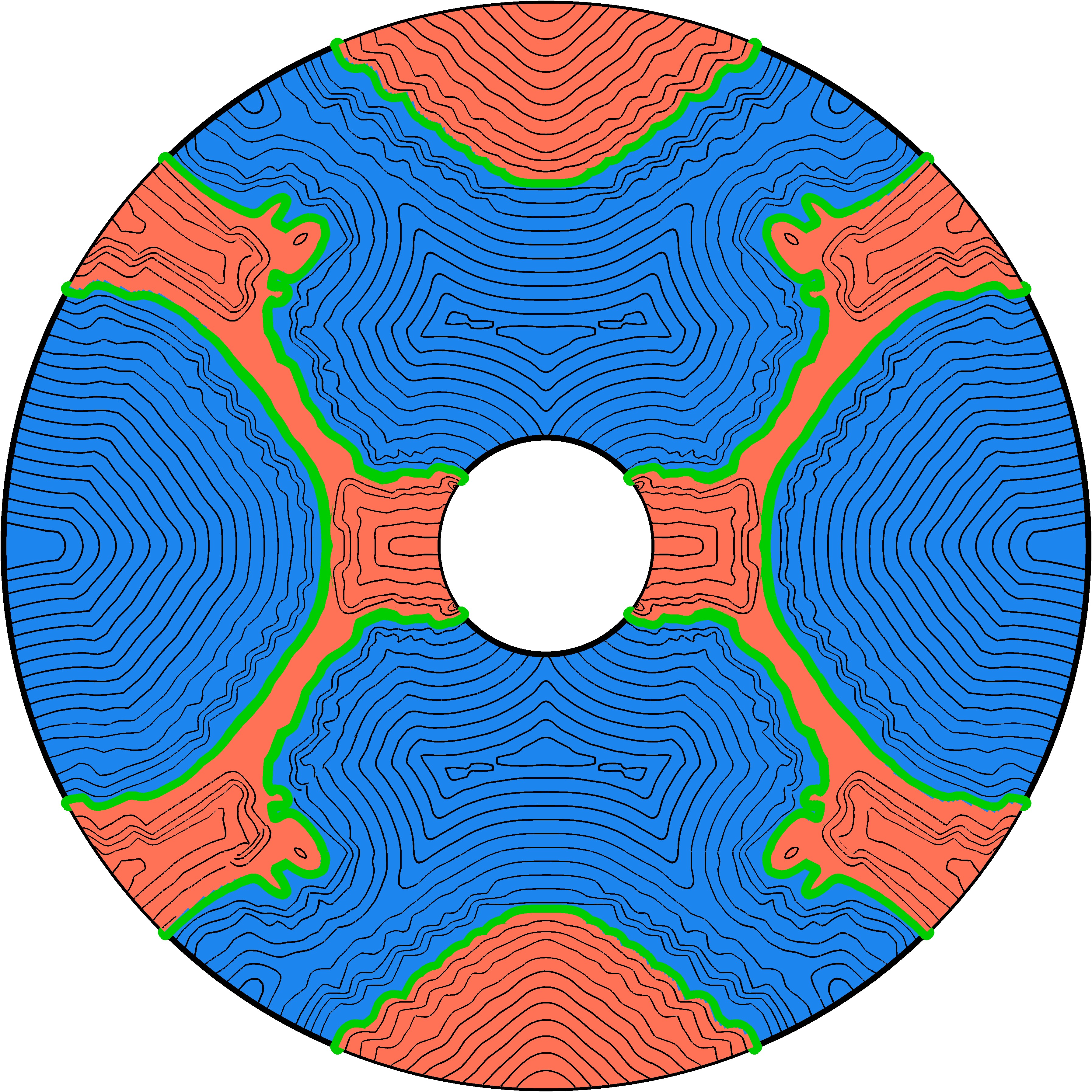}}
         \caption{\centering $\varPhi$ after reinitialization in 2D}
    \end{subfigure}
    \caption{The LSF before and after reinitialization. The green points (in (c)) represent the interface points before reinitialization, which lie exactly on the new interface.}
    \label{fig:levelset reinitialization}
\end{figure}

 Here, we utilize the geometry-based constrained reinitialization technique. \fref{fig:Geometry-based constrained LSF reinitialization} outlines the steps of the technique. Using the inverse of LSF parameterization, we find several points on the interface as shown in \fref{fig:Geometry-based constrained LSF reinitialization a}. For the sake of clarity, we call them the interface points. Then, the new LSF value at a point is defined as the distance to the closest interface point while keeping the sign from the original value (\fref{fig:Geometry-based constrained LSF reinitialization b}). The number of interface points decides the accuracy of the reinitialization. To find the new expansion coefficients, we solve the mass matrix system as shown in \eref{eq:Mass matrix system} by implementing the new LSF values at the integration/collocation points (\fref{fig:Geometry-based constrained LSF reinitialization c}). However, there is one difference from the initialization mass system. In this system, we apply the constraints which enforce the LSF values at interface points to be zero to preserve the location of the interface via the penalty method. An example of the LSF before and after reinitialization is shown in \fref{fig:levelset reinitialization}. 
\subsubsection{Tikhonov regularization}
\label{sec:Tikhonov regularization}
\par The effect of Tikhonov regularization~\cite{haber2004multilevel,tikhonov1995numerical} is similar to perimeter regularization, where the perimeter of the level-set interface is penalized. In Tikhonov regularization, an additional term penalizing the gradient of LSF is added to the main objective function. By penalizing the gradient of LSF, the optimization is directed towards smoother LSF and eventually smoother interface. between This Tikhonov term and corresponding sensitivity are written as,
\begin{equation}
\label{eq:Tikhonov regularization}
J_{\rm Tknv}= \int_{\Omega_{\rm design}} \nabla\varPhi^{\rm T}~\nabla\varPhi~d\Omega, \quad \text{and} \quad \dfrac{dJ_{\rm Tknv}}{d\varPhi_i}= \int_{\Omega_{\rm design}} 2\nabla\varPhi^{\rm T}~\nabla R_i~d\Omega.
\end{equation}
\par To implement Tikhonov regularization, the primary objective function and sensitivities are augmented by corresponding Tikhonov regularization terms using a weighing parameter $\chi$ as,
\begin{equation}
    J_{\rm total}= J + \chi J_{\mathrm{Tknv}}, \quad \text{and} \quad \dfrac{dJ_{\rm total}}{d\varPhi_i}=\dfrac{dJ}{d\varPhi_i}+\chi \dfrac{dJ_{\rm Tknv}}{d\varPhi_i}.
\end{equation}
\subsubsection{Volume regularization}
\label{sec:Volume regularization}
\par For heat manipulator optimization, we propose one more regularization, called volume regularization. In a metamaterial-based heat manipulator (made of two member materials), often one high $\kappa$ material and another low $\kappa$ material are used. The high $\kappa$ material works as the medium that allows the flow of the flux, while the low $\kappa$ material hinders the flow and guides it towards the required path. By taking into account these particular roles, the area where the flux is not flowing (and the temperature gradient is zero or negligibly small) can be filled by low $\kappa$ material. In other words, the areas, that do not contribute to the original objective, can be filled with low $\kappa$ material. By doing so, those areas are made as homogeneous as possible and free of unnecessary complex features. Mathematically, this objective is defined through a volume term as,
\begin{equation}
\label{eq:Volume regularization}
J_{\rm vol}= \int_{\Omega_{\rm design}} H(\varPhi)~d\Omega, \quad \text{and} \quad  \dfrac{dJ_{\rm vol}}{d\varPhi_i}= \int_{\Omega_{\rm design}} \delta(\varPhi)R_i~d\Omega.
\end{equation} 
\par The term calculates the volume of high $\kappa$ material, as we assign high $\kappa$ material on the positive side of the level set interface. As our optimization problem is a minimization problem, the term guides the optimization toward a larger volume of low $\kappa$ material. When this term is added to the main objective function through a weighing parameter, the combined objective of heat manipulation and volume filling can be achieved simultaneously. 
\par To implement volume regularization, the primary objective function and sensitivities are augmented by corresponding volume regularization terms using a weighing parameter $\rho$ as,
\begin{equation}
    J_{\rm total}= J + \rho J_{\mathrm{Tknv}}, \quad \text{and} \quad \dfrac{dJ_{\rm total}}{d\varPhi_i}=\dfrac{dJ}{d\varPhi_i}+\rho\dfrac{dJ_{\rm Tknv}}{d\varPhi_i}.
\end{equation}

\section{Numerical examples}
\label{sec:Numerical Examples}

In this section, we verify the proposed method, and test its efficiency and effectiveness for several numerical examples. We denote $p$ and $q$ as the order of NURBS approximation in two directions of a NURBS patch, which lies along circumferential direction and radial directions respectively for the given examples.
Since we are using `fmincon' optimization tool box in MATLAB, there are several inbuilt stopping criteria such as, `OptimalityTolerance' (tolerance value in first-order optimality measures), `StepTolerance' (tolerance value of the change in design variables' values), `ObjectiveLimit' (tolerance value of the objective function), `MaxFunctionEvaluations' (maximum number of function evaluations), `MaxIterations' (maximum number of iterations). We define `ObjectiveLimit'=$1\times 10^{-9}$, `StepTolerance'=$1\times 10^{-8}$, `OptimalityTolerance'=$1\times 10^{-6}$. If the LSF reinitialization is utilized, `MaxFunctionEvaluations' and  `MaxIterations' are employed to stop the optimization before reinitializing it. Also, the `StepTolerance' criterion often gets affected by poor LSF function. Therefore, the optimization is stopped when it fails by `StepTolerance' criterion for 4 consecutive times. Otherwise, the LSF is reinitialized and the optimization is run again. The stopping criteria, tolerance values, and applied regularizations differ slightly example-wise, and are mentioned in the descriptions of the examples. Most of the examples are symmetric along $x$ and $y$-axes. Therefore, $x$ and $y$-axes symmetry is applied in almost all numerical models, and we explicitly state when there is an exception. One more thing to note, from here onward, is that the value of the objective function for a particular test case will be mentioned in the caption.  
\subsection{Annular ring problem}
\label{sec:BenchAnncase}
\begin{figure}[!ht]
\centering
\includegraphics[width=0.5\textwidth]{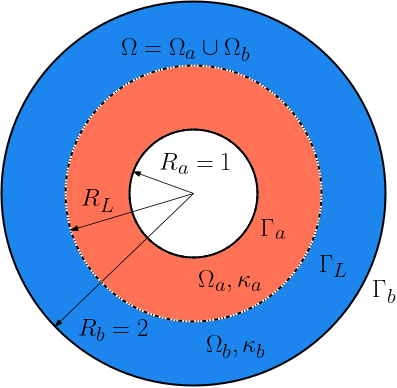}
\caption{Domain description of the annular ring problem.}\label{fig:BenchAnncase_domain}%
\end{figure} 
\par In this example, we consider a benchmark problem. As per authors' knowledge, there are not any benchmark problems of multi-material optimization with heat conduction phenomena and given interface conditions. The importance of this test case is to verify the proposed method with analytical solution. An annular ring with inner radius $R_{a}$ and outer radius $R_b$ is the domain $\mathrm{\Omega}$. The domain is divided into two regions $\mathrm{\Omega}_{a}$ and $\mathrm{\Omega}_{b}$ by an implicitly defined interface $\mathrm{\Gamma}_L$ (a circle with radius $R_L$) as shown in \fref{fig:BenchAnncase_domain}. Domains $\mathrm{\Omega}_{a}$ and $\mathrm{\Omega}_{b}$ are filled with isotropic materials of conductivity $\kappa_{a}$ and $\kappa_{b}$, respectively. For the state variable $T$, the boundary value problem is given as,
\begin{subequations}\label{eq:BenchAnn BVP}
\begin{align}
&\textrm{Governing equations}: \quad &\kappa_{a}\nabla^{2}T_{a}=0 \quad &\textrm{in} \quad \mathrm{\Omega}_{a},\\
& &\kappa_{b}\nabla^{2}T_{b}=0 \quad &\textrm{in} \quad \mathrm{\Omega}_{b},\\
&\textrm{Interface conditions}: \quad &
\kappa_{a} \frac{\partial T_{a}}{\partial r}=\kappa_{b} \frac{\partial T_{b}}{\partial r} \quad &\textrm{at} \quad r=R_L,\\
 & &T_{a}=T_{b} \quad &\textrm{at} \quad r=R_L,\\
&\textrm{Boundary conditions}: \quad & T_{a}=\overline{T}_{a} \quad &\textrm{at} \quad r=R_{a},\\
& &T_{b}=\overline{T}_{b} \quad &\textrm{at} \quad r=R_{b},
\end{align} 
\end{subequations}
and the objective is to maximize the functional $J(T)$,
\begin{equation} \label{eq:BenchAnn J}
    J(T) = \int_{\mathrm{\Omega}} T^2 d\mathrm{\Omega}.
\end{equation}
By comparing it with \eref{eq:Objective fun definition}, $\Omega_b=\mathrm{\Omega}$, $J_b=T^2$, and the surface term is absent.
\par In the simplest case, when the interface radius $R_{L}$ is the only design variable, the adjoint problem for the adjoint state $P$ can be constructed using \eref{eq:Adjoint BVP} as,
\begin{subequations}\label{eq:BenchAnn adjoint}
\begin{align}
&\textrm{Governing equations}: \quad &\kappa_{a}\nabla^{2}P_a=-2T_{a} \quad &\textrm{in} \quad \mathrm{\Omega}_{a},\\
& &\kappa_{b}\nabla^{2}P_b=-2T_{b} \quad &\textrm{in} \quad \mathrm{\Omega}_{b},\\
&\textrm{Interface conditions}: \quad &
\kappa_{a} \frac{\partial P_a}{\partial r}=\kappa_{b} \frac{\partial P_b}{\partial r} \quad &\textrm{at} \quad r=R_L,\\
 & &P_a=P_b \quad &\textrm{at} \quad r=R_L,\\
&\textrm{Boundary conditions}: \quad & P_a=0 \quad &\textrm{at} \quad r=R_{a},\\
& &P_b=0\quad &\textrm{at} \quad r=R_{b}.
\end{align} 
\end{subequations}
 For the given example, we take $R_{a}=1, R_{b}=2, T_{a}=0, T_{b}=100, \kappa_{a}=100$, and $\kappa_{b}=10$. The analytical solutions of the boundary value problem and adjoint problem are given in Appendix \ref{sec:Appendix A}. From the analytical solution, we can derive that the optimized value of $J(T)$, $J(T)=1.6094\times 10^4$, is achieved at $R_L\approx1.80612$. This is the only minimum in the given design space, and hence the unique solution to the optimization problem.

\begin{figure}[htbp!]
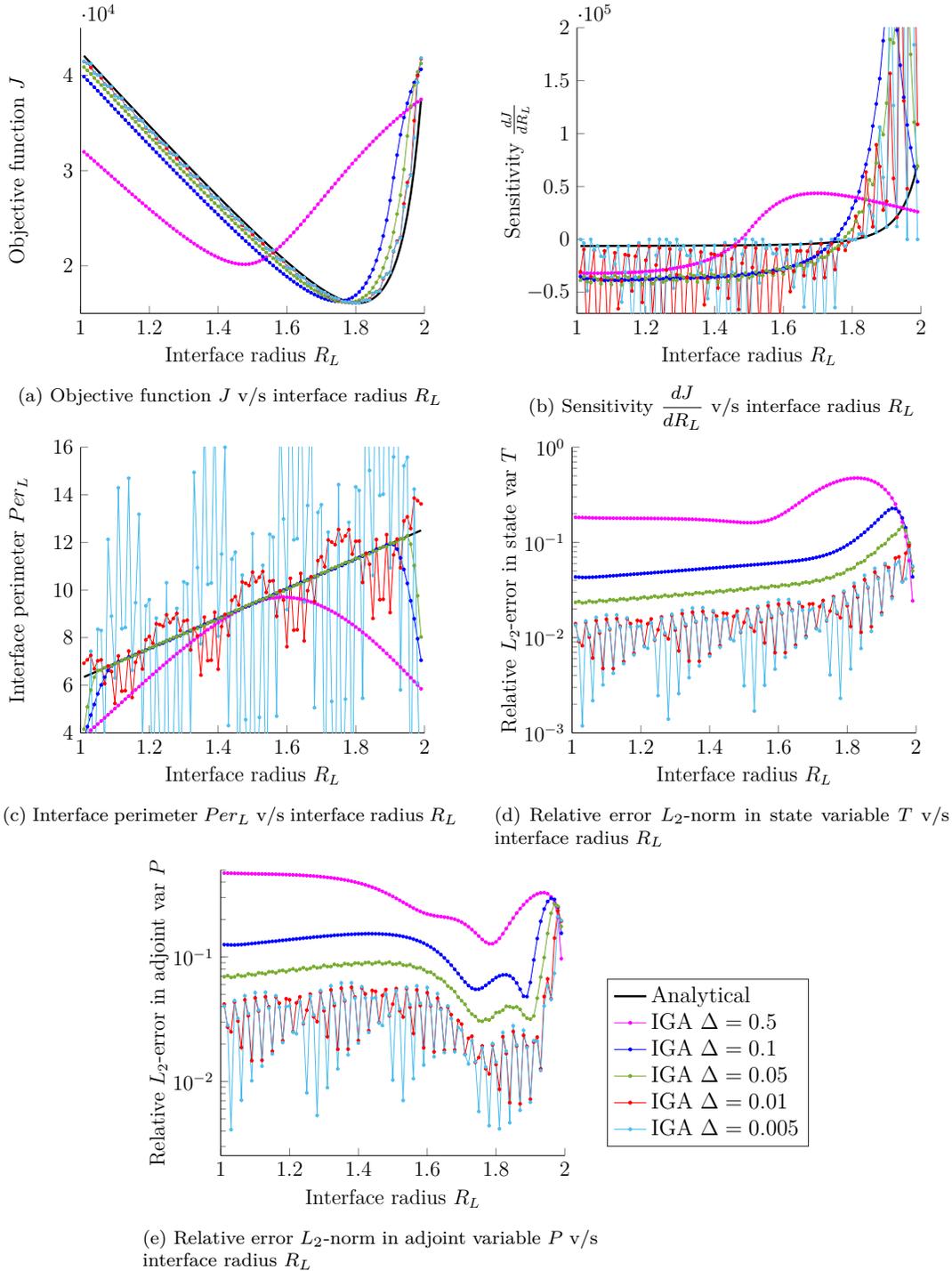

\centering
\begin{subfigure}[t]{0.45\textwidth}
\scalebox{0.75}{\input{Figures_benchAnnulus/benchAnncase_radeffect_J.tikz}}
        \caption{Objective function $J$ v/s interface radius $R_L$}
        \label{fig:BenchAnncase_J_wrt_R a}
    \end{subfigure}
    \quad
 \begin{subfigure}[t]{0.45\textwidth}
\scalebox{0.75}{\input{Figures_benchAnnulus/benchAnncase_radeffect_dJdR1.tikz}}
        \caption{Sensitivity $\dfrac{dJ}{dR_L}$ v/s interface radius $R_L$}
        \label{fig:BenchAnncase_dJdR_wrt_R b}
    \end{subfigure}
\\
\begin{subfigure}[t]{0.45\textwidth}
\scalebox{0.75}{\input{Figures_benchAnnulus/benchAnncase_radeffect_perimeter.tikz}}
        \caption{Interface perimeter $Per_L$ v/s interface radius $R_L$}
        \label{fig:BenchAnncase_peri_wrt_R c}
    \end{subfigure}
    \quad
 \begin{subfigure}[t]{0.45\textwidth}
\scalebox{0.75}{\input{Figures_benchAnnulus/benchAnncase_radeffect_error.tikz}}
        \caption{Relative error $L_2$-norm in state variable $T$ v/s interface radius $R_L$}
        \label{fig:BenchAnncase_err_wrt_R d}
    \end{subfigure}\\
    \begin{subfigure}[valign=t]{0.45\textwidth}
\scalebox{0.75}{\input{Figures_benchAnnulus/benchAnncase_radeffect_error_adjoint.tikz}}
        \caption{Relative error $L_2$-norm in adjoint variable $P$ v/s interface radius $R_L$}
        \label{fig:BenchAnncase_err_wrt_R e}
    \end{subfigure}
     \definecolor{mycolor1}{rgb}{1.00000,0.00000,1.00000}%
    \definecolor{mycolor2}{rgb}{0.46600,0.67400,0.18800}%
    \definecolor{mycolor3}{rgb}{0.30100,0.74500,0.93300}%
    \begin{subfigure}[valign=t]{0.2\textwidth}
    \scalebox{0.8}{\pgfplotslegendfromname{radeffleg}}
    \end{subfigure}

\caption{For the annular ring problem, variation of (a) objective function $J$, (b) sensitivity $\frac{dJ}{dR_L}$, (c) interface perimeter $Per_L$ and (d) relative $L_2$-error of state variable $T$ and (e) relative $L_2$-error of the adjoint variable $P$ with respect to interface radius $R_{L}$ for different values of bandwidth ($\Delta$) of approximate Heaviside function. The analytical results are shown in black color. It is observed that the IGA provides better accuracy in the
objective function with a smaller bandwidth. However, a bandwidth that is too small
can produce unstable results with oscillations in the optimization.}
    \label{fig:BenchAnncase_var_wrt_R}
\end{figure}

 \par At first, we compare the numerical results with analytical results for different support bandwidths $\Delta$ of the approximate Heaviside function in \eref{eq:Heavised-approx}. \fref{fig:BenchAnncase_var_wrt_R} shows the variation of quantities such as the objective function $J$, sensitivity $dJ/dR_L$, interface perimeter $Per_L$, error in the state and adjoint variables with respect to interface radius $R_L$ over range $[R_a,R_b]$. A constant mesh with 4389 degrees of freedom and 1089 expansion coefficients is exploited. For each value of $R_L$, the expansion coefficients are defined using \eref{eq:Mass matrix system}. The numerical sensitivity with respect to interface radius is calculated as follows,
 \begin{equation}
     \dfrac{dJ}{dR_L}=\dfrac{dJ}{d\boldsymbol{\Phi}}\cdot\dfrac{d\boldsymbol{\Phi}}{dR_L}, \quad \textrm{where} \quad \dfrac{d\boldsymbol{\Phi}}{dR_L}=\mathbf{M}^{-1} \dfrac{d\mathbf{\Psi}}{dR_L} \quad \textrm{(refer \eref{eq:Mass matrix system})},
 \end{equation}
and the interface perimeter $Per_L$ is calculated using the expansion coefficients as follows,
\begin{equation}
    Per_L=\int_{\Omega}\delta(\varPhi(\boldsymbol{\Phi}))~d\Omega.
\end{equation}
 The values of the support bandwidths $\Delta$ taken into account are $\Delta=0.5, 0.1, 0.05$, $0.01, 0.005$. From \fref{fig:BenchAnncase_var_wrt_R}, it is evident that the IGA provides better accuracy in the objective function with a smaller bandwidth. However, a bandwidth that is too small can produce unstable results with oscillations in the optimization. The reason behind it is inaccurate material information near the interface, as we are using density-based geometric mapping. Oscillations can stop the optimization prematurely or make it behave erratically. In addition, the mesh size has a direct relation with these oscillations. 
 \par
\begin{figure}[htbp!]
\centering
\begin{subfigure}[t]{0.44\textwidth}
\scalebox{0.85}{\input{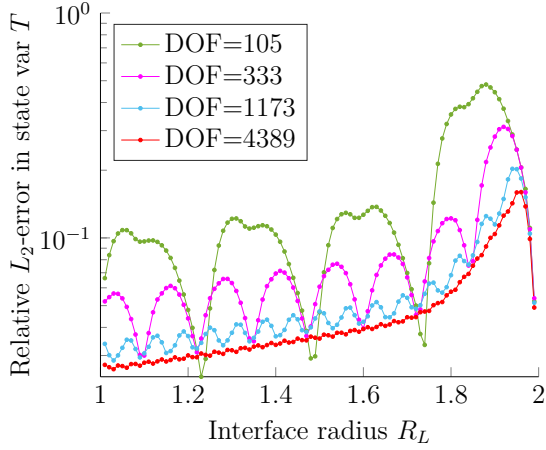}
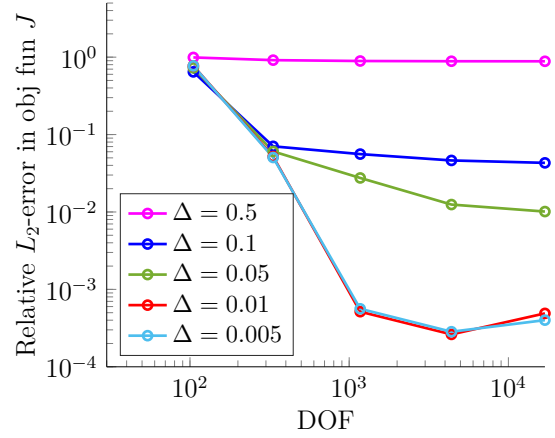
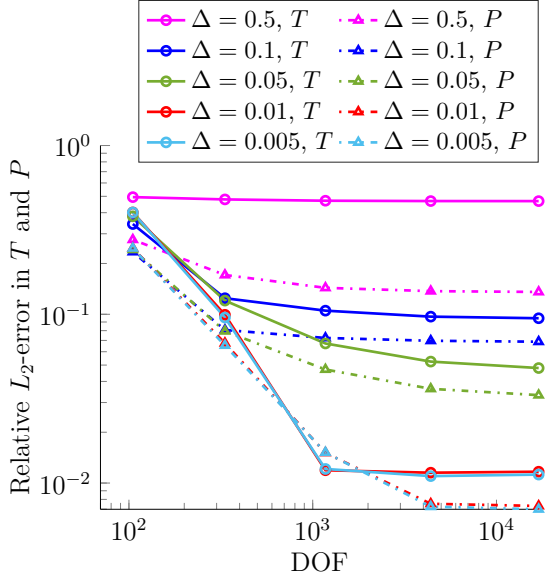
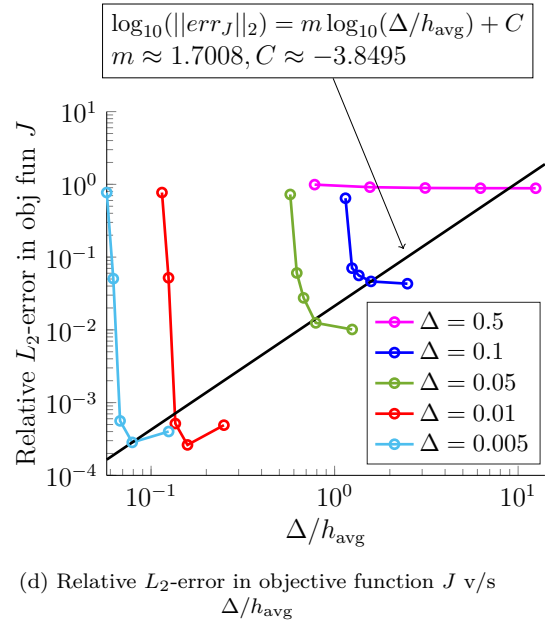}
        \caption{\centering Relative $L_2$-error in state variable $T$ v/s interface radius $R_{L}$}
        \label{fig:BenchAnncase_refeffect_a}
    \end{subfigure}
    \quad \quad
\begin{subfigure}[t]{0.44\textwidth}
\scalebox{0.85}{
%
%
\definecolor{mycolor1}{rgb}{1.00000,0.00000,1.00000}%
\definecolor{mycolor2}{rgb}{0.46600,0.67400,0.18800}%
\definecolor{mycolor3}{rgb}{0.30100,0.74500,0.93300}%
\begin{tikzpicture}

\begin{axis}[%
xmode=log,
xmin=30,
xmax=16965,
xminorticks=true,
xlabel={DOF},
ymode=log,
ymin=0.0001,
ymax=5,
yminorticks=true,
ylabel={Relative $L_{2}$-error in obj fun $J$},
axis x line*=bottom,
axis y line*=left,
legend style={at={(0.03,0.03)},anchor=south west,legend cell align=left,align=left,draw=white!15!black,font=\small}
]
\addplot [color=mycolor1,solid,line width=1.2pt,mark=o,mark options={solid}]
  table[row sep=crcr]{%
105	0.991401947177723\\
333	0.914796905379859\\
1173	0.891681533358828\\
4389	0.885596632030092\\
16965	0.884055689721597\\
};
\addlegendentry{$\Delta=0.5$};

\addplot [color=blue,solid,line width=1.2pt,mark=o,mark options={solid}]
  table[row sep=crcr]{%
105	0.644937944842772\\
333	0.0704654143581955\\
1173	0.0559454392531275\\
4389	0.0463197932849809\\
16965	0.0430985456463964\\
};
\addlegendentry{$\Delta=0.1$};

\addplot [color=mycolor2,solid,line width=1.2pt,mark=o,mark options={solid}]
  table[row sep=crcr]{%
105	0.726593804686611\\
333	0.0605368638818787\\
1173	0.0275128143112726\\
4389	0.0124672760350149\\
16965	0.0101233719146431\\
};
\addlegendentry{$\Delta=0.05$};

\addplot [color=red,solid,line width=1.2pt,mark=o,mark options={solid}]
  table[row sep=crcr]{%
105	0.7718275156325\\
333	0.0519712975988462\\
1173	0.000515120797784627\\
4389	0.000262124826146793\\
16965	0.000489755325121373\\
};
\addlegendentry{$\Delta=0.01$};

\addplot [color=mycolor3,solid,line width=1.2pt,mark=o,mark options={solid}]
  table[row sep=crcr]{%
105	0.7718275156325\\
333	0.0505518163019775\\
1173	0.000561032055968939\\
4389	0.000283790727754573\\
16965	0.000400690175234635\\
};
\addlegendentry{$\Delta=0.005$};

\end{axis}
\end{tikzpicture}%
}
        \caption{\centering Relative $L_2$-error in objective function $J$ v/s DOF}
        \label{fig:BenchAnncase_refeffect_b}
    \end{subfigure}\\
 \begin{subfigure}[t]{0.44\textwidth}
\scalebox{0.85}{
%
%
\definecolor{mycolor1}{rgb}{1.00000,0.00000,1.00000}%
\definecolor{mycolor2}{rgb}{0.46600,0.67400,0.18800}%
\definecolor{mycolor3}{rgb}{0.30100,0.74500,0.93300}%
\begin{tikzpicture}

\begin{axis}[%
xmode=log,
xmin=70,
xmax=16965,
xminorticks=true,
xlabel={DOF},
ymode=log,
ymin=0.0069920406490319,
ymax=1,
yminorticks=true,
ylabel={Relative $L_{2}$-error in $T$ and $P$},
axis x line*=bottom,
axis y line*=left,
legend style={at={(1.00,1.4)},anchor=north east,legend cell align=left,align=left,draw=white!15!black,font=\small,legend columns=2}
]
\addplot [color=mycolor1,solid,line width=1.2pt,mark=o,mark options={solid}]
  table[row sep=crcr]{%
105	0.494566522085533\\
333	0.479301964506503\\
1173	0.470855367136893\\
4389	0.468711562501805\\
16965	0.468390597496847\\
};
\addlegendentry{$\Delta=0.5$, $T$};

\addplot [color=mycolor1,dash pattern=on 1pt off 3pt on 3pt off 3pt,line width=1.2pt,mark=triangle,mark options={solid}]
  table[row sep=crcr]{%
105	0.276737832165034\\
333	0.170927750980757\\
1173	0.143457059370613\\
4389	0.137136910781103\\
16965	0.135616483370714\\
};
\addlegendentry{$\Delta=0.5$, $P$};

\addplot [color=blue,solid,line width=1.2pt,mark=o,mark options={solid}]
  table[row sep=crcr]{%
105	0.342624571670132\\
333	0.124517247150721\\
1173	0.105073888892516\\
4389	0.0968322443689322\\
16965	0.0947311106367554\\
};
\addlegendentry{$\Delta=0.1$, $T$};

\addplot [color=blue,dash pattern=on 1pt off 3pt on 3pt off 3pt,line width=1.2pt,mark=triangle,mark options={solid}]
  table[row sep=crcr]{%
105	0.234034791041442\\
333	0.0807159608875408\\
1173	0.0724121555882883\\
4389	0.0696750984548538\\
16965	0.0687802727536057\\
};
\addlegendentry{$\Delta=0.1$, $P$};

\addplot [color=mycolor2,solid,line width=1.2pt,mark=o,mark options={solid}]
  table[row sep=crcr]{%
105	0.380740339402373\\
333	0.120525892916596\\
1173	0.0671508409677285\\
4389	0.0524376492141602\\
16965	0.048022652286642\\
};
\addlegendentry{$\Delta=0.05$, $T$};

\addplot [color=mycolor2,dash pattern=on 1pt off 3pt on 3pt off 3pt,line width=1.2pt,mark=triangle,mark options={solid}]
  table[row sep=crcr]{%
105	0.241320900515912\\
333	0.0794015715873682\\
1173	0.0470837746399738\\
4389	0.0361723358335955\\
16965	0.0331844488806919\\
};
\addlegendentry{$\Delta=0.05$, $P$};

\addplot [color=red,solid,line width=1.2pt,mark=o,mark options={solid}]
  table[row sep=crcr]{%
105	0.401298682447853\\
333	0.0993174205122117\\
1173	0.0119067633292871\\
4389	0.0115153873570871\\
16965	0.0116709486999387\\
};
\addlegendentry{$\Delta=0.01$, $T$};

\addplot [color=red,dash pattern=on 1pt off 3pt on 3pt off 3pt,line width=1.2pt,mark=triangle,mark options={solid}]
  table[row sep=crcr]{%
105	0.245997398286278\\
333	0.0674522441214323\\
1173	0.0151069160834619\\
4389	0.00753464177271237\\
16965	0.00731478734499041\\
};
\addlegendentry{$\Delta=0.01$, $P$};

\addplot [color=mycolor3,solid,line width=1.2pt,mark=o,mark options={solid}]
  table[row sep=crcr]{%
105	0.401298682447853\\
333	0.0956421069655651\\
1173	0.0121183979194343\\
4389	0.0109937186028323\\
16965	0.011223231621505\\
};
\addlegendentry{$\Delta=0.005$, $T$};

\addplot [color=mycolor3,dash pattern=on 1pt off 3pt on 3pt off 3pt,line width=1.2pt,mark=triangle,mark options={solid}]
  table[row sep=crcr]{%
105	0.245997398286278\\
333	0.065366194432472\\
1173	0.0151686655005268\\
4389	0.00724475410331682\\
16965	0.0069920406490319\\
};
\addlegendentry{$\Delta=0.005$, $P$};

\end{axis}
\end{tikzpicture}%
}
        \caption{\centering Relative $L_2$-error in state variable $T$ and adjoint variable $P$ v/s DOF}
        \label{fig:BenchAnncase_refeffect_c}
    \end{subfigure} \quad \quad
    \begin{subfigure}[t]{0.44\textwidth}
\scalebox{0.85}{
%
%
\definecolor{mycolor1}{rgb}{1.00000,0.00000,1.00000}%
\definecolor{mycolor2}{rgb}{0.46600,0.67400,0.18800}%
\definecolor{mycolor3}{rgb}{0.30100,0.74500,0.93300}%
\begin{tikzpicture}

\begin{axis}[%
xmode=log,
xmin=0.057325,
xmax=14,
xminorticks=true,
xlabel={$\Delta/h_{\rm avg}$},
ymode=log,
ymin=0.0001,
ymax=10,
yminorticks=true,
ylabel={Relative $L_{2}$-error in obj fun $J$},
axis x line*=bottom,
axis y line*=left,
legend style={at={(0.99,0.03)},anchor=south east,legend cell align=left,align=left,draw=white!15!black,font=\small}
]
\addplot [color=mycolor1,solid,line width=1.2pt,mark=o,mark options={solid}]
  table[row sep=crcr]{%
0.778377498891037	0.991401947177723\\
1.55779188208883	0.914796905379859\\
3.11661507625843	0.891681533358828\\
6.2342572720716	0.885596632030092\\
12.469539221778	0.884055689721597\\
};
\addlegendentry{$\Delta=0.5$};

\addplot [color=blue,solid,line width=1.2pt,mark=o,mark options={solid}]
  table[row sep=crcr]{%
1.14650010994814	0.644937944842772\\
1.24487366727522	0.0704654143581955\\
1.35583449172977	0.0559454392531275\\
1.57938980624111	0.0463197932849809\\
2.49390784435561	0.0430985456463964\\
};
\addlegendentry{$\Delta=0.1$};

\addplot [color=mycolor2,solid,line width=1.2pt,mark=o,mark options={solid}]
  table[row sep=crcr]{%
0.57325005497407	0.726593804686611\\
0.622436833637609	0.0605368638818787\\
0.677917245864887	0.0275128143112726\\
0.789694903120555	0.0124672760350149\\
1.2469539221778	0.0101233719146431\\
};
\addlegendentry{$\Delta=0.05$};

\addplot [color=red,solid,line width=1.2pt,mark=o,mark options={solid}]
  table[row sep=crcr]{%
0.114650010994814	0.7718275156325\\
0.124487366727522	0.0519712975988462\\
0.135583449172977	0.000515120797784627\\
0.157938980624111	0.000262124826146793\\
0.249390784435561	0.000489755325121373\\
};
\addlegendentry{$\Delta=0.01$};

\addplot [color=mycolor3,solid,line width=1.2pt,mark=o,mark options={solid}]
  table[row sep=crcr]{%
0.057325005497407	0.7718275156325\\
0.0622436833637609	0.0505518163019775\\
0.0677917245864887	0.000561032055968939\\
0.0789694903120555	0.000283790727754573\\
0.12469539221778	0.000400690175234635\\
};
\addlegendentry{$\Delta=0.005$};

\addplot [color=black,solid,line width=1.2pt]
  table[row sep=crcr]{%
0.057325	0.00016458621222704\\
14	1.89449504564411\\
};
\end{axis}
\node[draw, align=left] at (3.3,6.8) (A){\small$\log_{10}(\vert \vert err_{J}\vert \vert_{2})=m\log_{10}(\Delta/h_{\rm avg})+C \vspace{0.1cm}$ \\  $m\approx 1.7008, C\approx-3.8495$};
\node[align=left] at (4.7,3.4) (B){};
\draw[->] (A) -- (B);
\end{tikzpicture}%
}
        \caption{\centering Relative $L_2$-error in objective function $J$ v/s $\Delta/h_{\rm avg}$}
        \label{fig:BenchAnncase_refeffect_d}
    \end{subfigure}
\caption{For the annular ring problem, the effect of DOF number on stability and accuracy (a) variation in the relative $L_2$-error of the state variable $T$ in the radius range as the mesh is refined, $\Delta=0.05$ (b) convergence of the objective function $J$, $\Delta=0.5,0.1,0.05,0.01,0.005$ (c) convergence of the state variable $T$ and adjoint variable $P$, $\Delta=0.5,0.1,0.05,0.01,0.005$ (d) relation between relative $L_2$-error in $J$ and $\Delta/h_{\rm avg}$, where $h_{\rm avg}$ is the average mesh size, $\Delta=0.5,0.1,0.05,0.01,0.005$. The errors in $J$, $T$, and $P$ reduce with mesh refinement bounded by the support bandwidth.}
    \label{fig:BenchAnncase_refeffect}
\end{figure}

 In \fref{fig:BenchAnncase_refeffect}, we show the effect of the mesh size on the numerical results. \fref{fig:BenchAnncase_refeffect_a} presents the fluctuation of relative $L_2$-error in state variable $T$ over the range of $R_L$ for four different mesh size with DOF=$105, 333, 1173, 4389$. $\Delta=0.05$ for all four cases. From the figure, it is evident that the accuracy and stability increase with the mesh refinement, which aligns with our expectations. Because we are using density-based geometry mapping, the finer mesh with a more precise material distribution improves the structural response. In addition, it captures the approximate Heaviside and Dirac delta functions with better accuracy to provide stable results. One point to note is that a finer mesh means higher computation effort in optimization and that provides a practical limit to the mesh size. 

 \par The relative $L_2$-error in objective function $J$ and the variables $T$ \& $P$ with respect to DOF are plotted in \fref{fig:BenchAnncase_refeffect_b} and \fref{fig:BenchAnncase_refeffect_c}, respectively. Here, we considered five values of $\Delta$ same as \fref{fig:BenchAnncase_var_wrt_R}. We observe a similar pattern, the errors in all three quantities reduce with mesh refinement bounded by the support bandwidth. In \fref{fig:BenchAnncase_refeffect_d}, the approximate relation between relative $L_2$-error in objective function $J$ and $\Delta/h_{\rm avg}$, where $h_{\rm avg}$ is the average mesh size, is developed. From the figure, it can be observed that the black line, $\log_{10}(\vert \vert err_{J}\vert \vert_{2})=m\log_{10}(\Delta/h_{\rm avg})+C$; $m\approx 1.7008, C\approx-3.8495$, represents an approximate upper bound on the improvement by refinement for a given value of $\Delta$. The improvement in accuracy via refinement is limited on the right side of the line. This empirical relation helps to find an appropriate mesh size (for a given $\Delta$) to use for optimization. \newcolumntype{L}[1]{>{\raggedright\let\newline\\\arraybackslash\hspace{0pt}}m{#1}}
\newcolumntype{C}[1]{>{\centering\let\newline\\\arraybackslash\hspace{0pt}}m{#1}}
\newcolumntype{R}[1]{>{\raggedleft\let\newline\\\arraybackslash\hspace{0pt}}mTable{#1}}
\renewcommand{\arraystretch}{1.25}   

\begin{figure}
\centering
\scalebox{0.9}{
\begin{tabular}[c]{|m{1em}|m{5.3em} m{5.3em}|m{5.3em} m{5.3em}| m{5.3em} m{5.3em}|}
\hline		
   & \multicolumn{2}{|c|}{$N_{\mathrm{var}}=25$} & \multicolumn{2}{|c|}{$N_{\mathrm{var}}=42$} & \multicolumn{2}{|c|}{$N_{\mathrm{var}}=1089$} \\
\hline 
\vspace{1.5cm}
   \rotatebox{90}{\centering Sample I}  &
    \begin{subfigure}[t]{0.15\textwidth}\begin{center}{
\includegraphics[width=1\textwidth]{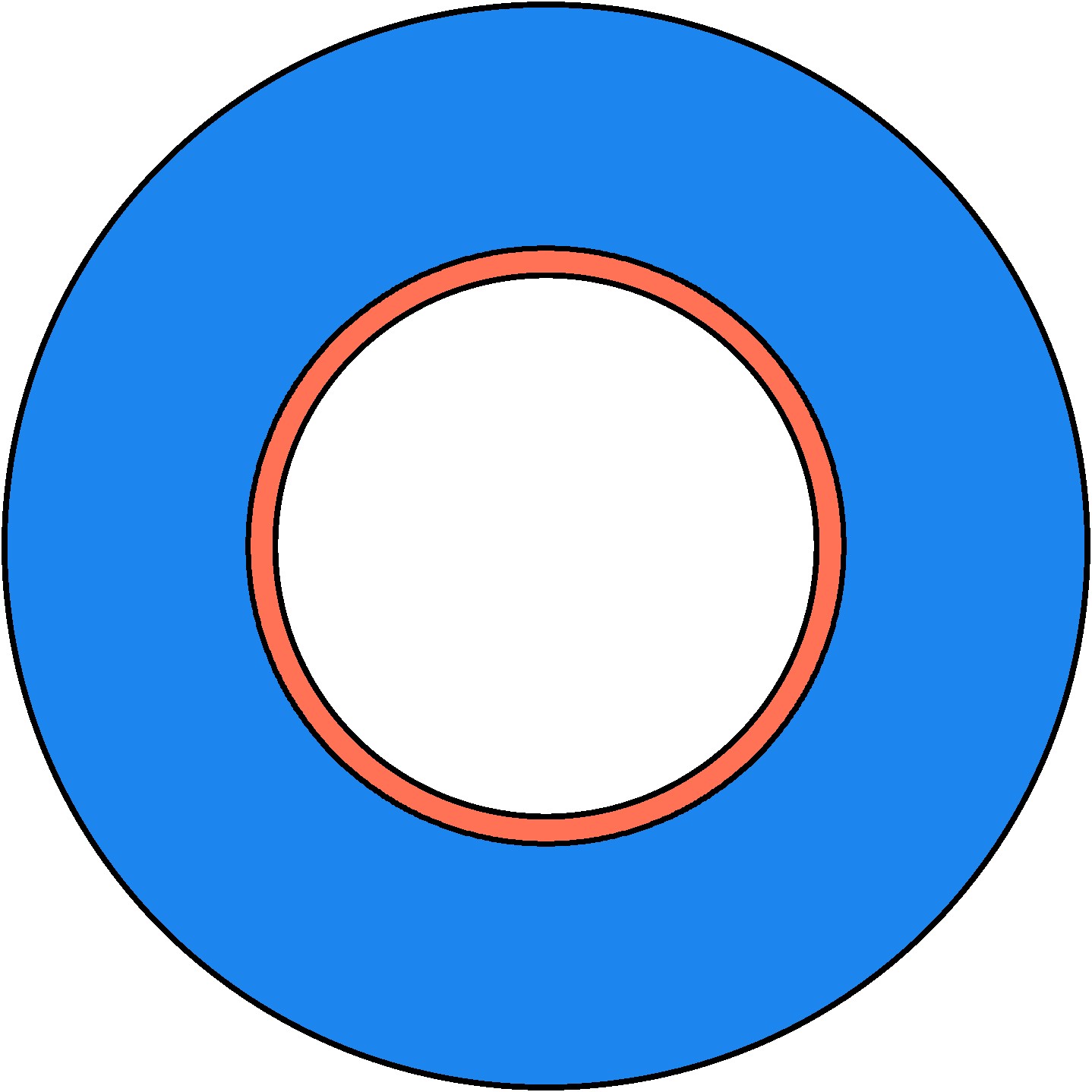}}
        \caption{\centering Initial topology}
         \label{fig:annular ring optimized geo a}        
    \end{center}
    \end{subfigure} &
    \begin{subfigure}[t]{0.15\textwidth}{\centering\includegraphics[width=1\textwidth]{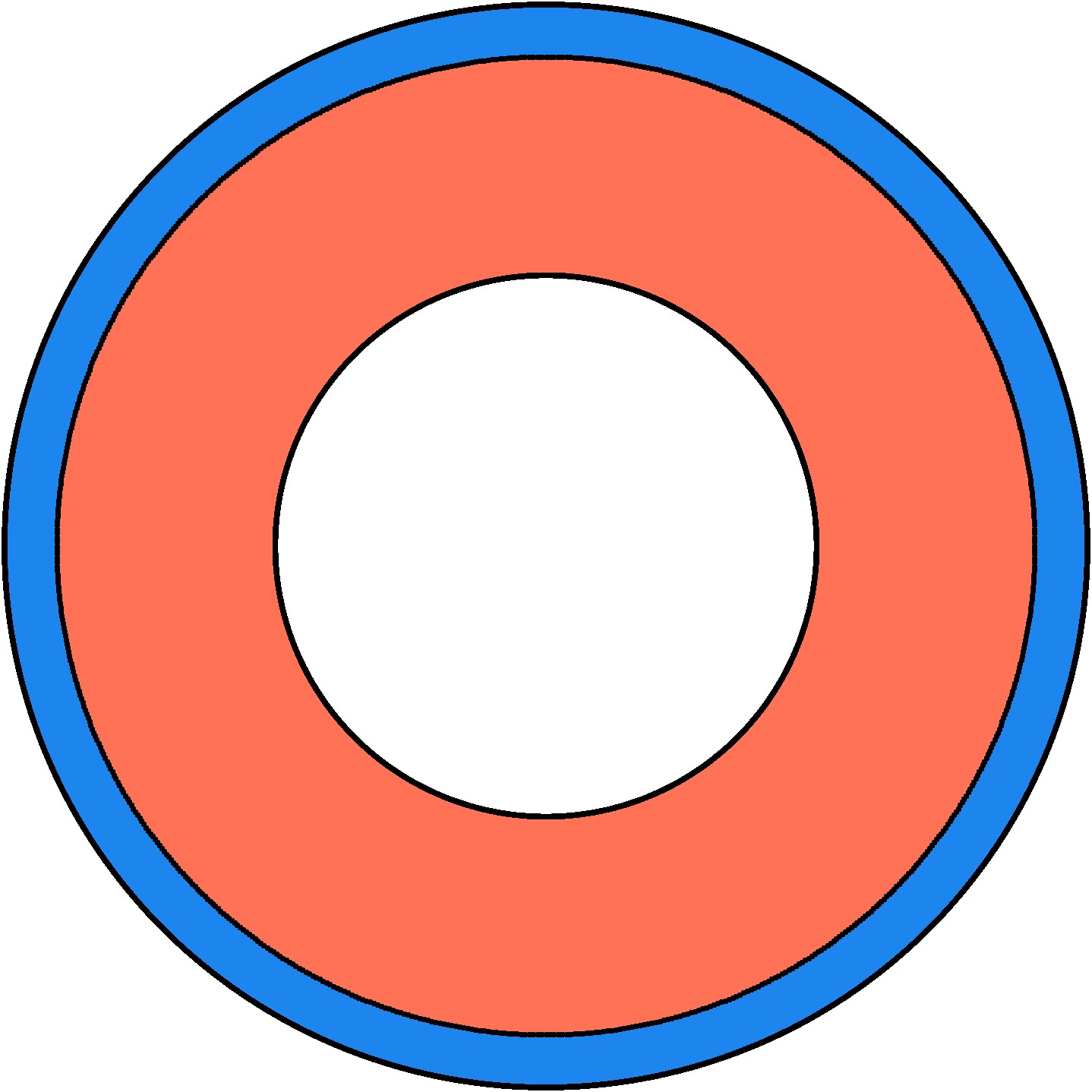}}
        \caption{\centering Optimized topology, $J=1.6099\times 10^4$}
         \label{fig:annular ring optimized geo b}
    \end{subfigure}&
    \begin{subfigure}[t]{0.15\textwidth}{\centering\includegraphics[width=1\textwidth]{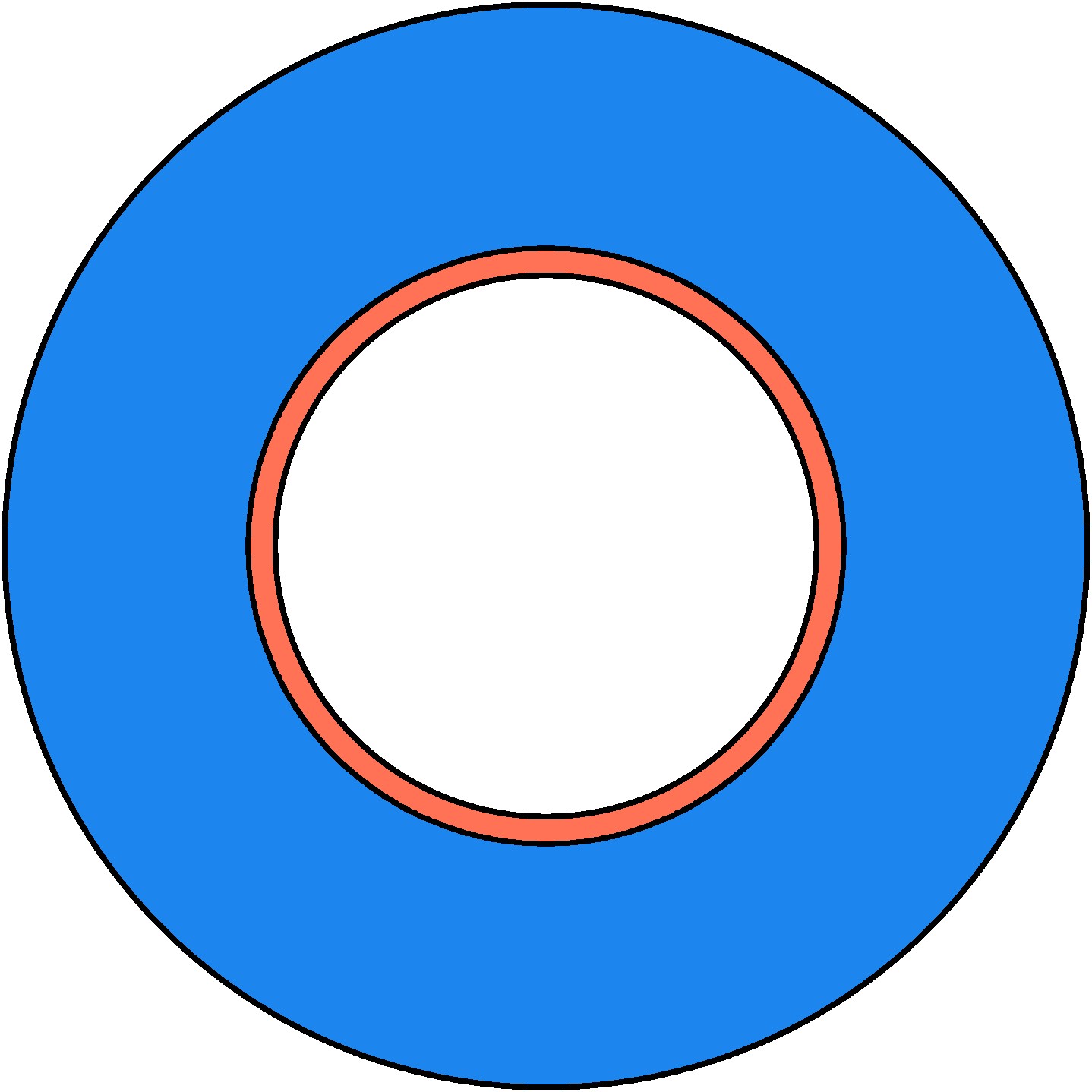}}
        \caption{\centering Initial topology}
         \label{fig:annular ring optimized geo c}
    \end{subfigure}&
    \begin{subfigure}[t]{0.15\textwidth}{\centering\includegraphics[width=1\textwidth]{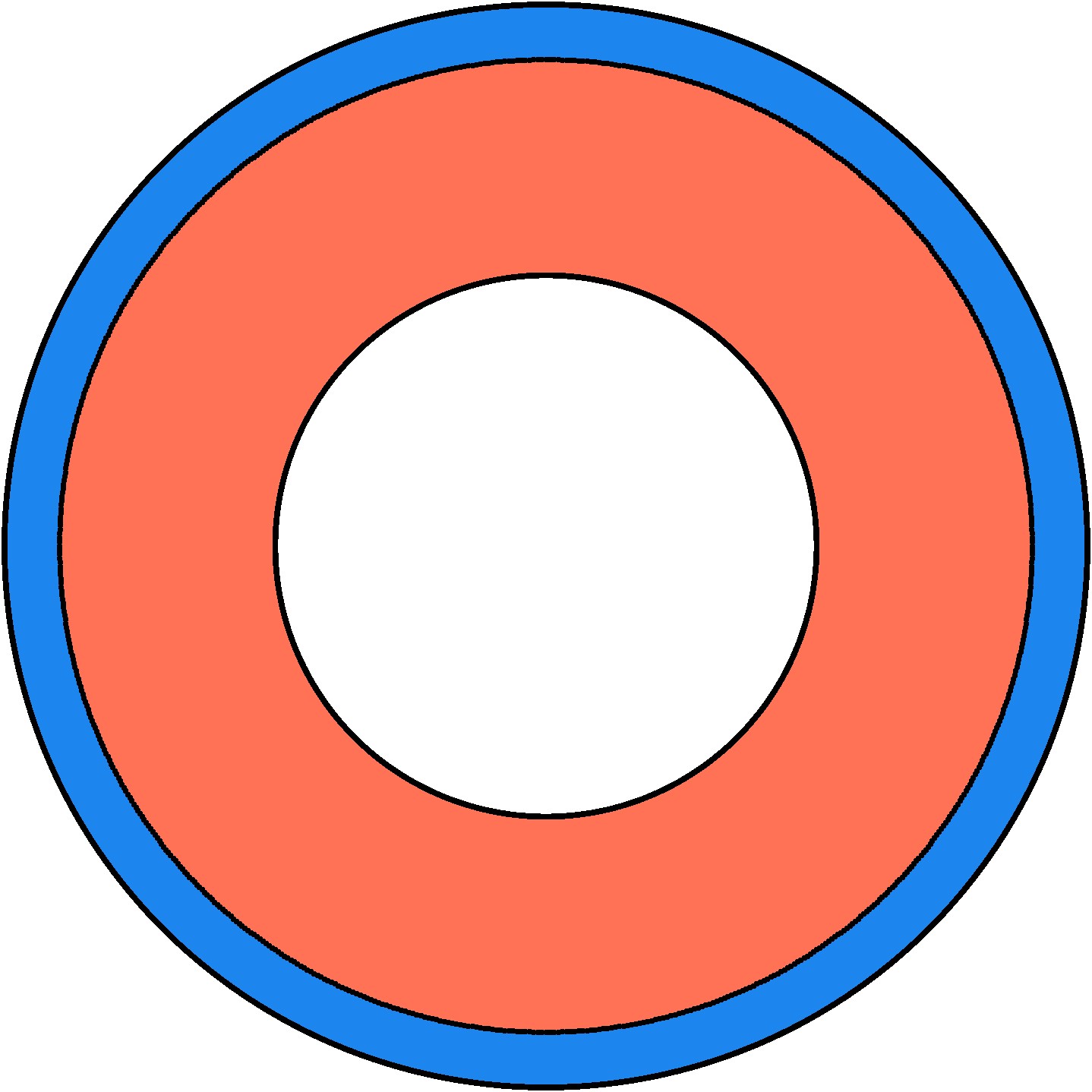}}
        \caption{\centering Optimized topology, $J=1.6099\times 10^4$}
         \label{fig:annular ring optimized geo d}
    \end{subfigure}&
    \begin{subfigure}[t]{0.15\textwidth}{\centering\includegraphics[width=1\textwidth]{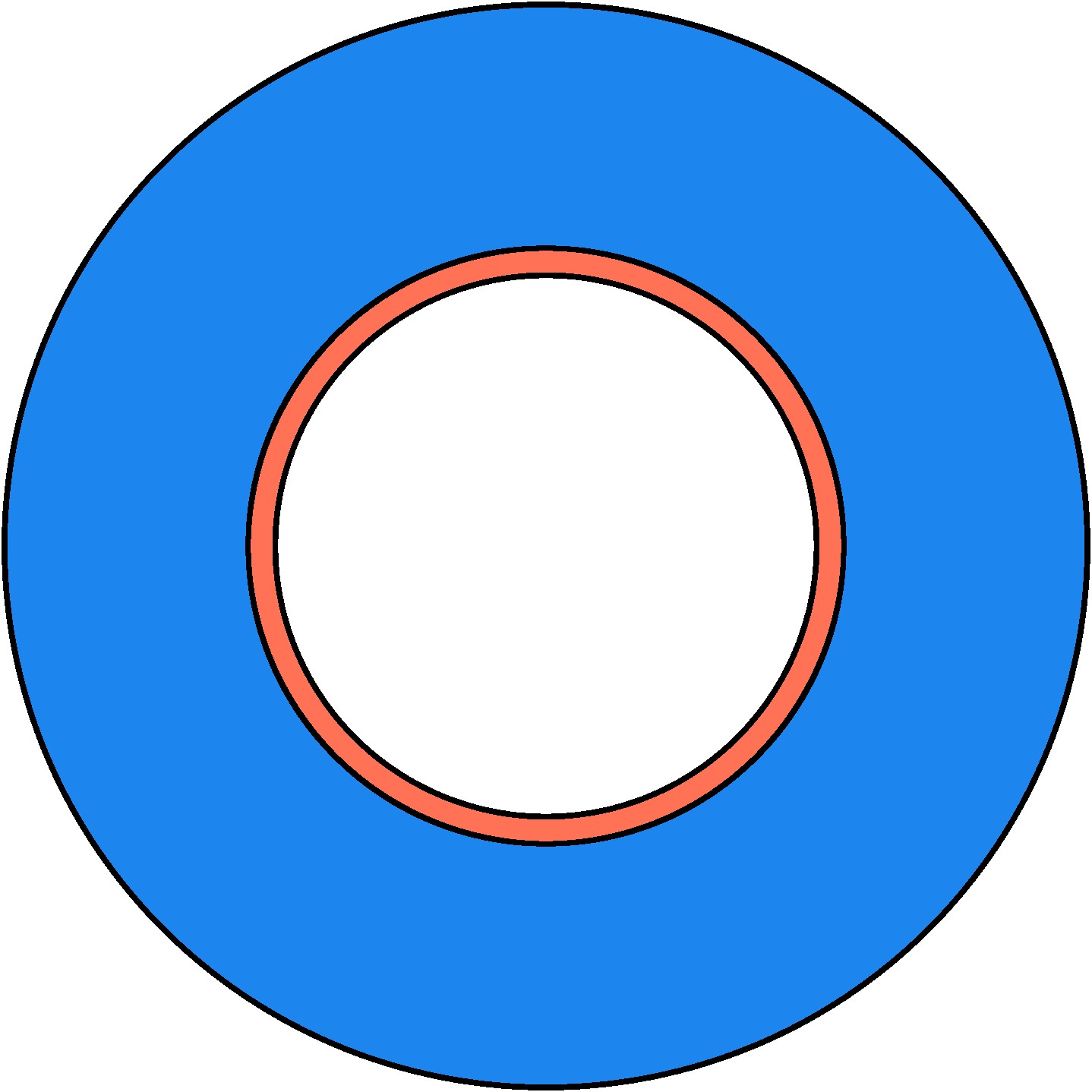}}
        \caption{\centering Initial topology}
         \label{fig:annular ring optimized geo e}
    \end{subfigure}&
    \begin{subfigure}[t]{0.15\textwidth}{\centering\includegraphics[width=1\textwidth]{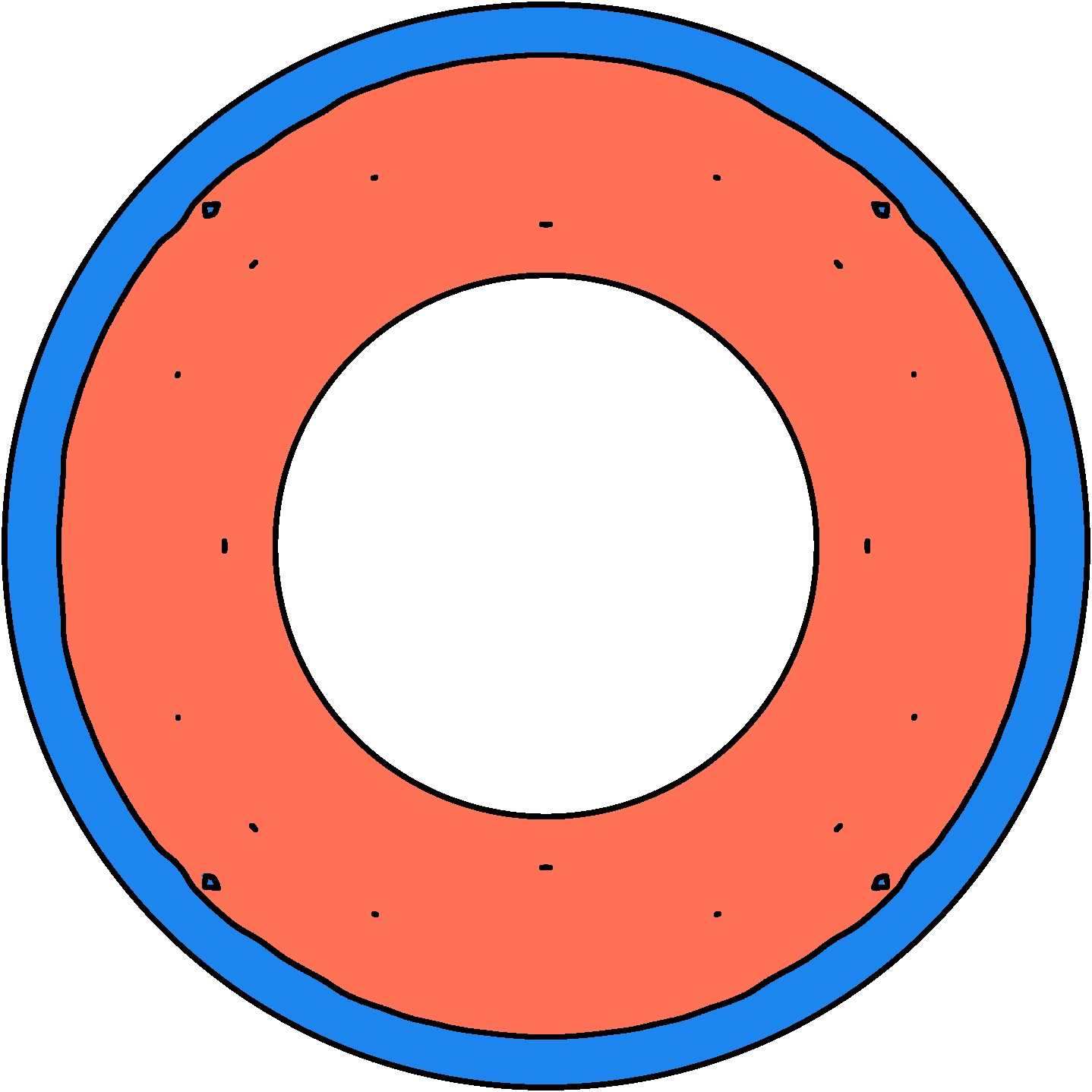}}
        \caption{\centering Optimized topology, $J=1.6099\times 10^4$}
         \label{fig:annular ring optimized geo f}
    \end{subfigure}\\
    \vspace{1cm}
\rotatebox{90}{\centering Sample II}  &\begin{subfigure}[t]{0.15\textwidth}{\centering\includegraphics[width=1\textwidth]{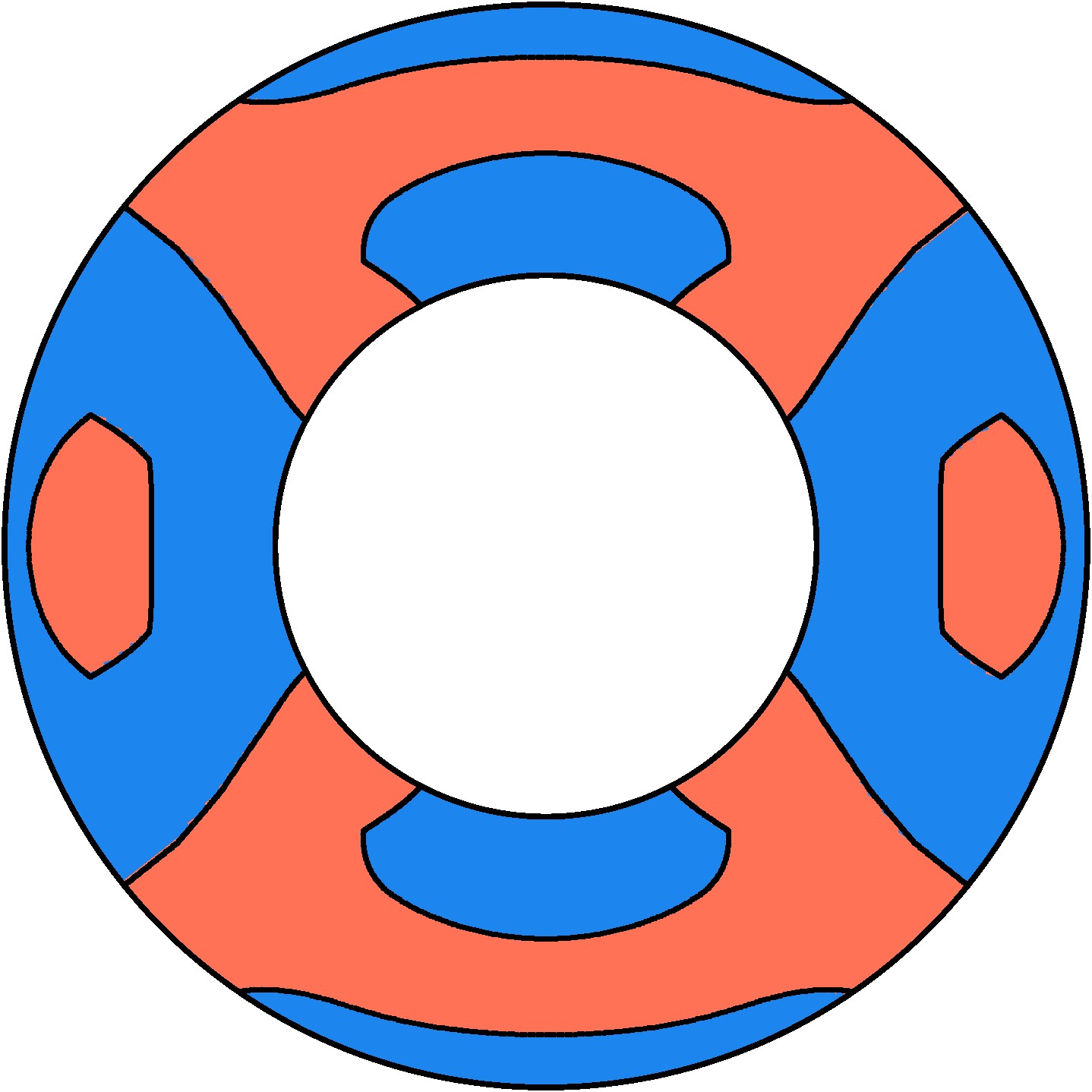}}
        \caption{\centering Initial topology}
         \label{fig:annular ring optimized geo g}
    \end{subfigure}&
    \begin{subfigure}[t]{0.15\textwidth}{\centering\includegraphics[width=1\textwidth]{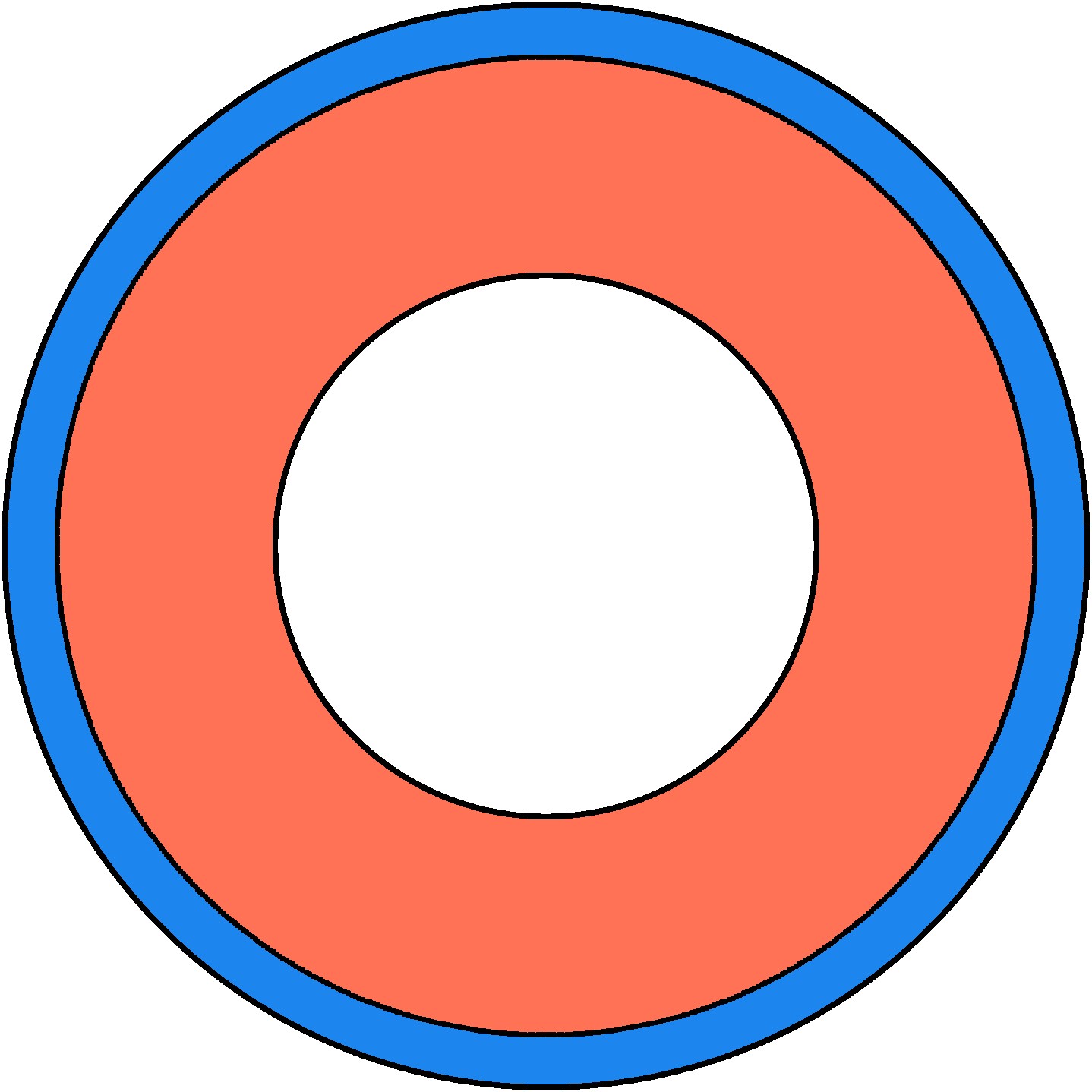}}
        \caption{\centering Optimized topology, $J=1.6099\times 10^4$}
         \label{fig:annular ring optimized geo h}
    \end{subfigure}&
    \begin{subfigure}[t]{0.15\textwidth}{\centering\includegraphics[width=1\textwidth]{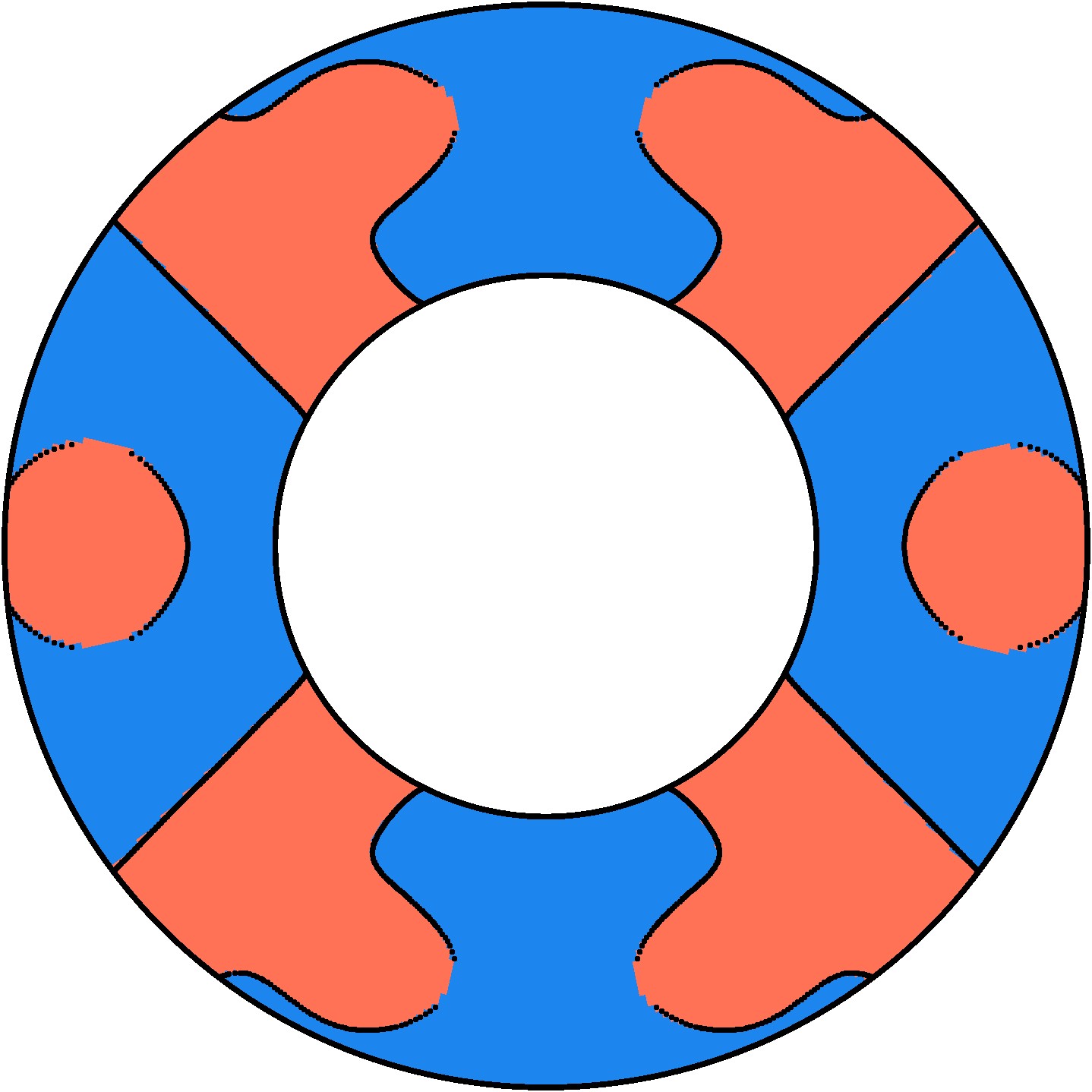}}
        \caption{\centering Initial topology}
         \label{fig:annular ring optimized geo i}
    \end{subfigure}&
    \begin{subfigure}[t]{0.15\textwidth}{\centering\includegraphics[width=1\textwidth]{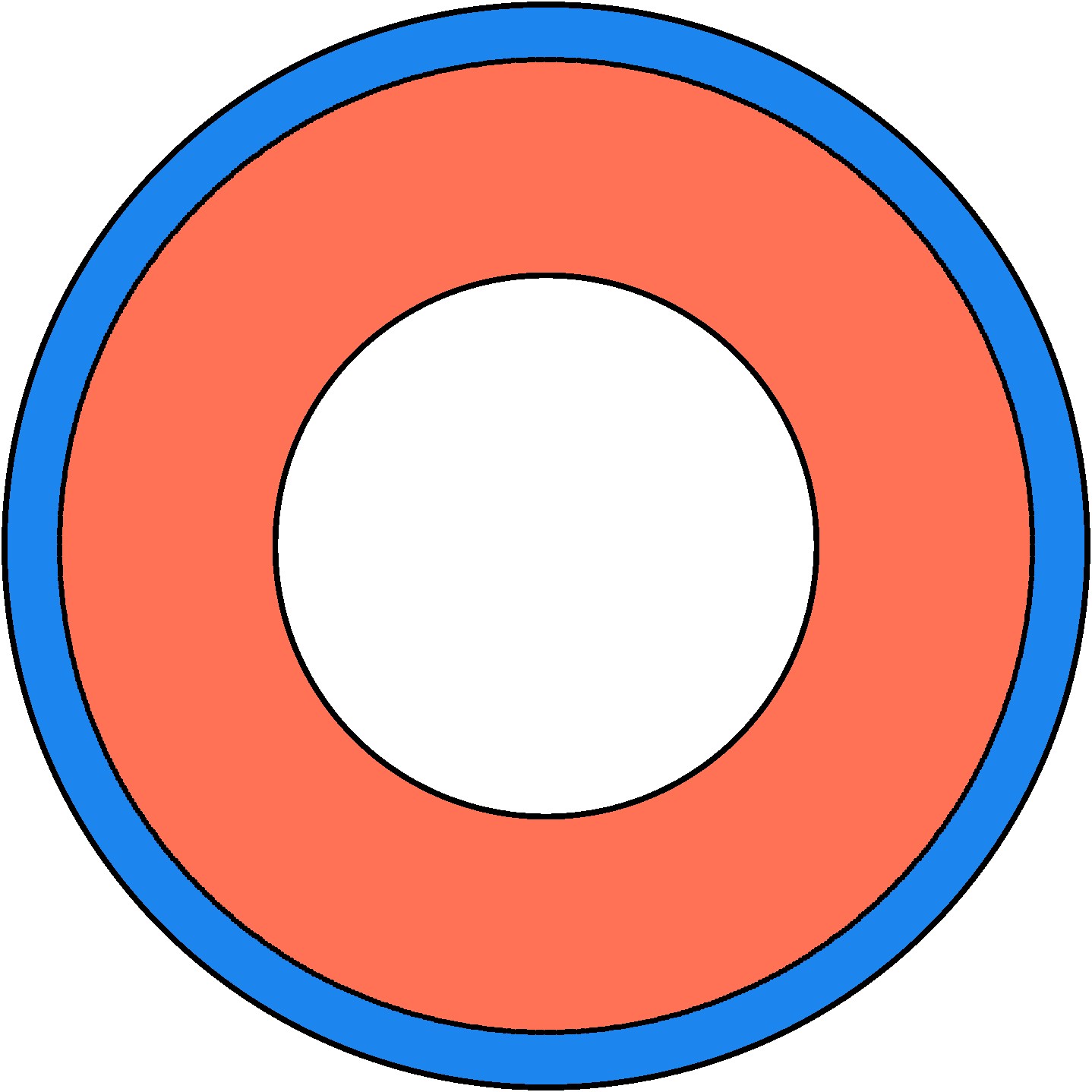}}
        \caption{\centering Optimized topology, $J=1.6099\times 10^4$}
         \label{fig:annular ring optimized geo j}
    \end{subfigure}&
    \begin{subfigure}[t]{0.15\textwidth}{\centering\includegraphics[width=1\textwidth]{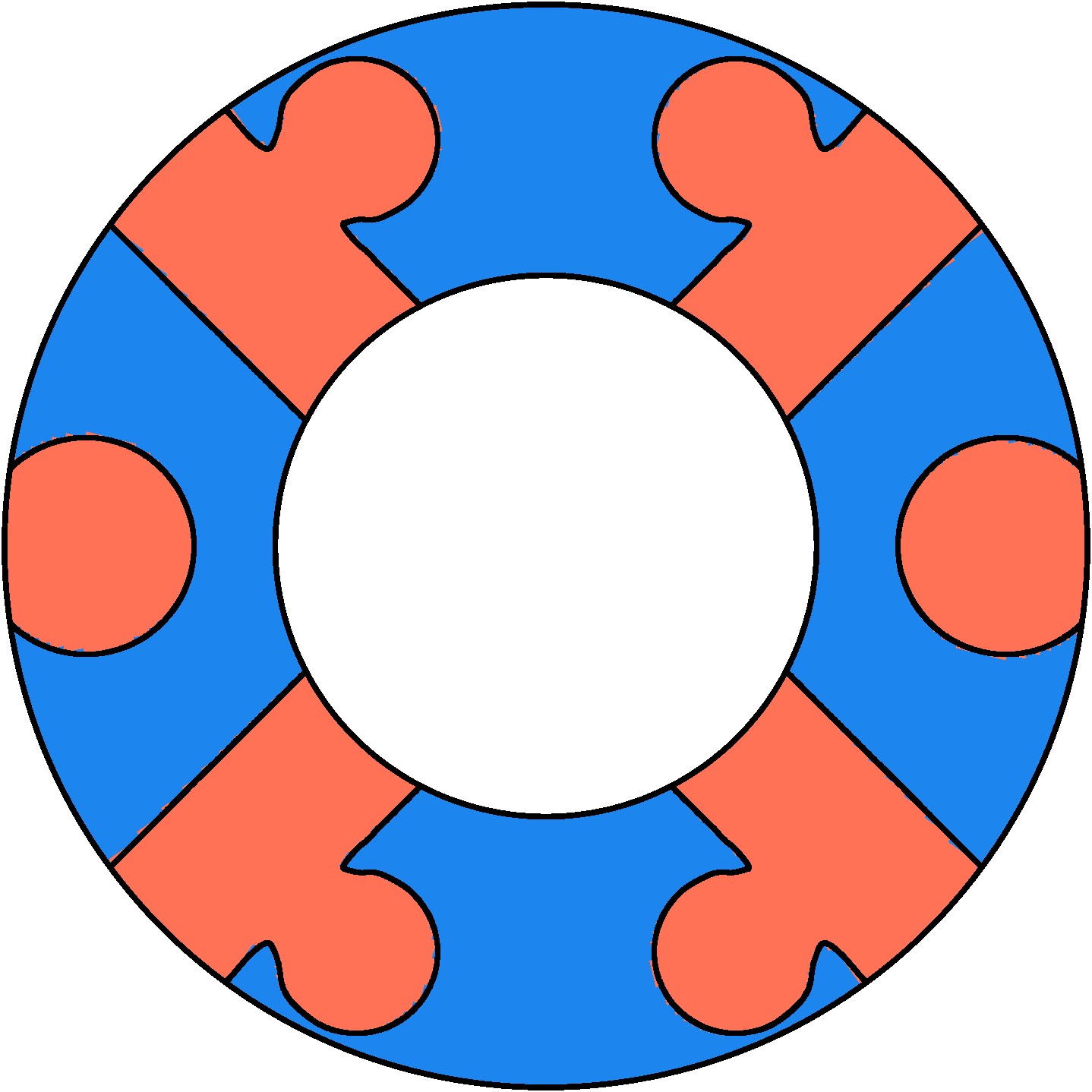}}
        \caption{\centering Initial topology}
         \label{fig:annular ring optimized geo k}
    \end{subfigure}&
    \begin{subfigure}[t]{0.15\textwidth}{\centering\includegraphics[width=1\textwidth]{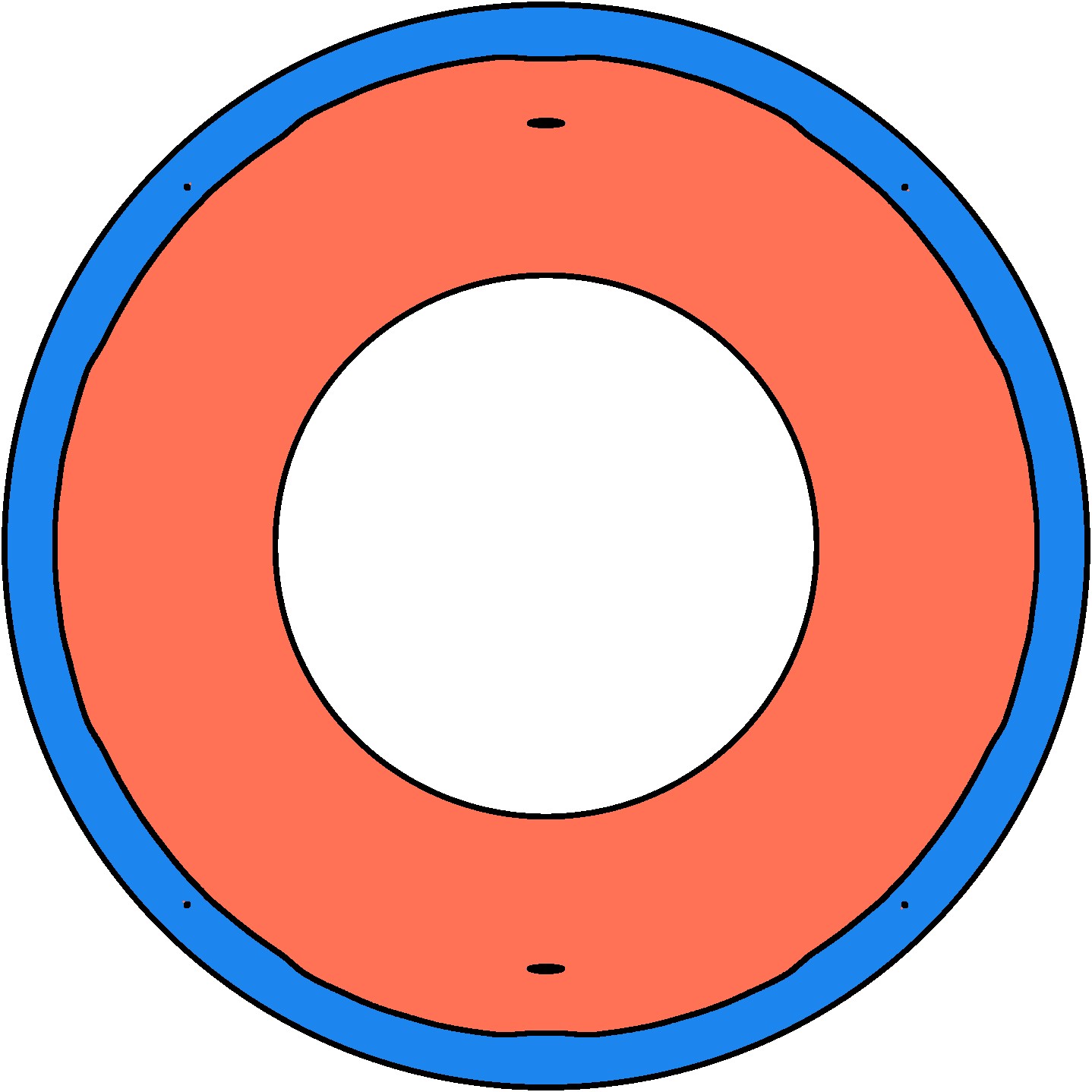}}
        \caption{\centering Optimized topology, $J=1.6099\times 10^4$}
         \label{fig:annular ring optimized geo l}
    \end{subfigure}\\
     \vspace{1cm}
    \rotatebox{90}{\centering Sample III} &   \begin{subfigure}[t]{0.15\textwidth}{\centering\includegraphics[width=1\textwidth]{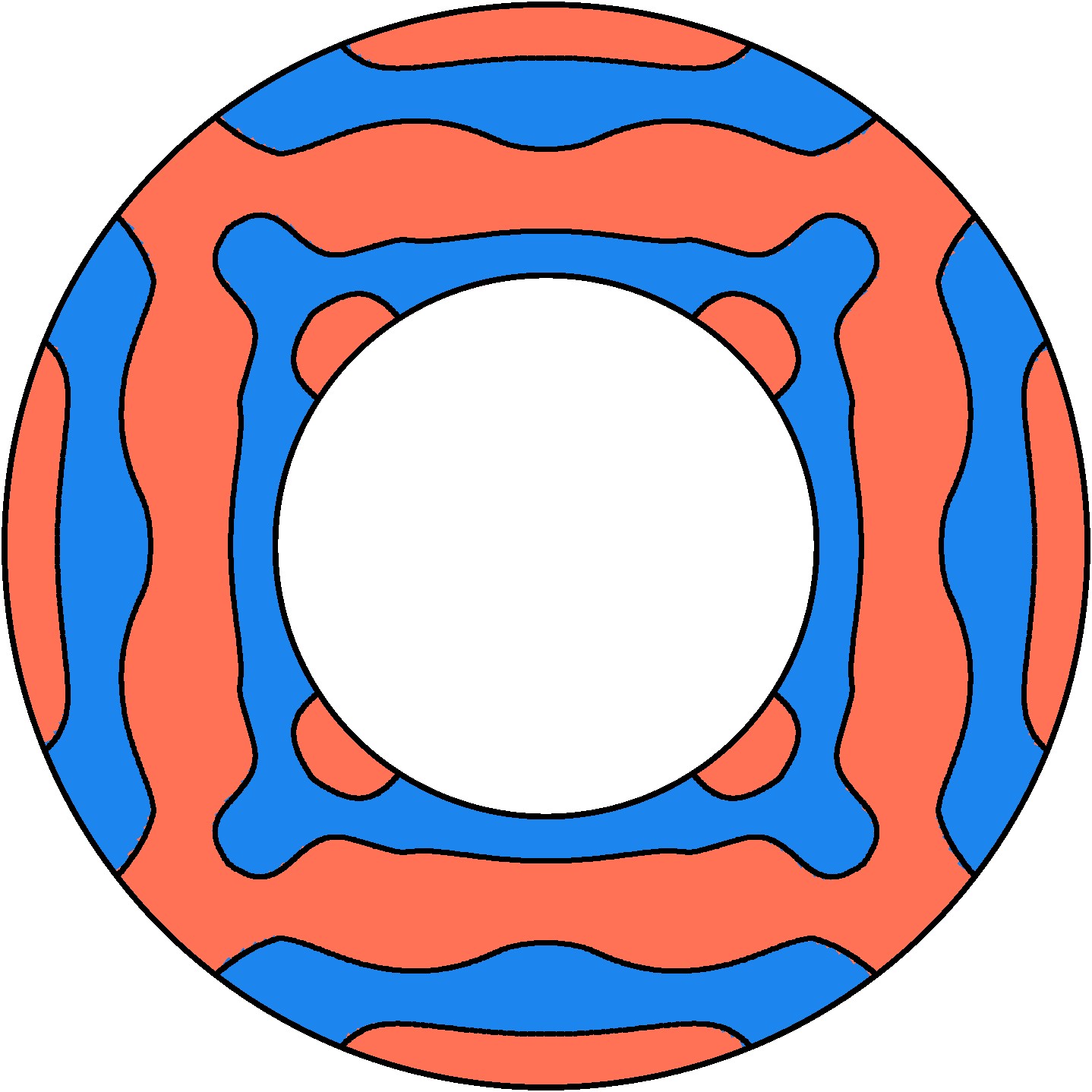}}
        \caption{\centering Initial topology}
         \label{fig:annular ring optimized geo m}
    \end{subfigure}&
    \begin{subfigure}[t]{0.15\textwidth}{\centering\includegraphics[width=1\textwidth]{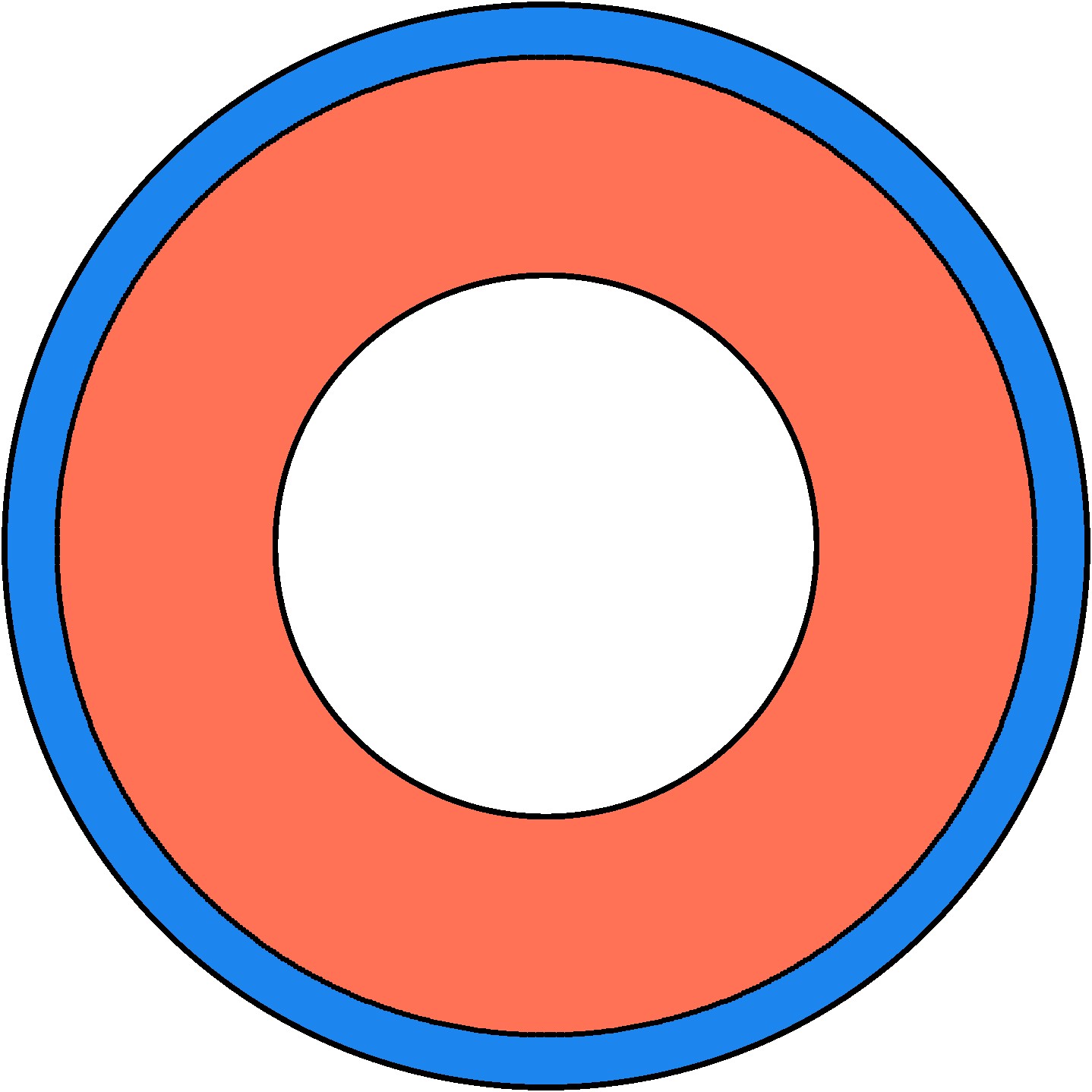}}
        \caption{\centering Optimized topology, $J=1.6099\times 10^4$}
         \label{fig:annular ring optimized geo n}
    \end{subfigure}&
    \begin{subfigure}[t]{0.15\textwidth}{\centering\includegraphics[width=1\textwidth]{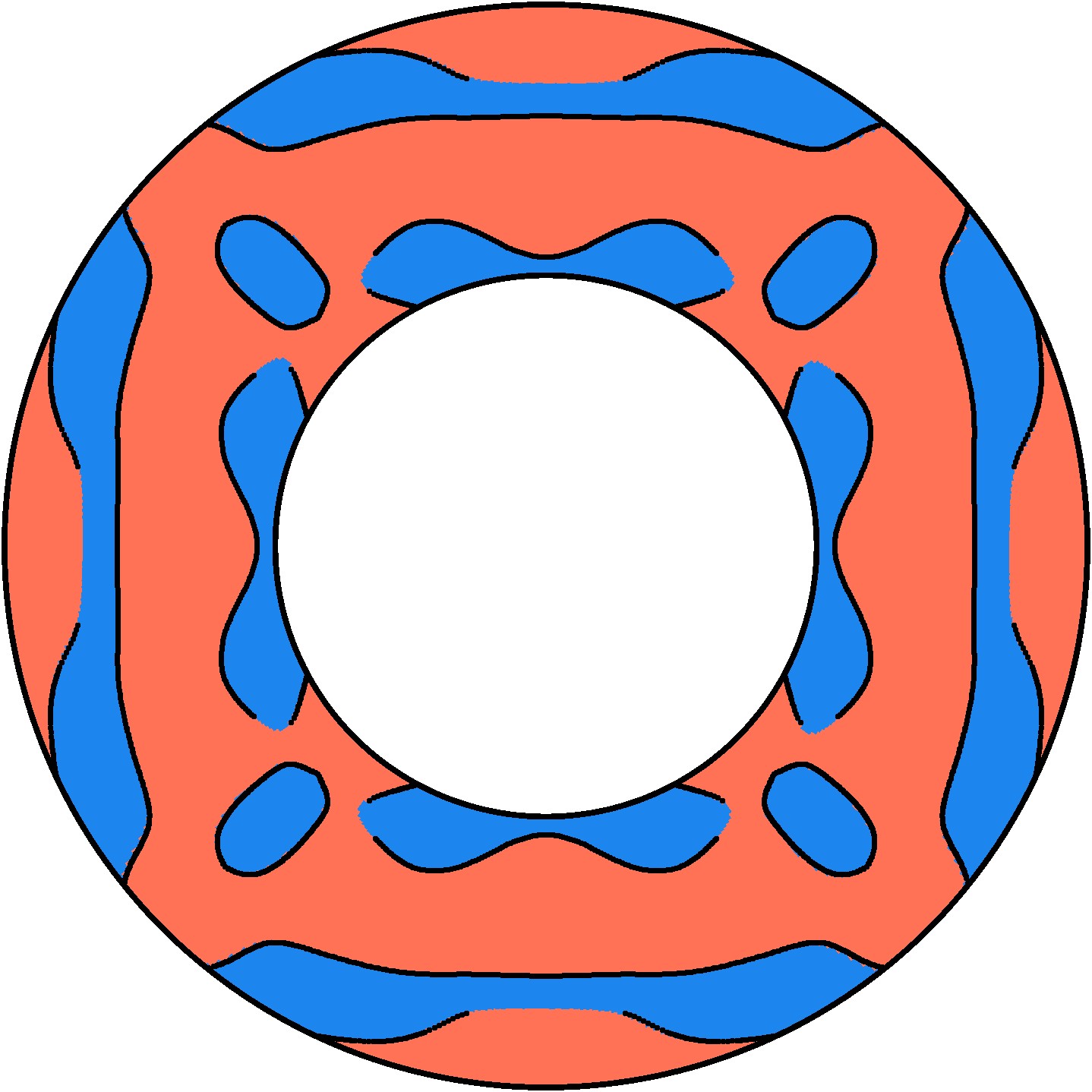}}
        \caption{\centering Initial topology}
         \label{fig:annular ring optimized geo o}
    \end{subfigure}&
    \begin{subfigure}[t]{0.15\textwidth}{\centering\includegraphics[width=1\textwidth]{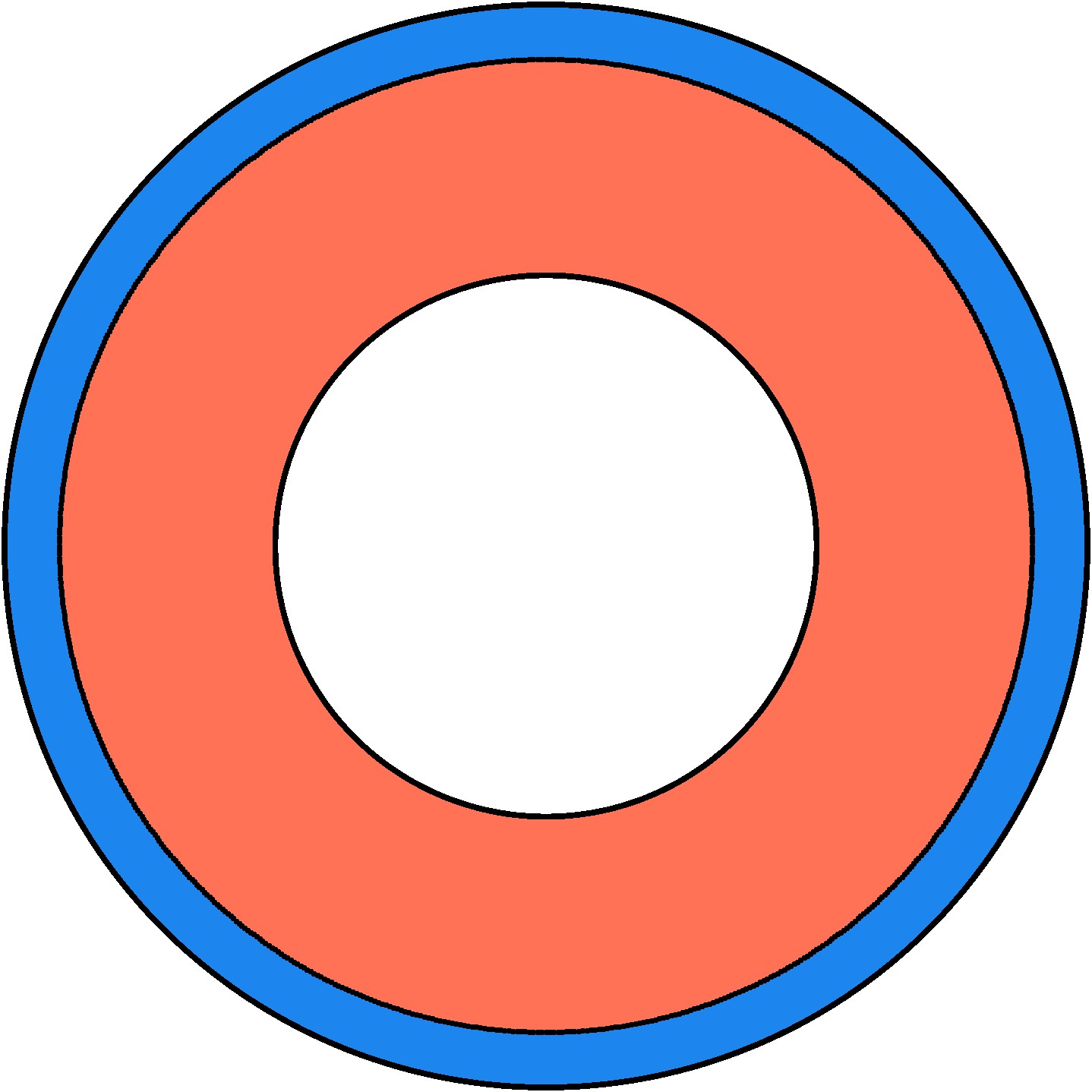}}
        \caption{\centering Optimized topology, $J=1.6099\times 10^4$}
         \label{fig:annular ring optimized geo p}
    \end{subfigure}&
    \begin{subfigure}[t]{0.15\textwidth}{\centering\includegraphics[width=1\textwidth]{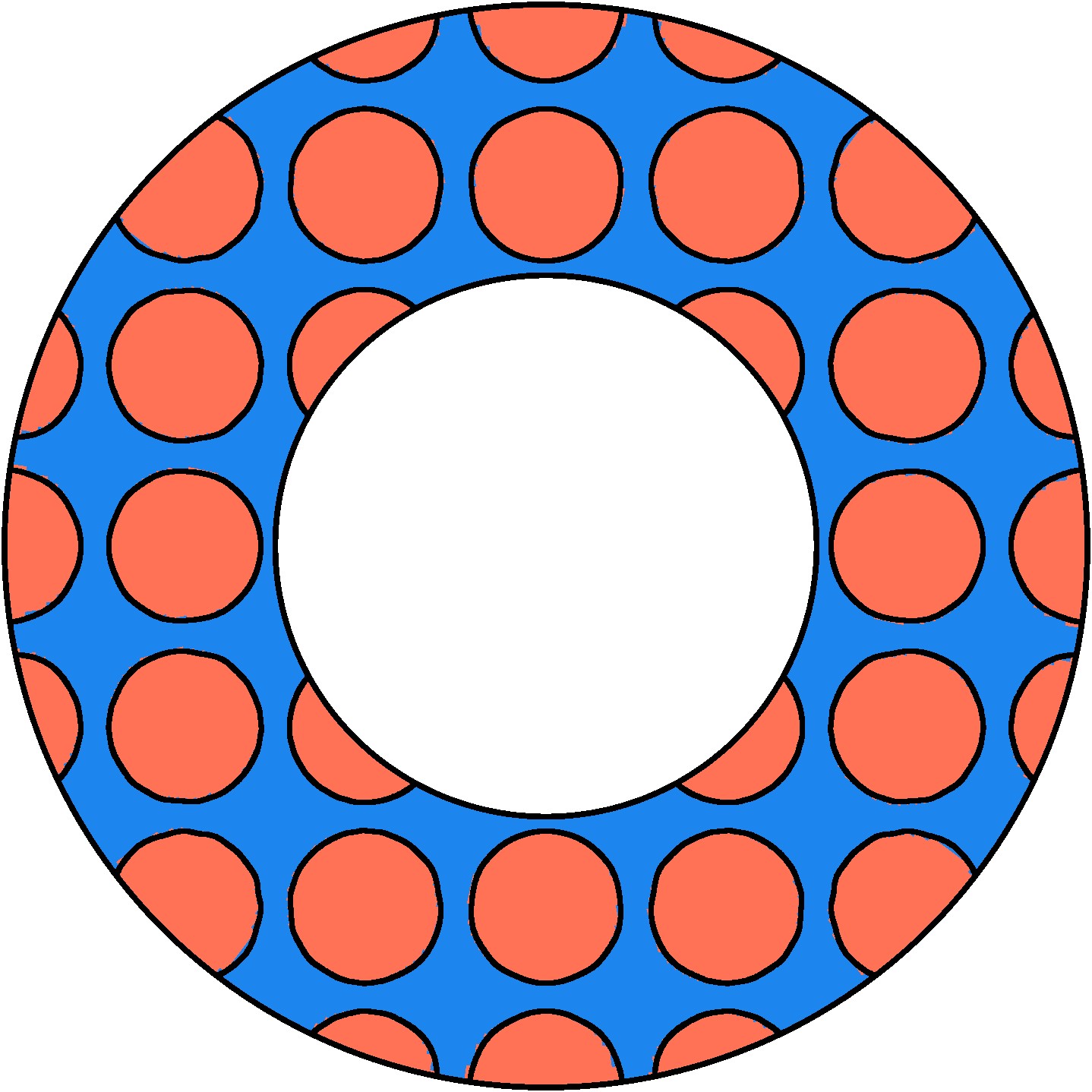}}
        \caption{\centering Initial topology}
         \label{fig:annular ring optimized geo q}
    \end{subfigure} &
    \begin{subfigure}[t]{0.15\textwidth}{\centering\includegraphics[width=1\textwidth]{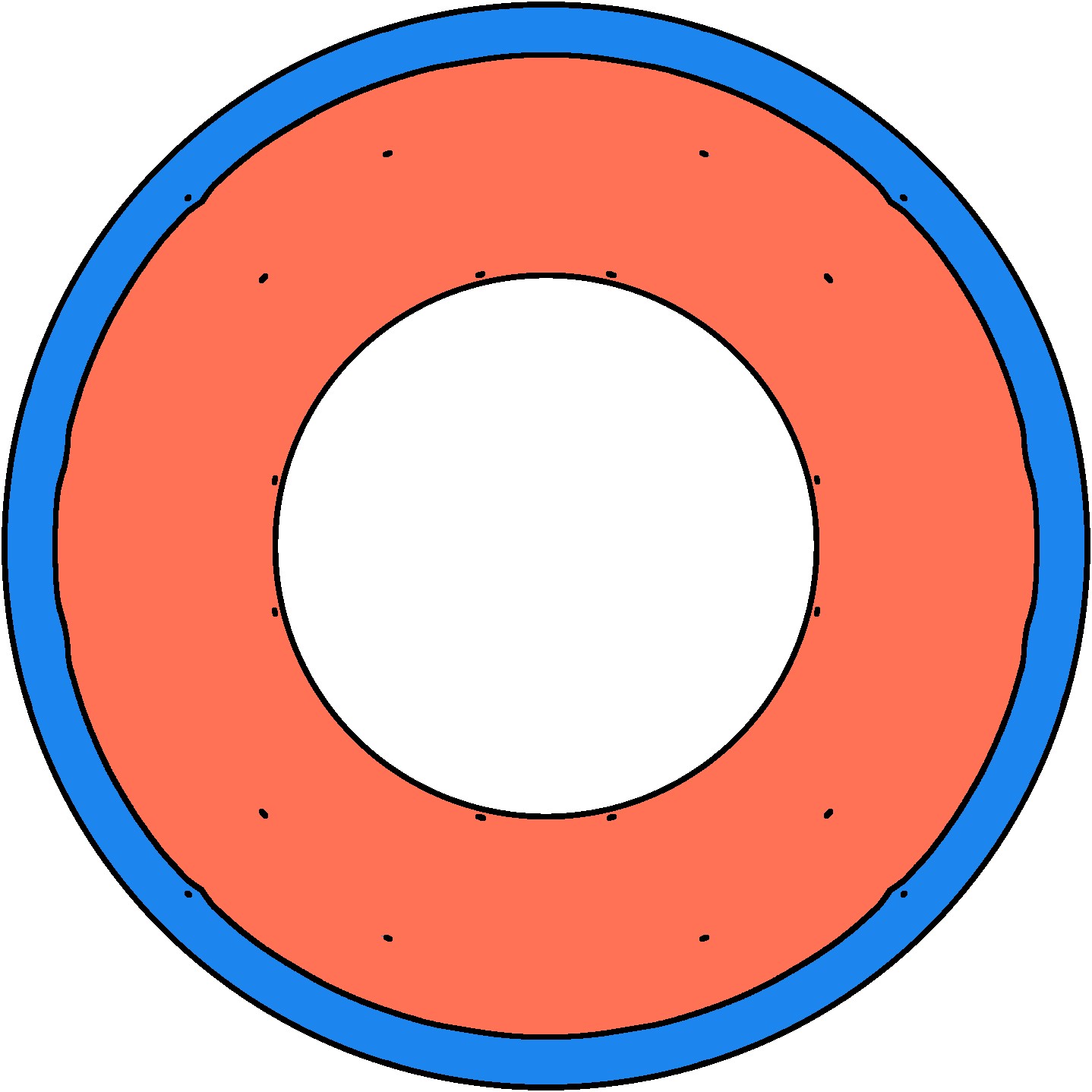}}
        \caption{\centering Optimized topology, $J=1.6099\times 10^4$}
         \label{fig:annular ring optimized geo r}
    \end{subfigure}
    \\
    \vspace{1cm}
  \rotatebox{90}{\centering Sample IV}  &     \begin{subfigure}[t]{0.15\textwidth}{\centering\includegraphics[width=1\textwidth]{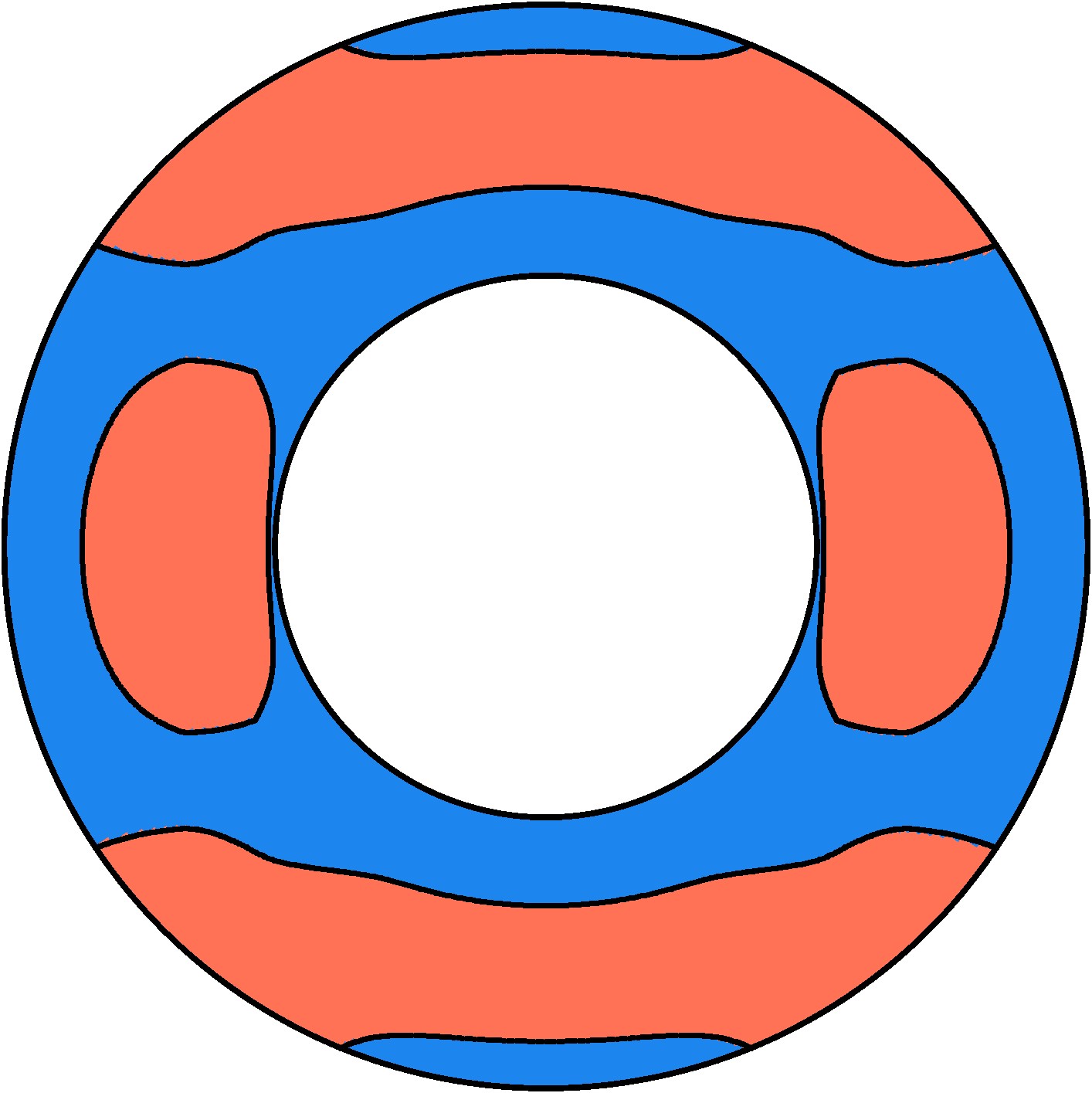}}
        \caption{\centering Initial topology}
         \label{fig:annular ring optimized geo s}
    \end{subfigure}&
    \begin{subfigure}[t]{0.15\textwidth}{\centering\includegraphics[width=1\textwidth]{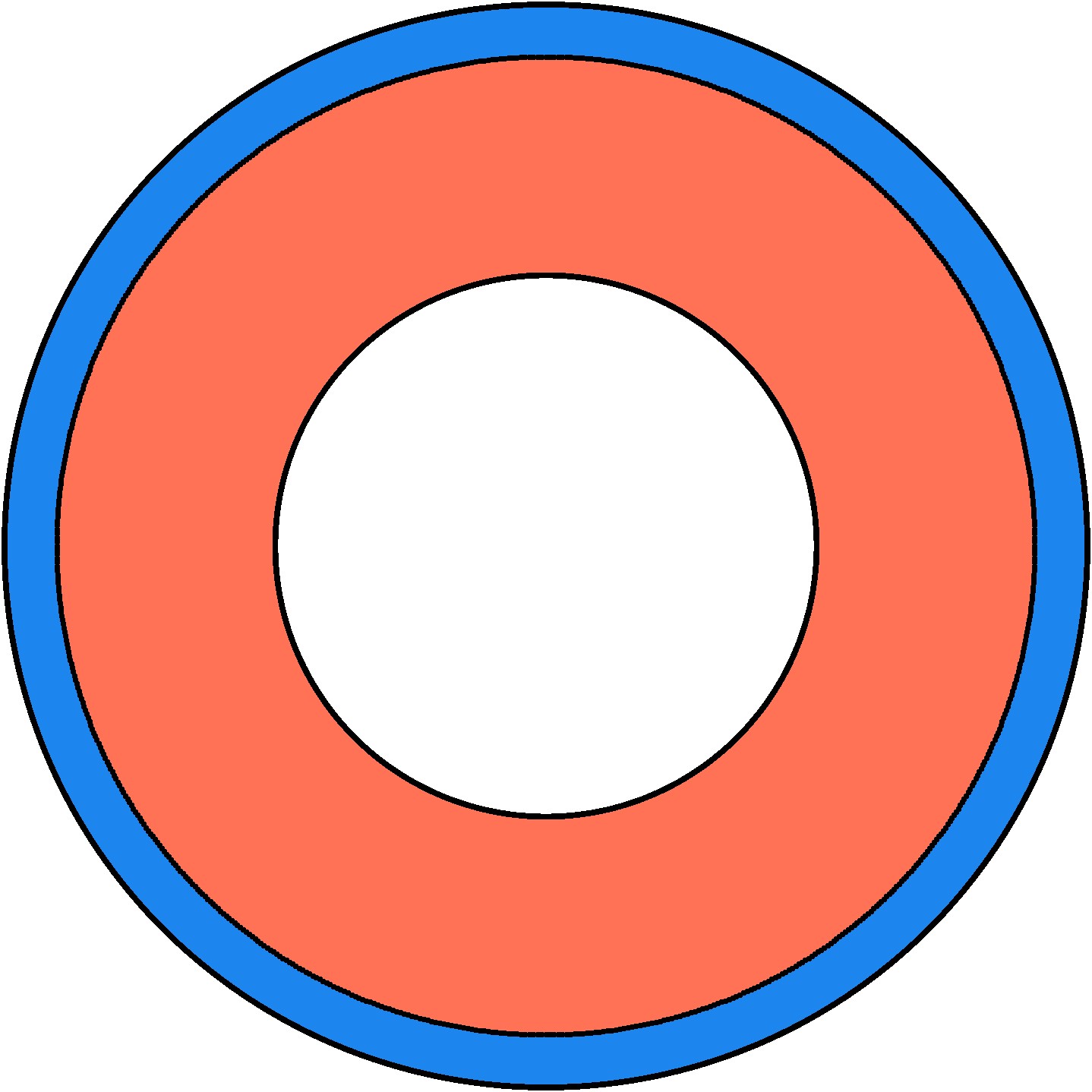}}
        \caption{\centering Optimized topology, $J=1.6099\times 10^4$}
         \label{fig:annular ring optimized geo t}
    \end{subfigure}&
    \begin{subfigure}[t]{0.15\textwidth}{\centering\includegraphics[width=1\textwidth]{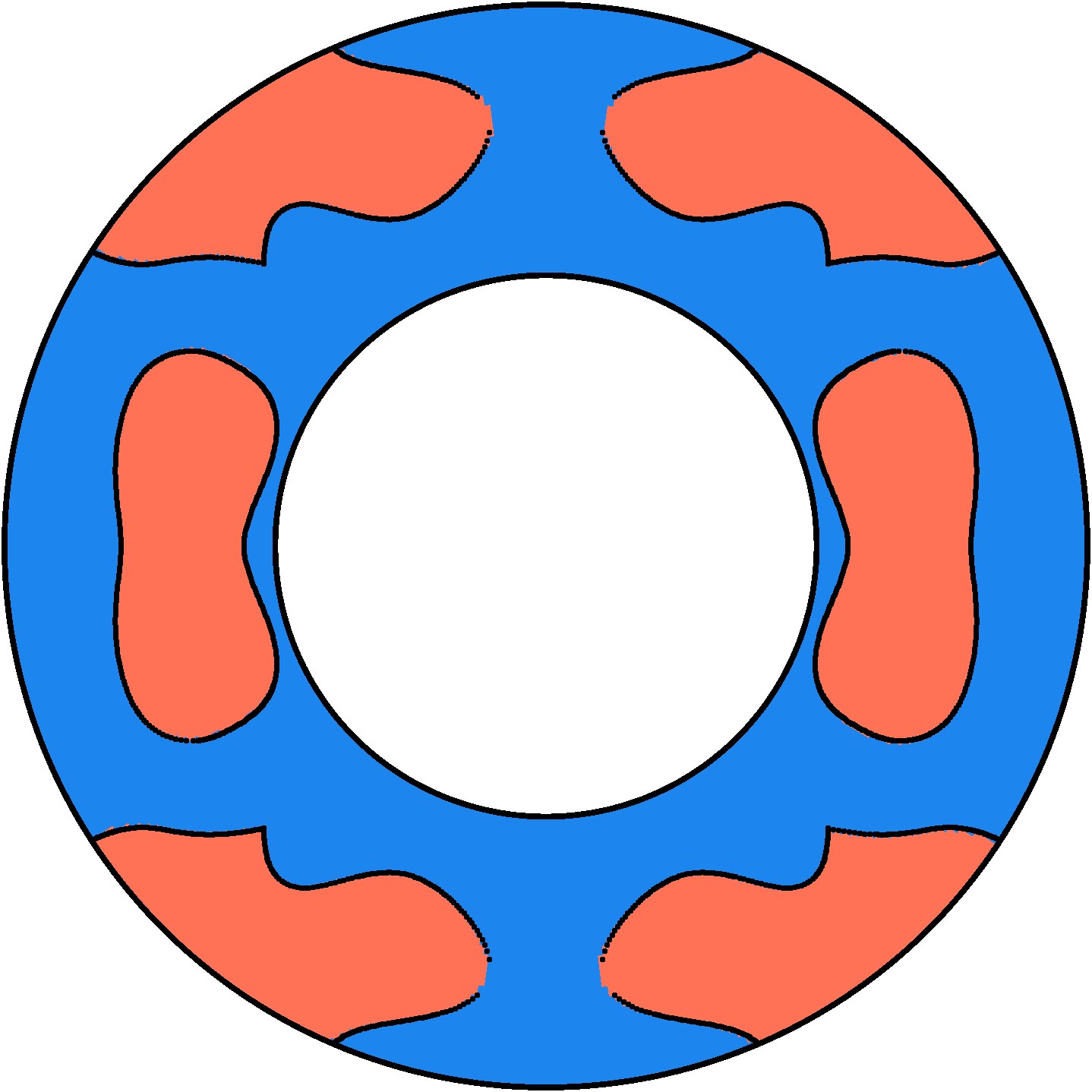}}
        \caption{\centering Initial topology}
         \label{fig:annular ring optimized geo u}
    \end{subfigure}&
    \begin{subfigure}[t]{0.15\textwidth}{\centering\includegraphics[width=1\textwidth]{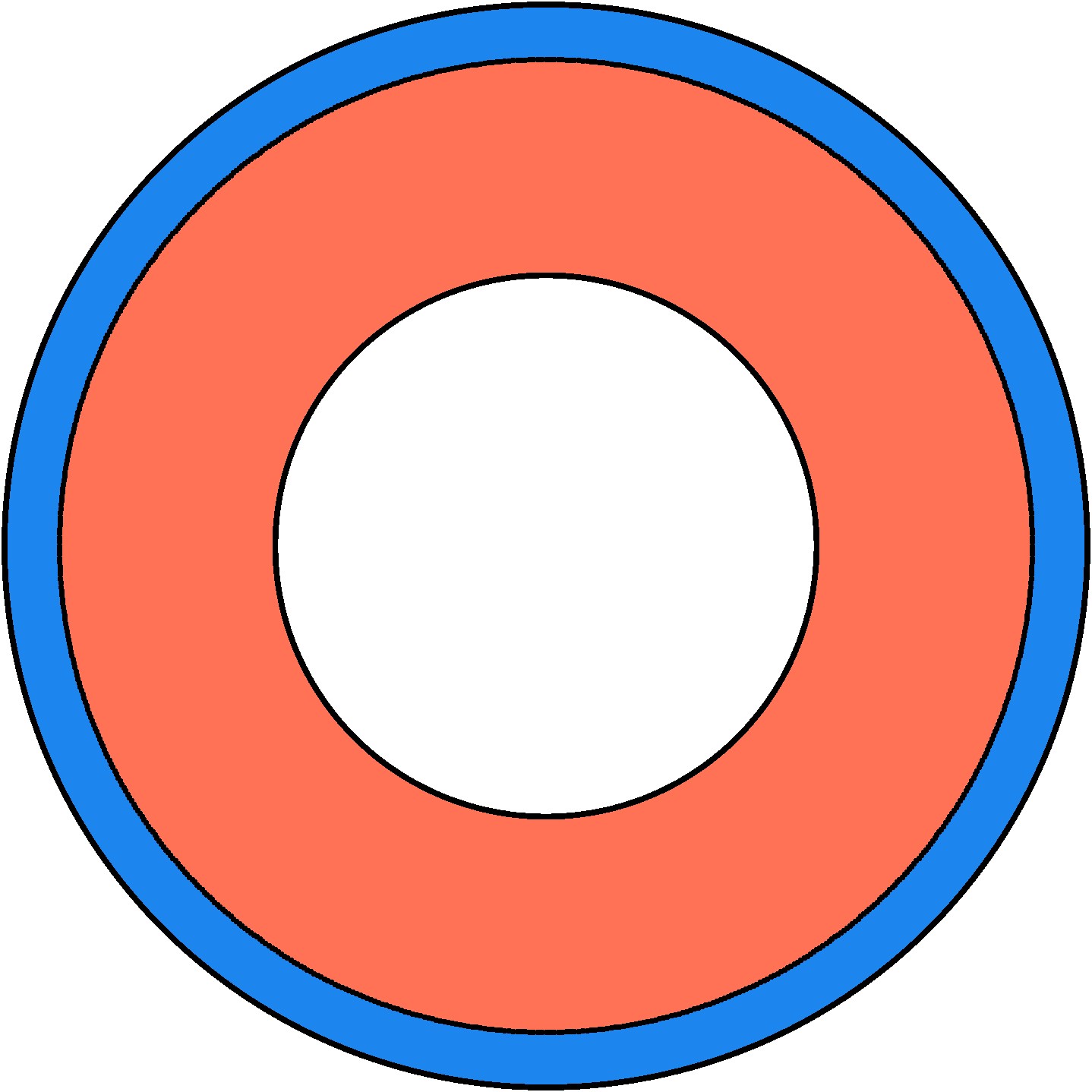}}
        \caption{\centering Optimized topology, $J=1.6099\times 10^4$}
         \label{fig:annular ring optimized geo v}
    \end{subfigure}&
    \begin{subfigure}[t]{0.15\textwidth}{\centering\includegraphics[width=1\textwidth]{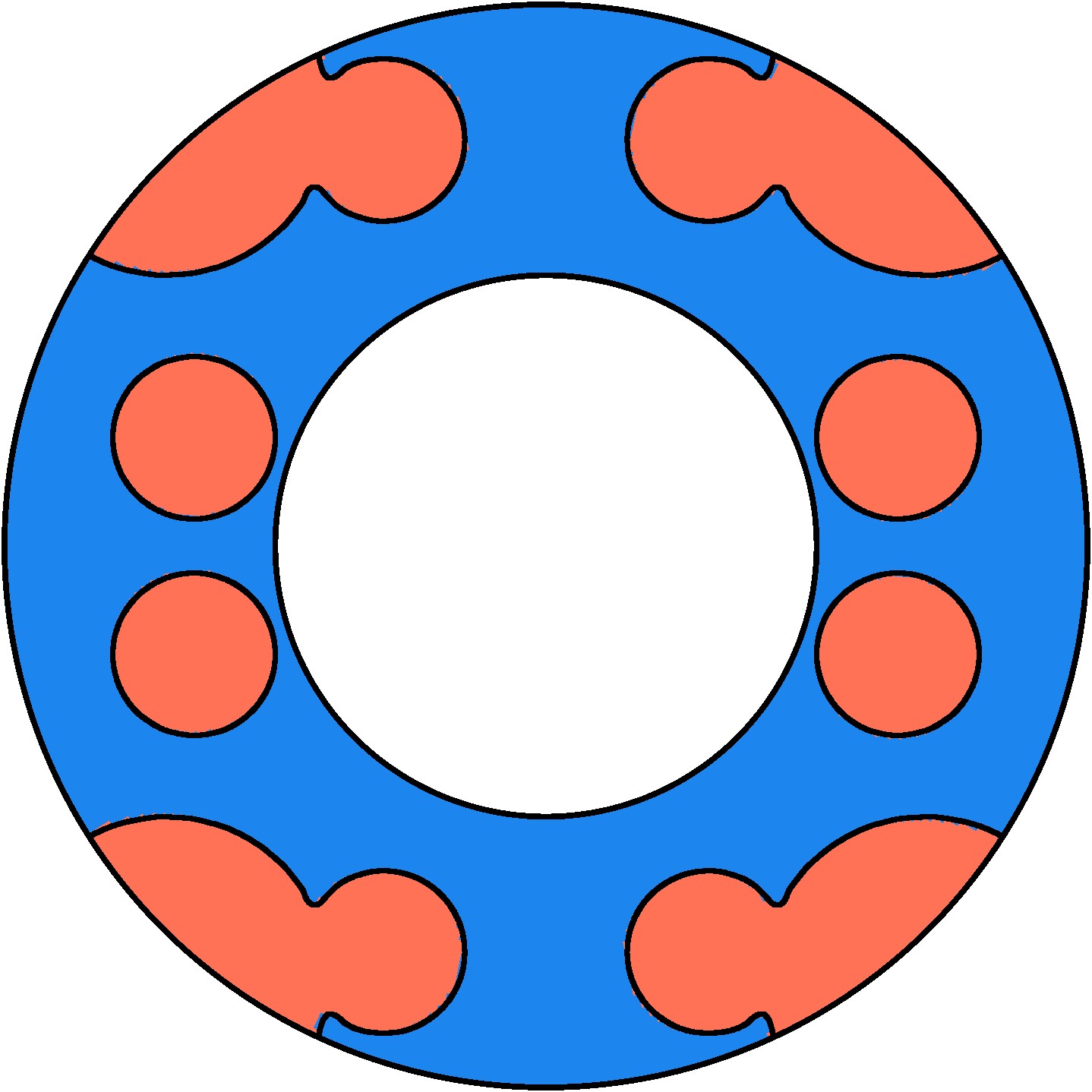}}
        \caption{\centering Initial topology}
         \label{fig:annular ring optimized geo w} 
    \end{subfigure}&
    \begin{subfigure}[t]{0.15\textwidth}{\centering\includegraphics[width=1\textwidth]{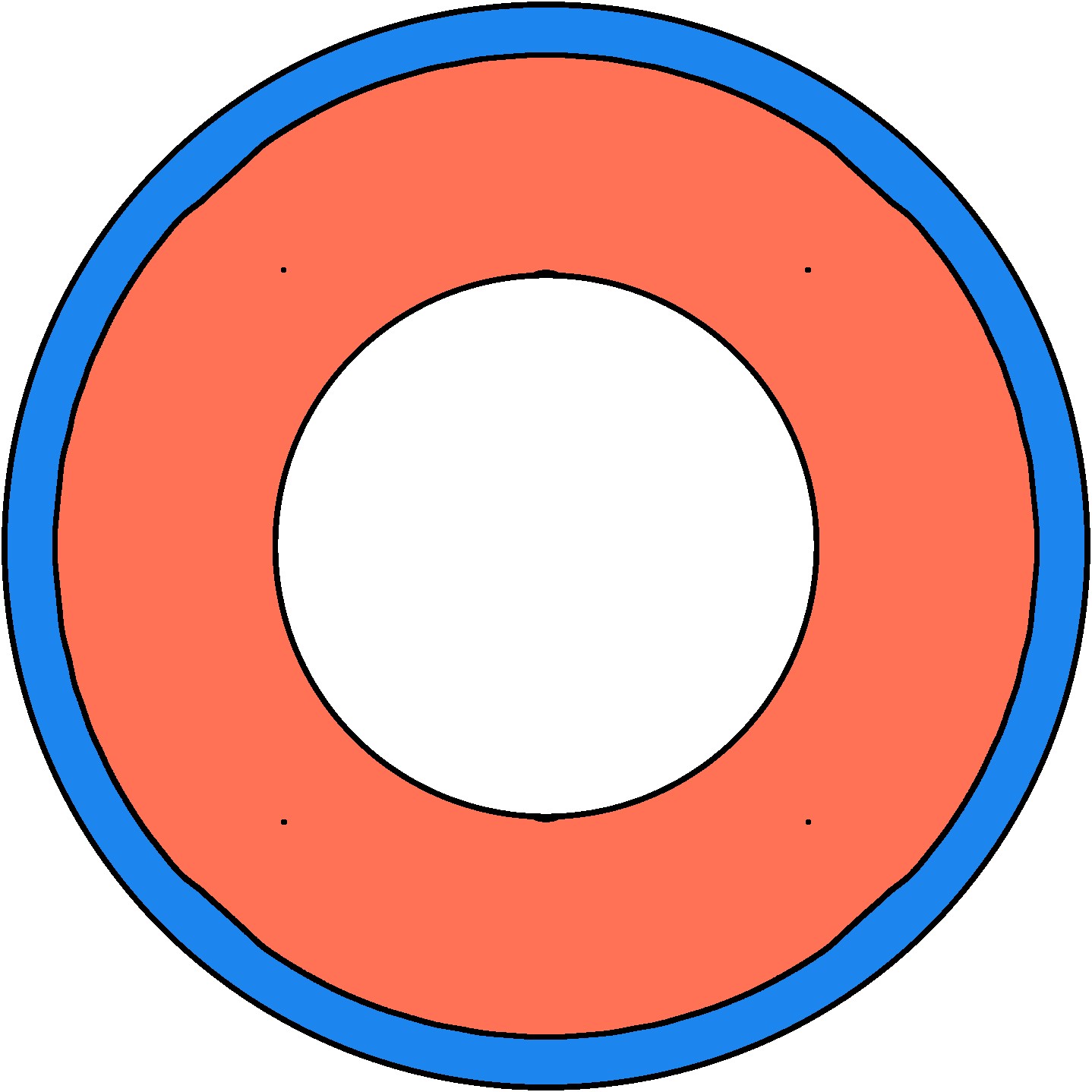}}
        \caption{\centering Optimized topology, $J=1.6099\times 10^4$}
        \label{figure:star_conv_plot_diff_kk x}\vspace{1mm} 
    \end{subfigure}\\ \hline
    
\end{tabular}

}
\caption{For the annular ring problem, initial and optimized topologies for three values of $N_{\rm var}=25, 42$ and $1089$ with $\Delta=0.05$. Four initial topologies (samples I, II, III, and IV) are discretized with the corresponding design basis. All optimized topologies are close to the optimal topology with a circular interface at $R_{L}\approx1.80612$ with the objective function $J=1.6098\times10^4$.}  
    \label{fig:annular ring optimized geo}
\end{figure}

 \par For optimization, we take the expansion coefficients as design variables instead of the interface radius.  \fref{fig:annular ring optimized geo} shows the initial and optimized topologies for three values of $N_{\rm var}$, $N_{\rm var}=25, 42$ and $1089$. The corresponding solution meshes have $4389$, $4658$, and $4389$ DOF, respectively. For $N_{\rm var}=25, 1089$, we use $p=2$ and $q=1$, while for $N_{\rm var}=42$, we use $p=3$ and $q=2$. We choose $\Delta=0.05$ to
trade off the accuracy, stability, and computational cost. We also consider four different LSFs (samples I, II, III, and IV) to define the initial topologies. For a particular LSF, the initial topology for each value of $N_{\rm var}$ can be slightly different due to their difference in parameterization. From the figure, we observe that each case generates a topology close to the optimal topology from the analytical solution regardless of the initial topology, which verifies the efficiency of the proposed method. The values of the objective function are also within $1\%$ variation of optimal value. It is interesting to observe that in all cases optimization converged to the minimum corresponding to the global minimum of $J$ in the design space with the unique parameter $R_L$. In the most general space, we only claim that it is a local minimum.

\subsection{Thermal cloak problem}
\label{sec:Chen2015case_cloak}
\subsubsection{Problem description}
\label{sec:Chen2015cloak Problem description}

 \begin{figure}[!htbp]
    \centering
    \setlength\figureheight{1\textwidth}
    \setlength\figurewidth{1\textwidth}
    \begin{subfigure}[b]{0.31\textwidth}{\centering\includegraphics[width=1\textwidth]{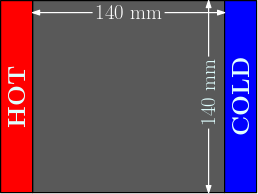}}
        \caption{A base material plate under constant heat flux.}
        \label{fig:Cloak problem schematics a}
    \end{subfigure}\quad
     \begin{subfigure}[b]{0.31\textwidth}{\centering\includegraphics[width=1\textwidth]{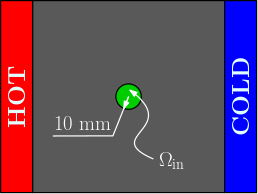}}
        \caption{An obstacle embedded in the plate.}
        \label{fig:Cloak problem schematics b}
    \end{subfigure}\quad
    \begin{subfigure}[b]{0.31\textwidth}{\centering\includegraphics[width=1\textwidth]{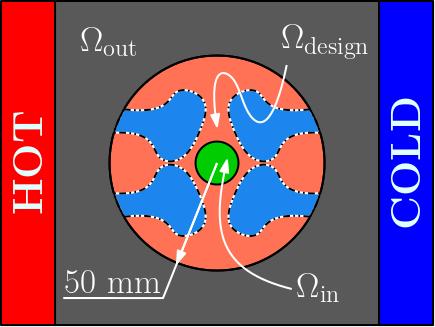}}
             \caption{A metamaterial-based cloak in the base material plate.}
             \label{fig:Cloak problem schematics c}
    \end{subfigure}
 \caption{Schematic design of (a) a base material (aluminium alloy) plate  under constant heat flux applied by the high-temperature source on the left side and low-temperature sink on the right side; (b) a circular insulated obstacle embedded in the base material plate; ($\mathrm{\Omega}_{\mathrm{in}}$ is the obstacle) (c) the obstacle and a surrounding metamaterial-based thermal cloak embedded in the base material plate; $\mathrm{\Omega}_{\mathrm{design}}$ is the domain of the cloak where the topology (the distribution of two materials, denoted by pink and blue colors) is optimized, $\mathrm{\Omega}_{\mathrm{out}}$ is the outside domain of remaining base material, where the temperature disturbance is sought to be reduced. $\mathrm{\Omega} = \mathrm{\Omega}_{\mathrm{in}} \cup \mathrm{\Omega}_{\mathrm{design}}\cup \mathrm{\Omega}_{\mathrm{out}}$.}
 \label{fig:Cloak problem schematics}
\end{figure}
\par In this example, a thermal cloak is optimized. The objective of a thermal cloak is to reduce the temperature disturbance created by an obstacle and produce the temperature distribution as if there were no obstacles. The geometry is referred from~\cite{Chen2015}, however, our focus is on a thermal cloak instead of a thermal concentrator as in~\cite{Chen2015}. We consider a square base material-aluminum alloy (6063) plate ($\kappa_{\rm Al}=200$~W/mK) with a side length of 140 mm. The plate is under constant temperature difference between the left side (at $300$ K) and the right side (at $200$ K) as shown in \fref{fig:Cloak problem schematics a}. The thermally insulated boundary condition, $\nabla T\cdot \boldsymbol{n}=0$, is applied on the remaining two sides. Now, the plate is embedded with a circular obstacle, which is a thermal insulator with very low thermal conductivity, $\kappa_{\rm insulator} = 0.0001$~W/mK (see \fref{fig:Cloak problem schematics b}). Due to the addition of an insulator, the temperature distribution is disturbed. Next, to reduce the temperature disturbance, a thermal cloak made of a metamaterial is added surrounding the obstacle (as shown in \fref{fig:Cloak problem schematics c}). All dimensions related to the given problem are shown in \fref{fig:Cloak problem schematics}. For the cloak, we chose a metamaterial made of copper and polydimethylsiloxane (PDMS), with thermal conductivities $\kappa_{\rm copper}=398$ W/mK and $\kappa_{\rm PDMS}=0.27$ W/mK, respectively. 
\subsubsection{Objective function}
\par The thermal cloak aims to reduce the temperature disturbance in $\mathrm{\Omega}_{\mathrm{out}}$ created by the inner obstacle $\mathrm{\Omega}_{\mathrm{in}}$. Mathematically, the cloak objective function is defined as,
\begin{equation}
    J_{\mathrm{cloak}}=\dfrac{1}{\widetilde{J}_{\mathrm{cloak}}} \int_{\mathrm{\Omega}_{\mathrm{out}}} \vert T - \overline{T} \vert^2~d\mathrm{\Omega},
    \end{equation}
with $\widetilde{J}_{\mathrm{cloak}}$ be the normalization value given as,
\begin{equation}
    \widetilde{J}_{\mathrm{cloak}}= \int_{\mathrm{\Omega}_{\mathrm{out}}} \vert \widetilde{T} - \overline{T} \vert^2~d\mathrm{\Omega},
    \end{equation}
Here, $\overline{T}$ represents the temperature field for the reference case, when the entire domain is filled with the base material, and $\widetilde{T}$  is the temperature field when $\mathrm{\Omega}_{\mathrm{design}}$ is entirely filled with the insulator. 
\par By comparison with \eref{eq:Objective fun definition}, $\Omega_b=\mathrm{\Omega}_{\rm out}$,  $J_b=\dfrac{1}{\widetilde{J}_{\mathrm{cloak}}}~\vert ~T~-~\overline{T}~\vert^2$, and the surface term is absent.

\subsubsection{Results and discussion}
 \begin{figure}[!htbp]
    \centering
    \setlength\figureheight{1\textwidth}
    \setlength\figurewidth{1\textwidth}
    \scalebox{0.8}{\input{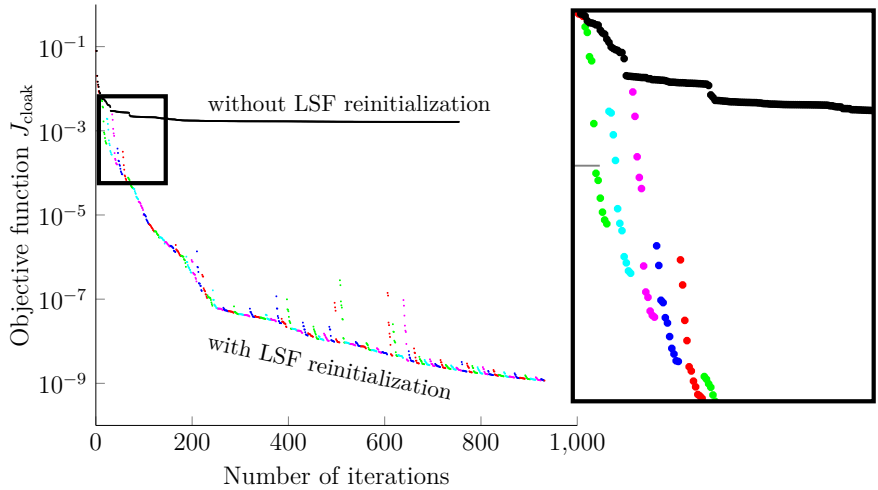}}
 \caption{For the thermal cloak problem, the convergence of the objective function $J_{\rm cloak}$ (with respect to the number of iterations) with and without LSF reinitialization. Black color dots show the case without LSF reinitialization, while the remaining colors represent the case with LSF reinitialization. The change in the color indicates that the LSF has been reinitialized at that iteration.}
 \label{fig:Chen2015cloak convergence}
\end{figure}

\renewcommand{\arraystretch}{1.5}   
\begin{figure}
\centering
\scalebox{0.9}{
\begin{tabular}[c]{|m{1em}|m{5.3em} m{5.3em}|m{5.3em} m{5.3em}| m{5.3em} m{5.3em}|}
\hline		
   & \multicolumn{2}{c|}{$N_{\mathrm{var}}=25$} & \multicolumn{2}{c|}{$N_{\mathrm{var}}=42$} & \multicolumn{2}{c|}{$N_{\mathrm{var}}=1089$} \\
\hline 
\vspace{0.2cm}
   \rotatebox{90}{\centering Sample I}  & \vspace{0.2cm}
    \begin{subfigure}[t]{0.15\textwidth}\begin{center}{
\includegraphics[width=1\textwidth]{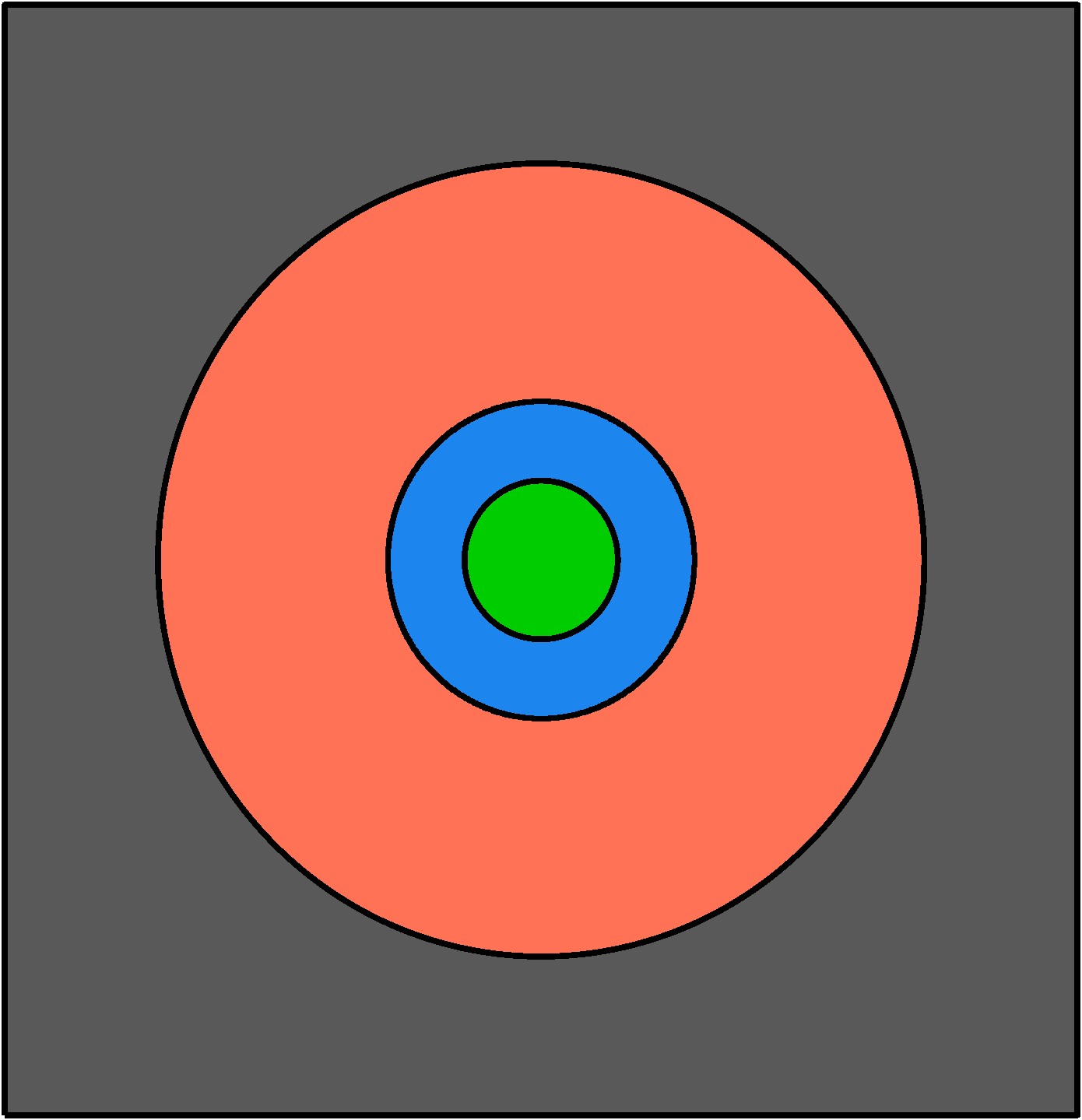}}
        \caption{\centering Initial topology}
        \label{fig:chen2015case optTop a}        
    \end{center}
    \end{subfigure} & \vspace{0.2cm}
    \begin{subfigure}[t]{0.15\textwidth}{\centering\includegraphics[width=1\textwidth]{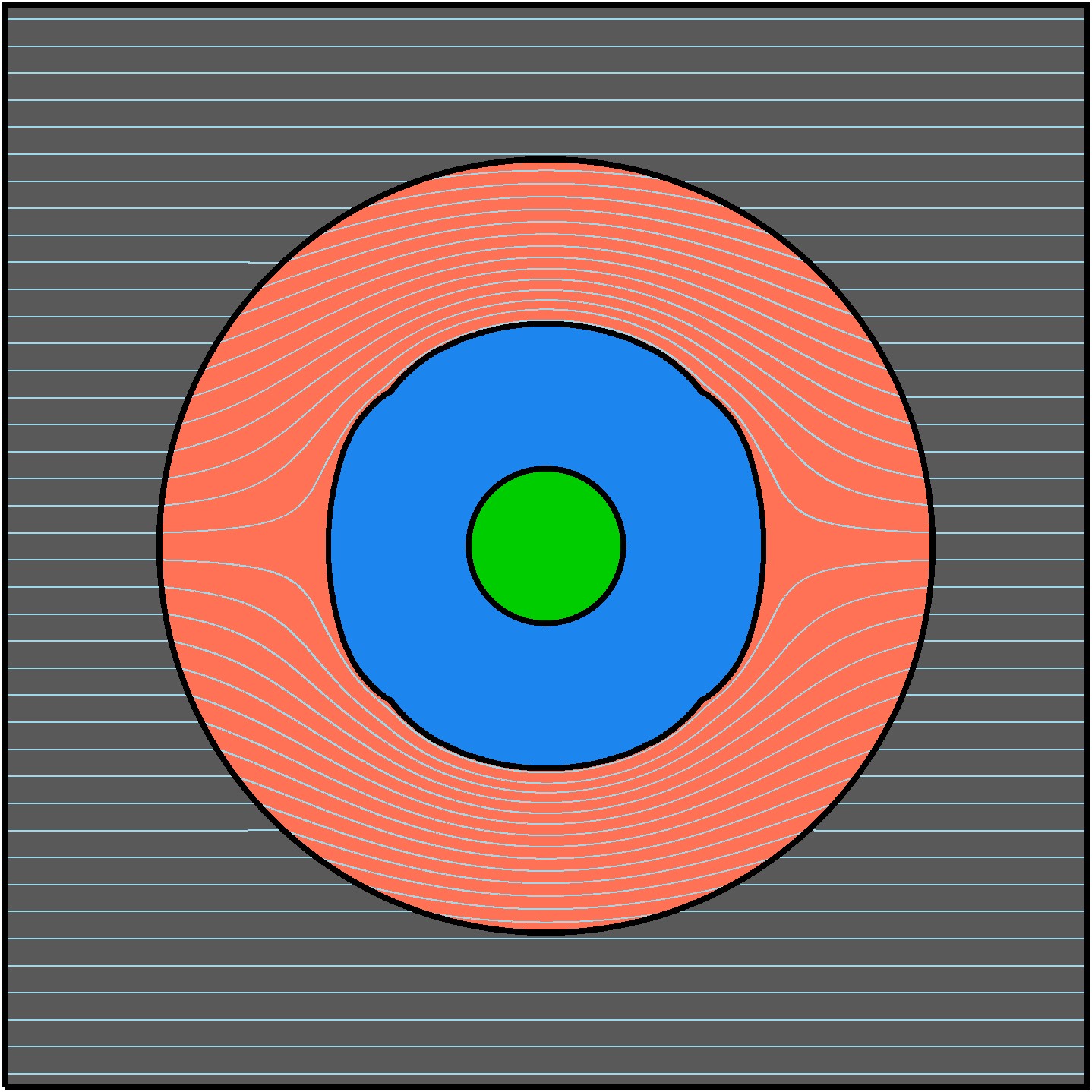}}
        \caption{\centering Optimized topology, $J=9.9321\times 10^{-10}$}
        \label{fig:chen2015case optTop b}
    \end{subfigure}& \vspace{0.2cm}
    \begin{subfigure}[t]{0.15\textwidth}{\centering\includegraphics[width=1\textwidth]{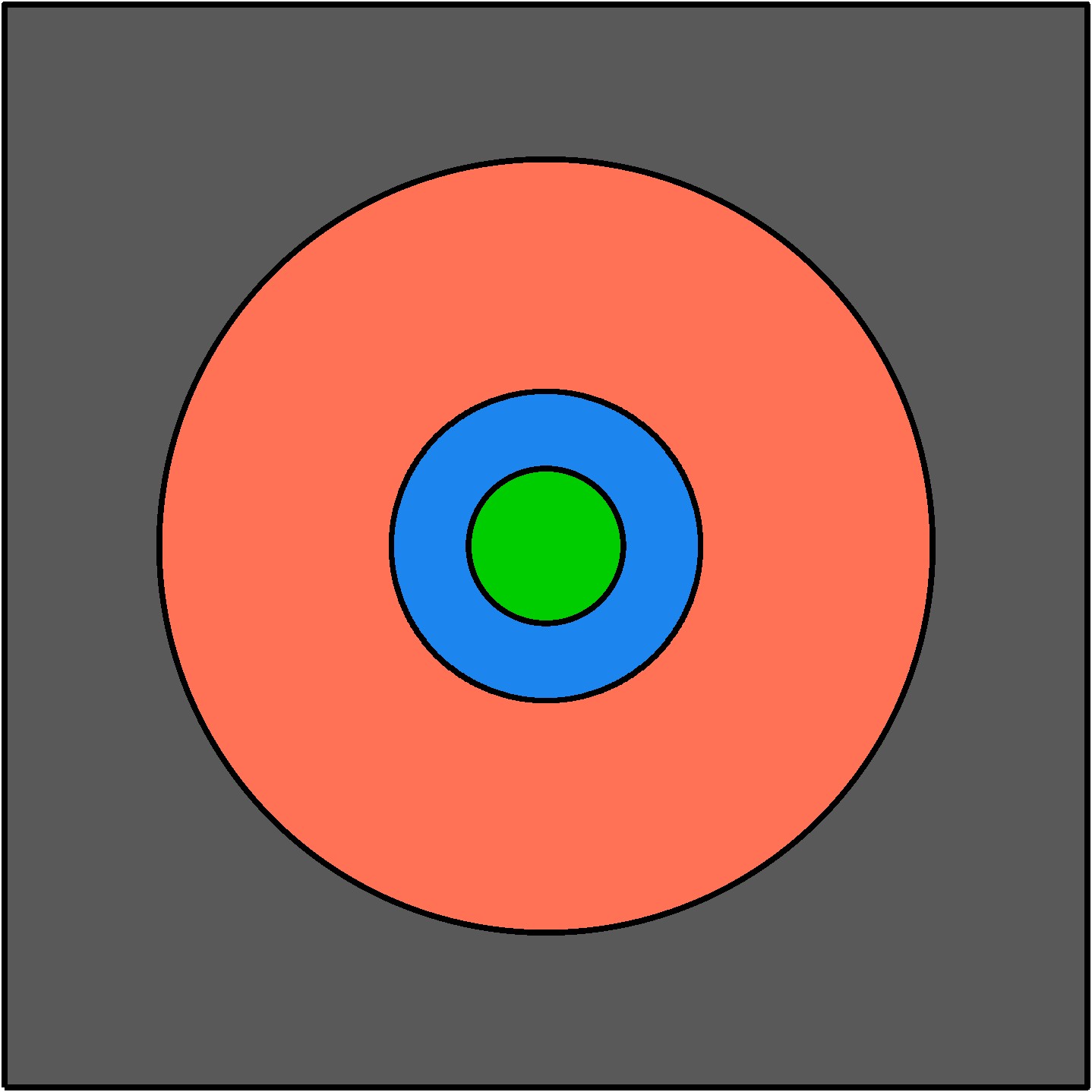}}
        \caption{\centering Initial topology}
        \label{fig:chen2015case optTop c}
    \end{subfigure}& \vspace{0.2cm}
    \begin{subfigure}[t]{0.15\textwidth}{\centering\includegraphics[width=1\textwidth]{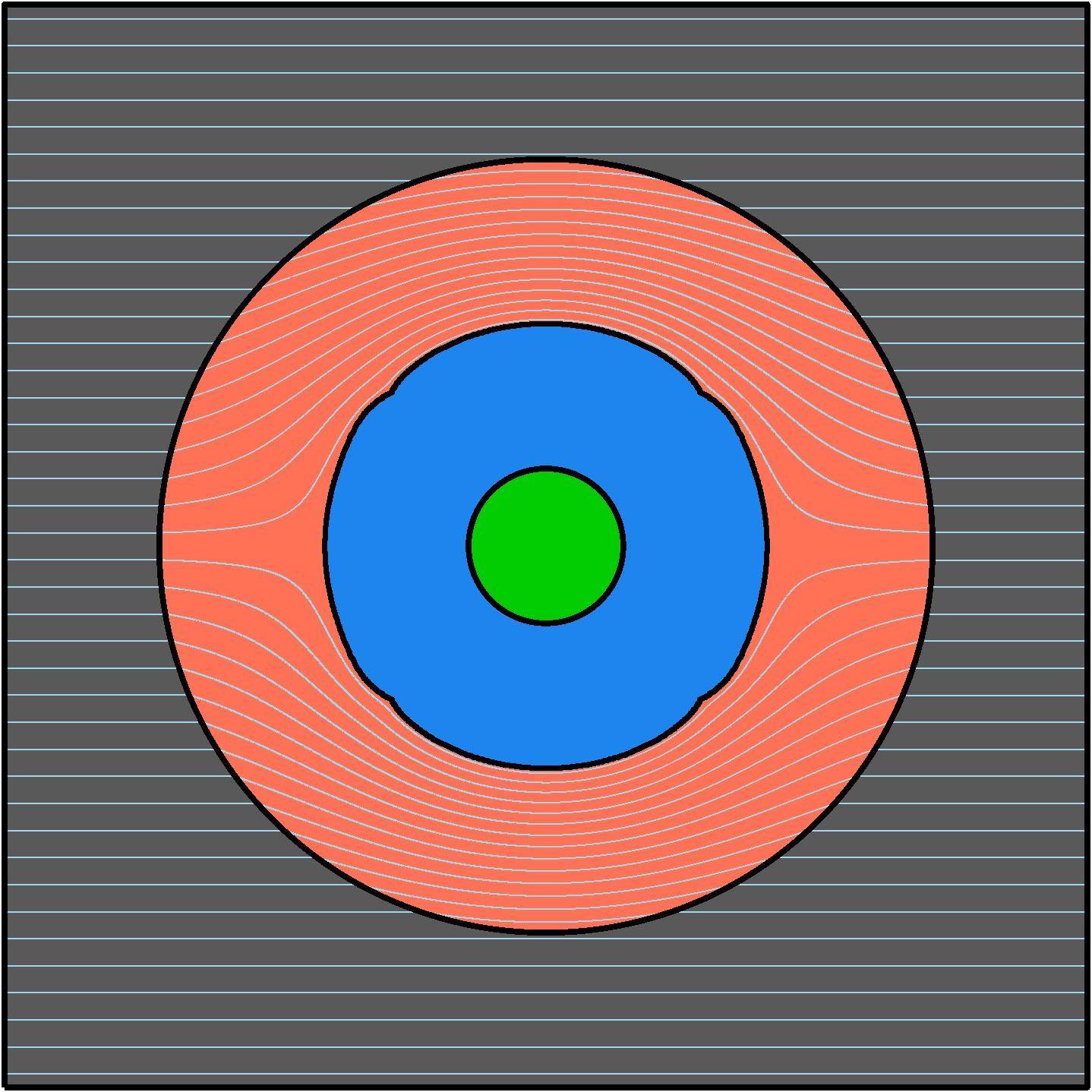}}
        \caption{\centering Optimized topology, $J=9.8838\times 10^{-10}$}
        \label{fig:chen2015case optTop d}
    \end{subfigure}&\vspace{0.2cm}
    \begin{subfigure}[t]{0.15\textwidth}{\centering\includegraphics[width=1\textwidth]{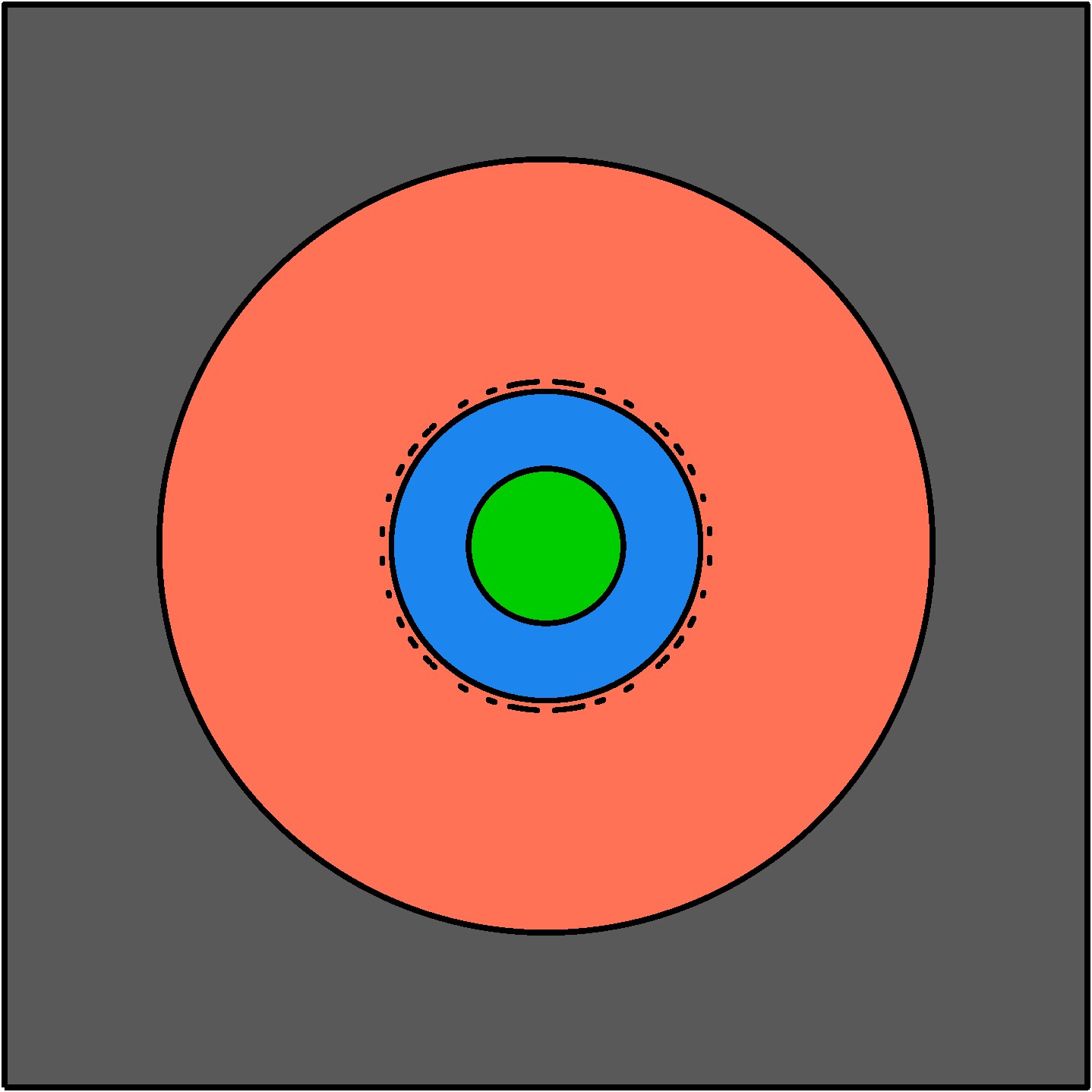}}
        \caption{\centering Initial topology}
        \label{fig:chen2015case optTop e}
    \end{subfigure}& \vspace{0.2cm}
    \begin{subfigure}[t]{0.15\textwidth}{\centering\includegraphics[width=1\textwidth]{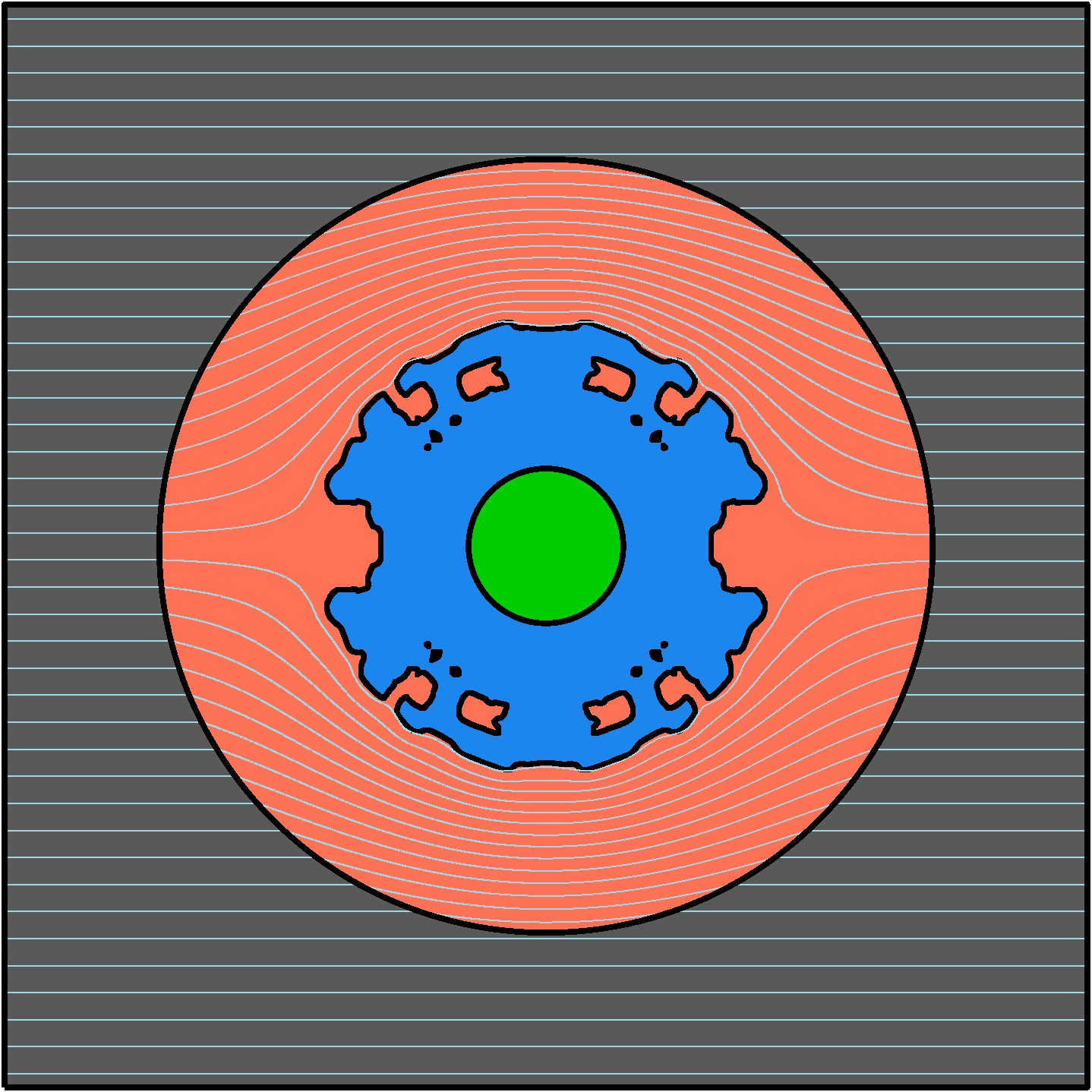}}
        \caption{\centering Optimized topology, $J=9.4025\times 10^{-10}$}
        \label{fig:chen2015case optTop f}
    \end{subfigure}\\
\rotatebox{90}{\centering Sample II}  &\begin{subfigure}[t]{0.15\textwidth}{\centering\includegraphics[width=1\textwidth]{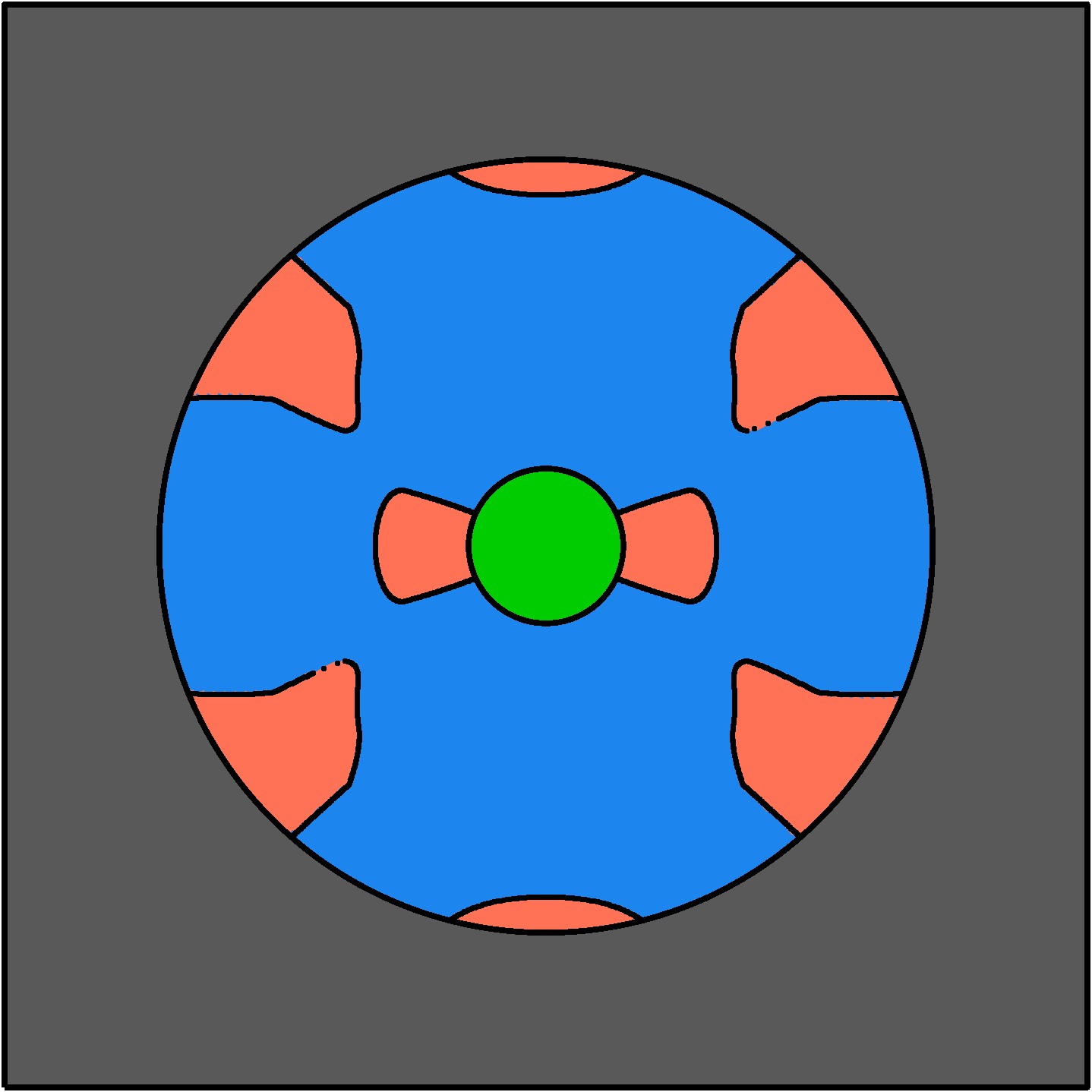}}
        \caption{\centering Initial topology}
        \label{fig:chen2015case optTop g}
    \end{subfigure}&
    \begin{subfigure}[t]{0.15\textwidth}{\centering\includegraphics[width=1\textwidth]{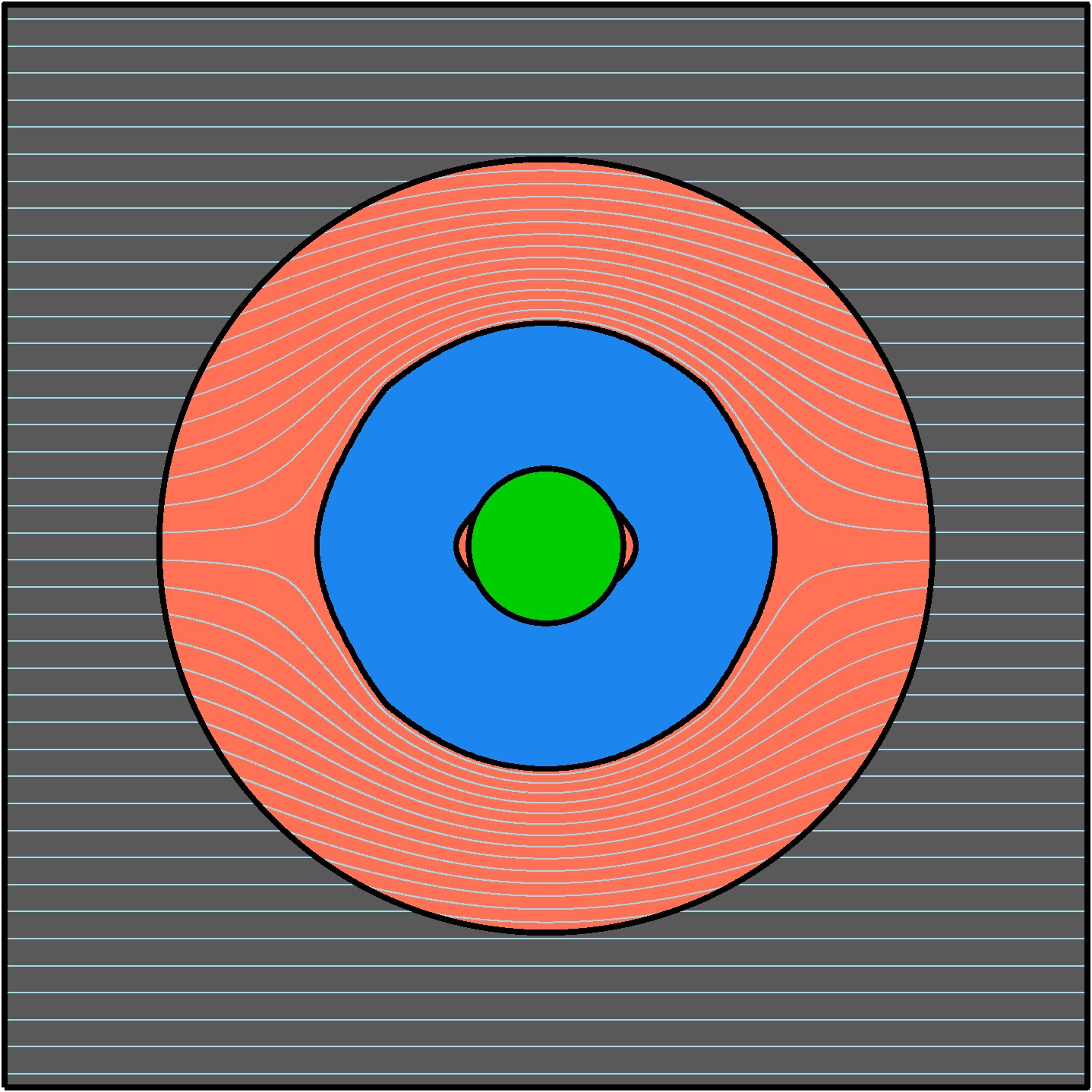}}
        \caption{\centering Optimized topology, $J=8.3342\times 10^{-10}$}
        \label{fig:chen2015case optTop h}
    \end{subfigure}&
    \begin{subfigure}[t]{0.15\textwidth}{\centering\includegraphics[width=1\textwidth]{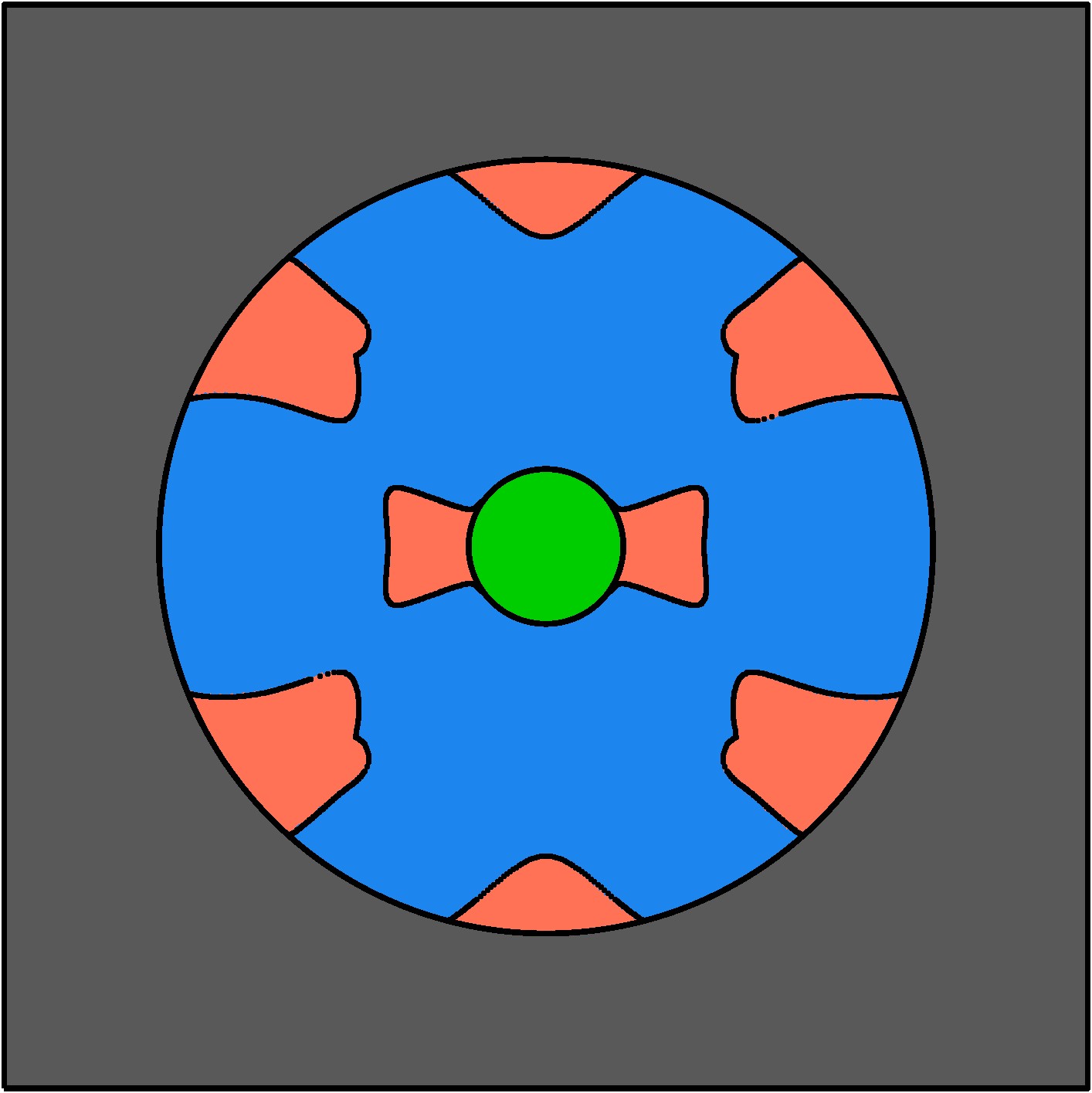}}
        \caption{\centering Initial topology}
        \label{fig:chen2015case optTop i}
    \end{subfigure}&
    \begin{subfigure}[t]{0.15\textwidth}{\centering\includegraphics[width=1\textwidth]{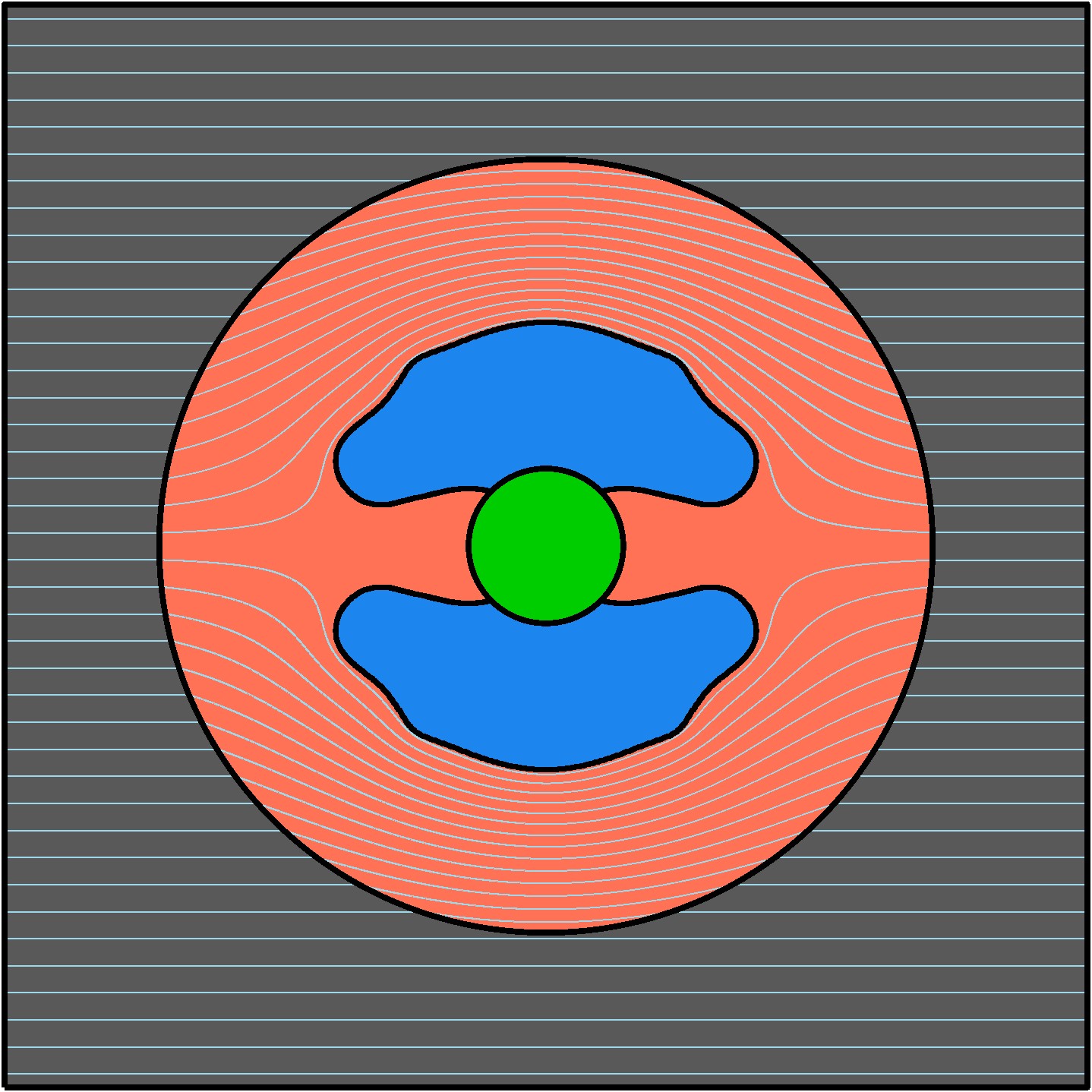}}
        \caption{\centering Optimized topology, $J=1.0162\times 10^{-9}$}
        \label{fig:chen2015case optTop j}
    \end{subfigure}&
    \begin{subfigure}[t]{0.15\textwidth}{\centering\includegraphics[width=1\textwidth]{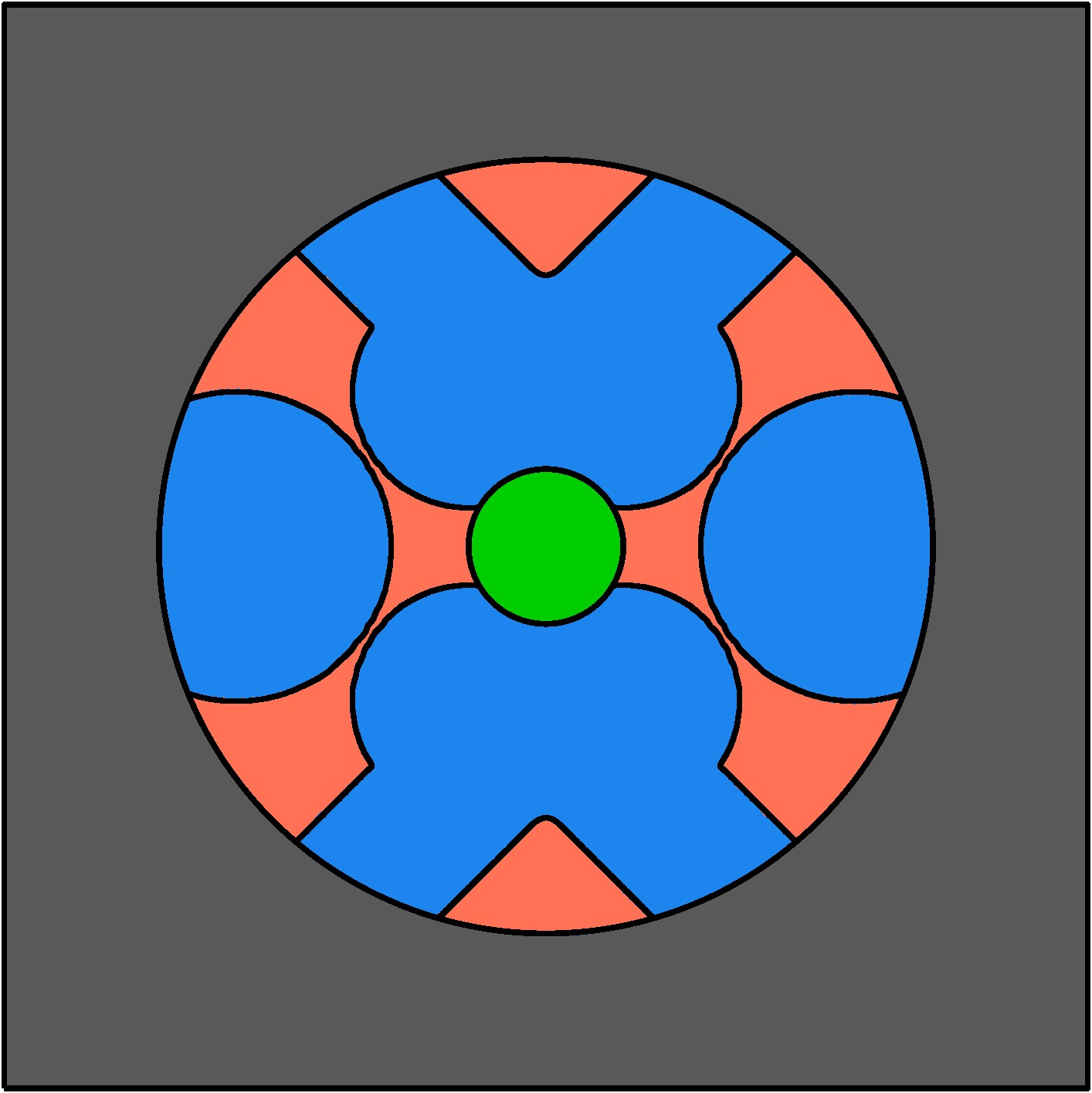}}
        \caption{\centering Initial topology}
        \label{fig:chen2015case optTop k}
    \end{subfigure}&
    \begin{subfigure}[t]{0.15\textwidth}{\centering\includegraphics[width=1\textwidth]{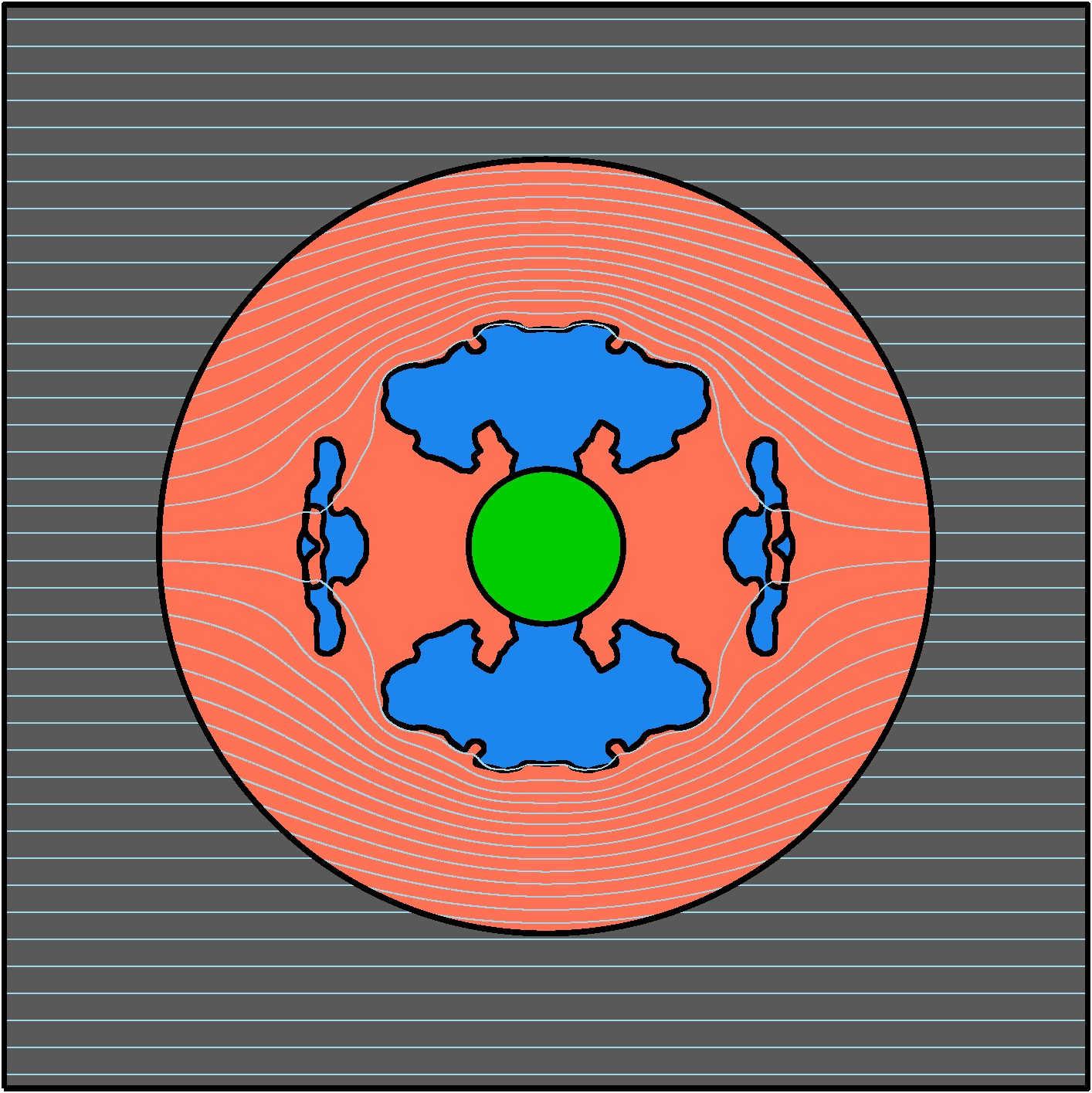}}
        \caption{\centering Optimized topology, $J=1.6808\times 10^{-9}$}
        \label{fig:chen2015case optTop l}
    \end{subfigure}\\
    \rotatebox{90}{\centering Sample III} &   \begin{subfigure}[t]{0.15\textwidth}{\centering\includegraphics[width=1\textwidth]{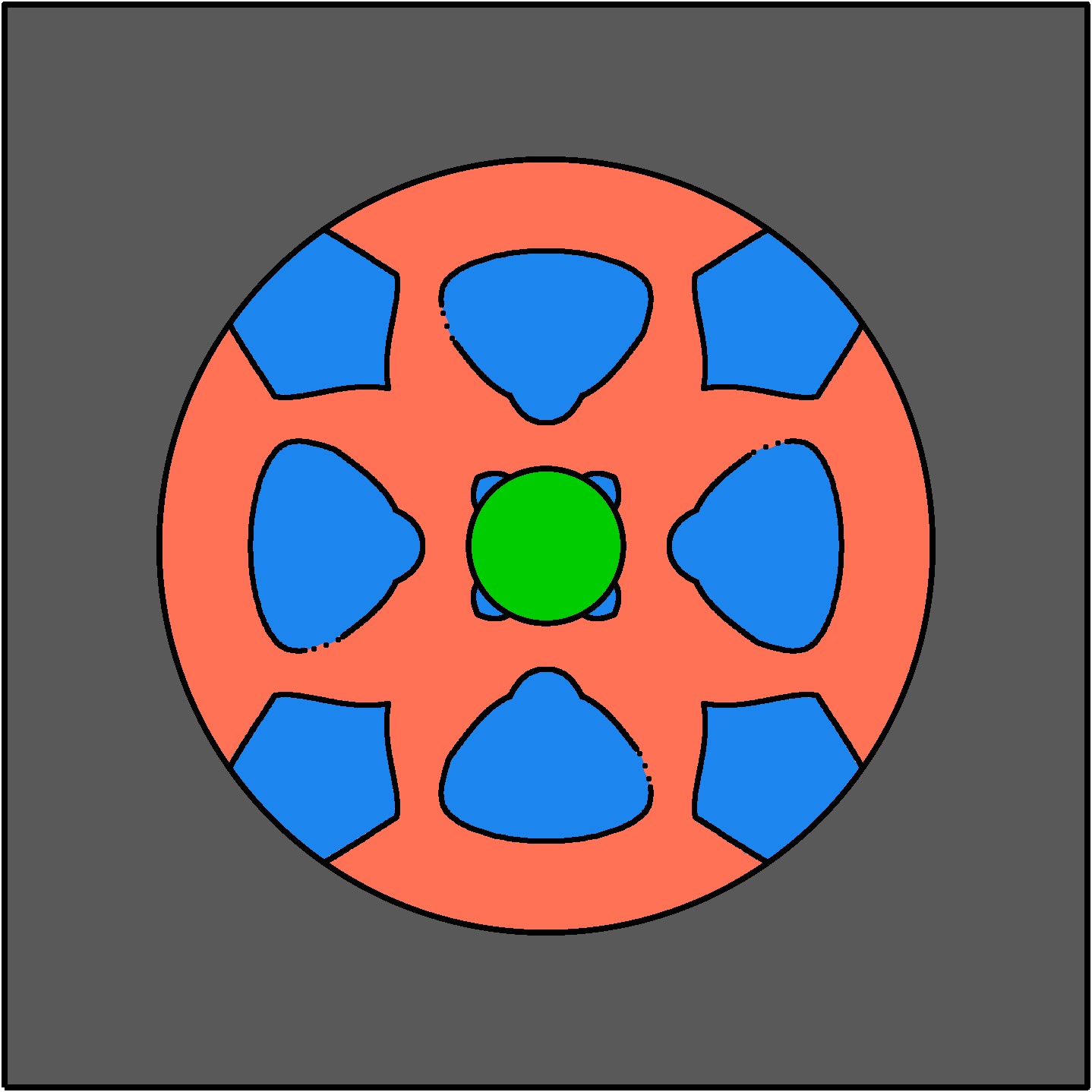}}
        \caption{\centering Initial topology}
        \label{fig:chen2015case optTop m}
    \end{subfigure}&
    \begin{subfigure}[t]{0.15\textwidth}{\centering\includegraphics[width=1\textwidth]{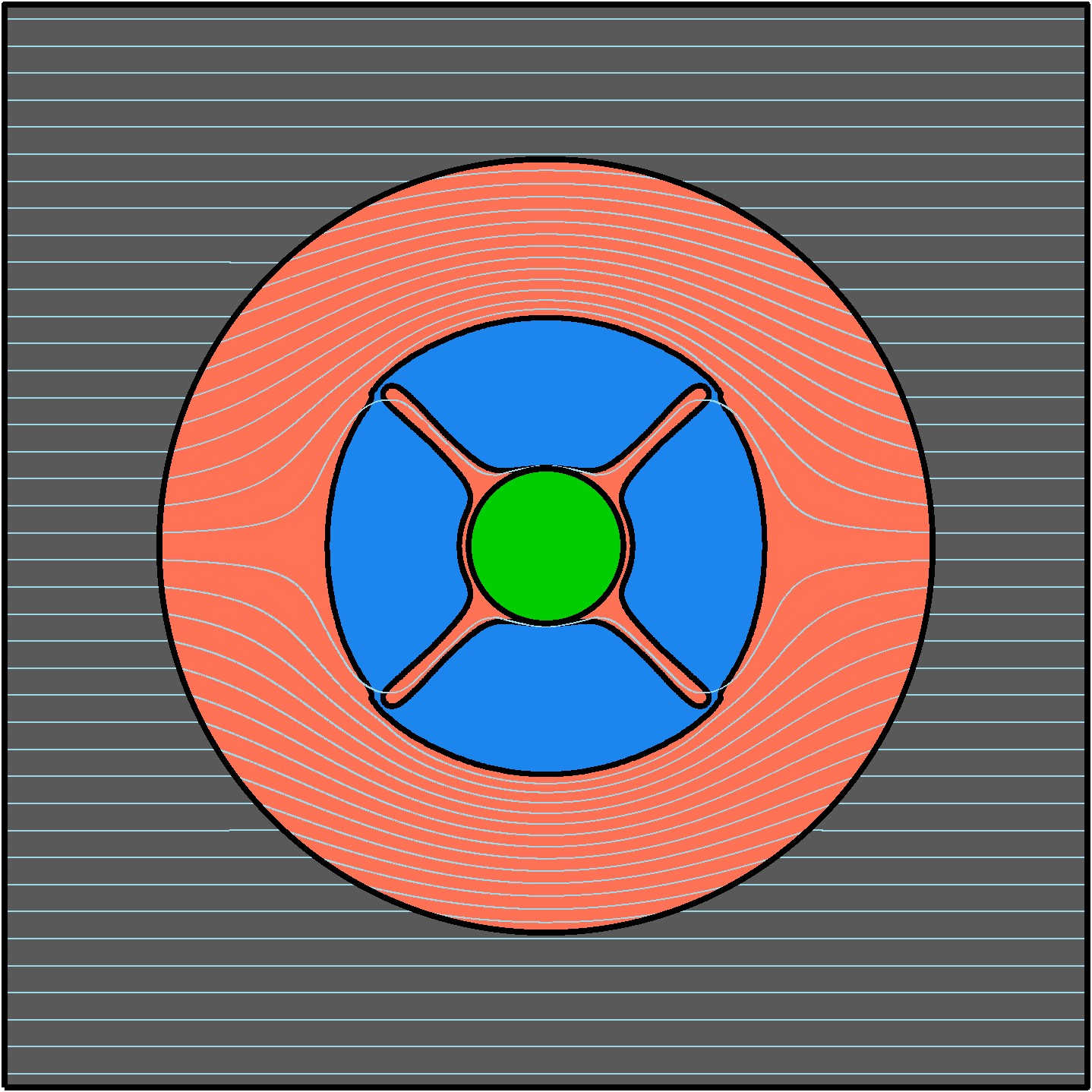}}
        \caption{\centering Optimized topology, $J=8.9454\times 10^{-10}$}
        \label{fig:chen2015case optTop n}
    \end{subfigure}&
    \begin{subfigure}[t]{0.15\textwidth}{\centering\includegraphics[width=1\textwidth]{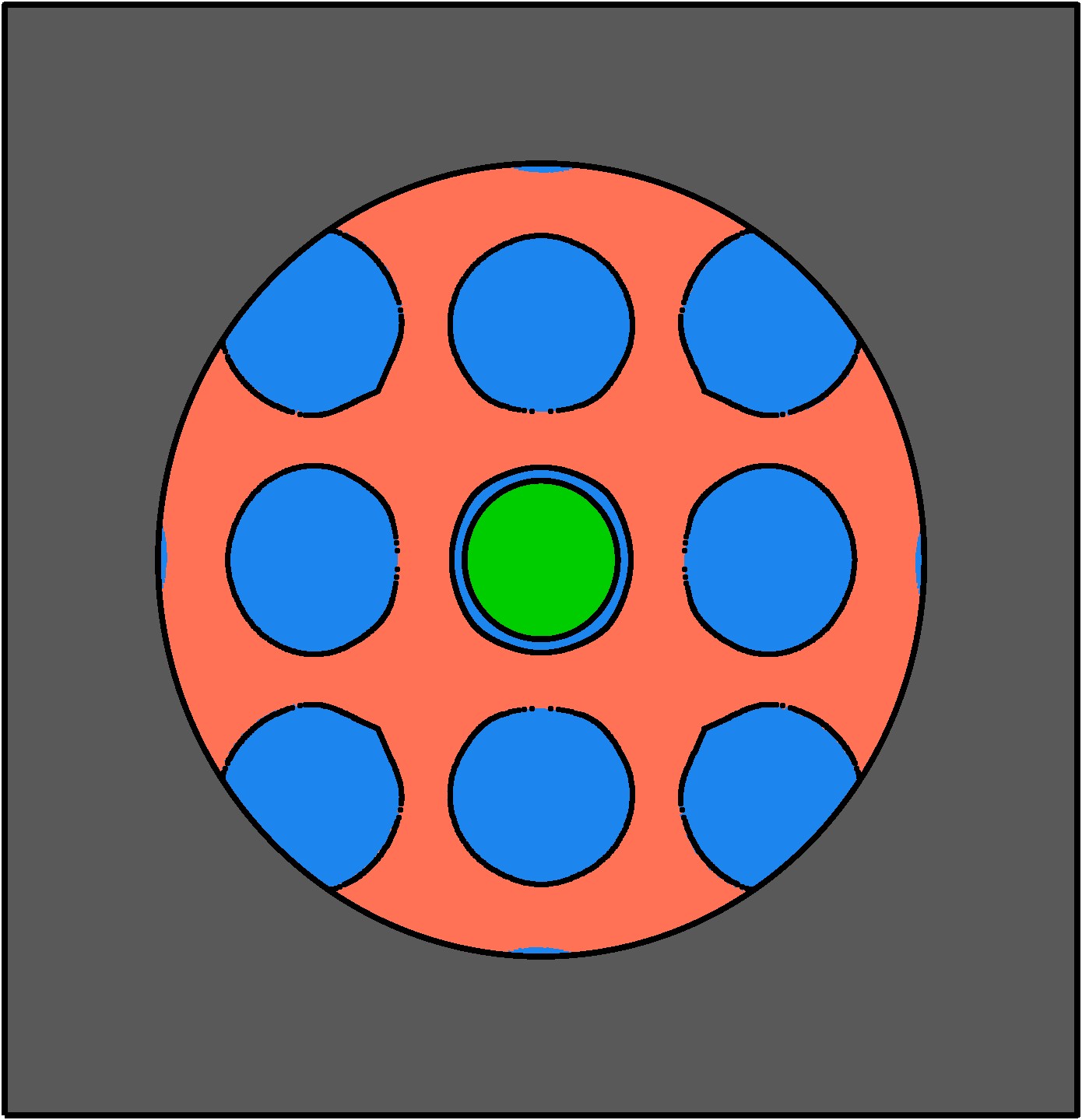}}
        \caption{\centering Initial topology}
        \label{fig:chen2015case optTop o}
    \end{subfigure}&
    \begin{subfigure}[t]{0.15\textwidth}{\centering\includegraphics[width=1\textwidth]{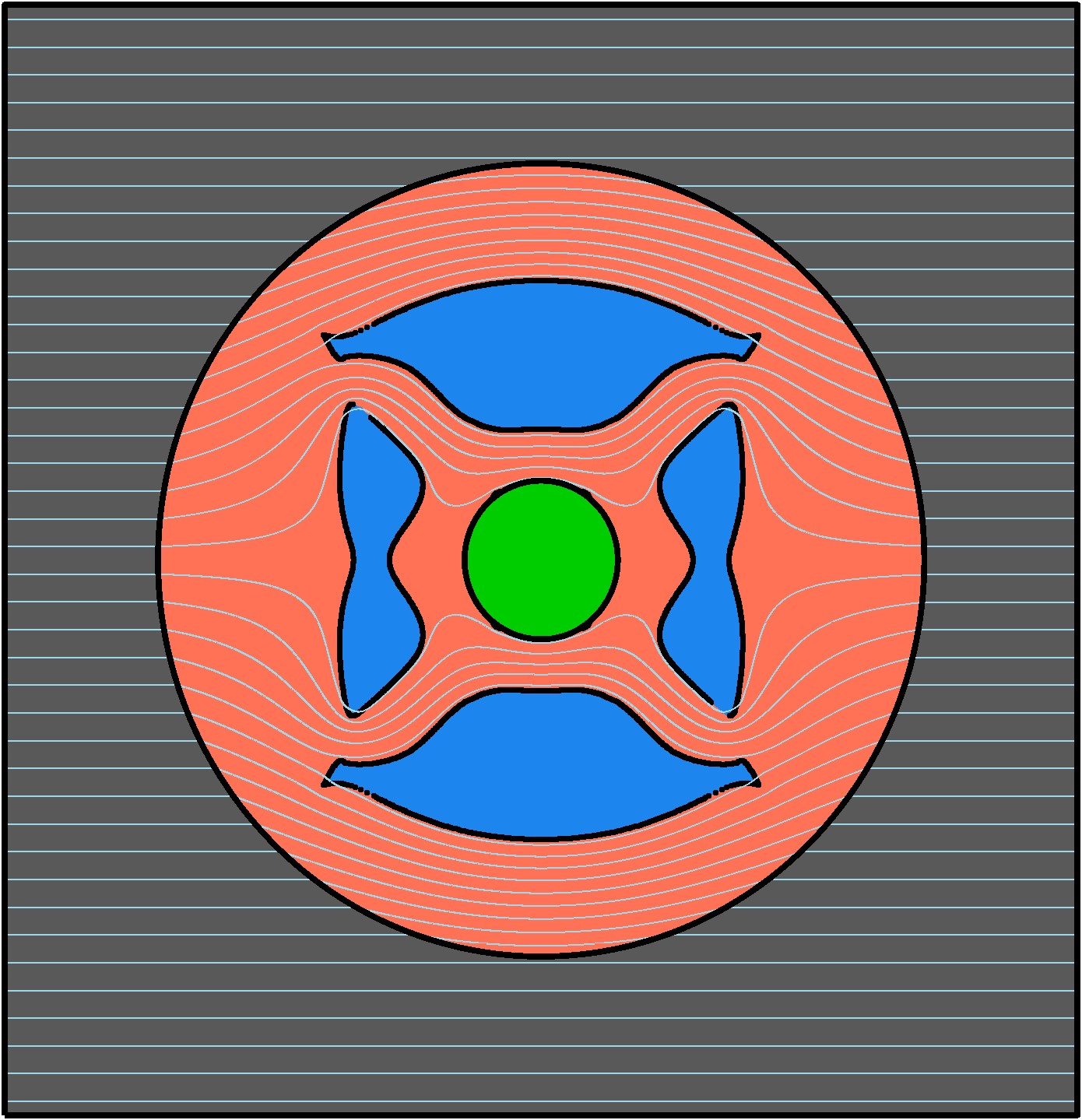}}
        \caption{\centering Optimized topology, $J=1.1647\times 10^{-9}$}
        \label{fig:chen2015case optTop p}
    \end{subfigure}&
    \begin{subfigure}[t]{0.15\textwidth}{\centering\includegraphics[width=1\textwidth]{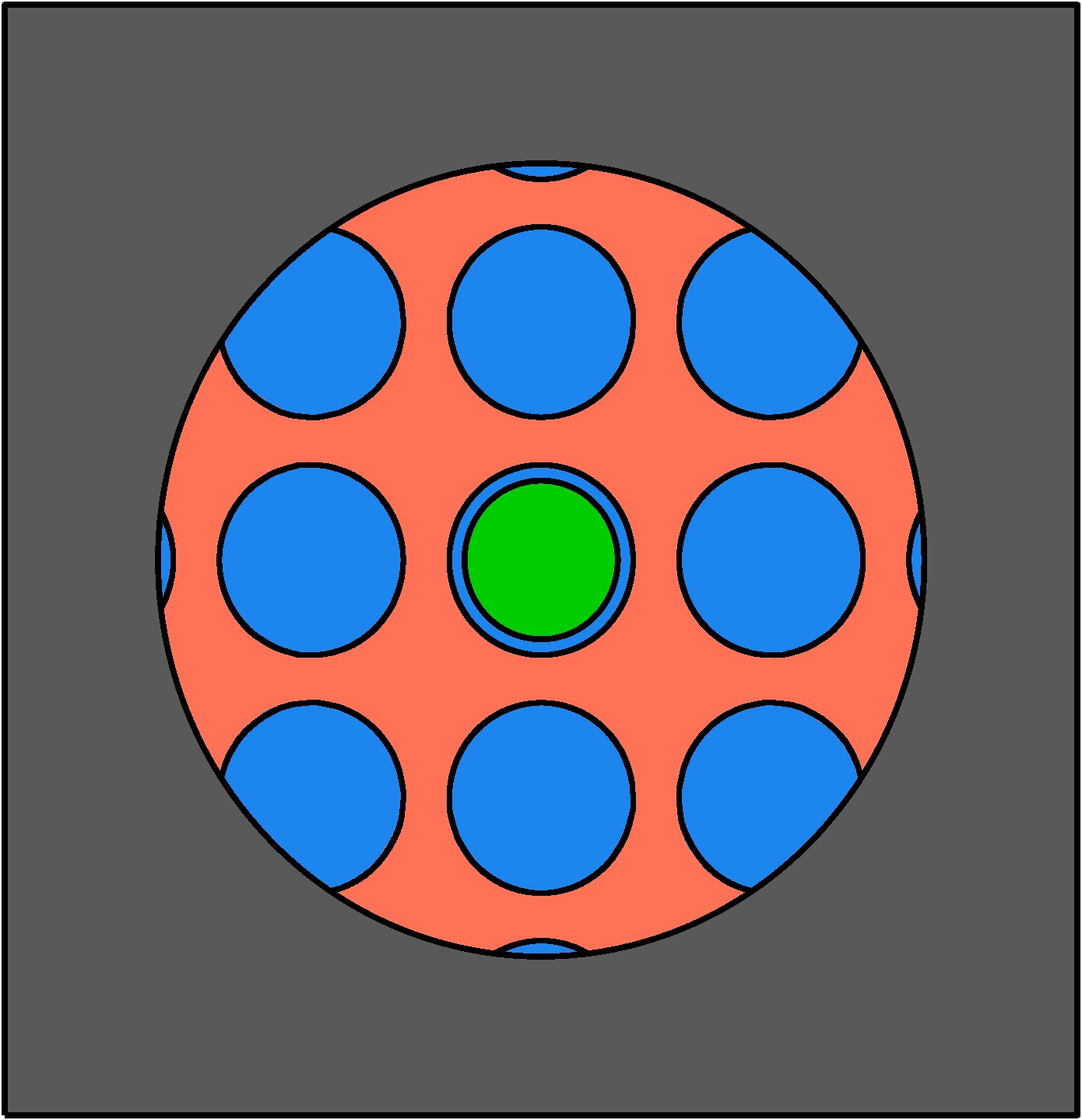}}
        \caption{\centering Initial topology}
        \label{fig:chen2015case optTop q}
    \end{subfigure} &
    \begin{subfigure}[t]{0.15\textwidth}{\centering\includegraphics[width=1\textwidth]{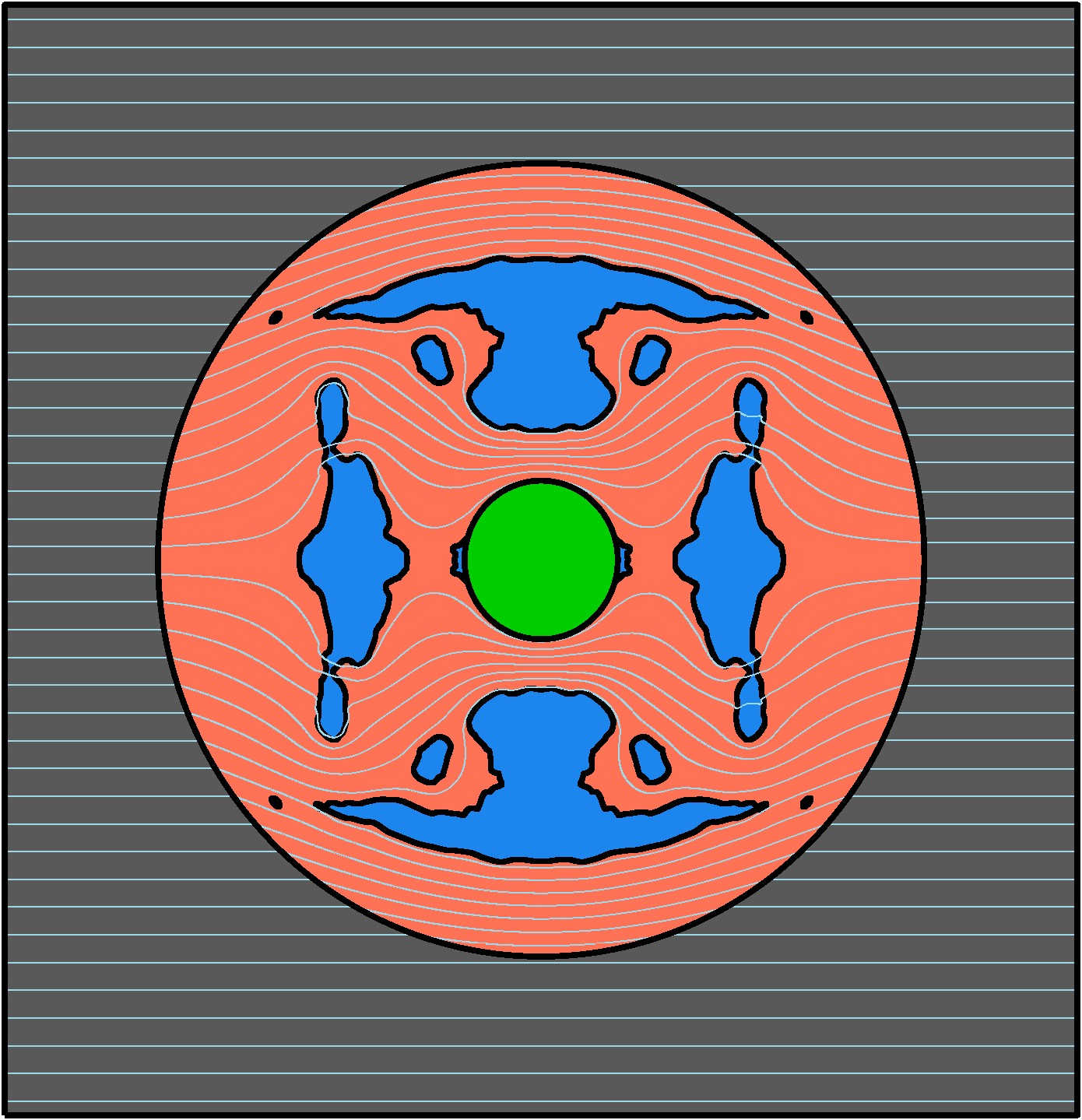}}
        \caption{\centering Optimized topology, $J=3.9496\times 10^{-9}$}
        \label{fig:chen2015case optTop r}
    \end{subfigure}
    \\
\hline
\end{tabular}

}
\caption{For the thermal cloak problem, initial and optimized topologies  for three values of $N_{\rm var}=25, 42$ and $1089$ with $\Delta=0.0005$. Three initial topologies (samples I, II, and III) are discretized with the corresponding design basis. All optimized topologies reach the objective function value of order $10^{-9}$-$10^{-10}$.}  
    \label{fig:chen2015case optTop}
\end{figure}

\par Here, $N_{\rm var}, p, q$ values are taken to be the same as in the previous example, and $\Delta=0.0005$. Three initial topologies (samples I, II, and III) are considered and the corresponding LSFs are discretized with $N_{\rm var}=25, 42$ and $1089$. 
The solution meshes corresponding to $N_{\rm var}=25, 42$ and $1089$ have $13167$, $13974$ and $13167$ DOF, respectively. Additionally, the LSF reinitialization is used to increase the convergence rate. For each knot span in a parameter direction, we utilize $20$ isoparameter lines to find the interface points (as described in \sref{sec:LSF reinitialization}). The reinitialization is performed every 10 iterations or 100 function evaluations. To illustrate the convergence pattern of the objective function with reinitialization, we show the convergence for sample III with $N_{\rm var}=1089$ in \fref{fig:Chen2015cloak convergence}. From \fref{fig:Chen2015cloak convergence}, it is observed that the reinitialization improves the convergence rate substantially. 
\par \fref{fig:chen2015case optTop} shows the initial and optimized topologies. From the figure, we see that the optimized topology depends on the initial topology and on the number of design variables. The objective function, however, successfully reaches the values of order $10^{-9}$-$10^{-10}$ for all cases. Larger $N_{\rm var}$ represents more design freedom, and that is evidently visible for the optimized topologies for $N_{\rm var}=1089$. Using a large $N_{\rm var}$ will allow exploring more detailed topologies, but at the same time, it can create unnecessary and complicated features. One of the solutions to the cloak problem is an exact bilayer cloak as proposed in~\cite{Han2014}, where the interface is a circle and the radius is uniquely defined by the conductivities at hand. From our optimization, we obtain several optimized topologies such as (\frefs{fig:chen2015case optTop b}, \ref{fig:chen2015case optTop d}, \ref{fig:chen2015case optTop f}, \ref{fig:chen2015case optTop h}, \ref{fig:chen2015case optTop n}) close to a bilayer cloak.

\renewcommand{\arraystretch}{1}   
\begin{figure}
\centering
\scalebox{0.9}{
\begin{tabular}[c]{|m{1em}|m{11.4em}|m{11.4em}|m{12.3em}|}
\hline		
   & \begin{center}
       {\small A homogeneous plate (Reference case)} 
   \end{center} & \begin{center}
       {\small A plate with an obstacle}
   \end{center} & \begin{center}
       {\small A plate with an obstacle and a thermal cloak (sample I and $N_{\rm var}=1089$)}
   \end{center} \\
\hline 
  \vspace{0.05cm}  
  \rotatebox{90}{\centering Flux flow}  & \vspace{0.2cm} 
    \begin{subfigure}[t]{0.23\textwidth}\begin{center}{
\includegraphics[width=1\textwidth]{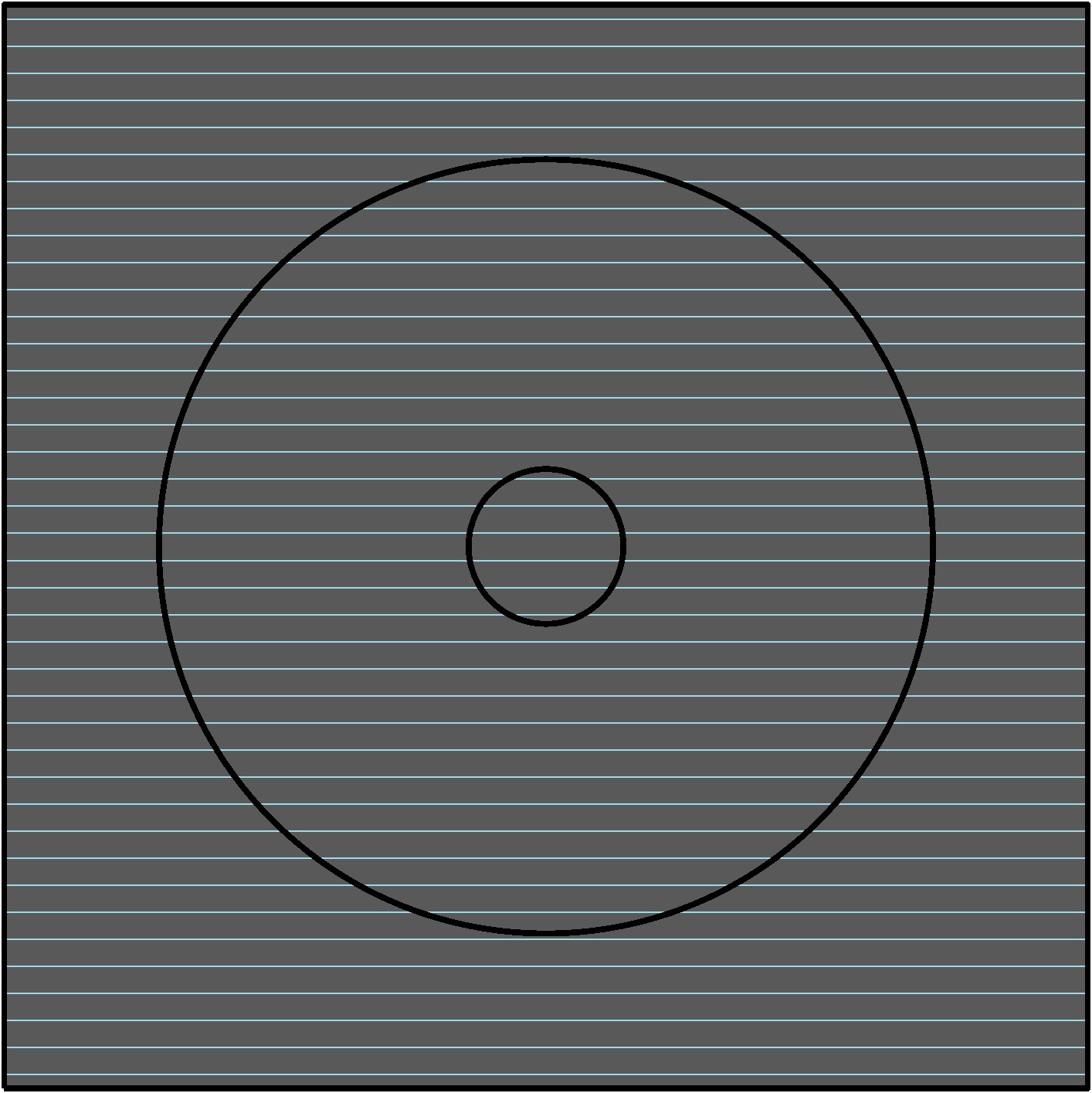}}
    \end{center}
    \end{subfigure} & \vspace{0.2cm} 
    \begin{subfigure}[t]{0.23\textwidth}\begin{center}{
\includegraphics[width=1\textwidth]{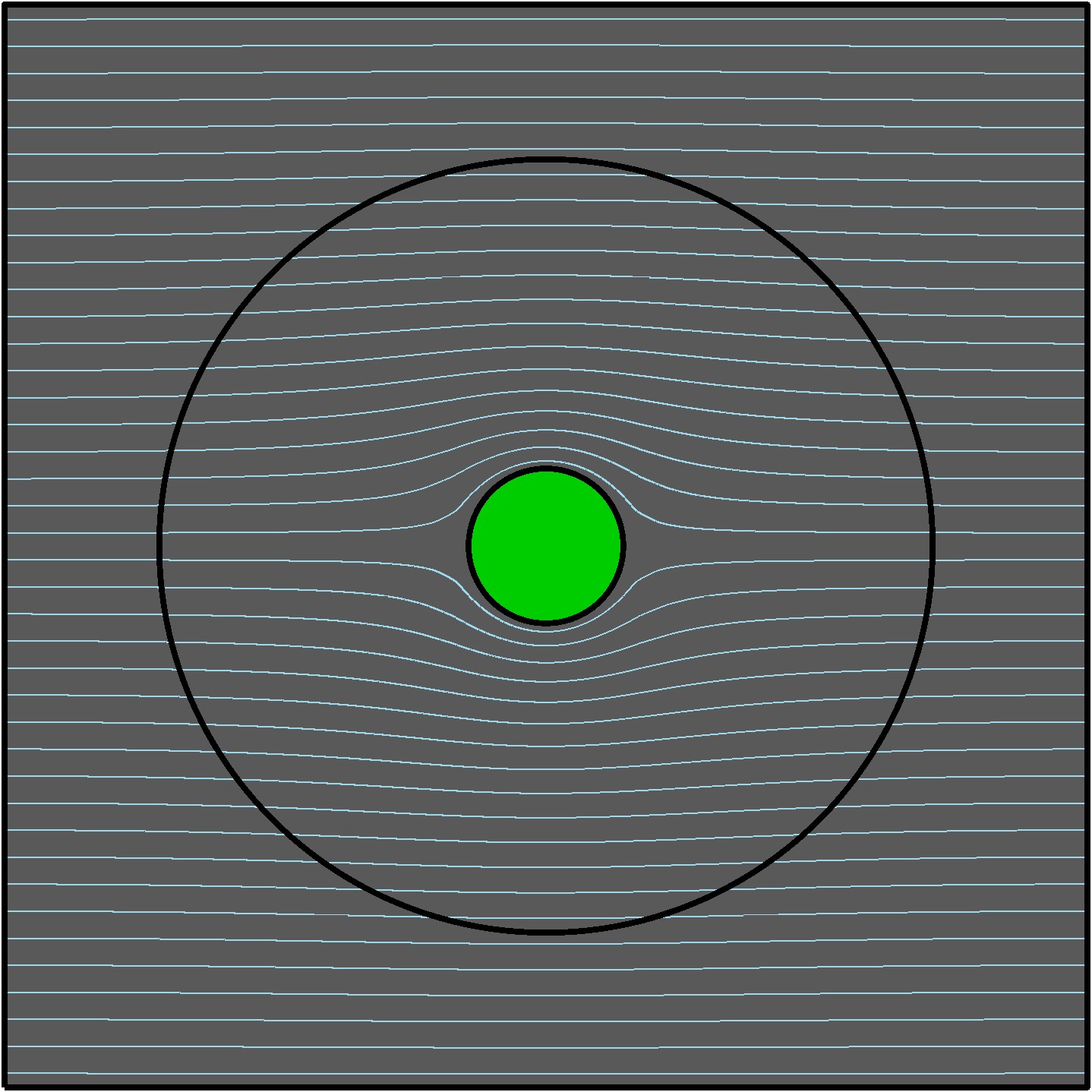}}
    \end{center}
    \end{subfigure} & \vspace{0.2cm} 
    \begin{subfigure}[t]{0.23\textwidth}\begin{center}{
\includegraphics[width=1\textwidth]{Figures_Chen2015cloak/Laplace_HT_LevelSetTop_Chen2015case_objT_2_ref_p00k00h00_DSNref_p00k55h00_sample1_inA_fluxPlot.jpg}}
    \end{center}
    \end{subfigure} \\  
    \rotatebox{90}{\centering Temp. $T$}  &
      \begin{subfigure}[t]{0.23\textwidth}\begin{center}{
\includegraphics[width=1\textwidth]{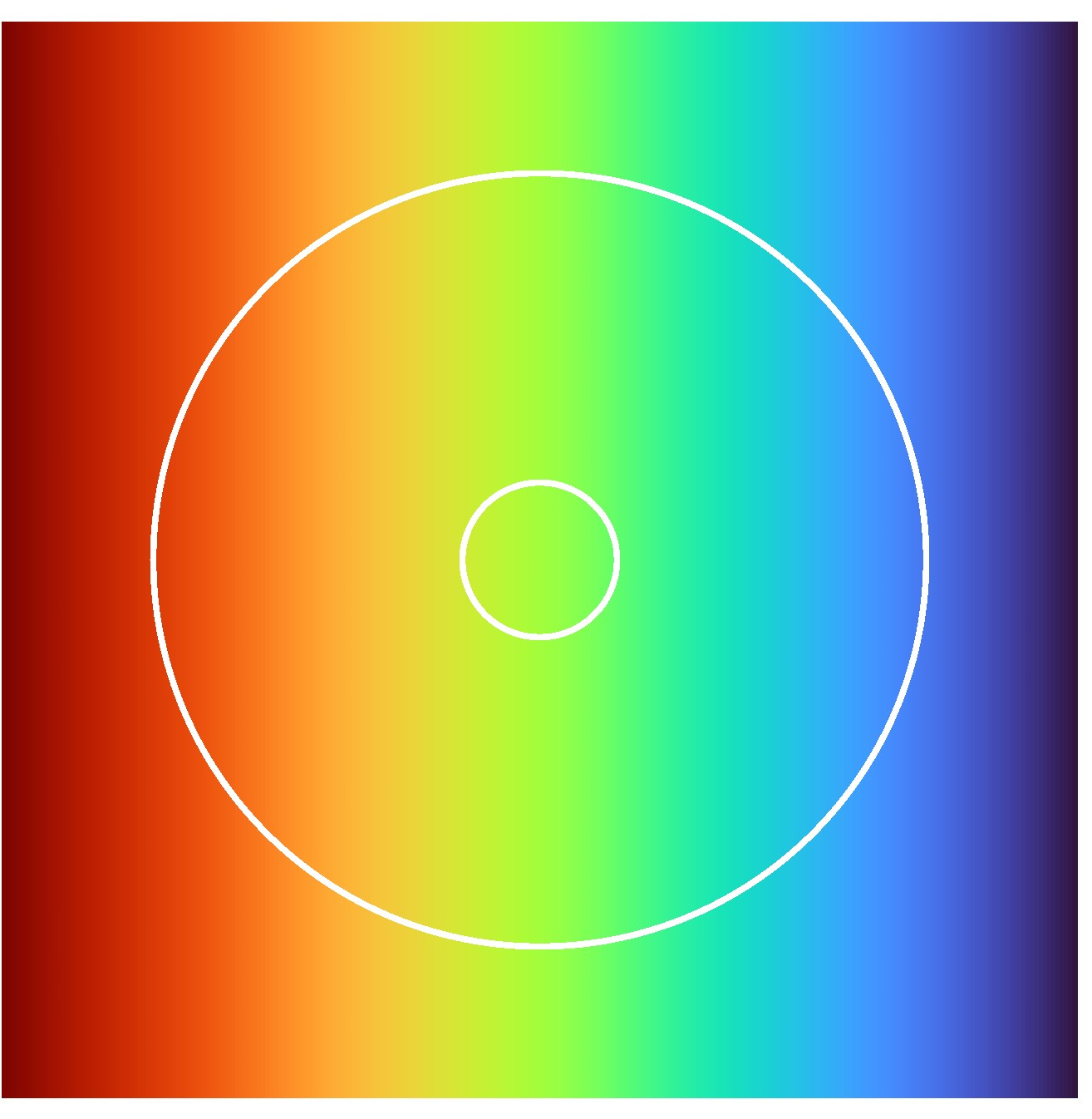}}
    \end{center}
    \end{subfigure} \begin{subfigure}[t]{0.071\textwidth}\begin{center}{
\includegraphics[width=1\textwidth]{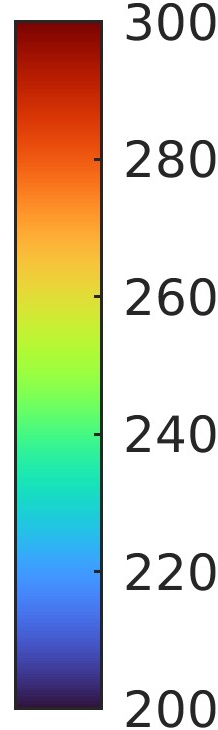}}
    \end{center}
    \end{subfigure} &
    \begin{subfigure}[t]{0.23\textwidth}\begin{center}{
\includegraphics[width=1\textwidth]{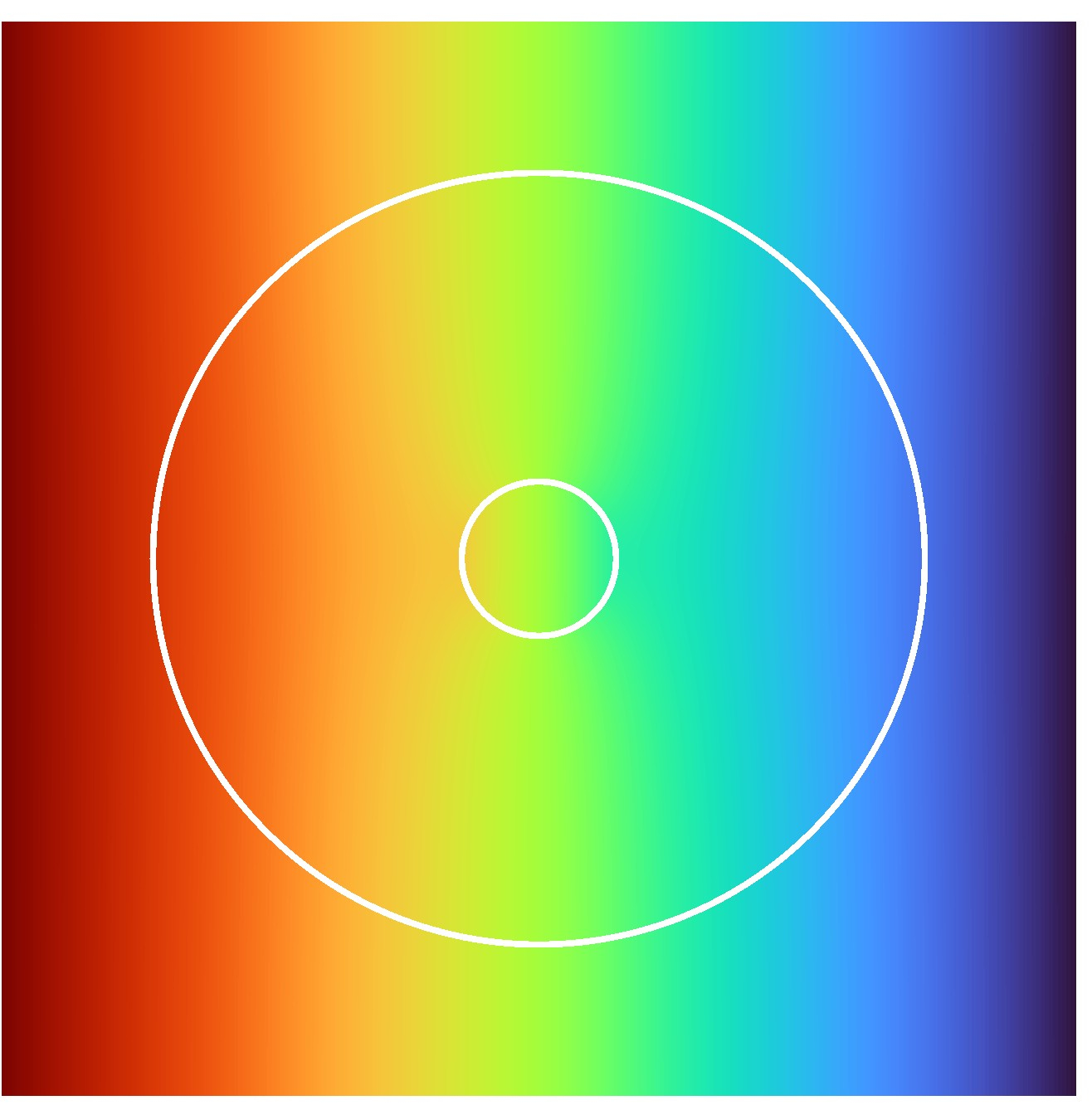}}
    \end{center}
    \end{subfigure} \begin{subfigure}[t]{0.071\textwidth}\begin{center}{
\includegraphics[width=1\textwidth]{Figures_Chen2015cloak/colorbar1.jpg}}
    \end{center}
    \end{subfigure} &
    \begin{subfigure}[t]{0.23\textwidth}\begin{center}{
\includegraphics[width=1\textwidth]{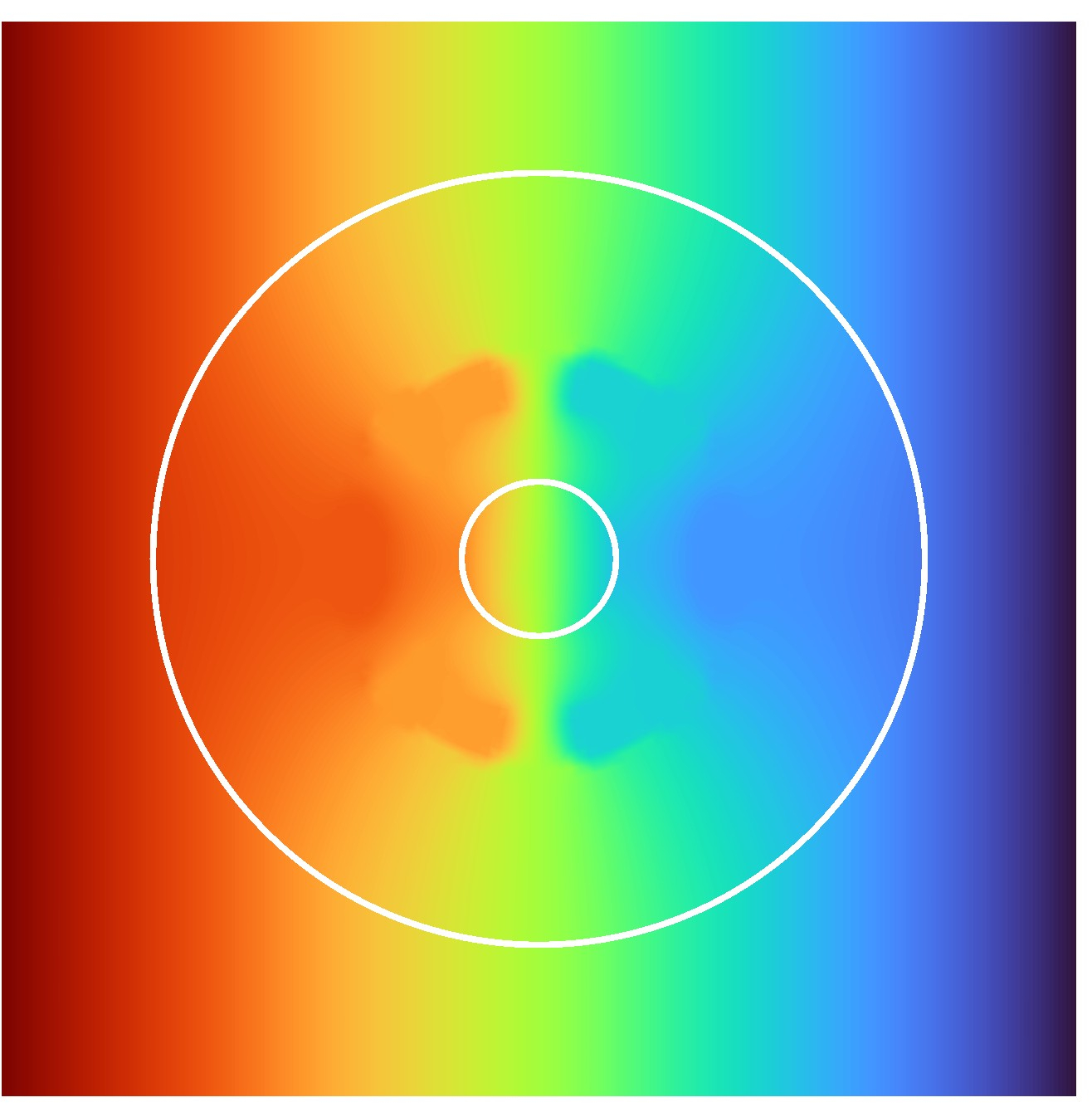}}
    \end{center}
    \end{subfigure} \begin{subfigure}[t]{0.071\textwidth}\begin{center}{
\includegraphics[width=1\textwidth]{Figures_Chen2015cloak/colorbar1.jpg}}
    \end{center}
    \end{subfigure} \\  
    \rotatebox{90}{\centering Temp. diff.~$T-\overline{T}$}  &
       &
    \begin{subfigure}[t]{0.23\textwidth}\begin{center}{
\includegraphics[width=1\textwidth]{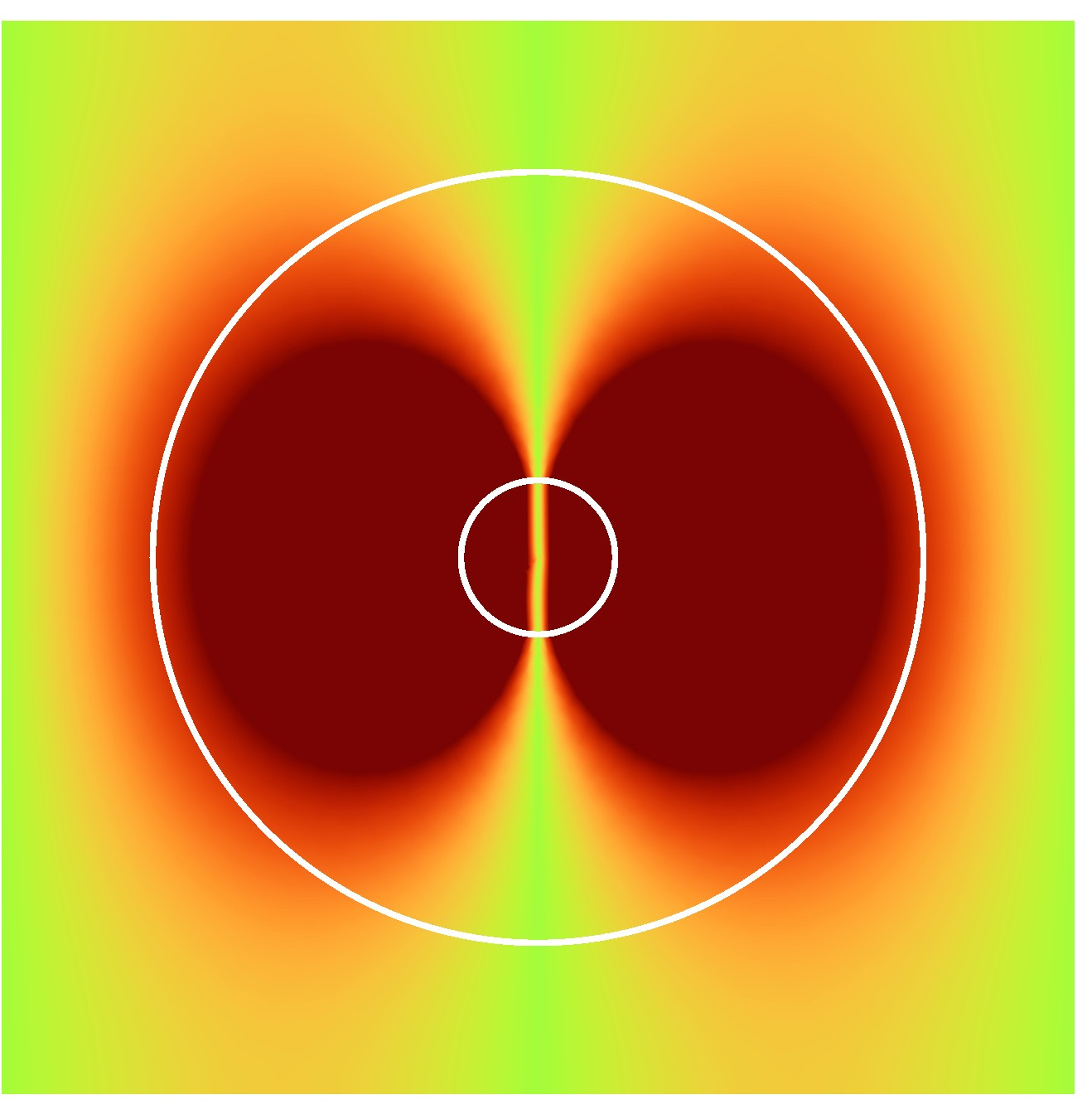}}
    \end{center}
    \end{subfigure} \begin{subfigure}[t]{0.068\textwidth}\begin{center}{
\includegraphics[width=1\textwidth]{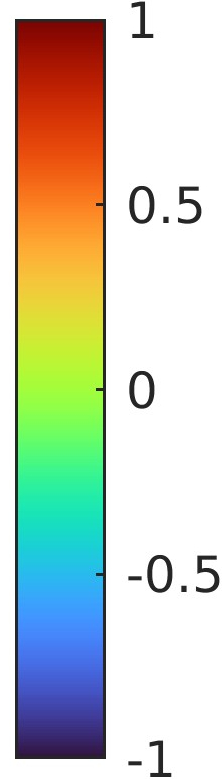}}
    \end{center}
    \end{subfigure} &
    \begin{subfigure}[t]{0.23
    \textwidth}\begin{center}{
\includegraphics[width=1\textwidth]{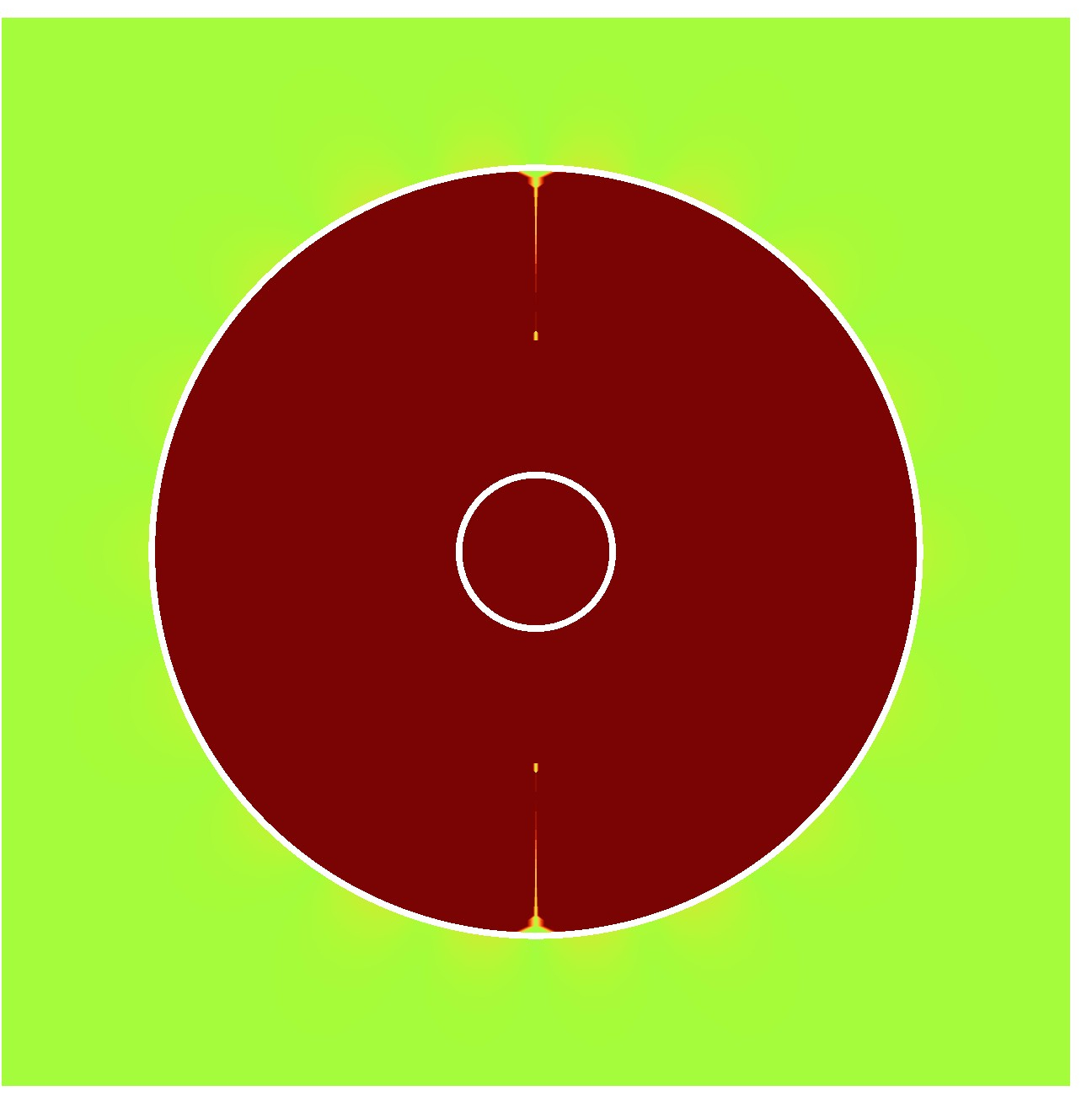}}
    \end{center}
    \end{subfigure} 
    \begin{subfigure}[t]{0.1\textwidth}\begin{center}{
\includegraphics[width=1\textwidth]{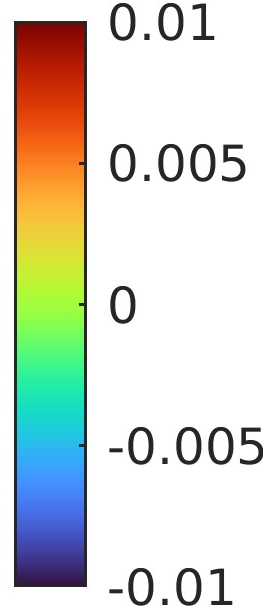}}
    \end{center}
    \end{subfigure} \\  
    \hline
    
\end{tabular}

}
\caption{For the thermal cloak problem, flux flow and temperature distribution for (left column) a homogeneous base material plate (reference case), (middle column) a base material plate embedded with a circular insulator obstacle, and (right column) a base material plate embedded with a circular insulator and surrounding thermal cloak (sample I and $N_{\rm var}=1089$). The temperature and heat flux distributions are shown. The thermal cloak reduces the temperature disturbance in $\mathrm{\Omega}_{\mathrm{out}}$.}
    \label{fig:chen2015case_cloak_tempDiff}
\end{figure}

To further illustrate the results of optimization, in \fref{fig:chen2015case_cloak_tempDiff}, we present an optimized thermal cloak for sample I with $N_{\rm var} = 1089$. We compare the flux flow and temperature distribution among the three cases shown in \fref{fig:Cloak problem schematics}: background plate under constant heat flux, a plate with an obstacle and a plate with an obstacle and the cloak. From the temperature difference, we can evidently see that the thermal cloak decreases the temperature disturbance created by the obstacle. In the $\Omega_{\rm out}$ region, the temperature distribution mimics the reference case, and the flux remains undisturbed.

\renewcommand{\arraystretch}{1.5}   
\begin{figure}
\centering
\scalebox{0.9}{
\begin{tabular}[c]{|m{0.6em}| m{5.6em} | m{5.6em}| m{5.3em}| m{5.3em}| m{5.3em}|m{5.3em} |}
\hline		
 & \multirow{2}{5.5em}{\centering Initial topology} & \multirow{2}{5.5em}{\centering W/o any reg. $J_{\rm total}=J_{\rm cloak}$} & \multicolumn{4}{c|}{Tikhonov regularization $J_{\rm total}=J_{\rm cloak} + \chi J_{\rm Tknv}$}\\
\cline{4-7}		
  &  & & $\chi=1 \times10^{-5}$ &$\chi=1 \times10^{-4}$ &  $\chi=1 \times10^{-3}$ &  $\chi=1 \times10^{-2}$  \\
\hline 
\vspace{0.2cm}
   \rotatebox{90}{\centering \small Sample I}  &
   \vspace{0.2cm}
   \begin{subfigure}[t]{0.15\textwidth}{\centering\includegraphics[width=1\textwidth]{Figures_Chen2015cloak/Laplace_HT_LevelSetTop_Chen2015case_objT_2_ref_p00k33h00_DSNref_p00k22h00_sample1_inA_initTop.jpg}}
        \caption{\centering Initial topology}
        \label{fig:chen2015case optTop TknvSmth a}
    \end{subfigure}  & \vspace{0.2cm}
   \begin{subfigure}[t]{0.15\textwidth}{\centering\includegraphics[width=1\textwidth]{Figures_Chen2015cloak/Laplace_HT_LevelSetTop_Chen2015case_objT_2_ref_p00k00h00_DSNref_p00k55h00_sample1_inA_fluxPlot.jpg}}
        \caption{\centering $J_{\rm cloak}=9.4025\times 10^{-10}$}
        \label{fig:chen2015case optTop TknvSmth b}
    \end{subfigure} &\vspace{0.2cm}
    \begin{subfigure}[t]{0.15\textwidth}{\centering\includegraphics[width=1\textwidth]{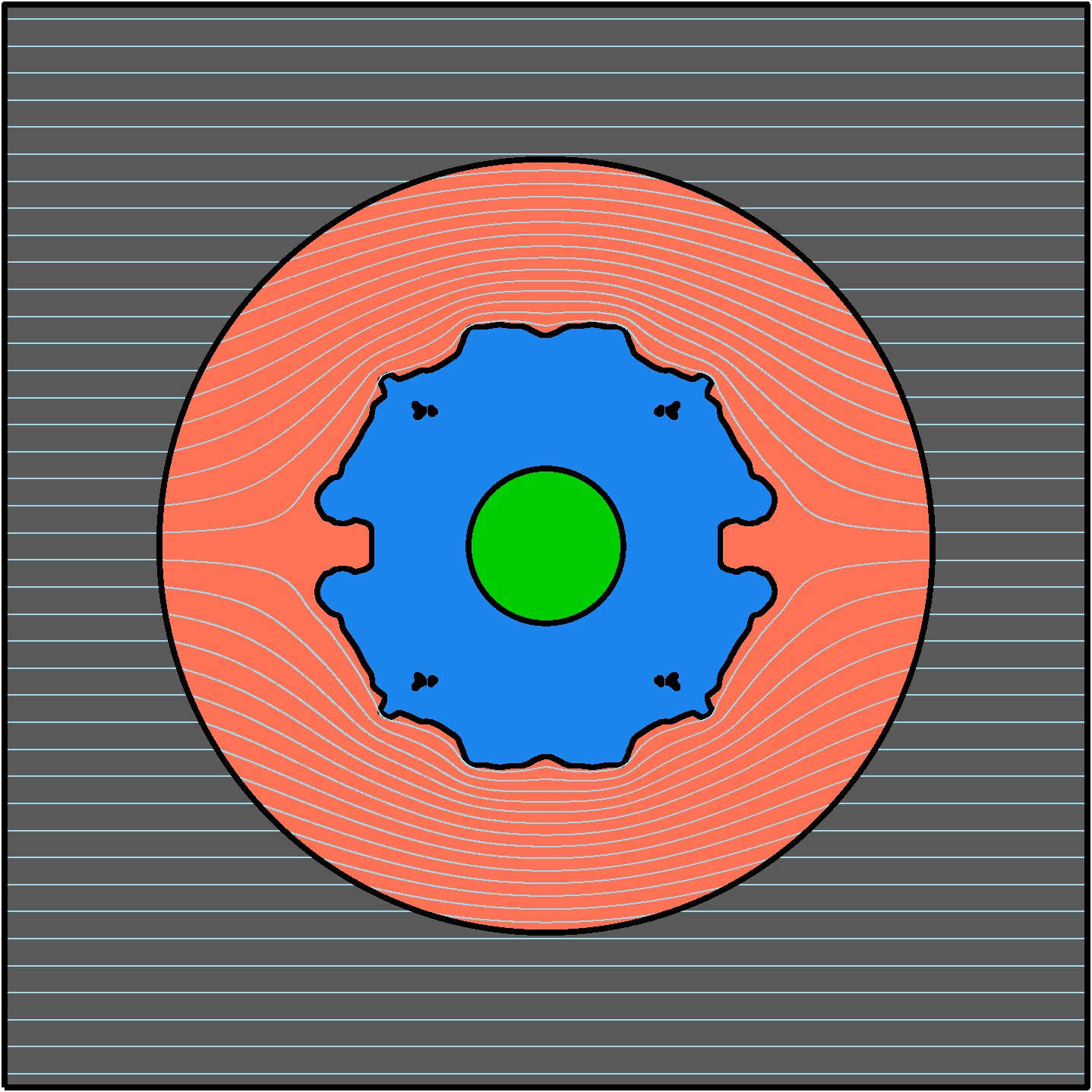}}
        \caption{\centering $J_{\rm cloak}=3.6926\times 10^{-9}$}
        \label{fig:chen2015case optTop TknvSmth c}
    \end{subfigure} & \vspace{0.2cm}
    \begin{subfigure}[t]{0.15\textwidth}{\centering\includegraphics[width=1\textwidth]{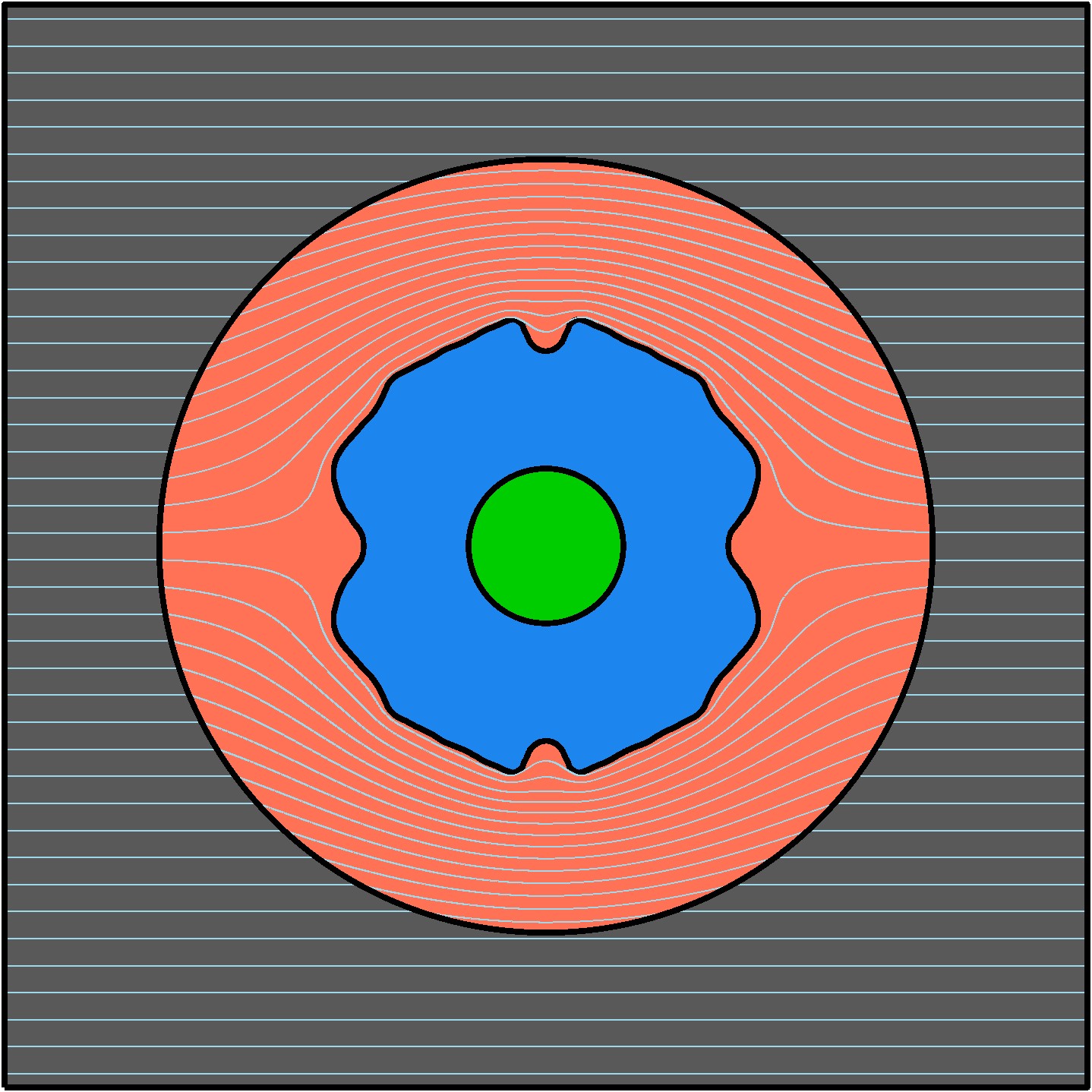}}
        \caption{\centering $J_{\rm cloak}=1.4965\times 10^{-9}$}
        \label{fig:chen2015case optTop TknvSmth d}
    \end{subfigure} & \vspace{0.2cm}
    \begin{subfigure}[t]{0.15\textwidth}{\centering\includegraphics[width=1\textwidth]{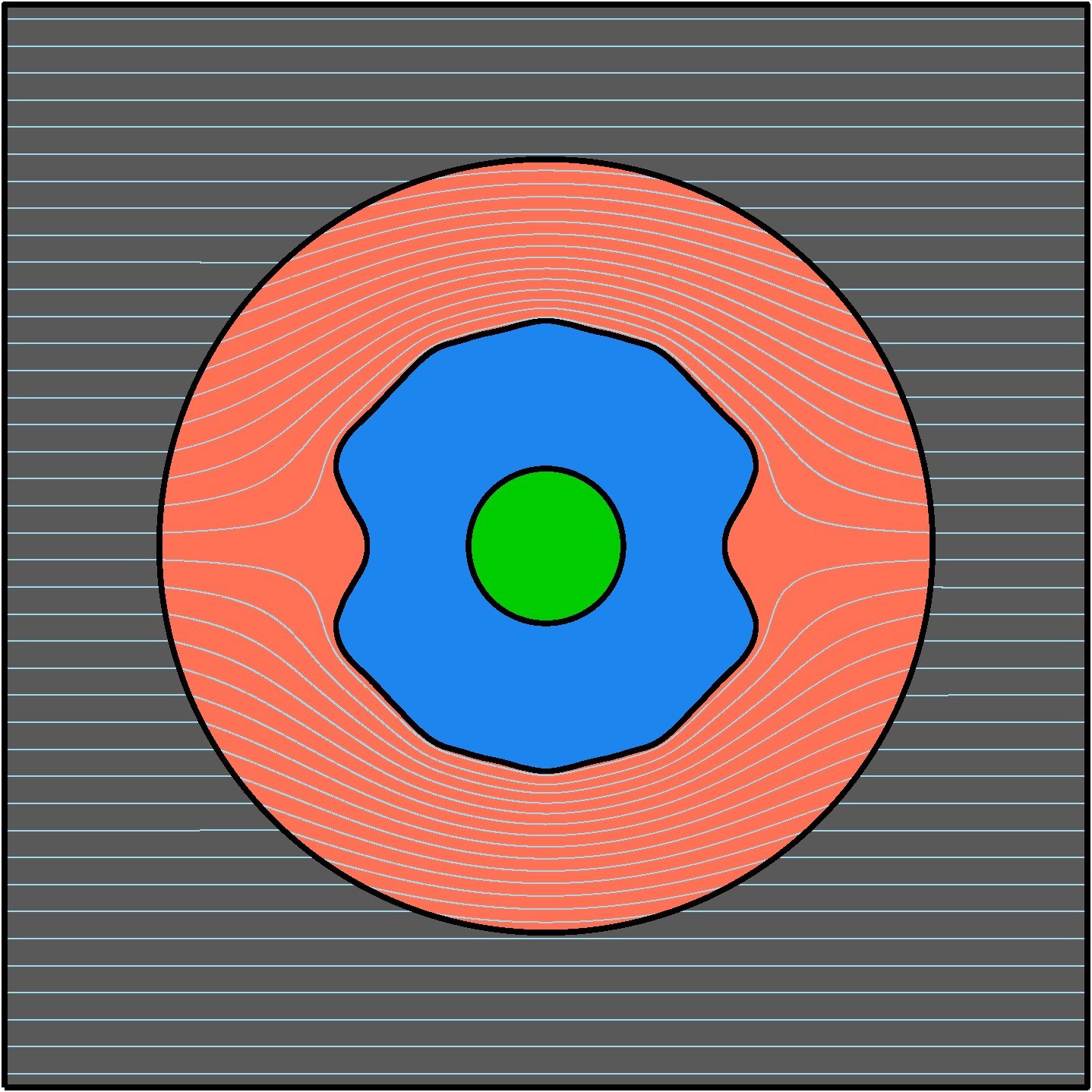}}
        \caption{\centering $J_{\rm cloak}=2.2925\times 10^{-9}$}
        \label{fig:chen2015case optTop TknvSmth e}
    \end{subfigure} & \vspace{0.2cm}
    \begin{subfigure}[t]{0.15\textwidth}{\centering\includegraphics[width=1\textwidth]{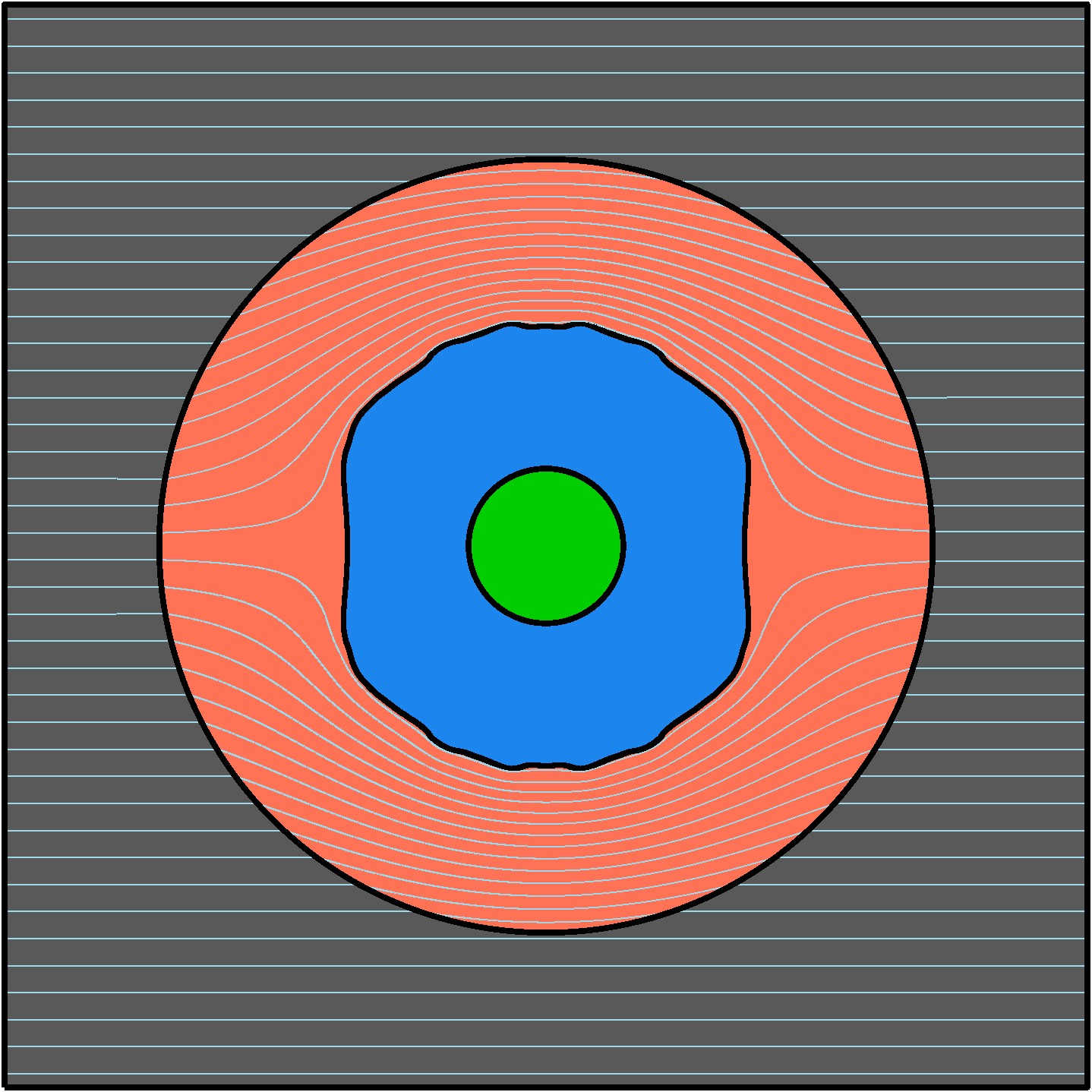}}
        \caption{\centering $J_{\rm cloak}=2.7410\times 10^{-7}$}
        \label{fig:chen2015case optTop TknvSmth f}
    \end{subfigure}
    \\
    \hline

   \rotatebox{90}{\centering \small Sample II}  &
   \vspace{0.2cm}
   \begin{subfigure}[t]{0.15\textwidth}{\centering\includegraphics[width=1\textwidth]{Figures_Chen2015cloak/Laplace_HT_LevelSetTop_Chen2015case_objT_2_ref_p00k33h00_DSNref_p00k22h00_sample1_inB_initTop.jpg}}
        \caption{\centering Initial topology}
        \label{fig:chen2015case optTop TknvSmth g}
    \end{subfigure}  & \vspace{0.2cm}
   \begin{subfigure}[t]{0.15\textwidth}{\centering\includegraphics[width=1\textwidth]{Figures_Chen2015cloak/Laplace_HT_LevelSetTop_Chen2015case_objT_2_ref_p00k00h00_DSNref_p00k55h00_sample1_inB_fluxPlot.jpg}}
        \caption{\centering $J_{\rm cloak}=1.6808\times 10^{-9}$}
        \label{fig:chen2015case optTop TknvSmth h}
    \end{subfigure}  &\vspace{0.2cm}
    \begin{subfigure}[t]{0.15\textwidth}{\centering\includegraphics[width=1\textwidth]{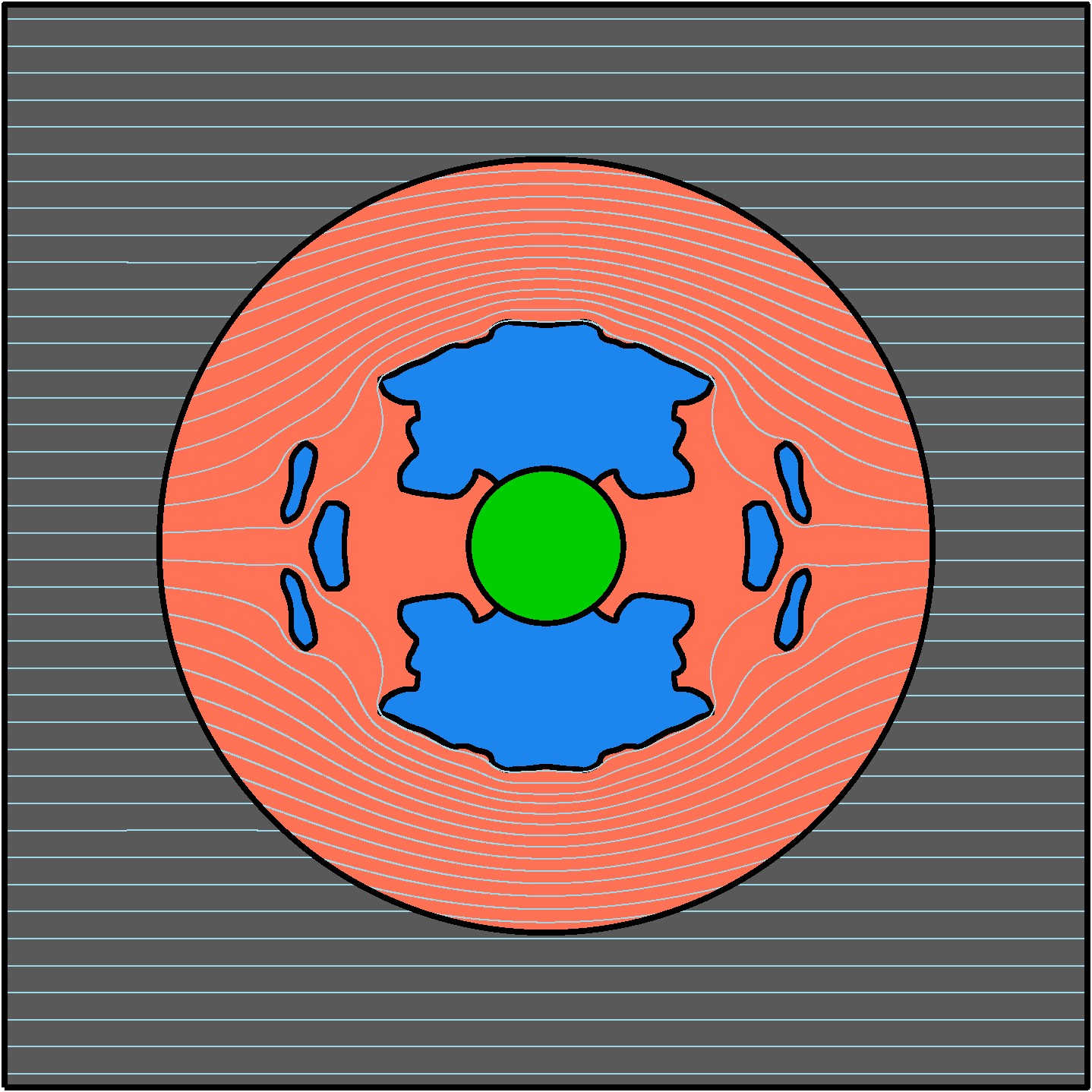}}
        \caption{\centering $J_{\rm cloak}=1.3970\times 10^{-8}$}
        \label{fig:chen2015case optTop TknvSmth i}
    \end{subfigure} & \vspace{0.2cm}
    \begin{subfigure}[t]{0.15\textwidth}{\centering\includegraphics[width=1\textwidth]{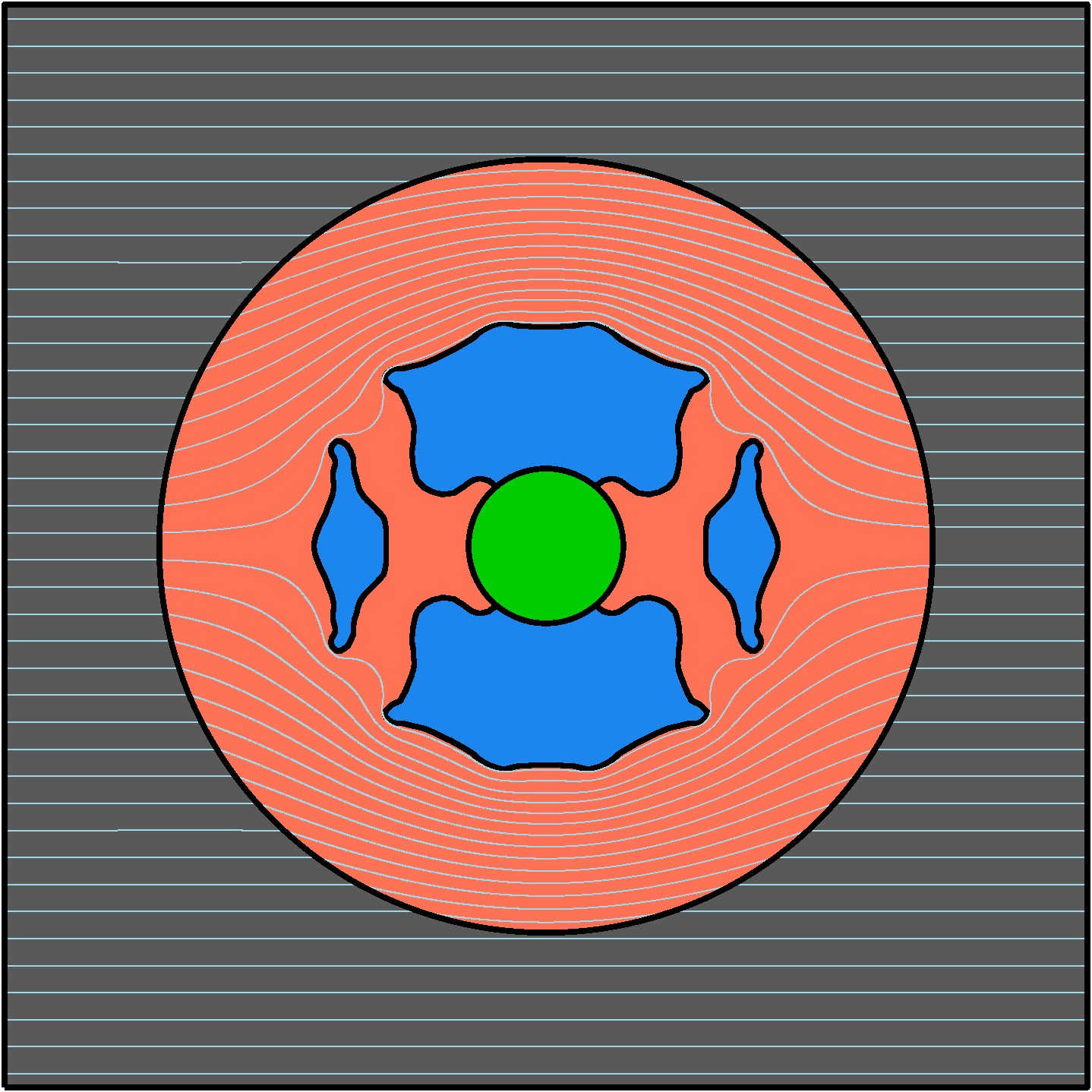}}
        \caption{\centering $J_{\rm cloak}=1.1734\times 10^{-8}$}
        \label{fig:chen2015case optTop TknvSmth j}
    \end{subfigure} & \vspace{0.2cm}
    \begin{subfigure}[t]{0.15\textwidth}{\centering\includegraphics[width=1\textwidth]{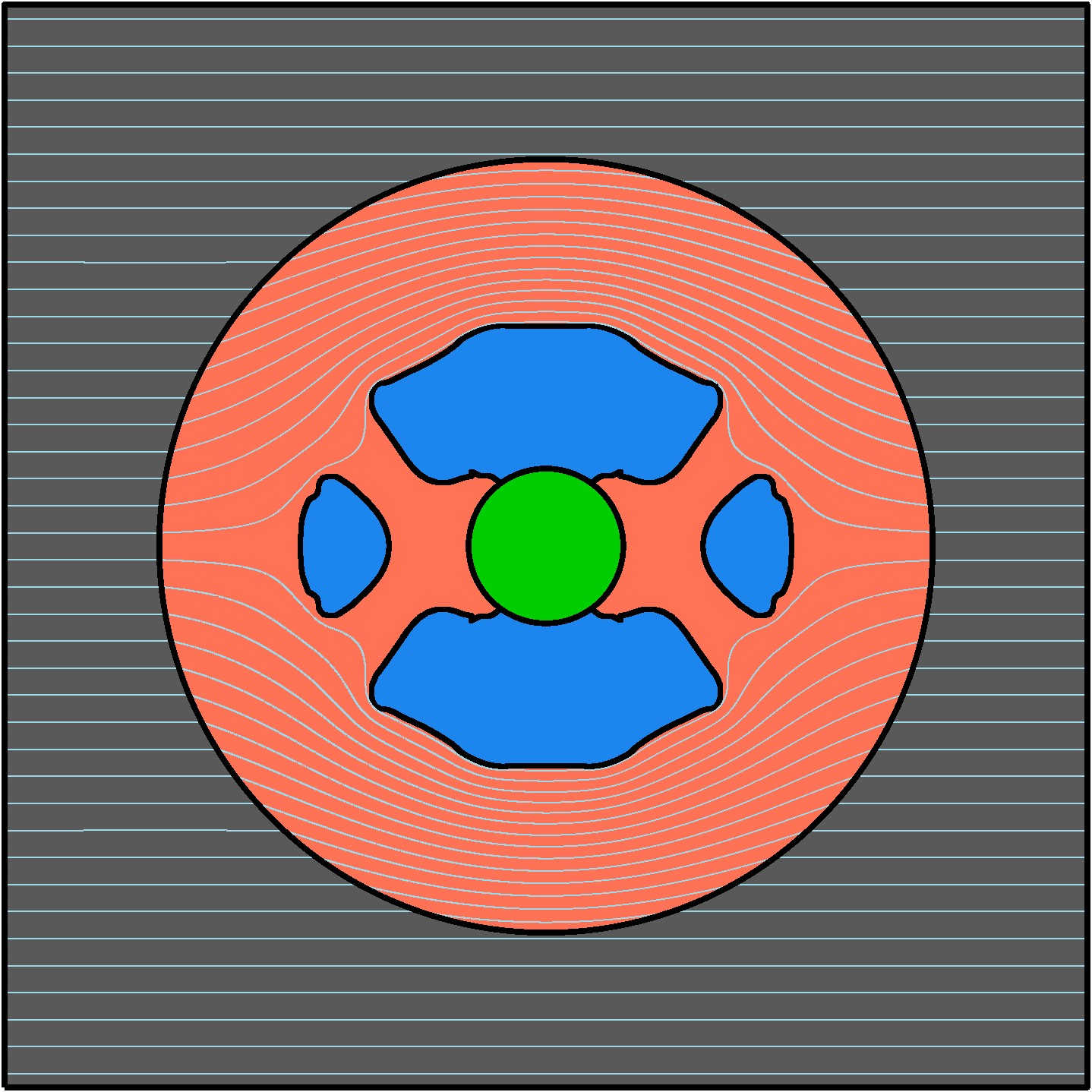}}
        \caption{\centering $J_{\rm cloak}=3.6621\times 10^{-8}$}
        \label{fig:chen2015case optTop TknvSmth k}
    \end{subfigure} & \vspace{0.2cm}
    \begin{subfigure}[t]{0.15\textwidth}{\centering\includegraphics[width=1\textwidth]{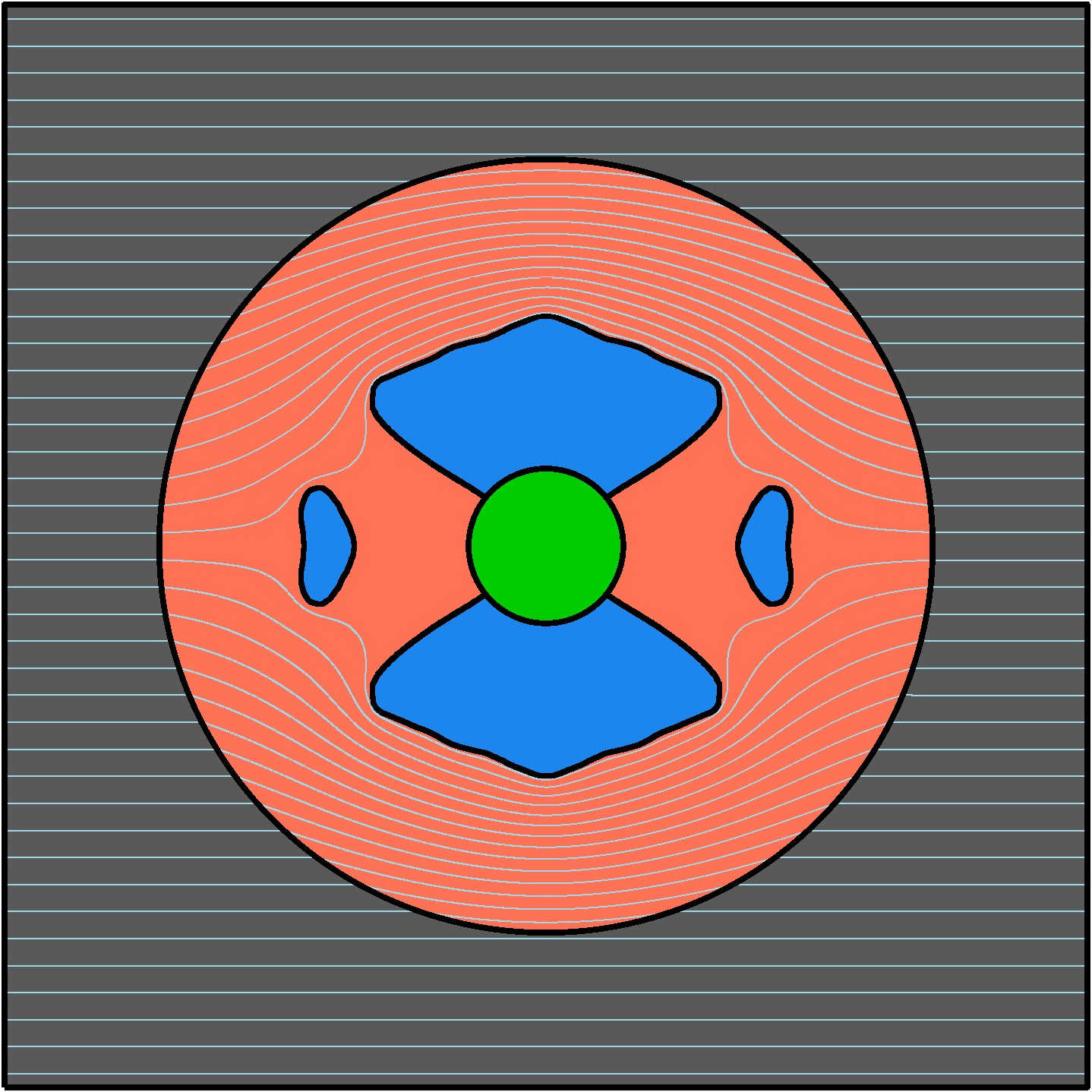}}
        \caption{\centering $J_{\rm cloak}=1.9286\times 10^{-7}$}
        \label{fig:chen2015case optTop TknvSmth l}
    \end{subfigure}   \\ 
\hline

   \rotatebox{90}{\centering \small Sample III} &
   \vspace{0.2cm}
   \begin{subfigure}[t]{0.15\textwidth}{\centering\includegraphics[width=1\textwidth]{Figures_Chen2015cloak/Laplace_HT_LevelSetTop_Chen2015case_objT_2_ref_p00k33h00_DSNref_p00k22h00_sample2_inB_initTop.jpg}}
        \caption{\centering Initial topology}
        \label{fig:chen2015case optTop TknvSmth m}
    \end{subfigure}   & \vspace{0.2cm}
   \begin{subfigure}[t]{0.15\textwidth}{\centering\includegraphics[width=1\textwidth]{Figures_Chen2015cloak/Laplace_HT_LevelSetTop_Chen2015case_objT_2_ref_p00k00h00_DSNref_p00k55h00_sample2_inB_fluxPlot.jpg}}
        \caption{\centering $J_{\rm cloak}=3.9218\times 10^{-9}$}
        \label{fig:chen2015case optTop TknvSmth n}
    \end{subfigure} &\vspace{0.2cm}
    \begin{subfigure}[t]{0.15\textwidth}{\centering\includegraphics[width=1\textwidth]{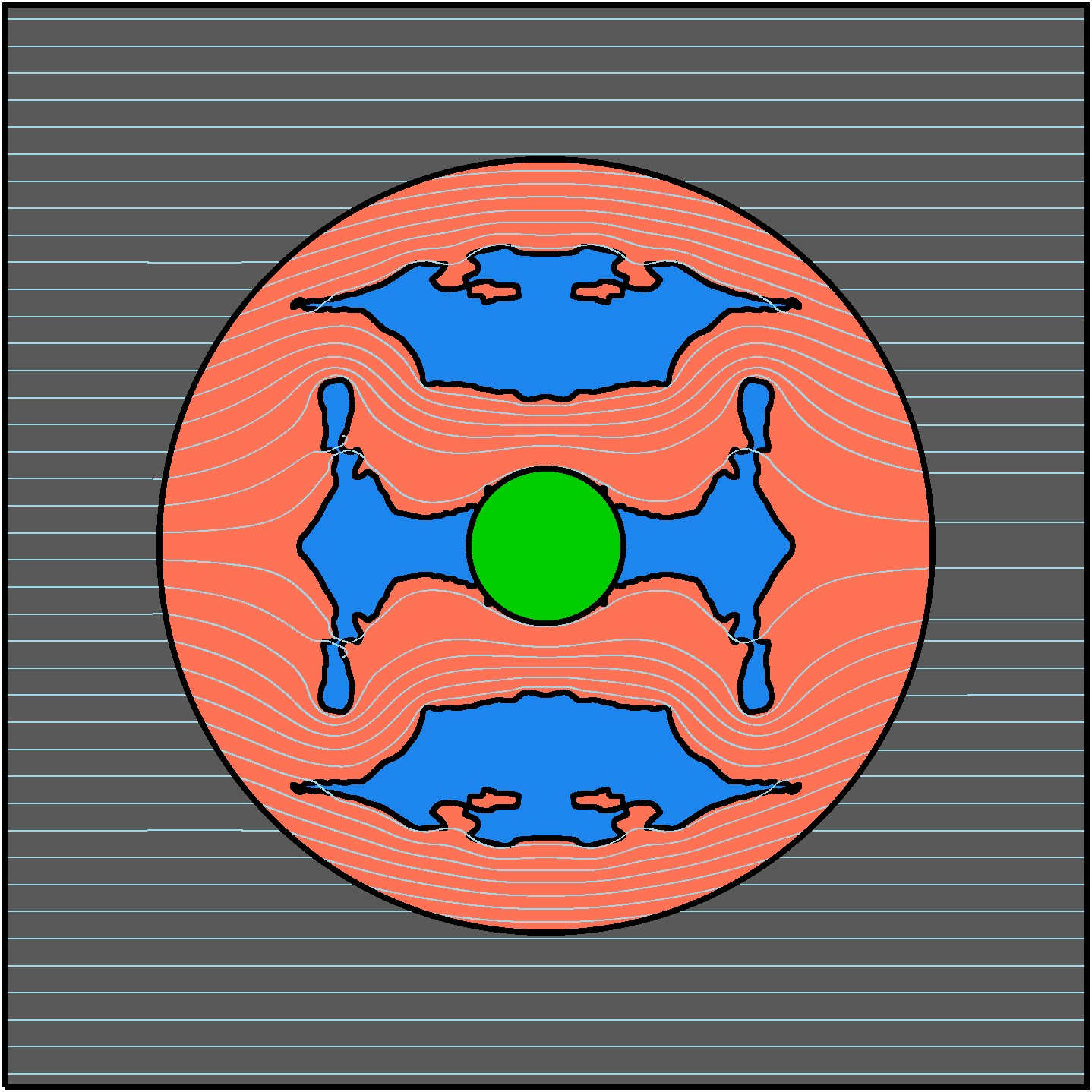}}
        \caption{\centering $J_{\rm cloak}=3.1303\times 10^{-8}$}
        \label{fig:chen2015case optTop TknvSmth o}
    \end{subfigure}& \vspace{0.2cm}
    \begin{subfigure}[t]{0.15\textwidth}{\centering\includegraphics[width=1\textwidth]{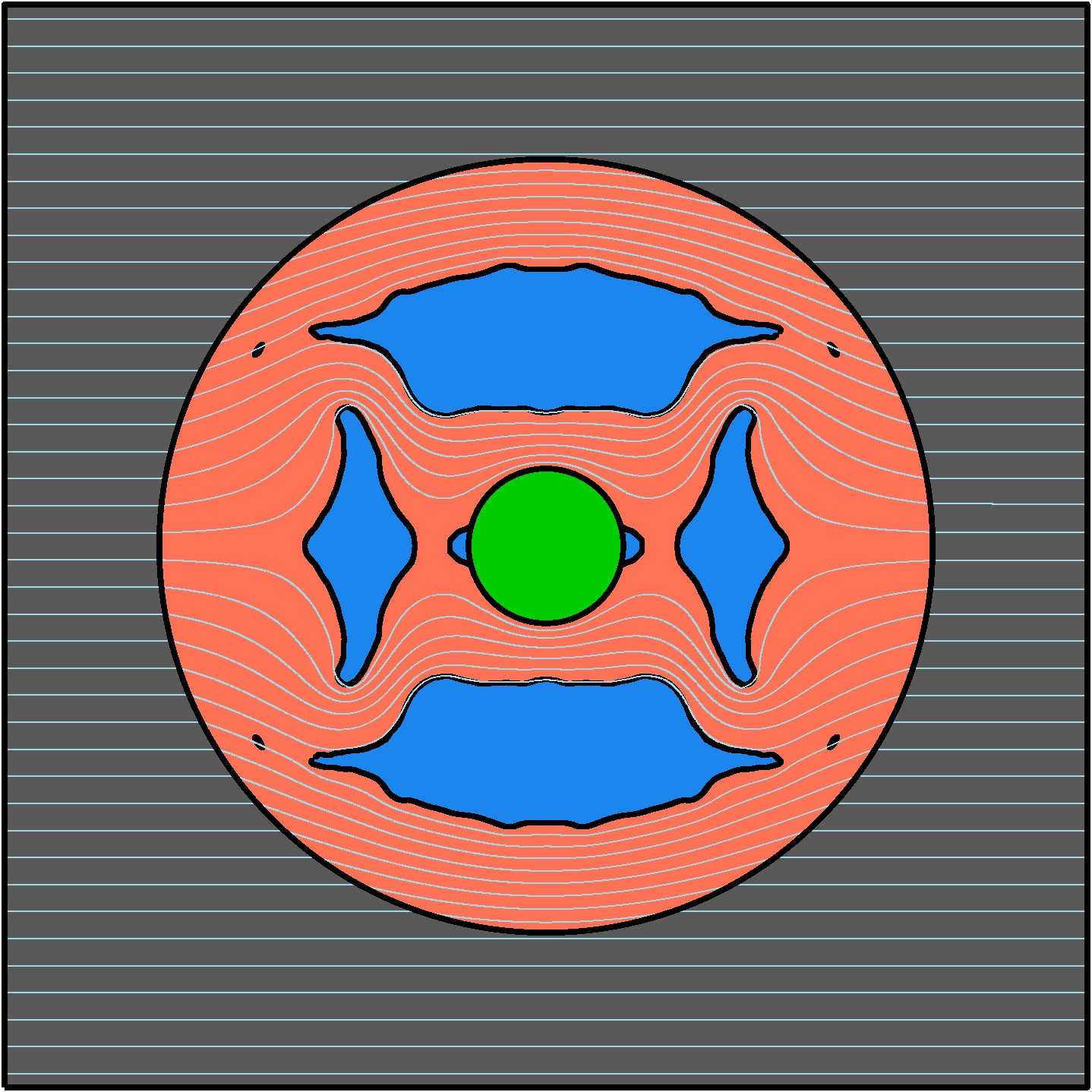}}
        \caption{\centering $J_{\rm cloak}=2.8539\times 10^{-8}$}
        \label{fig:chen2015case optTop TknvSmth p}
    \end{subfigure}& \vspace{0.2cm}
    \begin{subfigure}[t]{0.15\textwidth}{\centering\includegraphics[width=1\textwidth]{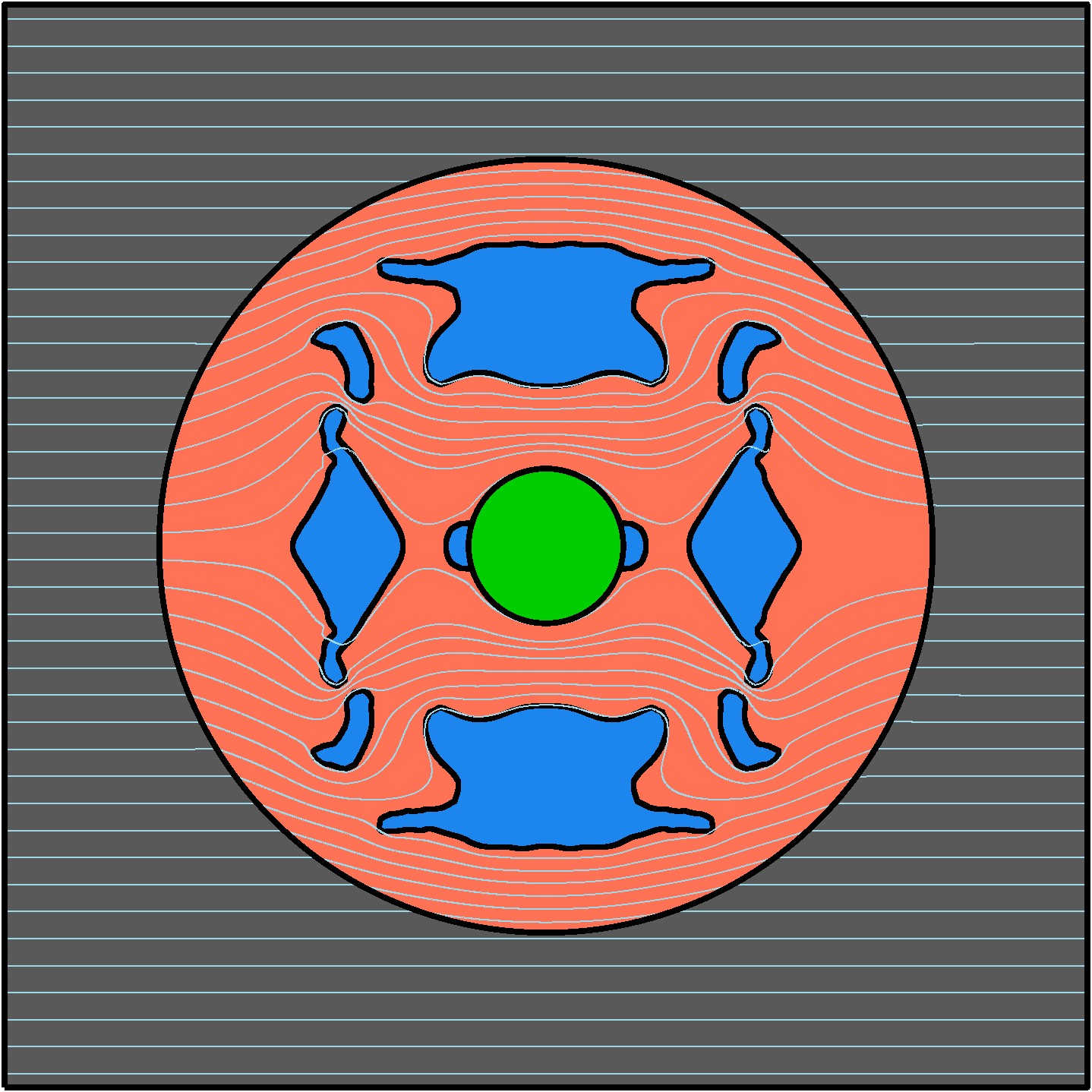}}
        \caption{\centering $J_{\rm cloak}=4.4867\times 10^{-7}$}
        \label{fig:chen2015case optTop TknvSmth q}
    \end{subfigure}  & \vspace{0.2cm}
    \begin{subfigure}[t]{0.15\textwidth}{\centering\includegraphics[width=1\textwidth]{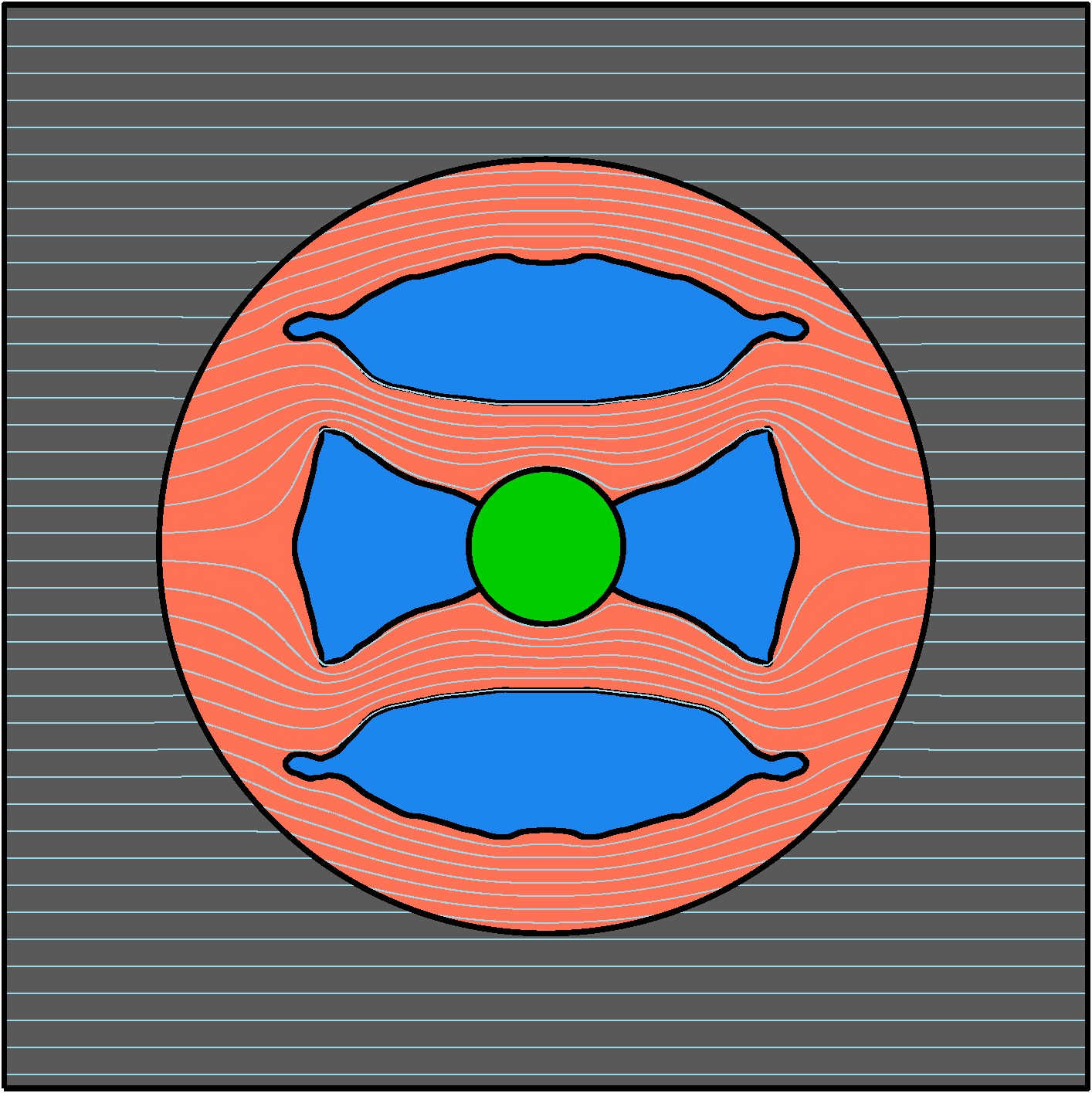}}
        \caption{\centering $J_{\rm cloak}=4.4867\times 10^{-7}$}
        \label{fig:chen2015case optTop TknvSmth r}
    \end{subfigure}  \\
    \hline
\end{tabular}

}
\caption{For thermal cloak problem, initial topologies, optimized topologies without any regularization and with Tikhonov regularization for $N_{\rm var}=1089$ with $\Delta=0.0005$. Three initial topologies (samples I, II and III) are discretized with the corresponding design basis. Four values of weighing parameter $\chi$ are considered. Tikhonov regularization provides smoother optimized topologies with a slight compromise on the $J_{\rm cloak}$-values.}  
    \label{fig:chen2015case optTop TknvSmth}
\end{figure}


\renewcommand{\arraystretch}{1.5}   
\begin{figure}
\centering
\scalebox{0.9}{
\begin{tabular}[c]{|m{0.6em}| m{5.8em} | m{5.3em}| m{5.3em}| m{5.3em}| m{5.3em}|m{5.3em} |}
\hline		
 & \multirow{2}{5.5em}{\centering Initial topology} & \multirow{2}{5.5em}{\centering W/o any reg. $J_{\rm total}=J_{\rm cloak}$} & \multicolumn{4}{c|}{Volume regularization $J_{\rm total}=J_{\rm cloak} + \rho J_{\rm vol}$}\\
\cline{4-7}		
  &  & & $\rho=1 \times10^{-5}$ &$\rho=1 \times10^{-4}$ &  $\rho=1 \times10^{-3}$ &  $\rho=1 \times10^{-2}$  \\
\hline 
\vspace{0.2cm}
   \rotatebox{90}{\centering \small Sample I}  &
   \vspace{0.2cm}
   \begin{subfigure}[t]{0.15\textwidth}{\centering\includegraphics[width=1\textwidth]{Figures_Chen2015cloak/Laplace_HT_LevelSetTop_Chen2015case_objT_2_ref_p00k33h00_DSNref_p00k22h00_sample1_inA_initTop.jpg}}
        \caption{\centering Initial topology}
        \label{fig:chen2015case optTop VolSmth a}
    \end{subfigure} 
    & \vspace{0.2cm}
   \begin{subfigure}[t]{0.15\textwidth}{\centering\includegraphics[width=1\textwidth]{Figures_Chen2015cloak/Laplace_HT_LevelSetTop_Chen2015case_objT_2_ref_p00k00h00_DSNref_p00k55h00_sample1_inA_fluxPlot.jpg}}
        \caption{\centering $J_{\rm cloak}=9.4025\times 10^{-10}$}
        \label{fig:chen2015case optTop VolSmth b}
    \end{subfigure} &\vspace{0.2cm}
    \begin{subfigure}[t]{0.15\textwidth}{\centering\includegraphics[width=1\textwidth]{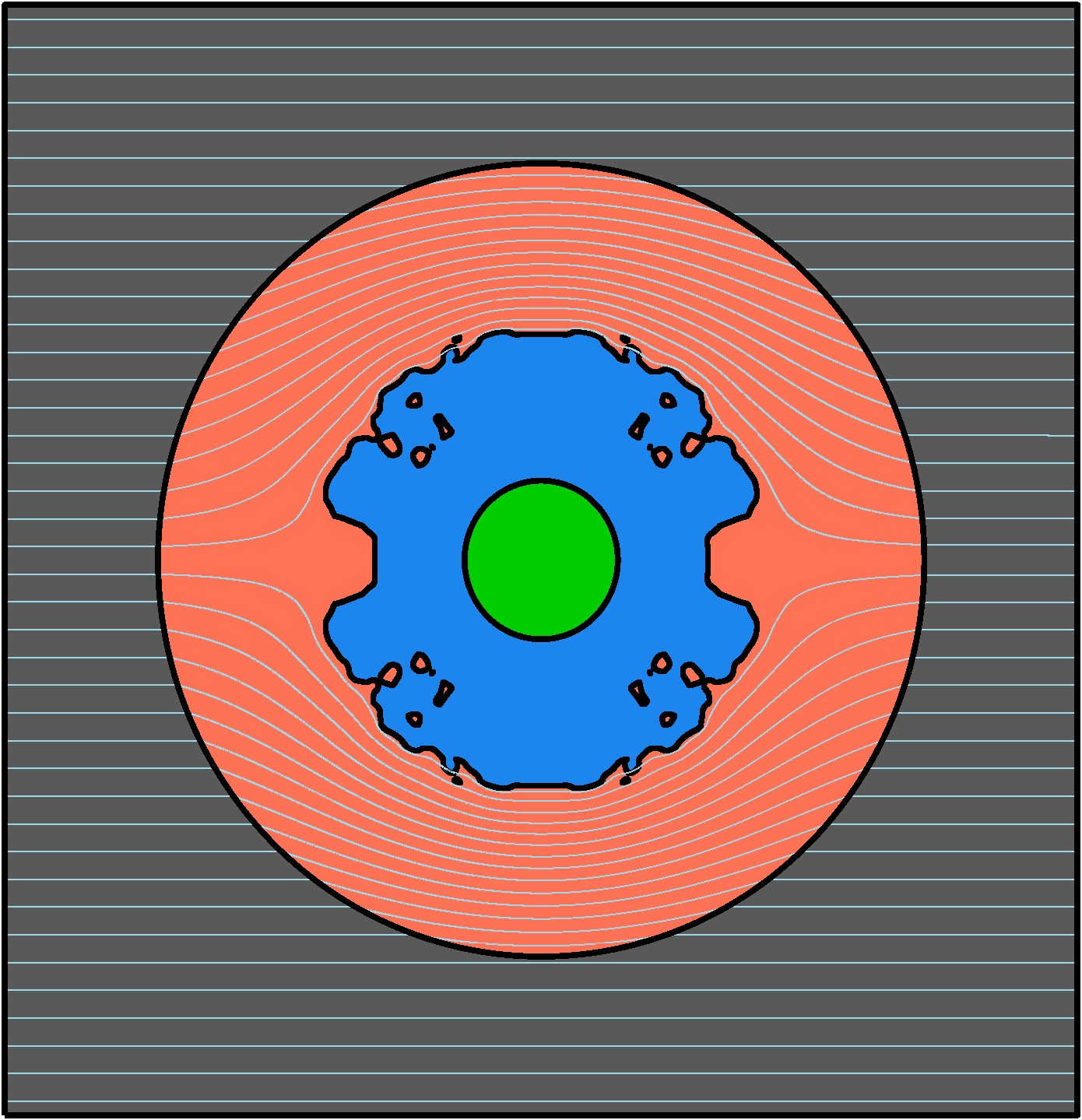}}
        \caption{\centering $J_{\rm cloak}=1.0381\times 10^{-9}$}
        \label{fig:chen2015case optTop VolSmth f}
    \end{subfigure}& \vspace{0.2cm}
    \begin{subfigure}[t]{0.15\textwidth}{\centering\includegraphics[width=1\textwidth]{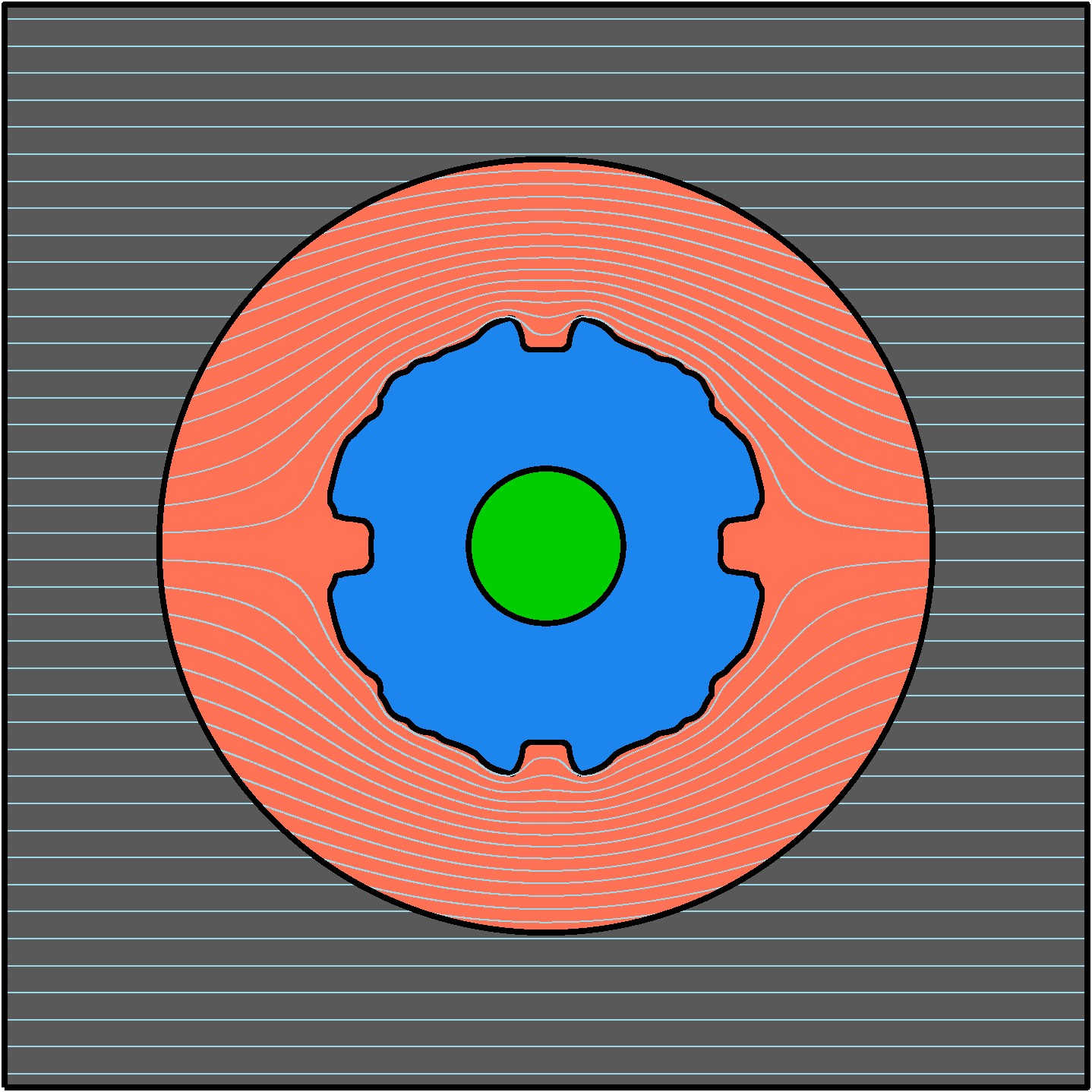}}Initial
        \caption{\centering $J_{\rm cloak}=6.0961\times 10^{-10}$}
        \label{fig:chen2015case optTop VolSmth e}
    \end{subfigure} & \vspace{0.2cm}
    \begin{subfigure}[t]{0.15\textwidth}{\centering\includegraphics[width=1\textwidth]{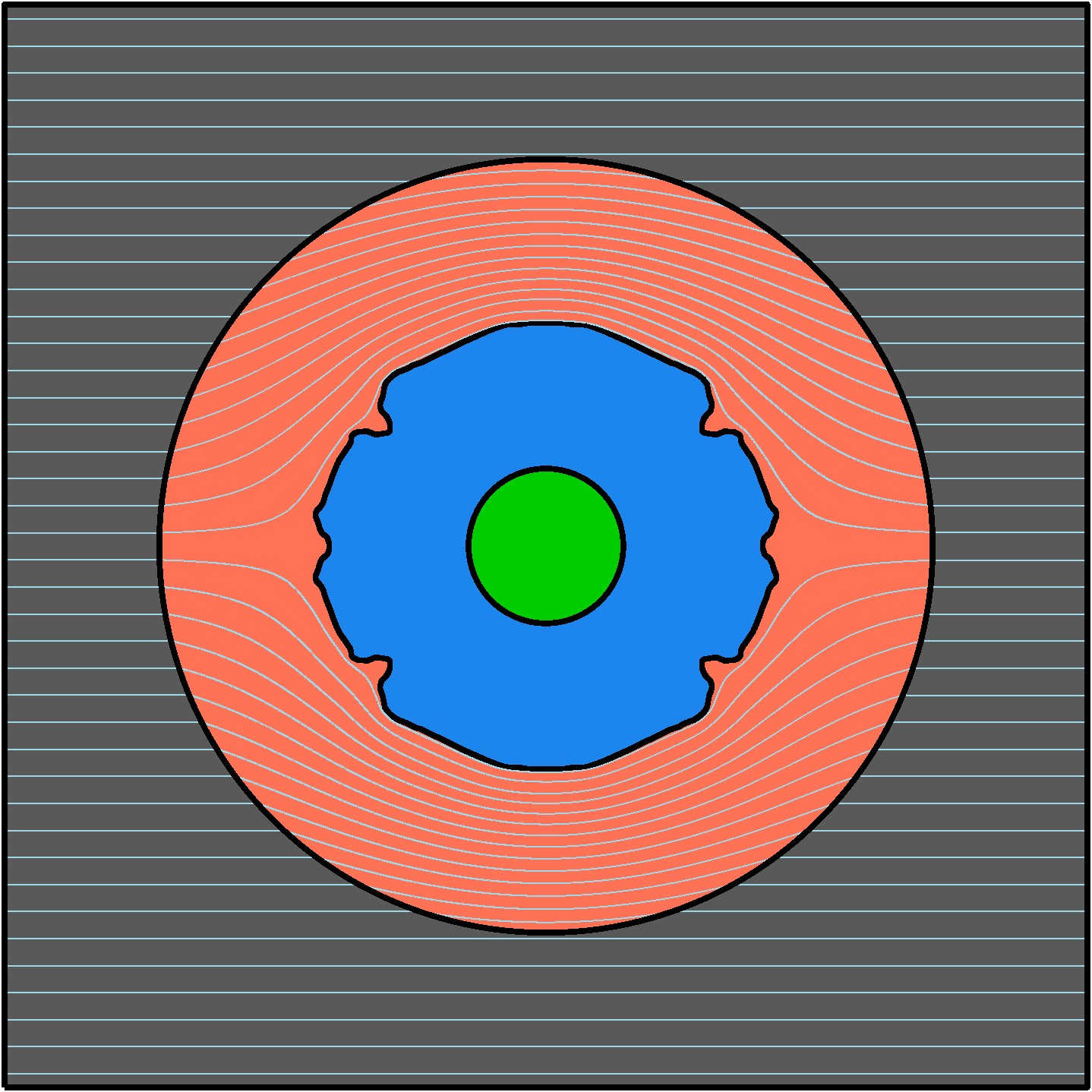}}Initial
        \caption{\centering $J_{\rm cloak}=8.2470\times 10^{-9}$}
        \label{fig:chen2015case optTop VolSmth d} 
    \end{subfigure}& \vspace{0.2cm}
    \begin{subfigure}[t]{0.15\textwidth}{\centering\includegraphics[width=1\textwidth]{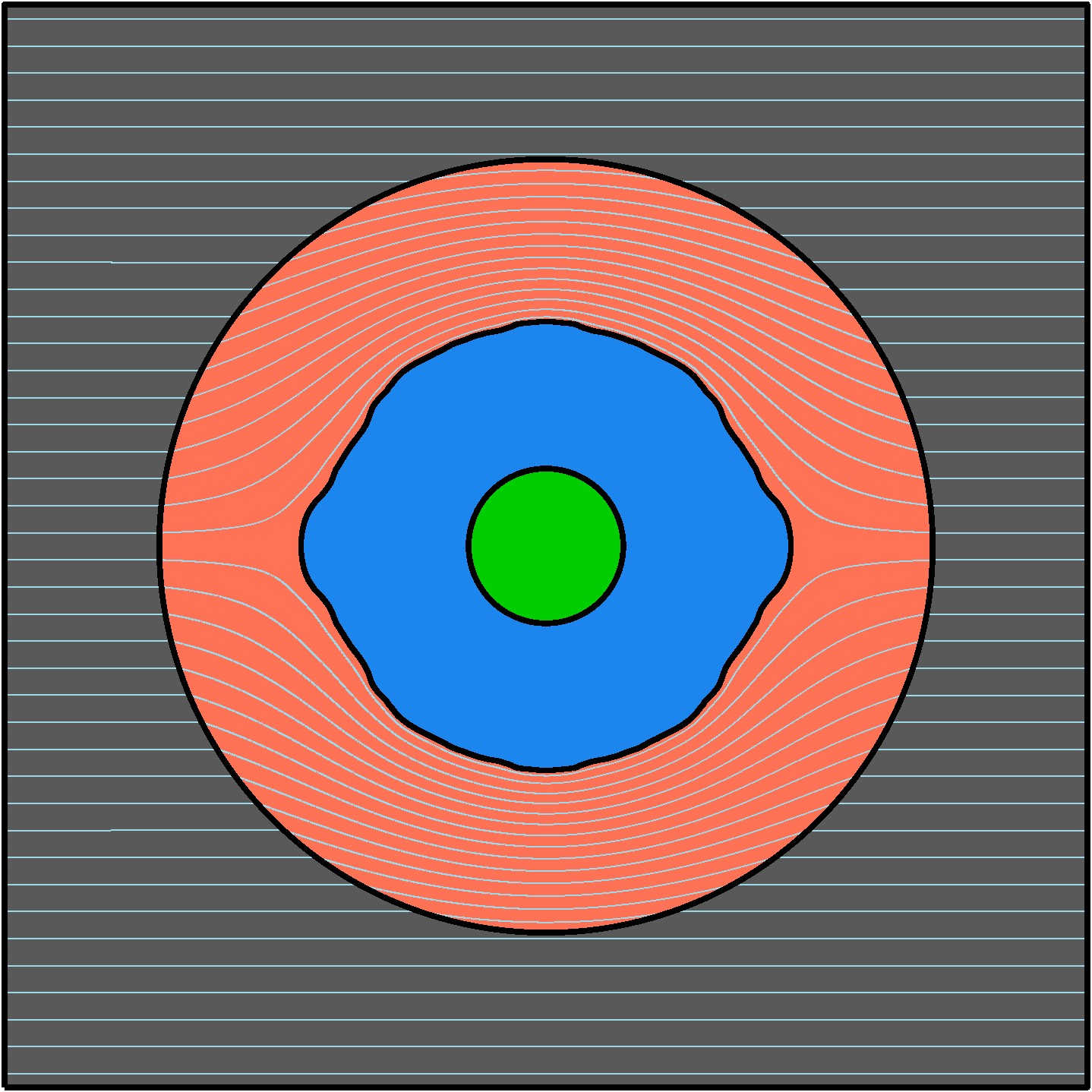}}
        \caption{\centering $J_{\rm cloak}=4.2666\times 10^{-8}$}
        \label{fig:chen2015case optTop VolSmth c}
    \end{subfigure}
    \\

    \hline
   \rotatebox{90}{\centering \small Sample II} &
   \vspace{0.2cm}
   \begin{subfigure}[t]{0.15\textwidth}{\centering\includegraphics[width=1\textwidth]{Figures_Chen2015cloak/Laplace_HT_LevelSetTop_Chen2015case_objT_2_ref_p00k33h00_DSNref_p00k22h00_sample1_inB_initTop.jpg}}
        \caption{\centering Initial topology}
        \label{fig:chen2015case optTop VolSmth g}
    \end{subfigure}   & \vspace{0.2cm}
   \begin{subfigure}[t]{0.15\textwidth}{\centering\includegraphics[width=1\textwidth]{Figures_Chen2015cloak/Laplace_HT_LevelSetTop_Chen2015case_objT_2_ref_p00k00h00_DSNref_p00k55h00_sample1_inB_fluxPlot.jpg}}
        \caption{\centering $J_{\rm cloak}=1.6808\times 10^{-9}$}
        \label{fig:chen2015case optTop VolSmth h}
    \end{subfigure}&\vspace{0.2cm}
    \begin{subfigure}[t]{0.15\textwidth}{\centering\includegraphics[width=1\textwidth]{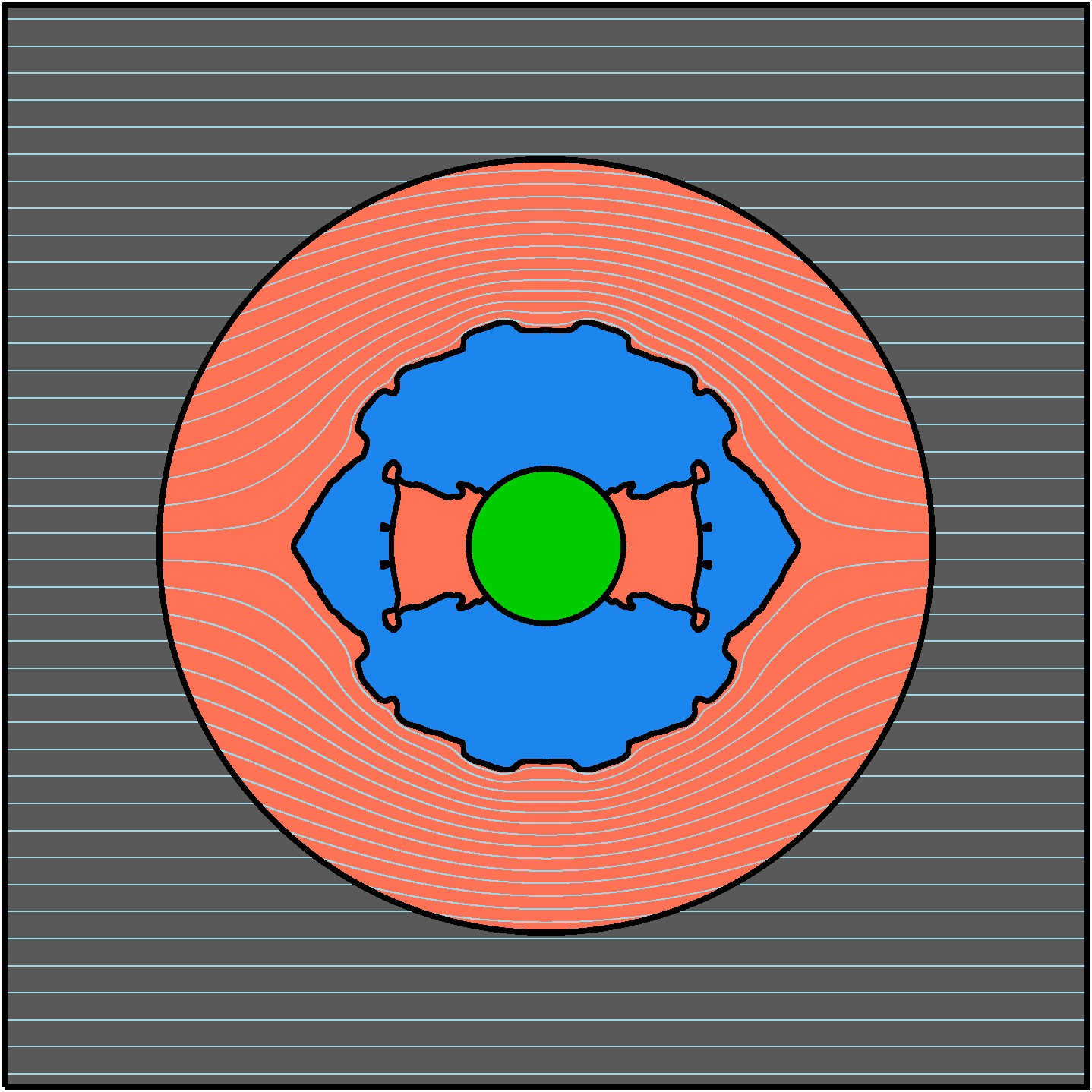}}
        \caption{\centering $J_{Initial\rm cloak}=2.5519\times 10^{-9}$}
        \label{fig:chen2015case optTop VolSmth l}
    \end{subfigure}& \vspace{0.2cm}
    \begin{subfigure}[t]{0.15\textwidth}{\centering\includegraphics[width=1\textwidth]{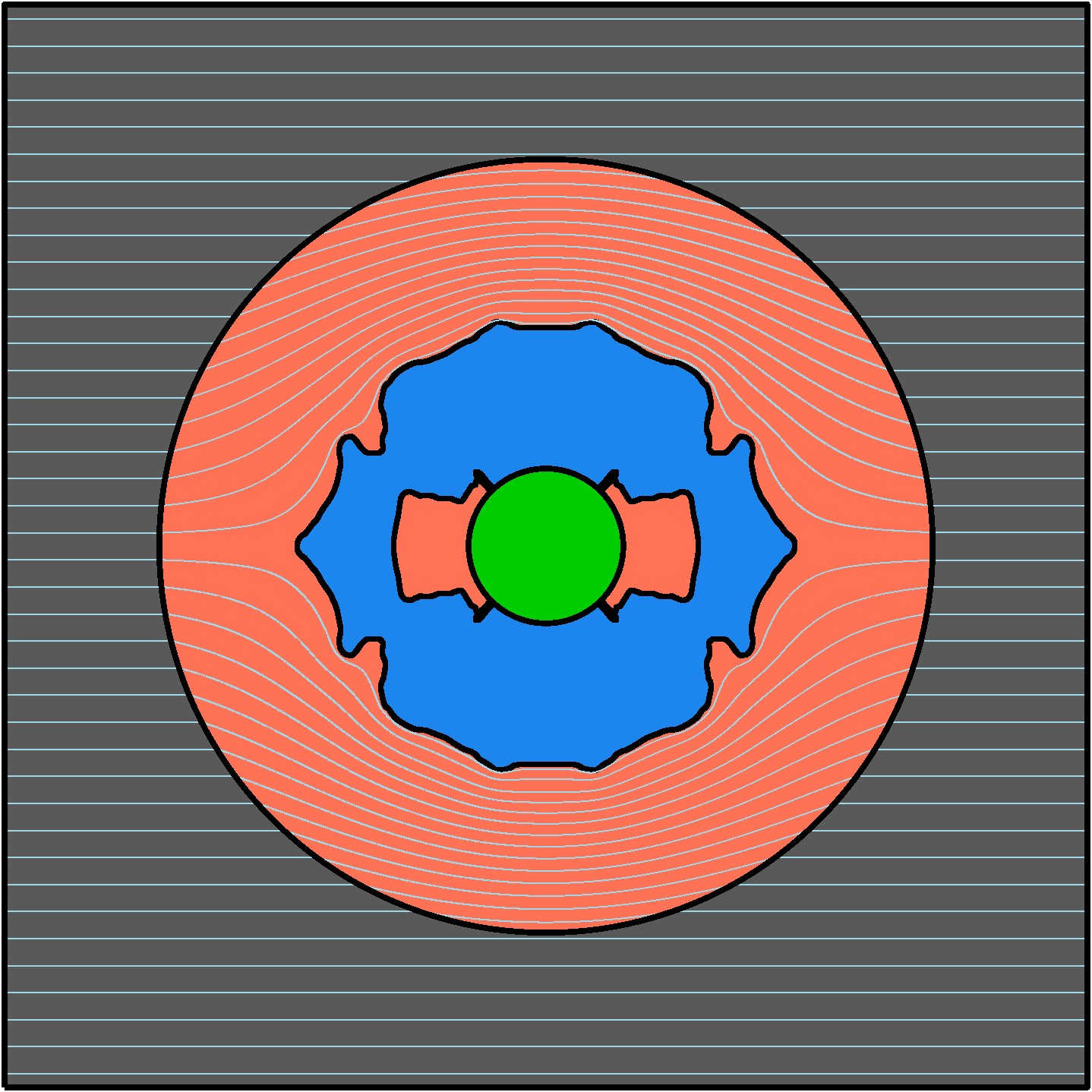}}
        \caption{\centering $J_{\rm cloak}=2.6986\times 10^{-9}$}
        \label{fig:chen2015case optTop VolSmth k}
    \end{subfigure} & \vspace{0.2cm}
    \begin{subfigure}[t]{0.15\textwidth}{\centering\includegraphics[width=1\textwidth]{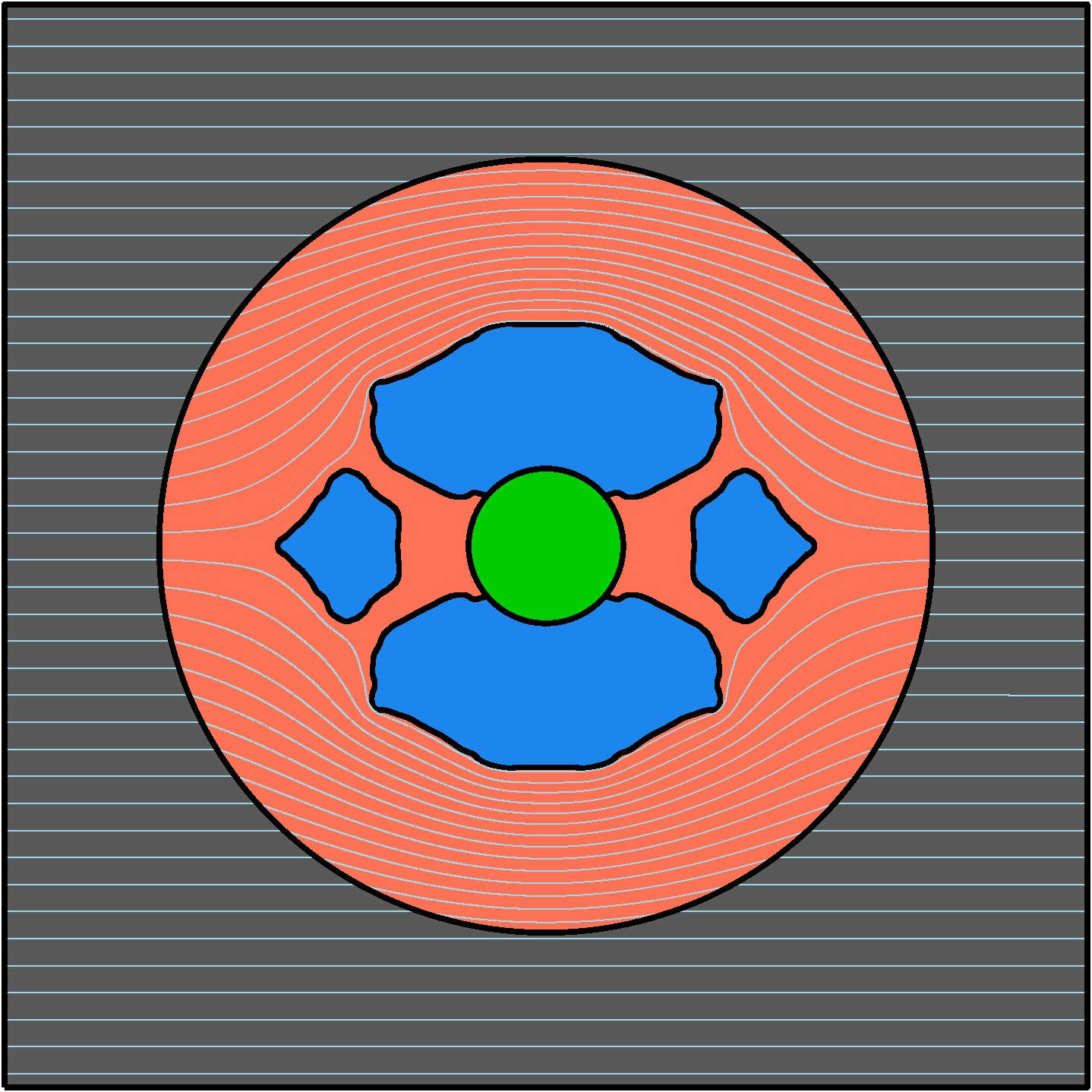}}
        \caption{\centering $J_{\rm cloak}=4.2448\times 10^{-8}$}
        \label{fig:chen2015case optTop VolSmth j}
    \end{subfigure} & \vspace{0.2cm}
    \begin{subfigure}[t]{0.15\textwidth}{\centering\includegraphics[width=1\textwidth]{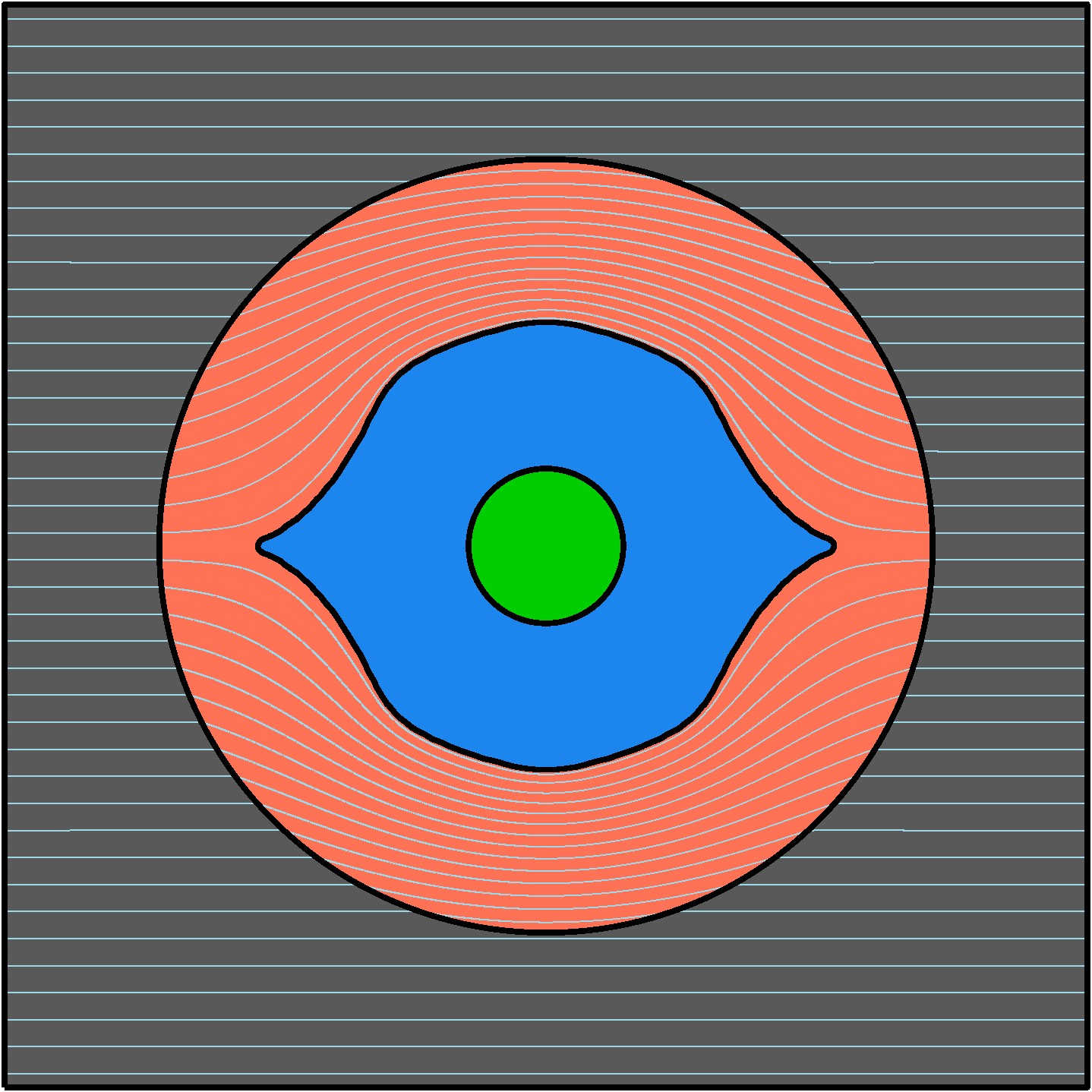}}
        \caption{\centering $J_{\rm cloak}=1.5302\times 10^{-7}$}
        \label{fig:chen2015case optTop VolSmth i}
    \end{subfigure}
    \\ 
\hline
   \rotatebox{90}{\centering \small Sample III} &
   \vspace{0.2cm}
   \begin{subfigure}[t]{0.15\textwidth}{\centering\includegraphics[width=1\textwidth]{Figures_Chen2015cloak/Laplace_HT_LevelSetTop_Chen2015case_objT_2_ref_p00k33h00_DSNref_p00k22h00_sample2_inB_initTop.jpg}}
        \caption{\centering Initial topology}
        \label{fig:chen2015case optTop VolSmth m}
    \end{subfigure}   & \vspace{0.2cm}
   \begin{subfigure}[t]{0.15\textwidth}{\centering\includegraphics[width=1\textwidth]{Figures_Chen2015cloak/Laplace_HT_LevelSetTop_Chen2015case_objT_2_ref_p00k00h00_DSNref_p00k55h00_sample2_inB_fluxPlot.jpg}}
        \caption{\centering $J_{\rm cloak}=3.9496\times 10^{-9}$}
        \label{fig:chen2015case optTop VolSmth n}
    \end{subfigure} &\vspace{0.2cm}
    \begin{subfigure}[t]{0.15\textwidth}{\centering\includegraphics[width=1\textwidth]{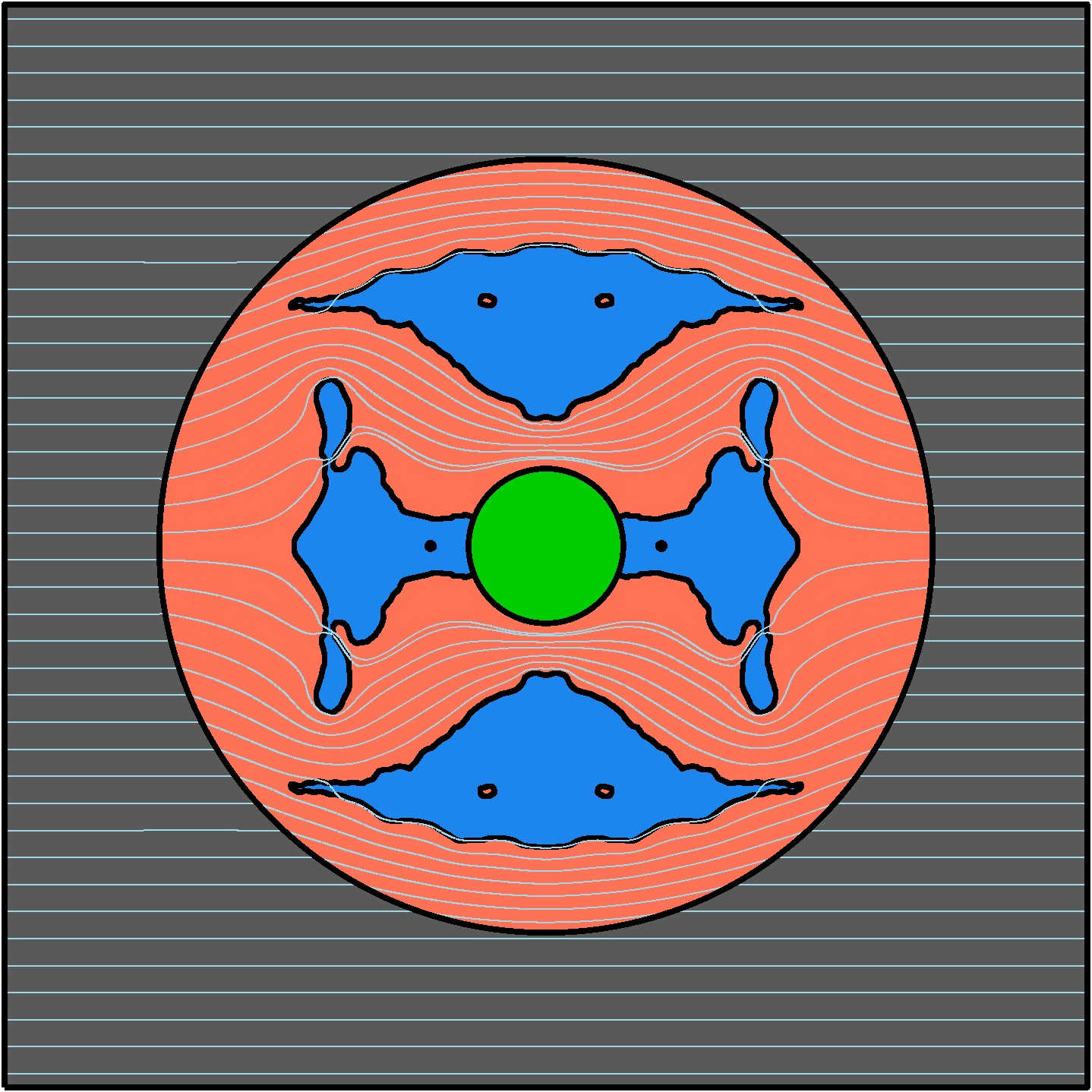}}
        \caption{\centering $J_{\rm cloak}=2.8493\times 10^{-8}$}
        \label{fig:chen2015case optTop VolSmth r}
    \end{subfigure}& \vspace{0.2cm}
    \begin{subfigure}[t]{0.15\textwidth}{\centering\includegraphics[width=1\textwidth]{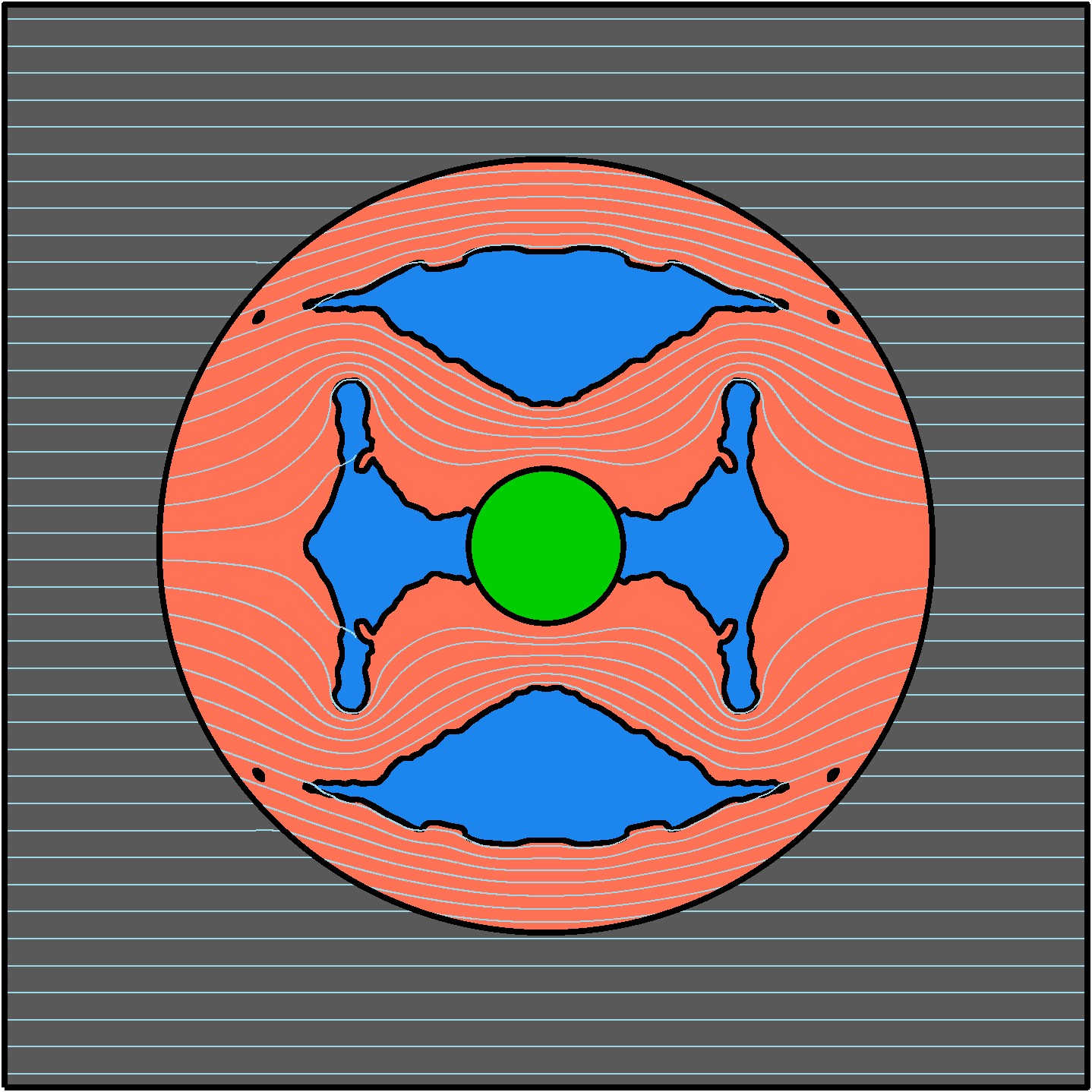}}
        \caption{\centering $J_{\rm cloak}=8.3947\times 10^{-9}$}
        \label{fig:chen2015case optTop VolSmth q}
    \end{subfigure} & \vspace{0.2cm}
    \begin{subfigure}[t]{0.15\textwidth}{\centering\includegraphics[width=1\textwidth]{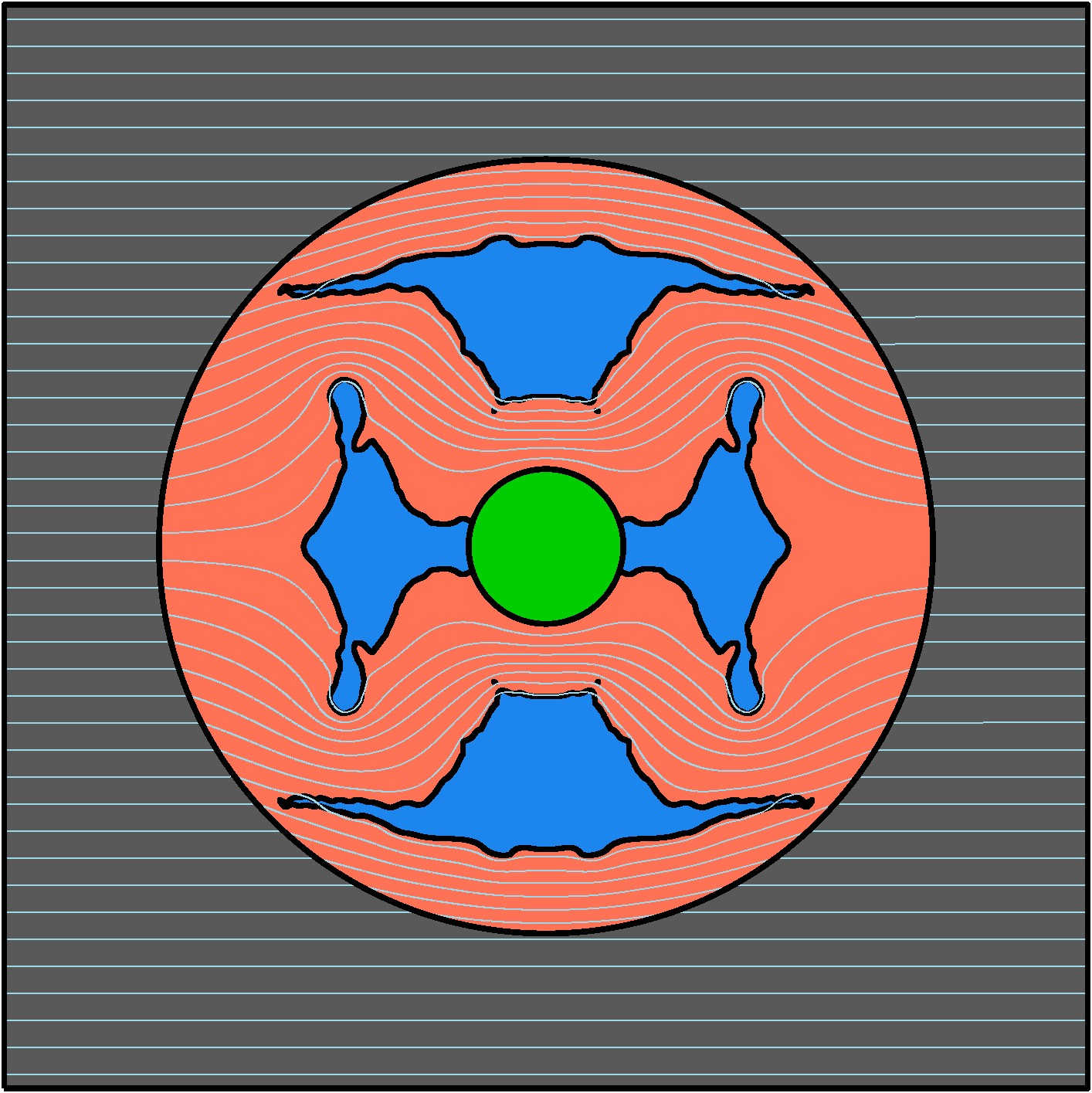}}
        \caption{\centering $J_{\rm cloak}=3.2274\times 10^{-7}$}
        \label{fig:chen2015case optTop VolSmth p}
    \end{subfigure}& \vspace{0.2cm}
    \begin{subfigure}[t]{0.15\textwidth}{\centering\includegraphics[width=1\textwidth]{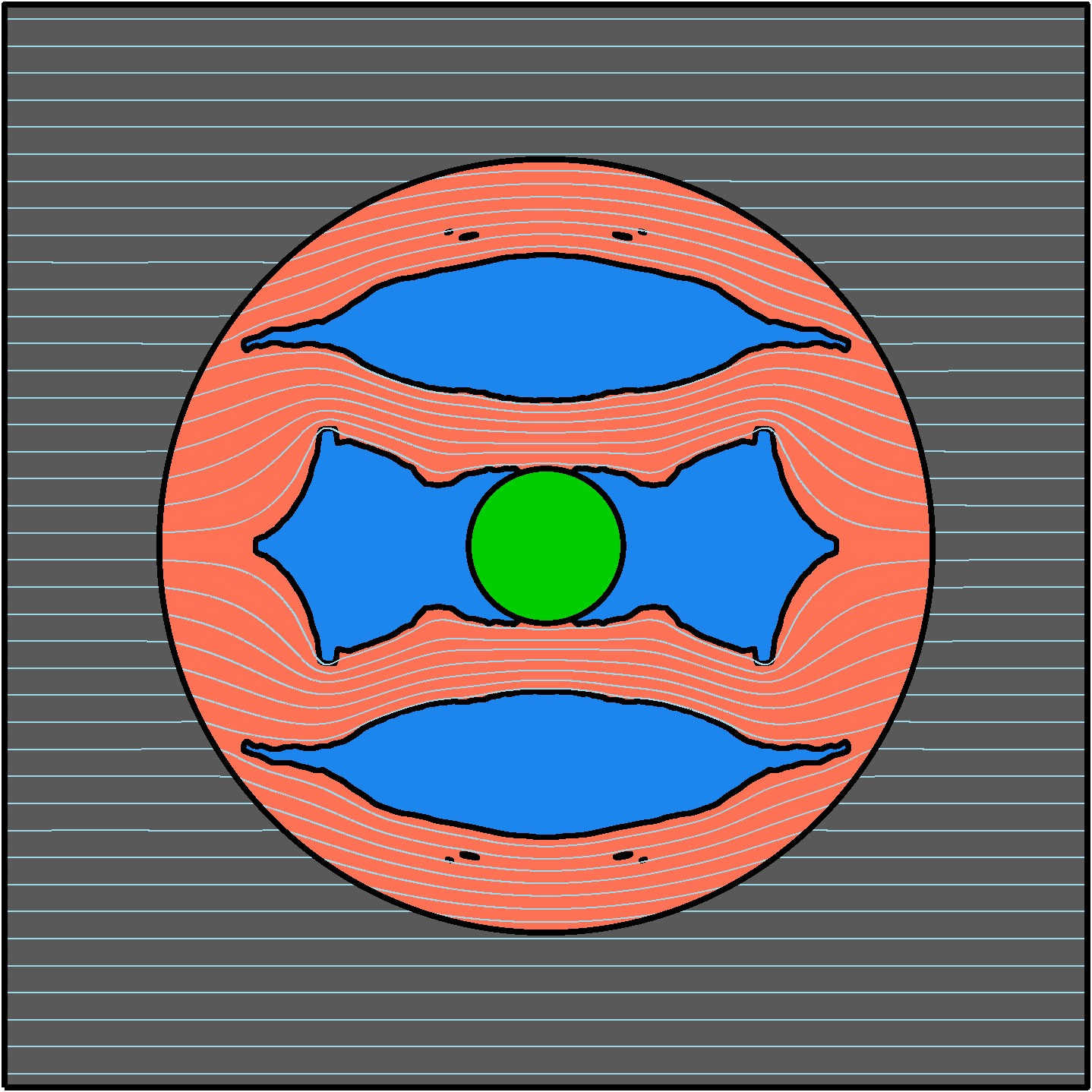}}
        \caption{\centering $J_{\rm cloak}=1.6057\times 10^{-6}$}
        \label{fig:chen2015case optTop VolSmth o}
    \end{subfigure}
    \\
    \hline
\end{tabular}

}
\caption{For the thermal cloak problem, initial topologies, optimized topologies without any regularization and with volume regularization for $N_{\rm var}=1089$ with $\Delta=0.0005$. Three initial topologies (samples I, II and III) are discretized with the corresponding design basis. Four values of weighing parameter $\rho$ are considered. Volume regularization fills the zero-flux area with PDMS material with a slight compromise on the $J_{\rm cloak}$-values.}  
    \label{fig:chen2015case optTop VolSmth}
\end{figure}

\newcolumntype{U}{>{\centering\arraybackslash}m{5.6em}}
\newcolumntype{V}{>{\centering\arraybackslash}m{5.3em}}
\newcolumntype{W}{>{\centering\arraybackslash}m{0.6em}}
\renewcommand{\arraystretch}{1.5}   
\begin{figure}
\centering
\scalebox{0.9}{
\begin{tabular}[c]{| W | U | U | V | V | V | V |}
\hline		
 & \multirow{2}{5.5em}{\centering Initial topology} & \multirow{2}{5.5em}{\centering W/o any regularization $J_{\rm total}=J_{\rm cloak}$} & \multicolumn{4}{c|}{Tikhonov +Volume reg. $J_{\rm total}=J_{\rm cloak} + \chi J_{\rm Tknv} + \rho J_{\rm vol}$ }\\
\cline{4-7}		
  &  & & \centering
     Set-A $\chi=1 \times10^{-4}$ $\rho=1\times10^{-4}$ & \centering Set-B $\chi=1 \times10^{-3}$ $\rho=1\times10^{-4}$ & \centering Set-C $\chi=1 \times10^{-4}$ $\rho=1\times10^{-2}$ & {\centering Set-D \newline$\chi=1 \times10^{-3}$ $\rho=1\times10^{-2}$}\\
\hline 
\vspace{0.2cm}
   \rotatebox{90}{\centering \small Sample I}  &
   \vspace{0.2cm}
   \begin{subfigure}[t]{0.15\textwidth}{\centering\includegraphics[width=1\textwidth]{Figures_Chen2015cloak/Laplace_HT_LevelSetTop_Chen2015case_objT_2_ref_p00k33h00_DSNref_p00k22h00_sample1_inA_initTop.jpg}}
        \caption{\centering Initial topology}
         \label{fig:chen2015case optTop TVSmth a}
    \end{subfigure}  & \vspace{0.2cm}
   \begin{subfigure}[t]{0.15\textwidth}{\centering\includegraphics[width=1\textwidth]{Figures_Chen2015cloak/Laplace_HT_LevelSetTop_Chen2015case_objT_2_ref_p00k00h00_DSNref_p00k55h00_sample1_inA_fluxPlot.jpg}}
        \caption{\centering $J_{\rm cloak}=9.4025\times 10^{-10}$}
         \label{fig:chen2015case optTop TVSmth b}
    \end{subfigure} & \vspace{0.2cm}
    \begin{subfigure}[t]{0.15\textwidth}{\centering\includegraphics[width=1\textwidth]{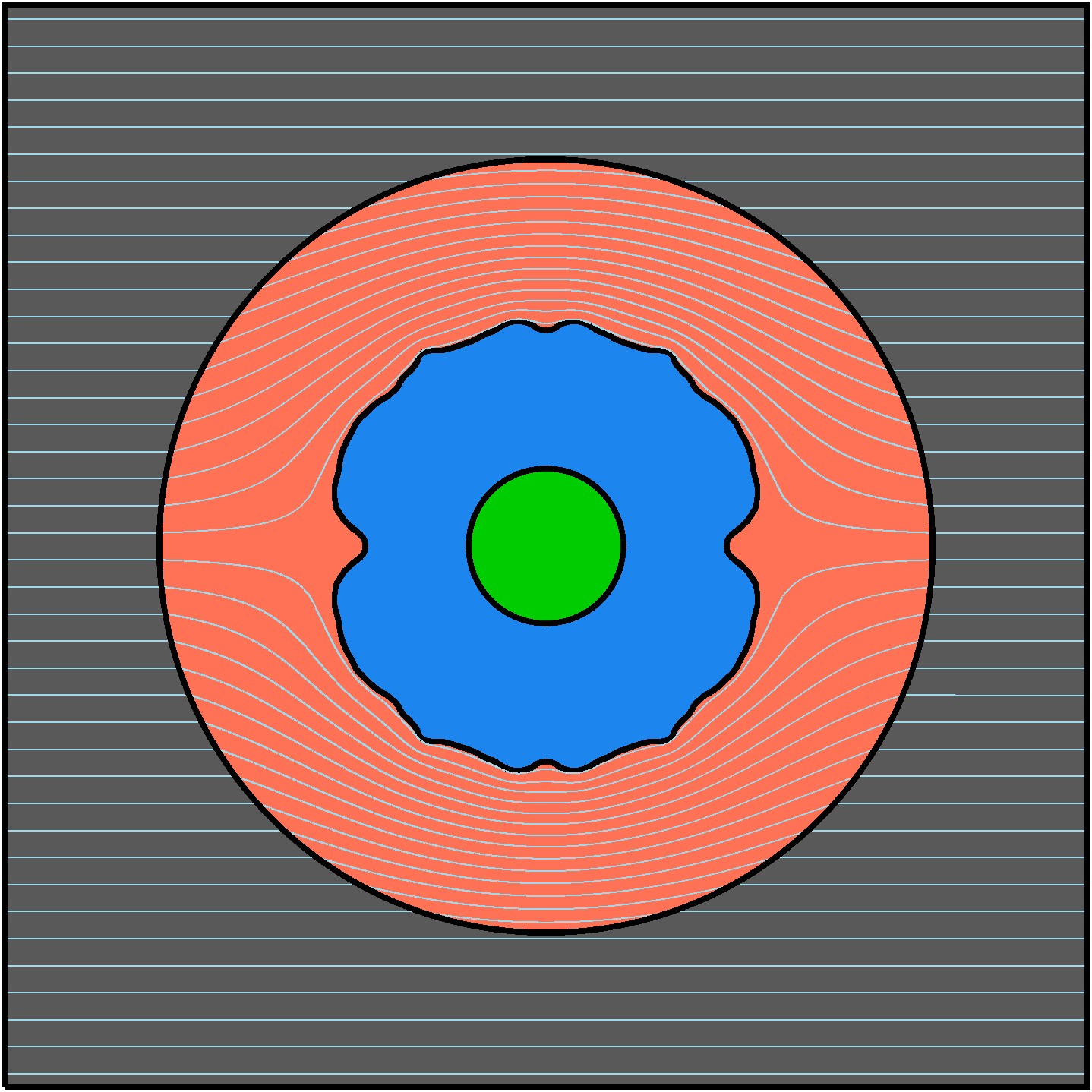}}
        \caption{\centering $J_{\rm cloak}=4.7079\times 10^{-9}$}
         \label{fig:chen2015case optTop TVSmth c}
    \end{subfigure} &\vspace{0.2cm}
    \begin{subfigure}[t]{0.15\textwidth}{\centering\includegraphics[width=1\textwidth]{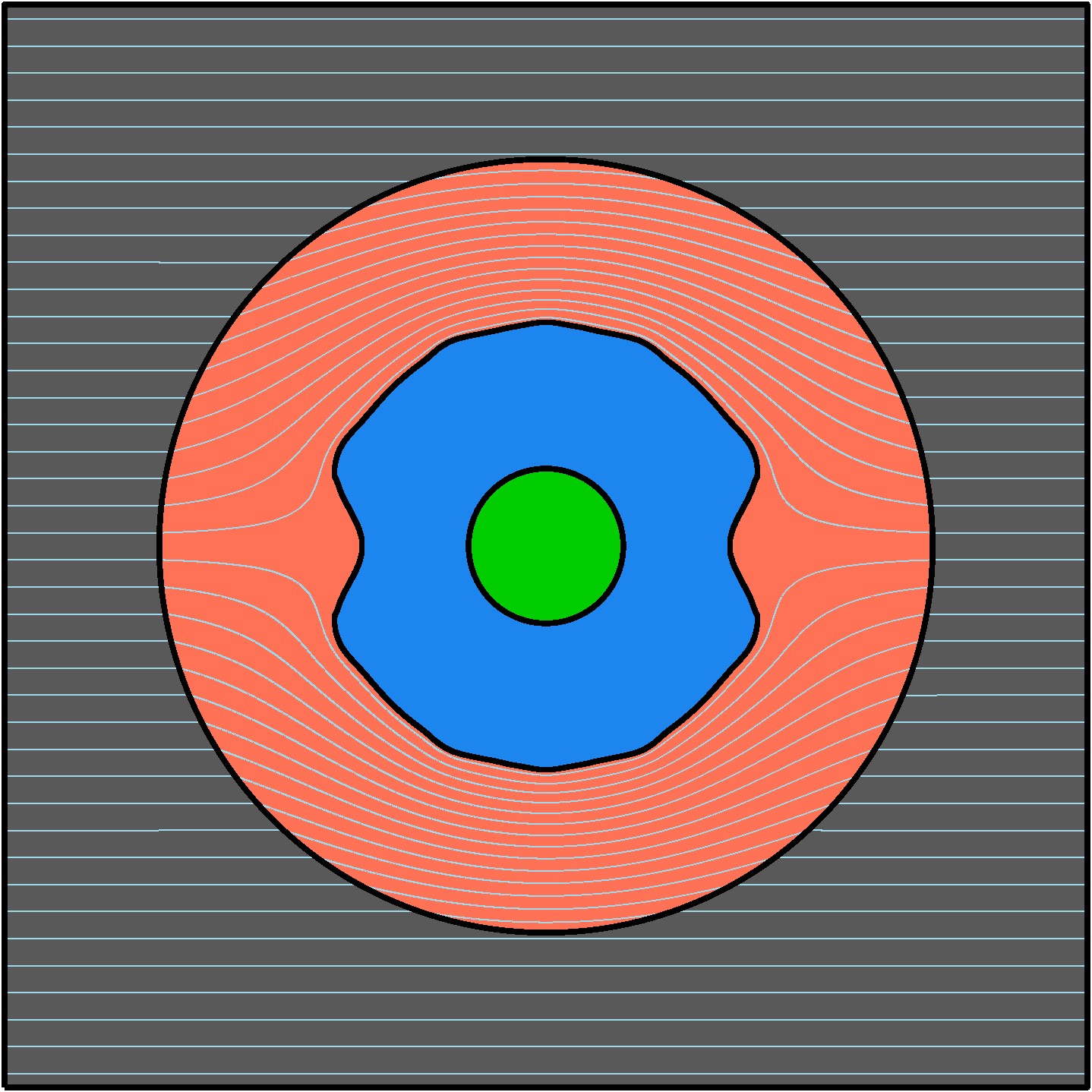}}
        \caption{\centering $J_{\rm cloak}=6.6095\times 10^{-9}$}
         \label{fig:chen2015case optTop TVSmth d}
    \end{subfigure}& \vspace{0.2cm}
    \begin{subfigure}[t]{0.15\textwidth}{\centering\includegraphics[width=1\textwidth]{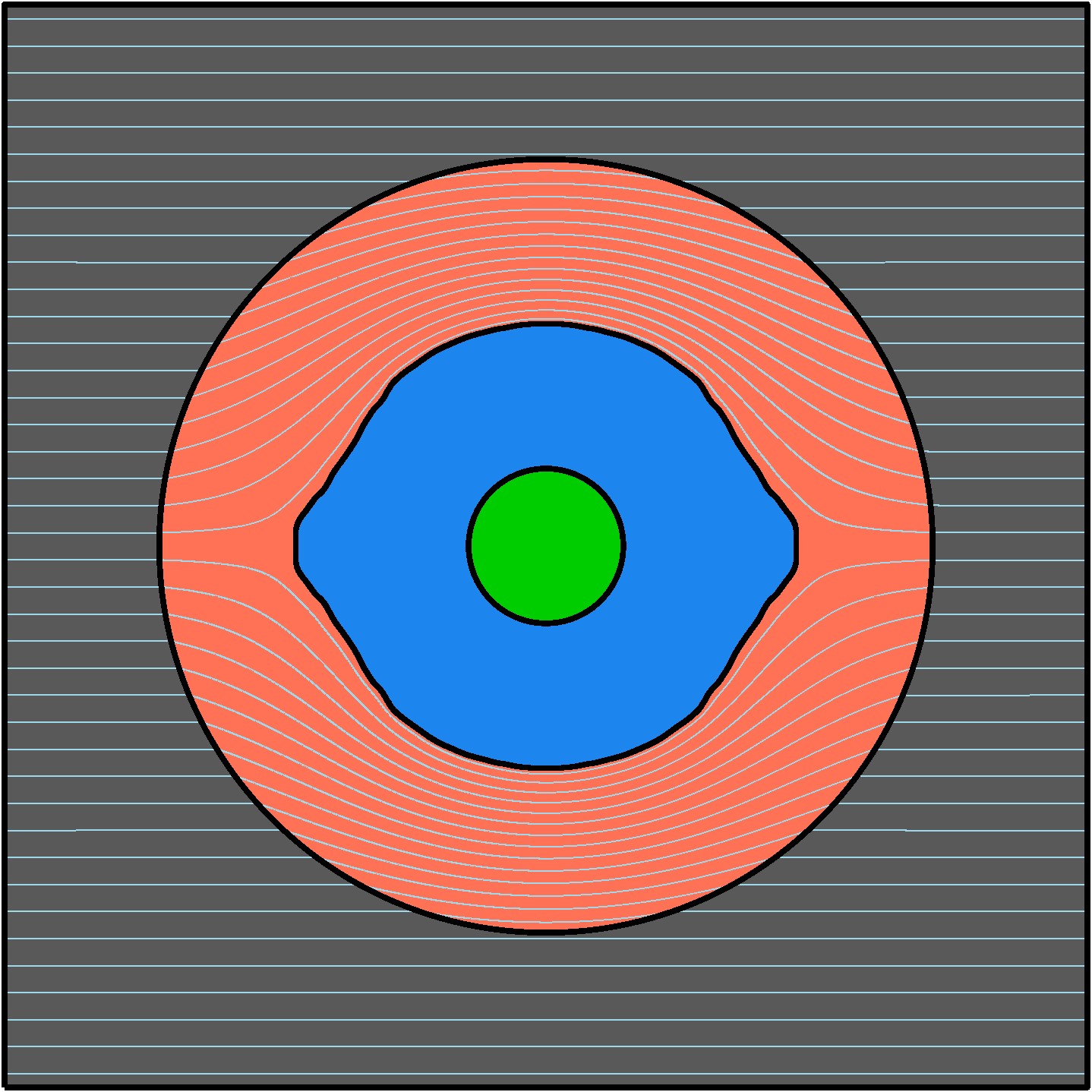}}
        \caption{\centering $J_{\rm cloak}=1.3305\times 10^{-7}$}
         \label{fig:chen2015case optTop TVSmth e}
    \end{subfigure}&  \vspace{0.2cm}
    \begin{subfigure}[t]{0.15\textwidth}{\centering\includegraphics[width=1\textwidth]{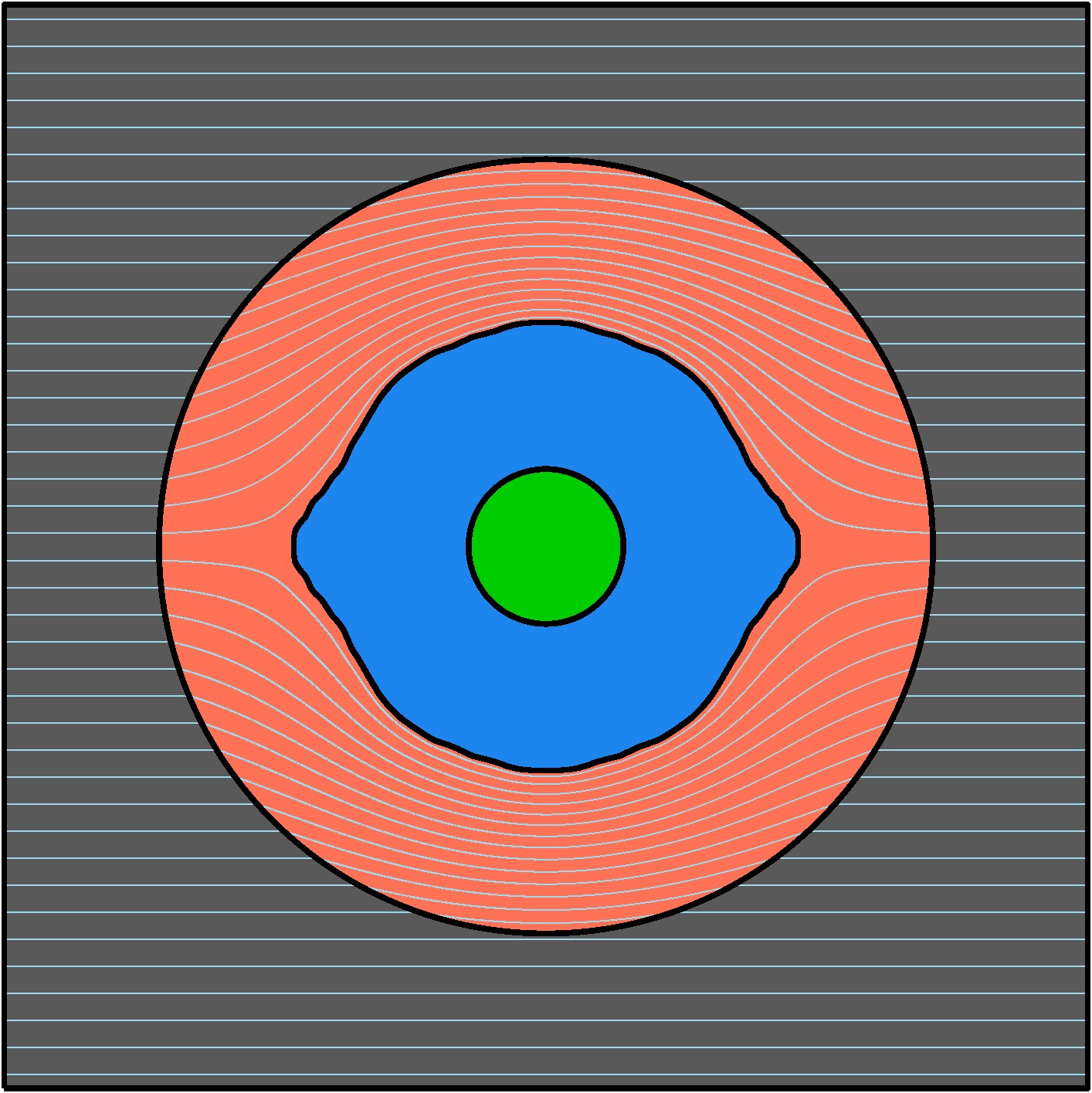}}
        \caption{\centering $J_{\rm cloak}=1.0838\times 10^{-7}$}
         \label{fig:chen2015case optTop TVSmth f}
    \end{subfigure}\\
    \hline
   \rotatebox{90}{\centering \small Sample II}  &
   \vspace{0.2cm}
   \begin{subfigure}[t]{0.15\textwidth}{\centering\includegraphics[width=1\textwidth]{Figures_Chen2015cloak/Laplace_HT_LevelSetTop_Chen2015case_objT_2_ref_p00k33h00_DSNref_p00k22h00_sample1_inB_initTop.jpg}}
        \caption{\centering Initial topology}
         \label{fig:chen2015case optTop TVSmth g}
    \end{subfigure}  & \vspace{0.2cm}
   \begin{subfigure}[t]{0.15\textwidth}{\centering\includegraphics[width=1\textwidth]{Figures_Chen2015cloak/Laplace_HT_LevelSetTop_Chen2015case_objT_2_ref_p00k00h00_DSNref_p00k55h00_sample1_inB_fluxPlot.jpg}}
        \caption{\centering $J_{\rm cloak}=1.6808\times 10^{-9}$}
         \label{fig:chen2015case optTop TVSmth h}
    \end{subfigure} & \vspace{0.2cm}
    \begin{subfigure}[t]{0.15\textwidth}{\centering\includegraphics[width=1\textwidth]{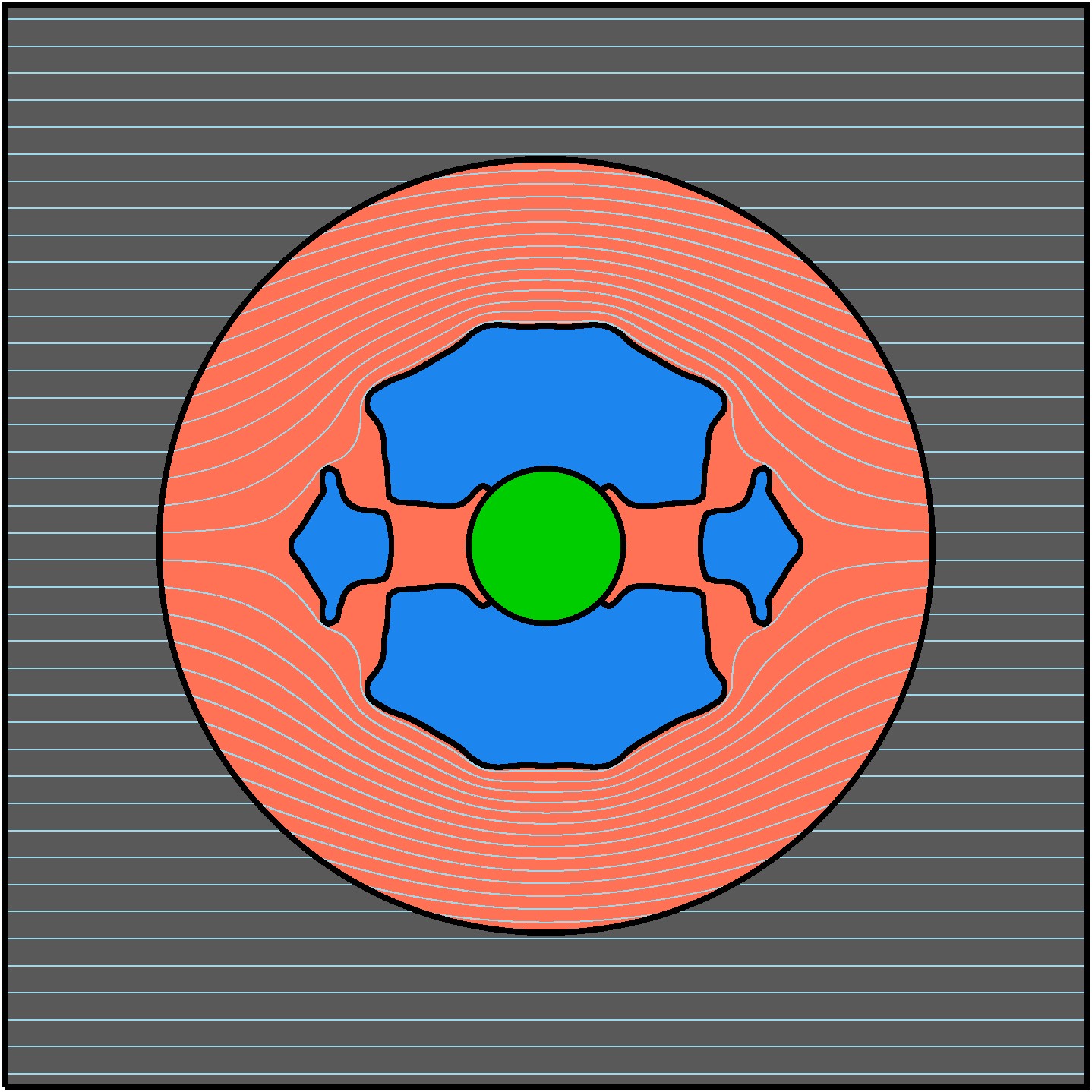}}
        \caption{\centering $J_{\rm cloak}=1.0086\times 10^{-8}$}
         \label{fig:chen2015case optTop TVSmth i}
    \end{subfigure} &\vspace{0.2cm}
    \begin{subfigure}[t]{0.15\textwidth}{\centering\includegraphics[width=1\textwidth]{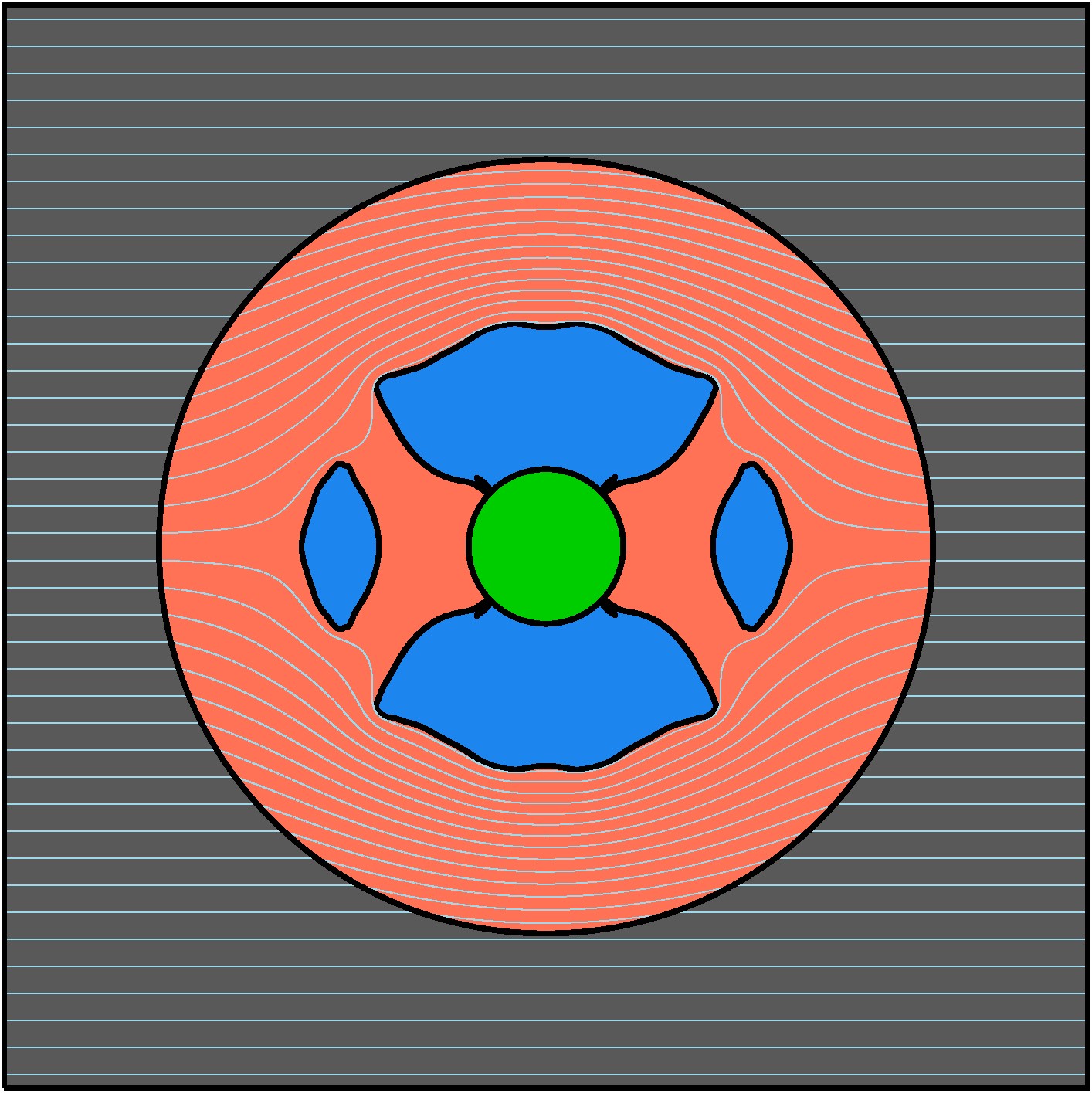}}
        \caption{\centering $J_{\rm cloak}=6.2857\times 10^{-8}$}
         \label{fig:chen2015case optTop TVSmth j}
    \end{subfigure}& \vspace{0.2cm}
    \begin{subfigure}[t]{0.15\textwidth}{\centering\includegraphics[width=1\textwidth]{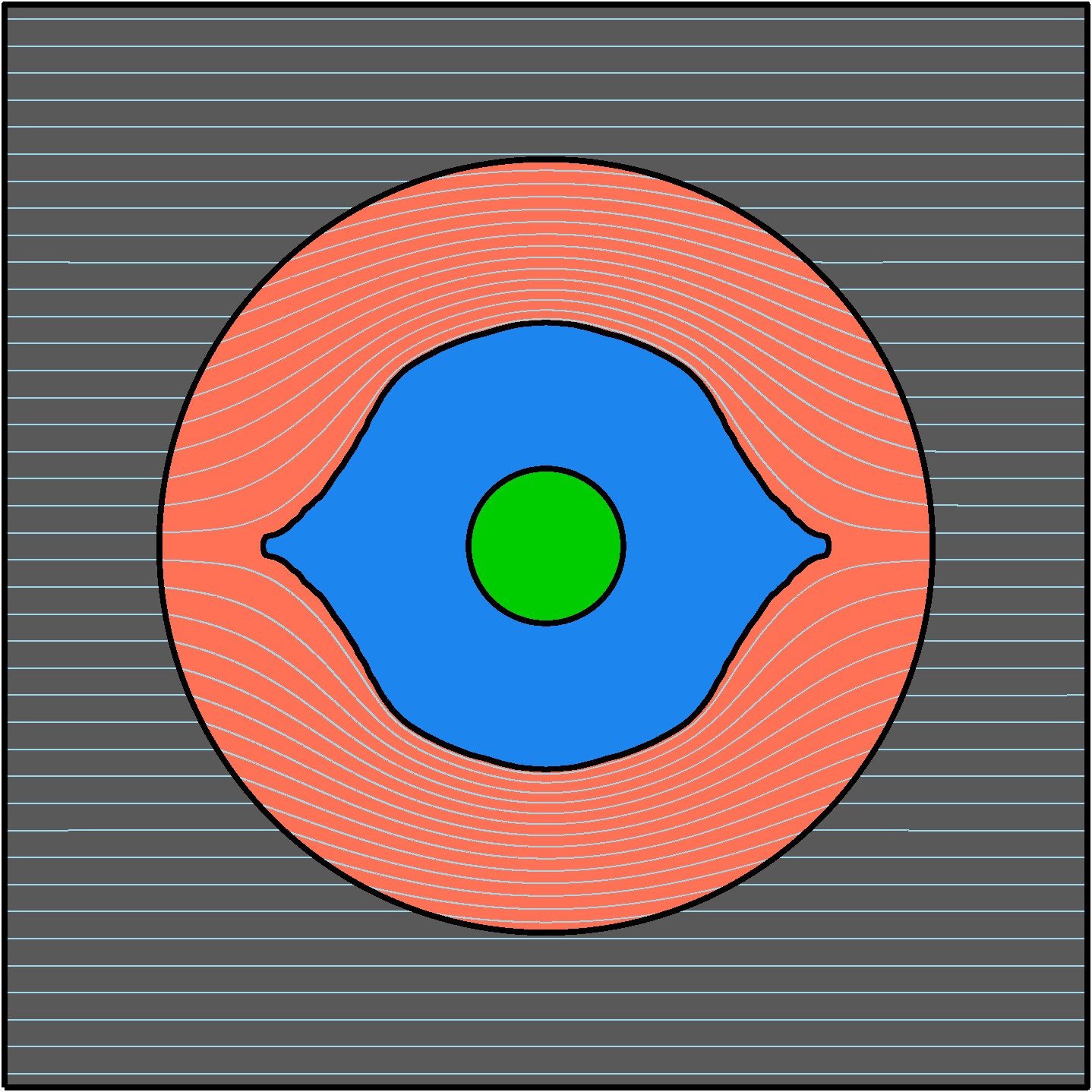}}
        \caption{\centering $J_{\rm cloak}=1.5262\times 10^{-7}$}
         \label{fig:chen2015case optTop TVSmth k}
    \end{subfigure}& \vspace{0.2cm}
    \begin{subfigure}[t]{0.15\textwidth}{\centering\includegraphics[width=1\textwidth]{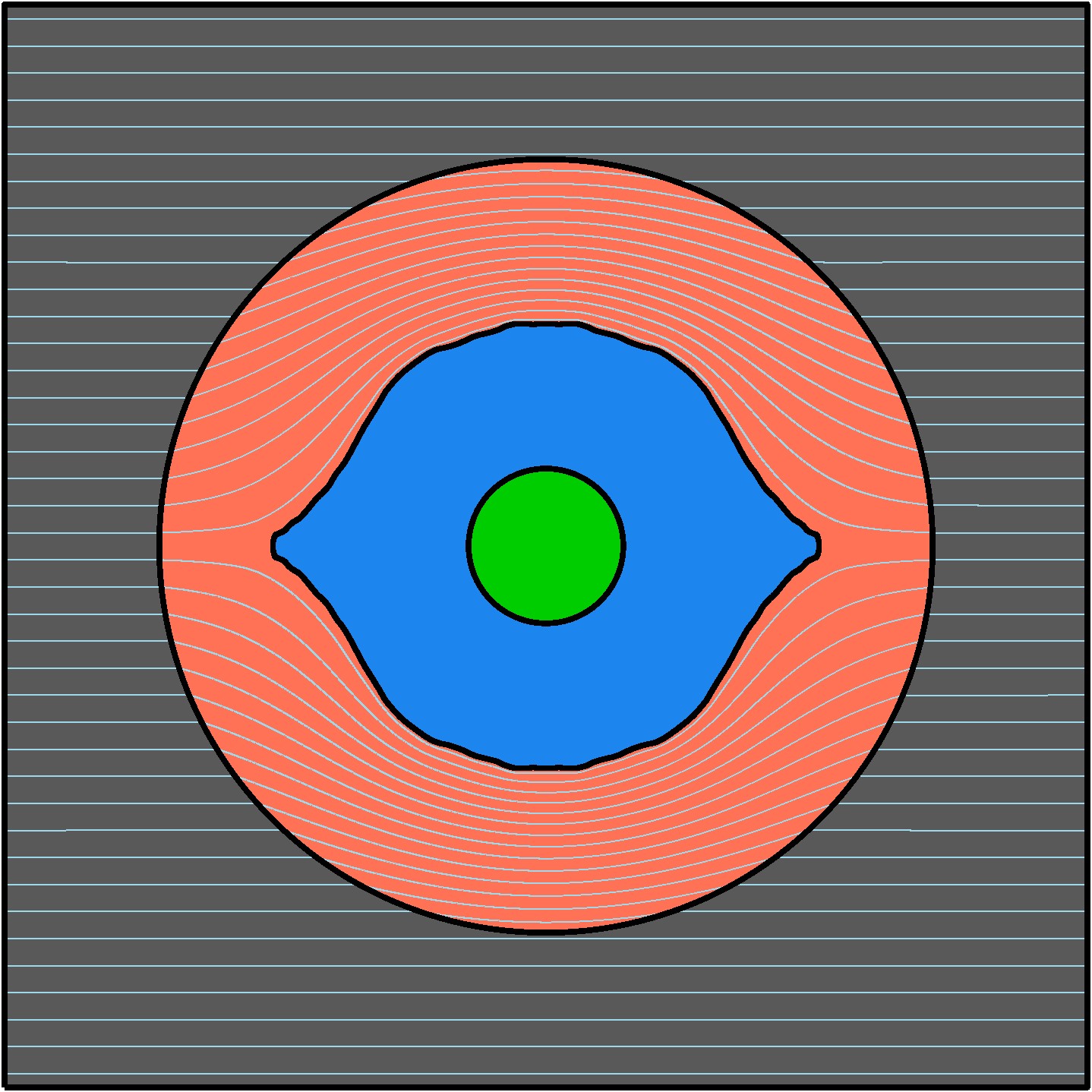}}
        \caption{\centering $J_{\rm cloak}=1.3121\times 10^{-7}$}
         \label{fig:chen2015case optTop TVSmth l}
    \end{subfigure}\\ 
\hline
   \rotatebox{90}{\centering \small Sample III} &
   \vspace{0.2cm}
   \begin{subfigure}[t]{0.15\textwidth}{\centering\includegraphics[width=1\textwidth]{Figures_Chen2015cloak/Laplace_HT_LevelSetTop_Chen2015case_objT_2_ref_p00k33h00_DSNref_p00k22h00_sample2_inB_initTop.jpg}}
        \caption{\centering Initial topology}
         \label{fig:chen2015case optTop TVSmth m}
    \end{subfigure}   & \vspace{0.2cm}
   \begin{subfigure}[t]{0.15\textwidth}{\centering\includegraphics[width=1\textwidth]{Figures_Chen2015cloak/Laplace_HT_LevelSetTop_Chen2015case_objT_2_ref_p00k00h00_DSNref_p00k55h00_sample2_inB_fluxPlot.jpg}}
        \caption{\centering $J_{\rm cloak}=3.9496\times 10^{-9}$}
         \label{fig:chen2015case optTop TVSmth n}
    \end{subfigure} & \vspace{0.2cm}
    \begin{subfigure}[t]{0.15\textwidth}{\centering\includegraphics[width=1\textwidth]{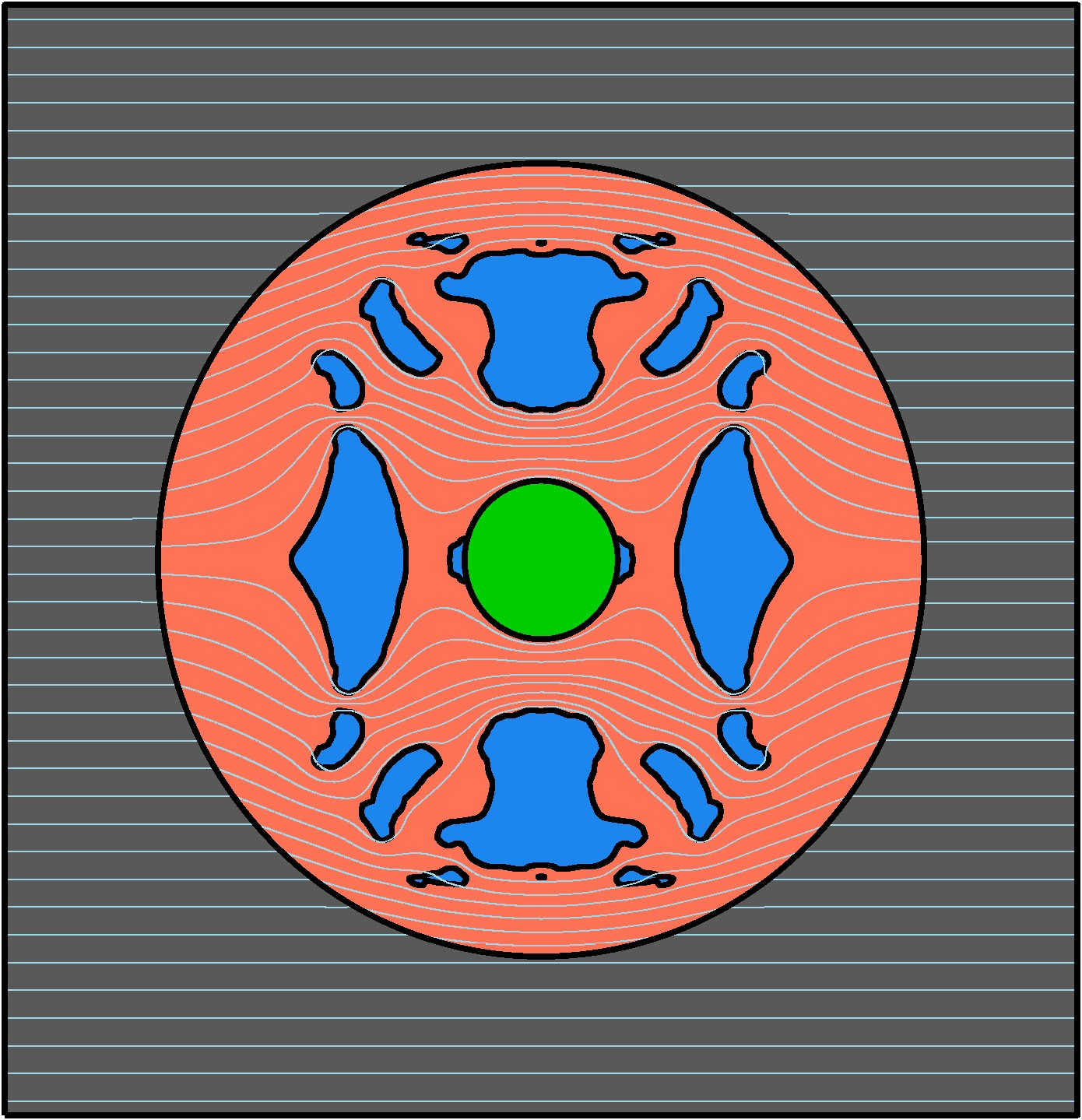}}
        \caption{\centering $J_{\rm cloak}=4.1915\times 10^{-8}$}
         \label{fig:chen2015case optTop TVSmth o}
    \end{subfigure}&\vspace{0.2cm}
    \begin{subfigure}[t]{0.15\textwidth}{\centering\includegraphics[width=1\textwidth]{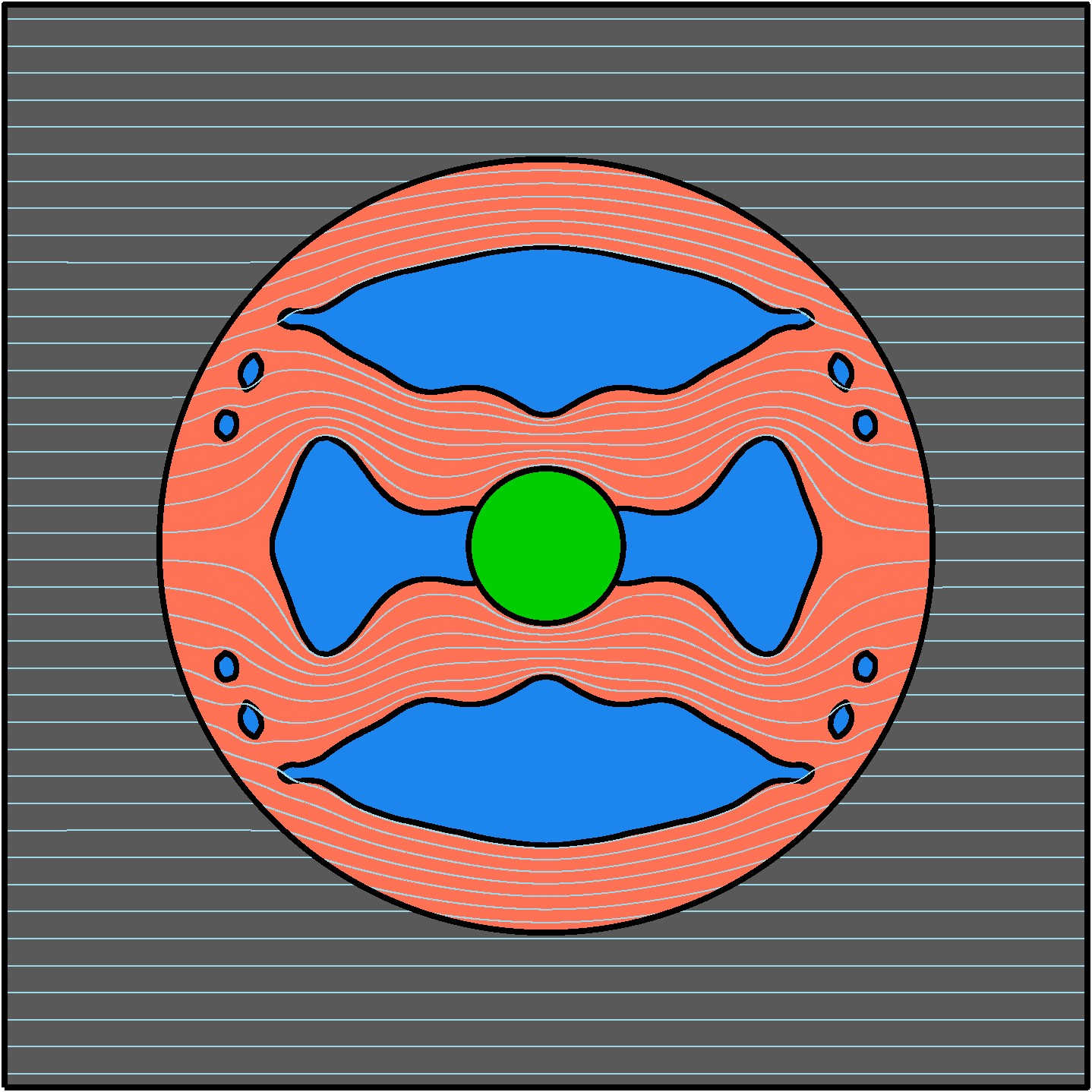}}
        \caption{\centering $J_{\rm cloak}=2.1605\times 10^{-7}$}
         \label{fig:chen2015case optTop TVSmth p}
    \end{subfigure}& \vspace{0.2cm}
    \begin{subfigure}[t]{0.15\textwidth}{\centering\includegraphics[width=1\textwidth]{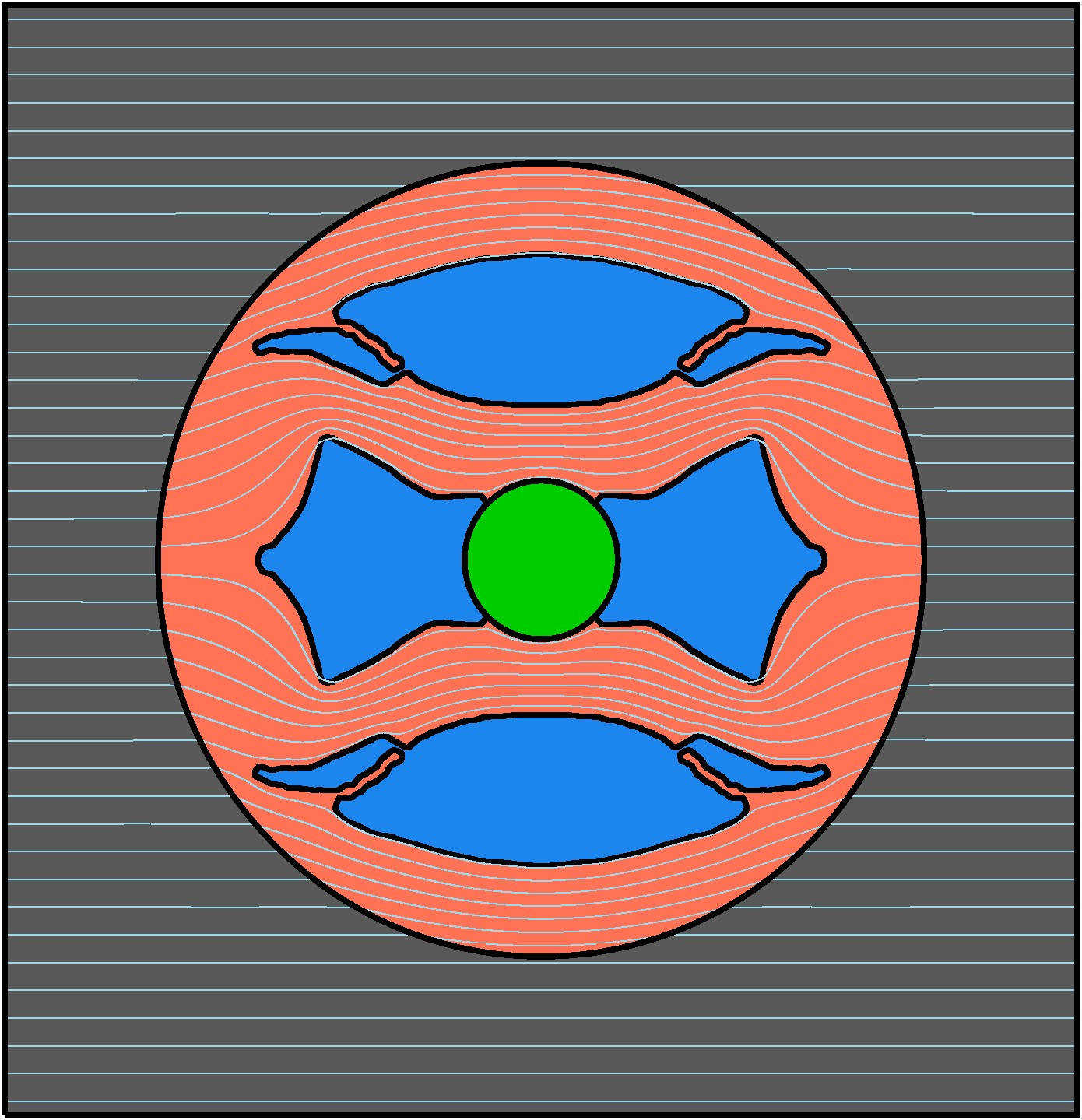}}
        \caption{\centering $J_{\rm cloak}=1.6656\times 10^{-6}$}
         \label{fig:chen2015case optTop TVSmth q}
    \end{subfigure} & \vspace{0.2cm}
    \begin{subfigure}[t]{0.15\textwidth}{\centering\includegraphics[width=1\textwidth]{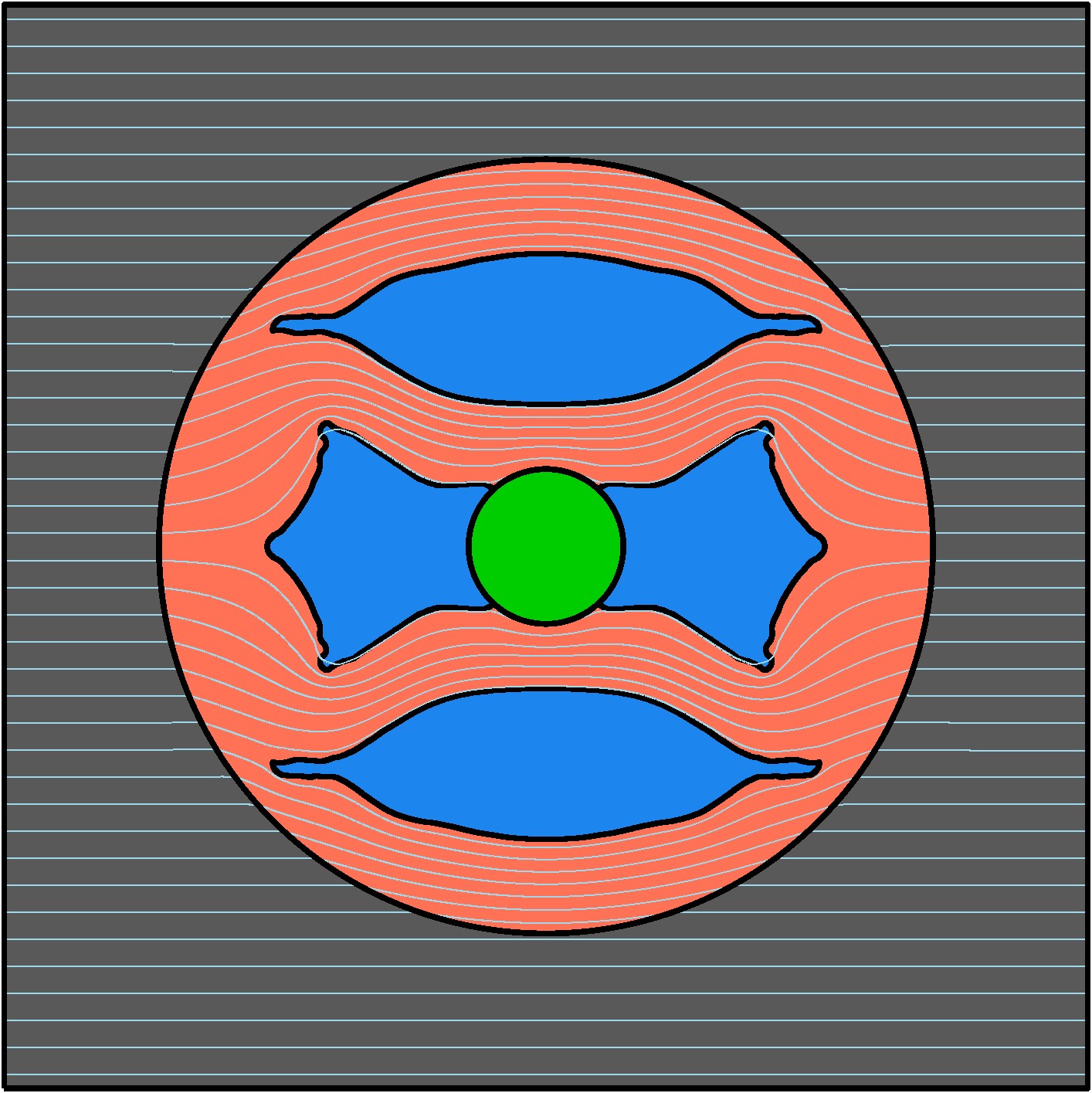}}
        \caption{\centering $J_{\rm cloak}=1.7590\times 10^{-6}$}
         \label{fig:chen2015case optTop TVSmth r}
    \end{subfigure}\\
    \hline
\end{tabular}

}
\caption{For the thermal cloak problem, initial topologies, optimized topologies  without any regularization and with combined Tikhonov and volume regularization for $N_{\rm var}=1089$ with $\Delta=0.0005$. Three initial topologies (samples I, II and III) are discretized with the corresponding design basis. Four sets of weighing parameters $\chi$ and $\rho$ are considered. Combined regularization can fill the zero-flux area with PDMS material as well as produce smoother topologies with a slight compromise on the $J_{\rm cloak}$-values.}  
    \label{fig:chen2015case optTop TVSmth}
\end{figure}


\par In this paragraph, we discuss how to avoid unnecessary complex features in optimized topologies as shown in \frefs{fig:chen2015case optTop f}, \ref{fig:chen2015case optTop l} and \ref{fig:chen2015case optTop r}. The goal is to use regularization techniques to produce smooth and less-complex topologies. First, we explore the Tikhonov regularization. With the Tikhonov regularization, the total objective function $J_{\rm total}= J_{\mathrm{cloak}} + \chi J_{\mathrm{Tknv}}$.
The optimization is performed with four values of $\chi$: $10^{-5}$, $10^{-4}$, $10^{-3}$, and $10^{-2}$. The optimization results are presented in \fref{fig:chen2015case optTop TknvSmth}. For $\chi=10^{-5}$, $10^{-4}$, the effect of regularization is very minor, and consequently, we can see that some of the small features are still present. For larger values of $\chi$ ($10^{-3}$, $10^{-2}$), the optimized topologies are smoother and with fewer complex features, as expected. However, regularization provides a slight constraint on the $J_{\mathrm{cloak}}$-values. The values, which are within the order of $10^{-7}$, are still capable of providing cloaking effect with enough accuracy.
\par Next, we explore the effect of another type of regularization, volume regularization. With the volume regularization, the total objective function $J_{\rm total}= J_{\mathrm{cloak}} + \rho J_{\mathrm{vol}}$.  Here, we use four values of $\rho$: $10^{-5}$, $10^{-4}$, $10^{-3}$ and $10^{-2}$. \fref{fig:chen2015case optTop VolSmth} shows the optimization results. From the figure, it is evident that volume regularization with higher $\rho$ fills the zero-flux area with PDMS material. For lower values of $\rho$, the effect is not dominant. Similarly to the last regularization, $J_{\mathrm{cloak}}$-values suffer up to some extent, however, still lie within order $10^{-6}$. Also, for sample II, the optimized topology without any regularization (as shown in \fref{fig:chen2015case optTop VolSmth h}) is quite different from the bilayer. However, after volume regularization, \frefs{fig:chen2015case optTop VolSmth i}-\ref{fig:chen2015case optTop VolSmth j} are close to the bilayer cloak.
\par Thereupon, we check the combination of these two regularizations mentioned in the last two paragraphs. We consider four combinations of $\chi=10^{-4}, 10^{-3}$ and $\rho=10^{-4}, 10^{-2}$, denoted by sets A to D. The optimization results are shown in \fref{fig:chen2015case optTop TVSmth}. For sample I, all sets give similar results with very negligible differences in topology. For samples II and III, sets C and D provide smoother topologies with smaller perimeters and less complex features compared to sets A and B. $J_{\mathrm{cloak}}$-values, however, are slightly higher for set-C and D. Therefore, we can say that $\chi$ and $\rho$ are decided based on trade-off between the smoothness of the topologies and fulfillment of the cloaking objective. However, it is very difficult to predict the exact values of $\chi$ and $\rho$ a priori and should be decided on the basis of the trial and error method. Other regularization techniques such as perimeter regularization and sensitivity smoothing can also be applied to get smoother geometries. Nonetheless, the issue of lack of apriori knowledge of the appropriate value of weighing parameters  remains. Another way to avoid complex topologies is by applying geometric constraints such as minimum length scale in the optimization problem. Geometry-constrained optimization, however, is beyond the scope of the current work, and it will be explored in future research.
\par

 \begin{figure}[!htbp]
    \centering
    \setlength\figureheight{1\textwidth}
    \setlength\figurewidth{1\textwidth}
    \begin{subfigure}[t]{0.30\textwidth}{\centering\includegraphics[width=1\textwidth]{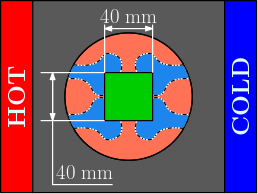}}
        \caption{\centering Config. I - square obstacle}
         \label{fig:Cloak obsctacle schematic a}
    \end{subfigure}\quad
     \begin{subfigure}[t]{0.30\textwidth}{\centering\includegraphics[width=1\textwidth]{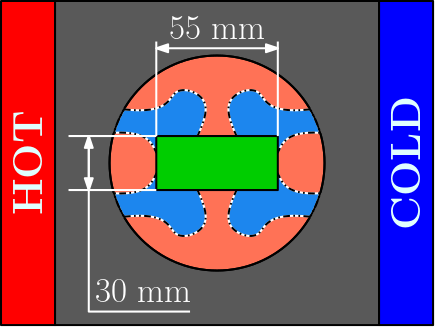}}
        \caption{\centering Config. II - horizontal rectangular obstacle}
         \label{fig:Cloak obsctacle schematic b}
    \end{subfigure}\quad
    \begin{subfigure}[t]{0.30\textwidth}{\centering\includegraphics[width=1\textwidth]{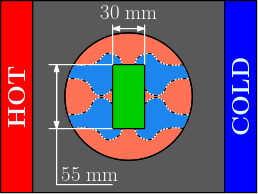}}
        \caption{\centering Config. III - vertical rectangular obstacle}
         \label{fig:Cloak obsctacle schematic c}
    \end{subfigure}\\
    \begin{subfigure}[t]{0.30\textwidth}{\centering\includegraphics[width=1\textwidth]{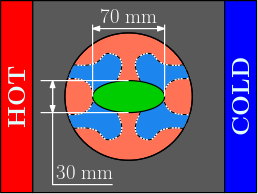}}
             \caption{\centering Config. IV - horizontal ellipsoidal obstacle}
              \label{fig:Cloak obsctacle schematic d}
    \end{subfigure} \quad
    \begin{subfigure}[t]{0.30\textwidth}{\centering\includegraphics[width=1\textwidth]{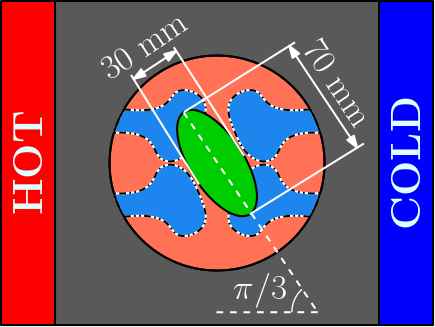}}
             \caption{\centering Config. V - inclined ellipsoidal obstacle}
              \label{fig:Cloak obsctacle schematic e}
    \end{subfigure}\\ 
    \begin{subfigure}[t]{0.30\textwidth}{\centering\includegraphics[width=1\textwidth]{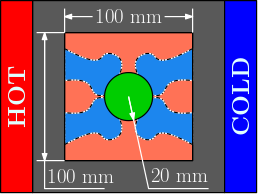}}
             \caption{\centering Config. VI - square cloak}
              \label{fig:Cloak obsctacle schematic f}
    \end{subfigure} \quad
    \begin{subfigure}[t]{0.30\textwidth}{\centering\includegraphics[width=1\textwidth]{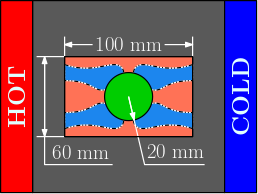}}
             \caption{\centering Config. VII - horintal rectangular cloak}
              \label{fig:Cloak obsctacle schematic g}
    \end{subfigure}\quad
    \begin{subfigure}[t]{0.30\textwidth}{\centering\includegraphics[width=1\textwidth]{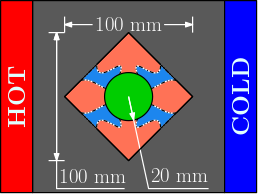}}
             \caption{\centering Config. VII - Inclined square cloak}
              \label{fig:Cloak obsctacle schematic h}
    \end{subfigure}
 \caption{Configurations and dimensions of different types geometries of a thermal cloak problem: Config. I - square obstacle, Config. II - horizontal rectangular obstacle, Config. III - vertical rectangular obstacle, Config. IV - horizontal ellipsoidal obstacle, Config. V - inclined ellipsoidal obstacle, Config. VI - square cloak, Config. VII - horizontal rectangular cloak, Config. VIII - inclined square cloak.}
 \label{fig:Cloak obsctacle schematic}
\end{figure}

\renewcommand{\arraystretch}{1.5}   
\begin{figure}
\centering
\scalebox{0.89}{
\begin{tabular}[c]{| m{5.6em} | m{5.6em}|m{5.6em}|| m{5.6em}| m{5.6em}| m{5.6em} |}
\hline	\vspace{-0.3cm}	
 \begin{center}
    Initial topology \vspace{-0.5cm}
 \end{center}  & \vspace{-0.3cm}	\begin{center}
    W/o any reg.  \vspace{-0.5cm}
 \end{center}  & \vspace{-0.3cm}	\begin{center}
    Volume + Tikhonov reg. \vspace{-0.5cm}
 \end{center}  & \vspace{-0.3cm}	\begin{center}
    Initial topology \vspace{-0.5cm}
 \end{center}  & \vspace{-0.3cm}	\begin{center}
    W/o any reg. \vspace{-0.5cm}
 \end{center}  & \vspace{-0.3cm}	\begin{center}
    Volume + Tikhonov reg. \vspace{-0.5cm}
 \end{center} \\
 \hline
\multicolumn{3}{|c||}{\small \hspace{30mm} Config. I \hspace{3mm}\hfill \scriptsize$(\chi,\rho)=(10^{-3},10^{-2})$} &  \multicolumn{3}{c|}{\small \hspace{28mm} Config. V \hfill \scriptsize$(\chi,\rho)=(10^{-2},10^{-4})$} \\
 \hline
   \vspace{0.2cm}
   \begin{subfigure}[t]{0.15\textwidth}{\centering\includegraphics[width=1\textwidth]{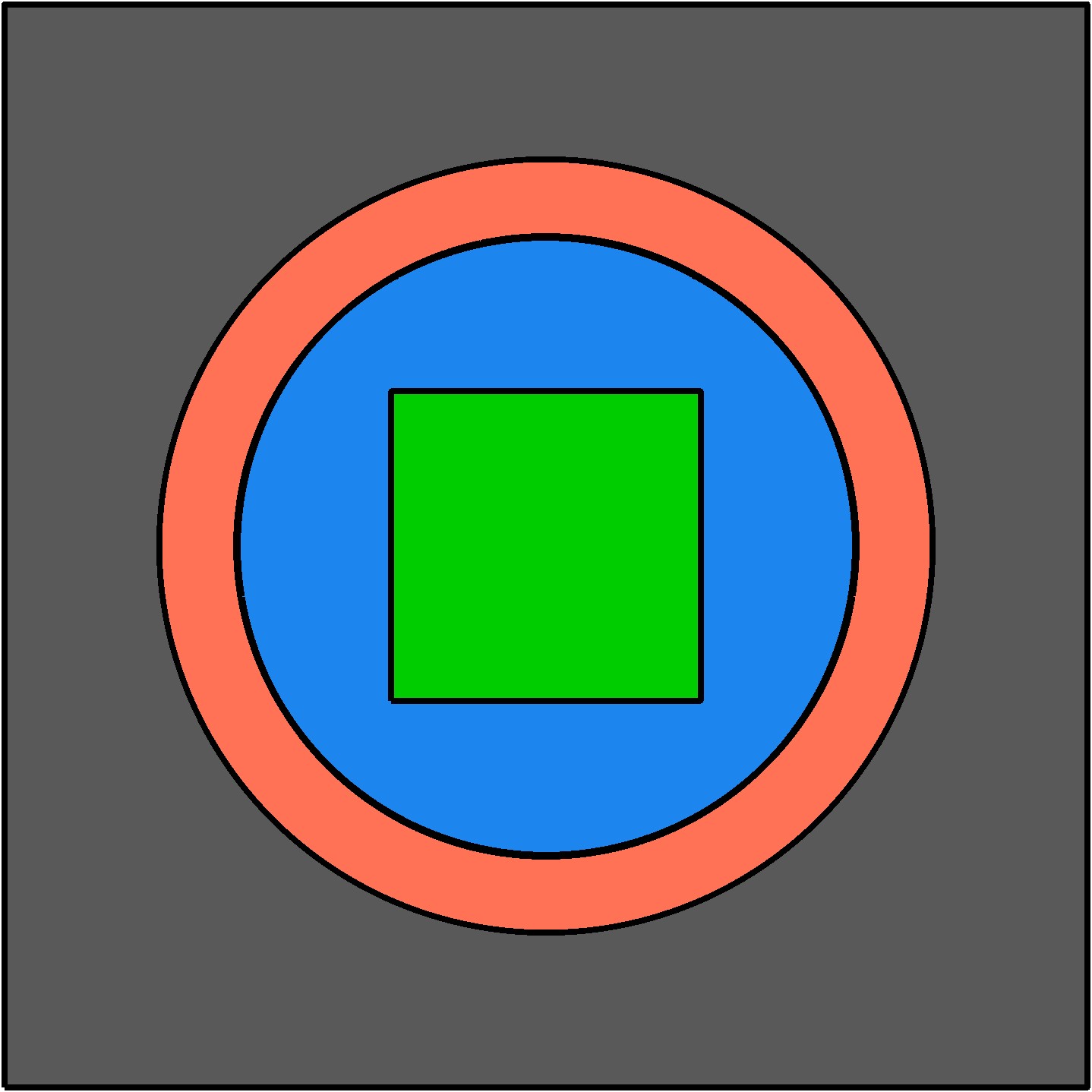}}
        \caption{\centering Initial topology}
       \label{fig:Mchen2015case optTop a}
    \end{subfigure}  & \vspace{0.2cm}
   \begin{subfigure}[t]{0.15\textwidth}{\centering\includegraphics[width=1\textwidth]{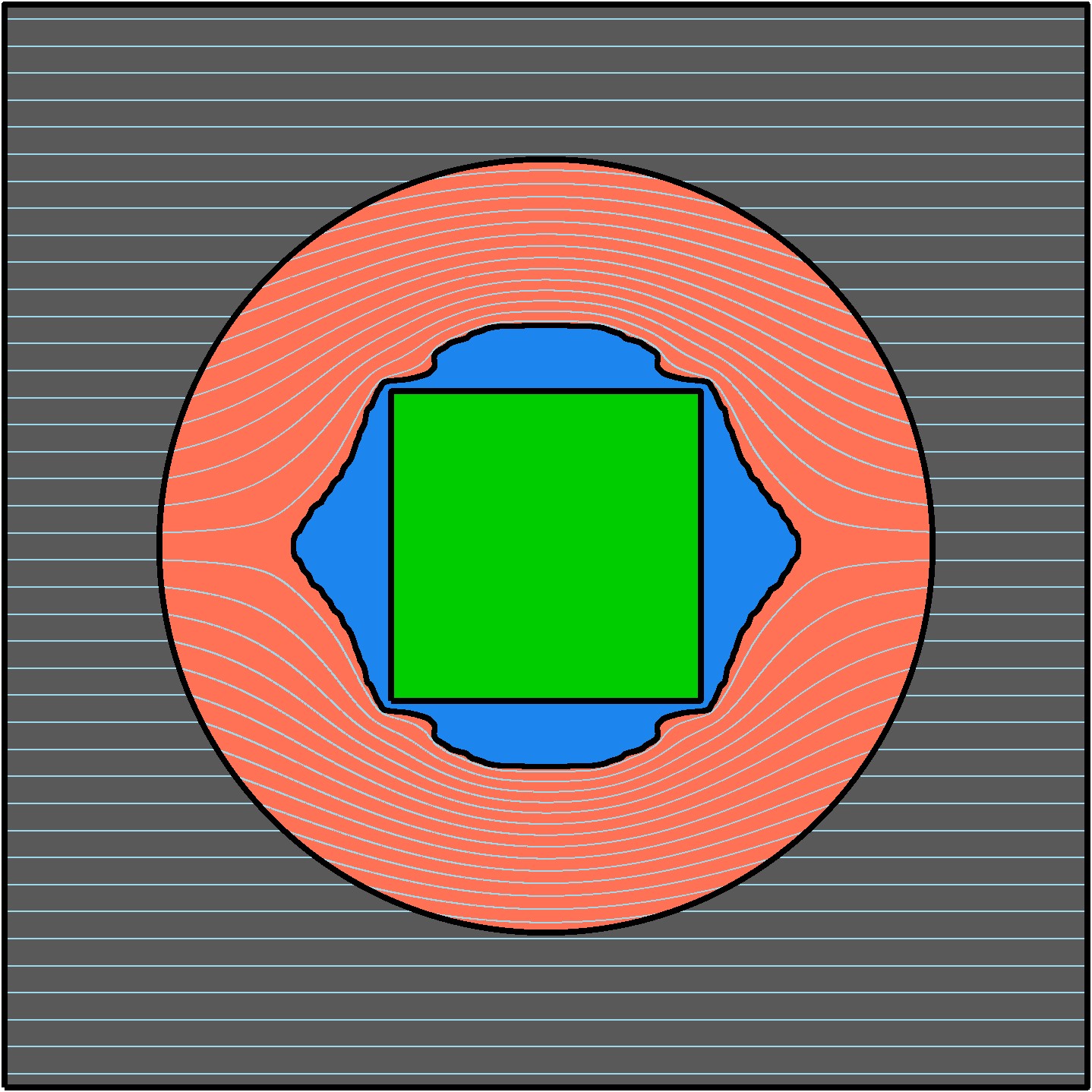}}
        \caption{\centering $J_{\rm cloak}=2.8844\times 10^{-9}$}
       \label{fig:Mchen2015case optTop b}
    \end{subfigure} & \vspace{0.2cm}
    \begin{subfigure}[t]{0.15\textwidth}{\centering\includegraphics[width=1\textwidth]{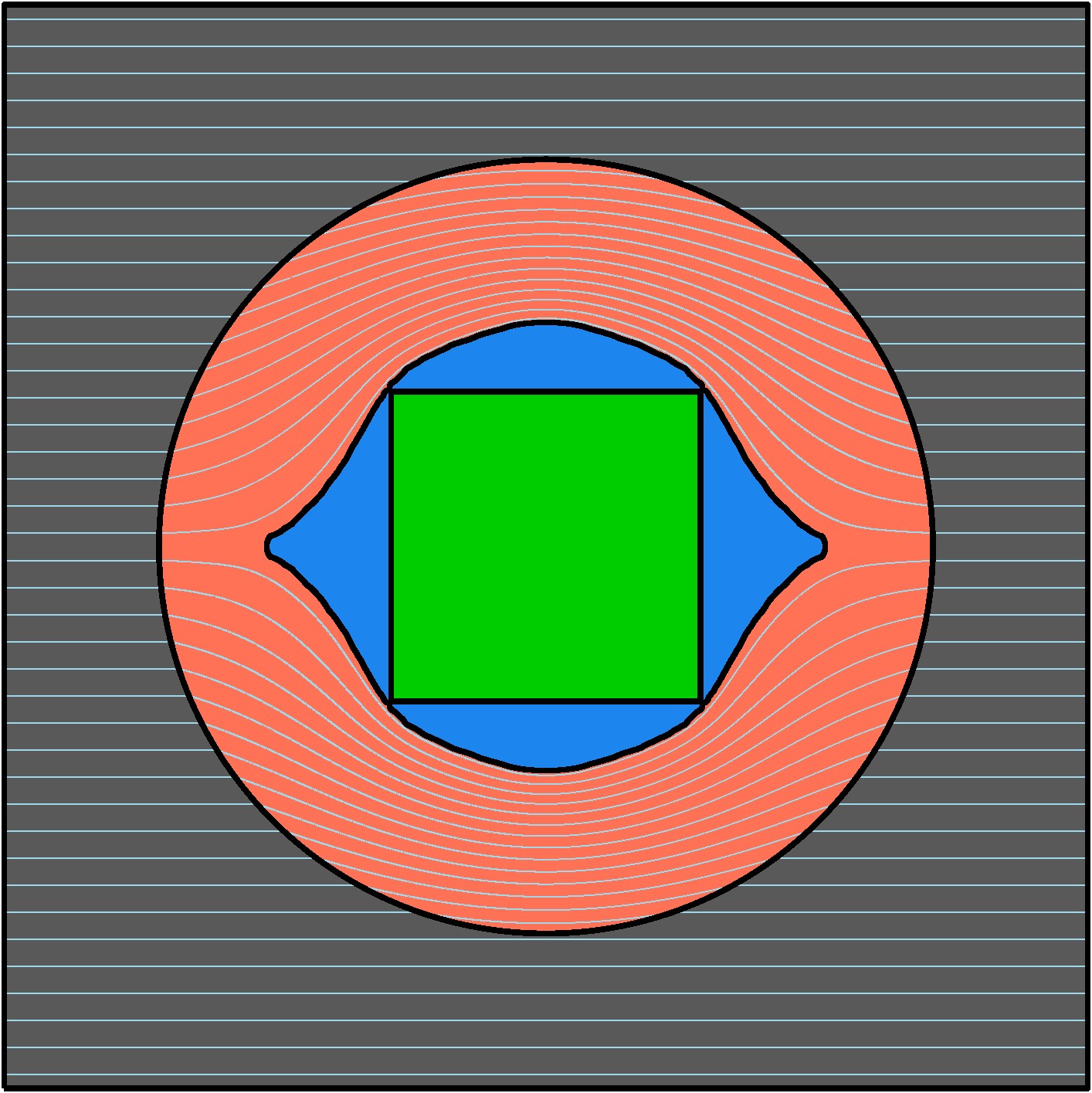}}
        \caption{\centering $J_{\rm cloak}=2.5868\times 10^{-7}$}
       \label{fig:Mchen2015case optTop c}
    \end{subfigure} & 
    
   \vspace{0.2cm}
   \begin{subfigure}[t]{0.15\textwidth}{\centering\includegraphics[width=1\textwidth]{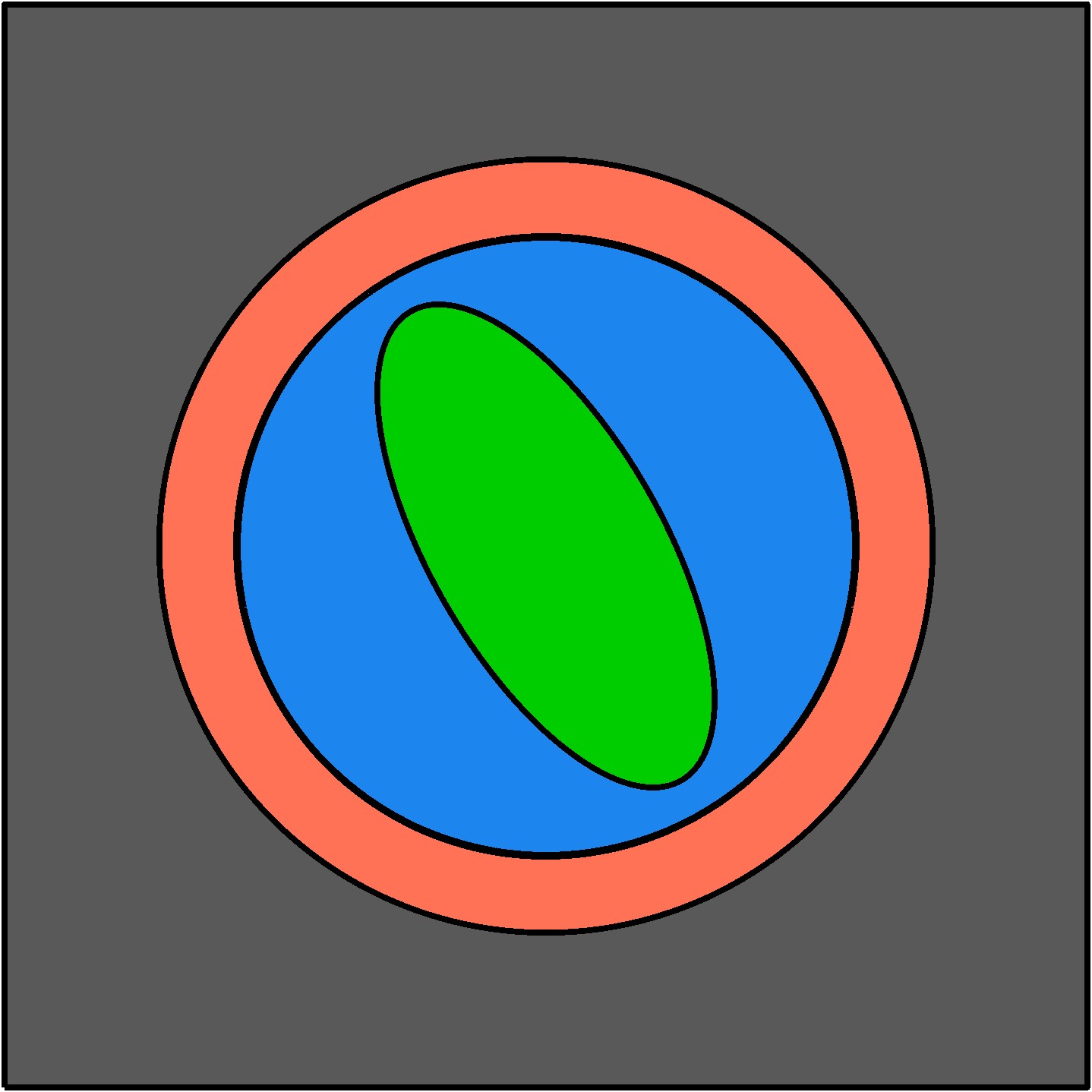}}
        \caption{\centering Initial topology}
       \label{fig:Mchen2015case optTop d}
    \end{subfigure}  & \vspace{0.2cm}
   \begin{subfigure}[t]{0.15\textwidth}{\centering\includegraphics[width=1\textwidth]{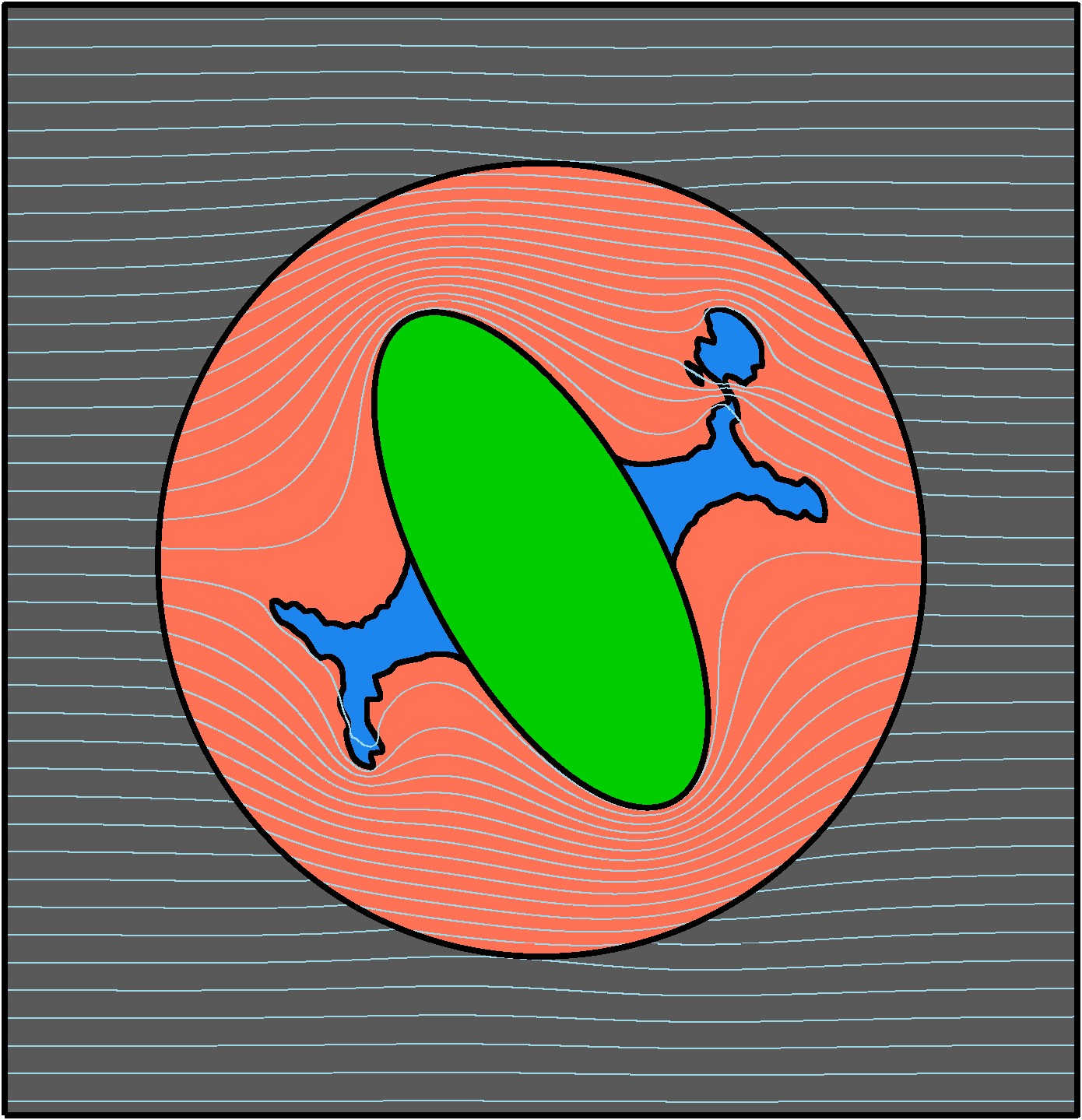}}
        \caption{\centering $J_{\rm cloak}=4.2623\times 10^{-3}$}
       \label{fig:Mchen2015case optTop e}
    \end{subfigure} & \vspace{0.2cm}
    \begin{subfigure}[t]{0.15\textwidth}{\centering\includegraphics[width=1\textwidth]{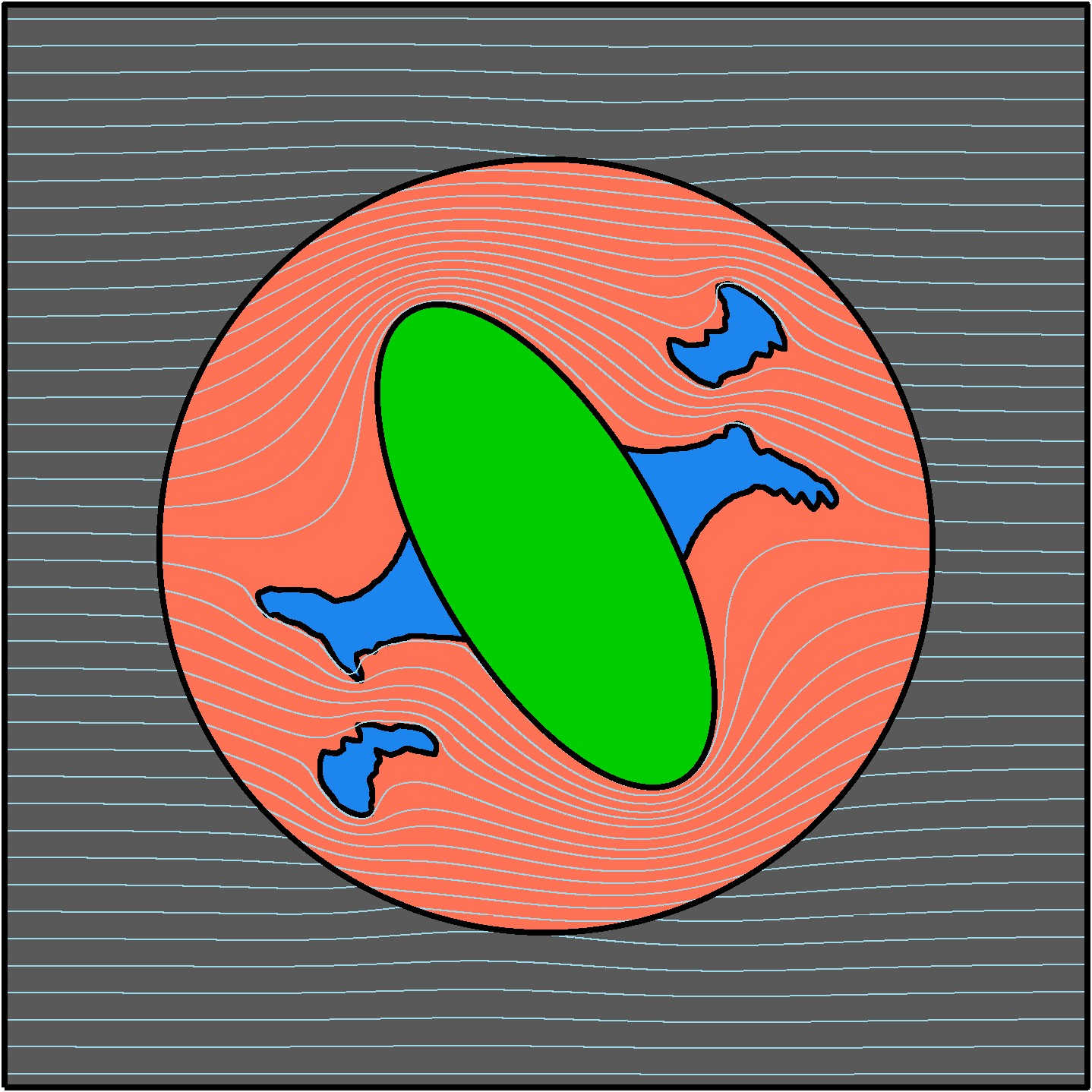}}
        \caption{\centering $J_{\rm cloak}=3.9801\times 10^{-3}$}
       \label{fig:Mchen2015case optTop f}
    \end{subfigure}\\
    \hline

\multicolumn{3}{|c||}{\small \hspace{30mm} Config. II \hspace{2mm}\hfill \scriptsize$(\chi,\rho)=(10^{-4},10^{-2})$} &  \multicolumn{3}{c|}{\small \hspace{28mm} Config. VI \hfill \scriptsize$(\chi,\rho)=(10^{-3},10^{-4})$}
\\
 \hline
   \vspace{0.2cm}
   \begin{subfigure}[t]{0.15\textwidth}{\centering\includegraphics[width=1\textwidth]{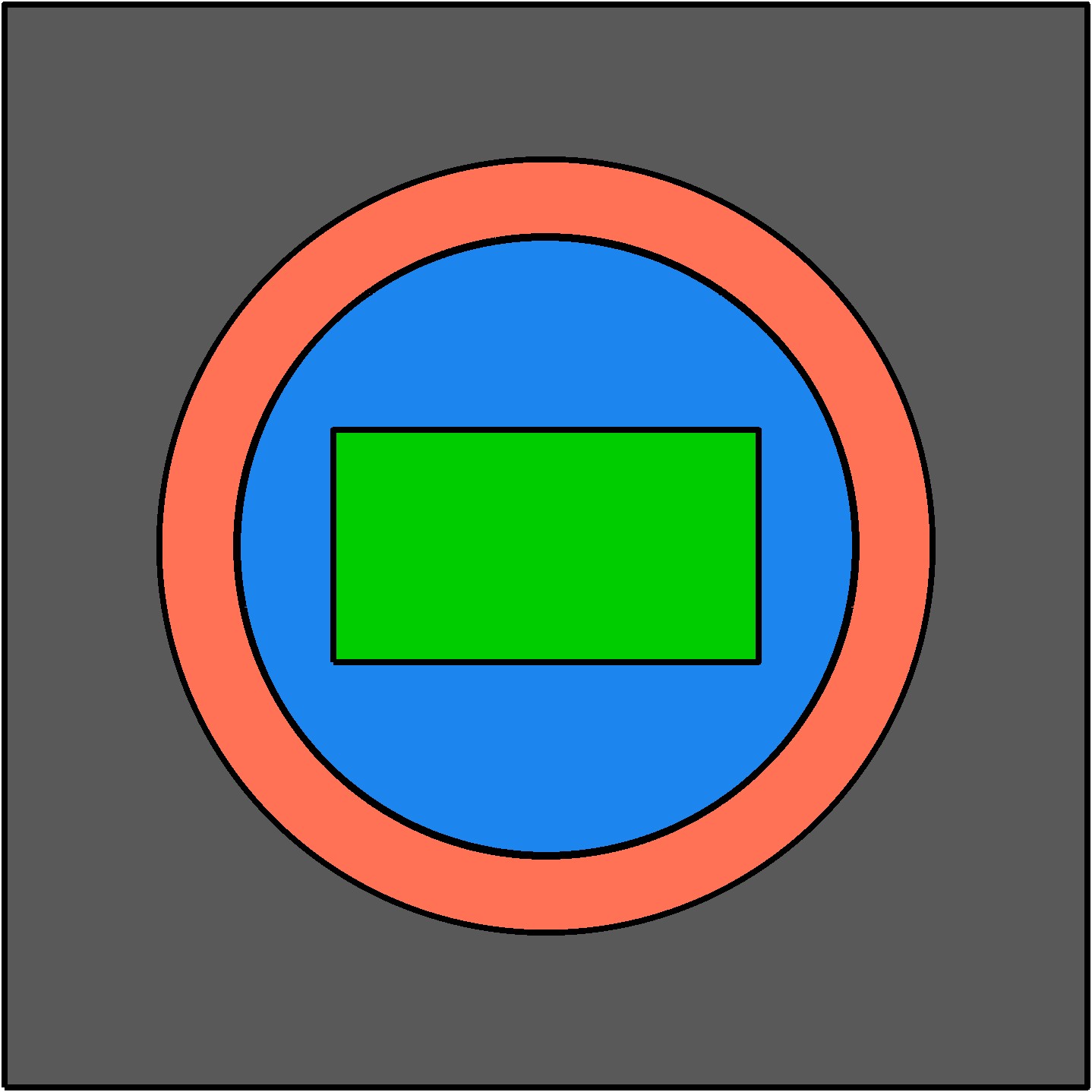}}
        \caption{\centering Initial topology}
       \label{fig:Mchen2015case optTop g}
    \end{subfigure}  & \vspace{0.2cm}
   \begin{subfigure}[t]{0.15\textwidth}{\centering\includegraphics[width=1\textwidth]{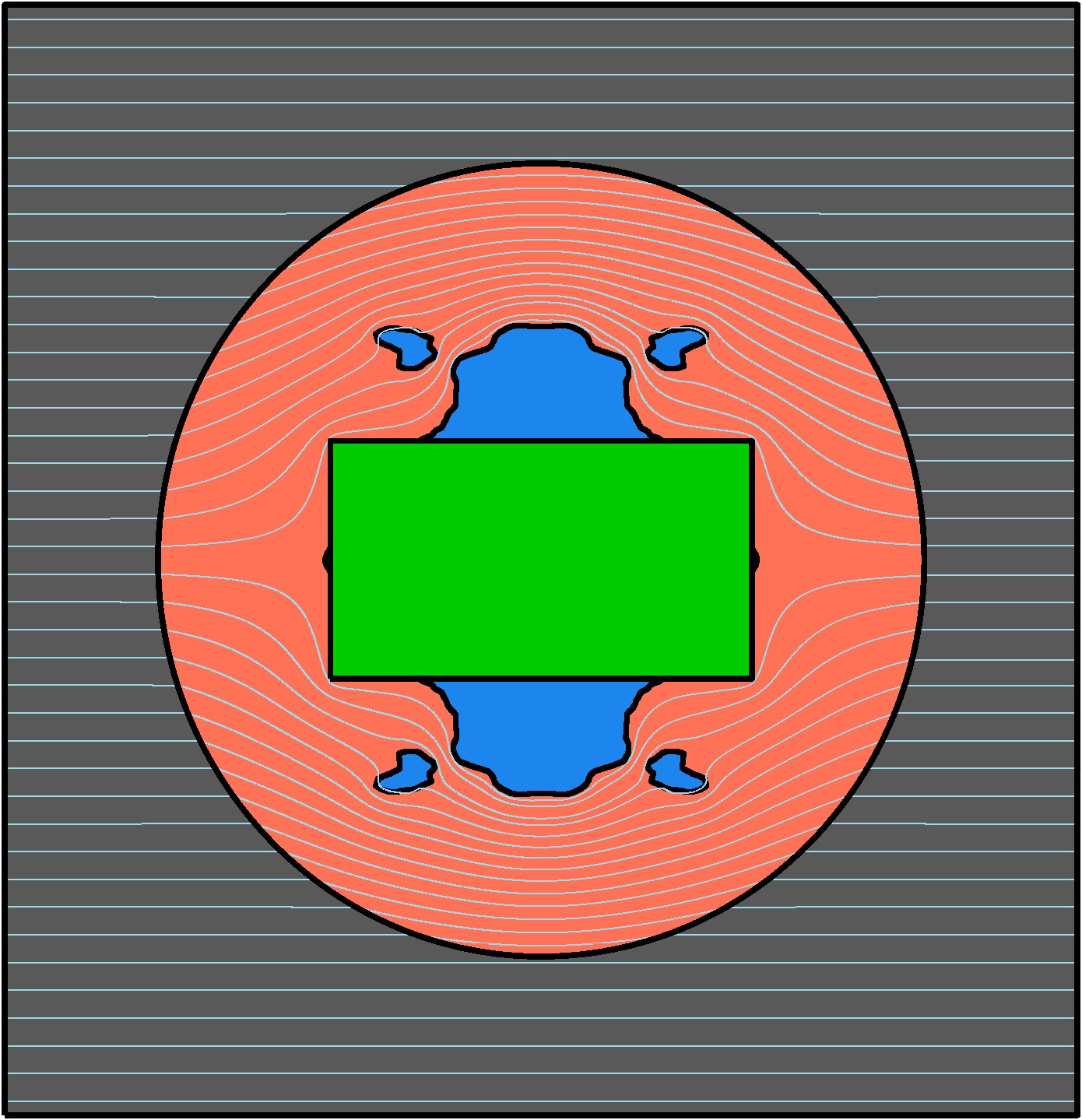}}
        \caption{\centering $J_{\rm cloak}=3.6564\times 10^{-7}$}
       \label{fig:Mchen2015case optTop h}
    \end{subfigure} & \vspace{0.2cm}
    \begin{subfigure}[t]{0.15\textwidth}{\centering\includegraphics[width=1\textwidth]{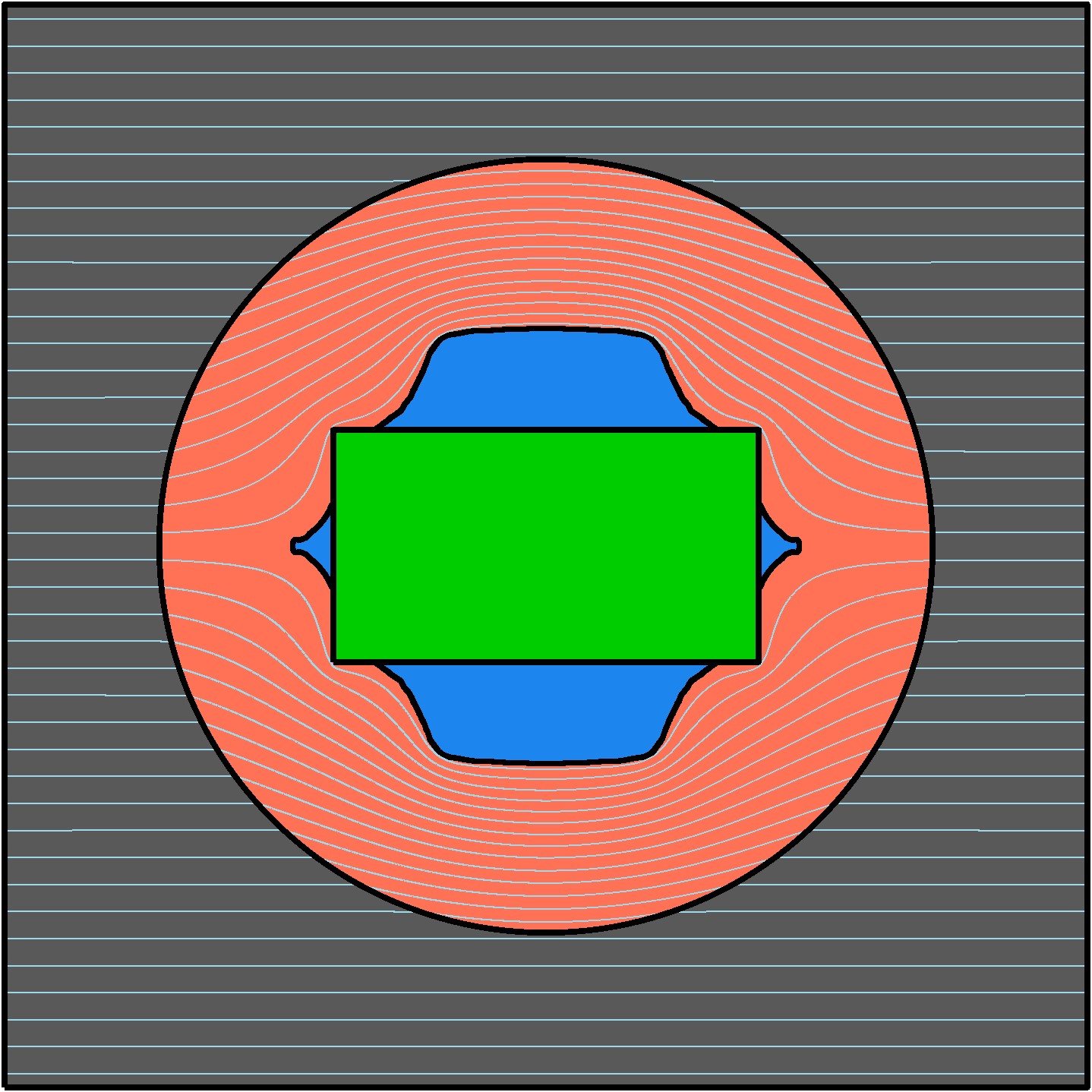}}
        \caption{\centering $J_{\rm cloak}=1.4318\times 10^{-6}$}
       \label{fig:Mchen2015case optTop i}
    \end{subfigure}& 
    
   \vspace{0.2cm}
   \begin{subfigure}[t]{0.15\textwidth}{\centering\includegraphics[width=1\textwidth]{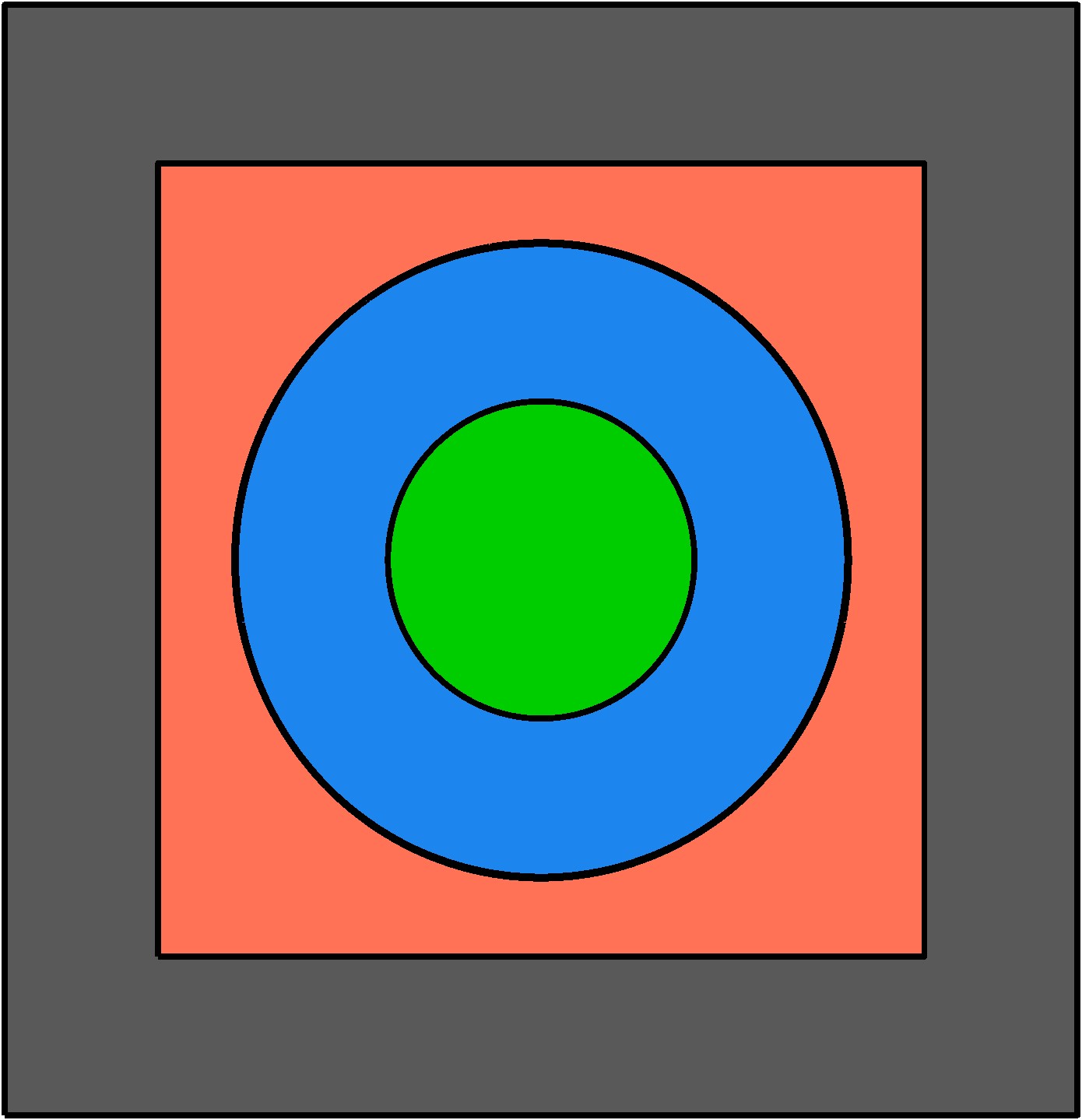}}
        \caption{\centering Initial topology}
       \label{fig:Mchen2015case optTop j}
    \end{subfigure}  & \vspace{0.2cm}
   \begin{subfigure}[t]{0.15\textwidth}{\centering\includegraphics[width=1\textwidth]{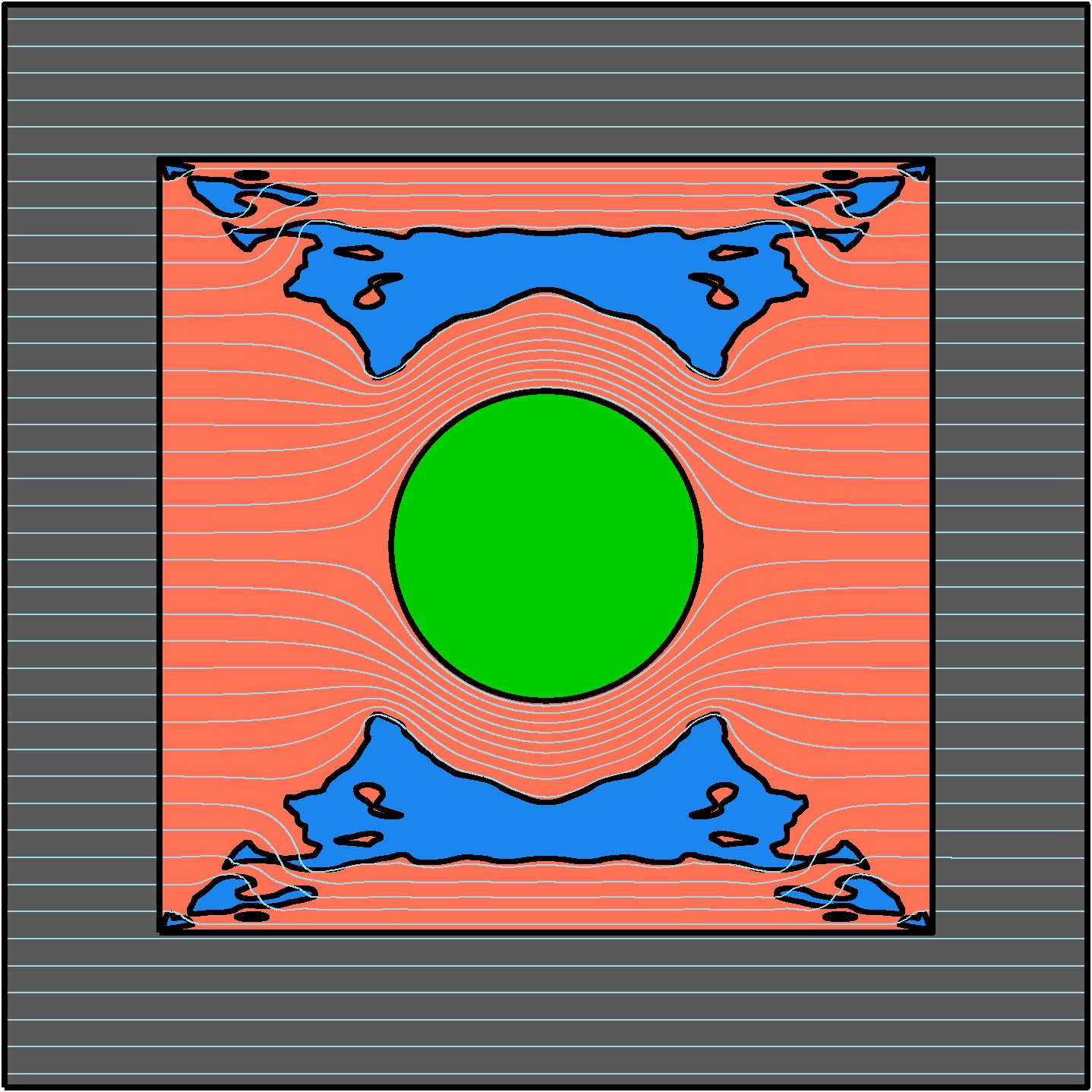}}
        \caption{\centering $J_{\rm cloak}=2.6690\times 10^{-8}$}
       \label{fig:Mchen2015case optTop k}
    \end{subfigure} & \vspace{0.2cm}
    \begin{subfigure}[t]{0.15\textwidth}{\centering\includegraphics[width=1\textwidth]{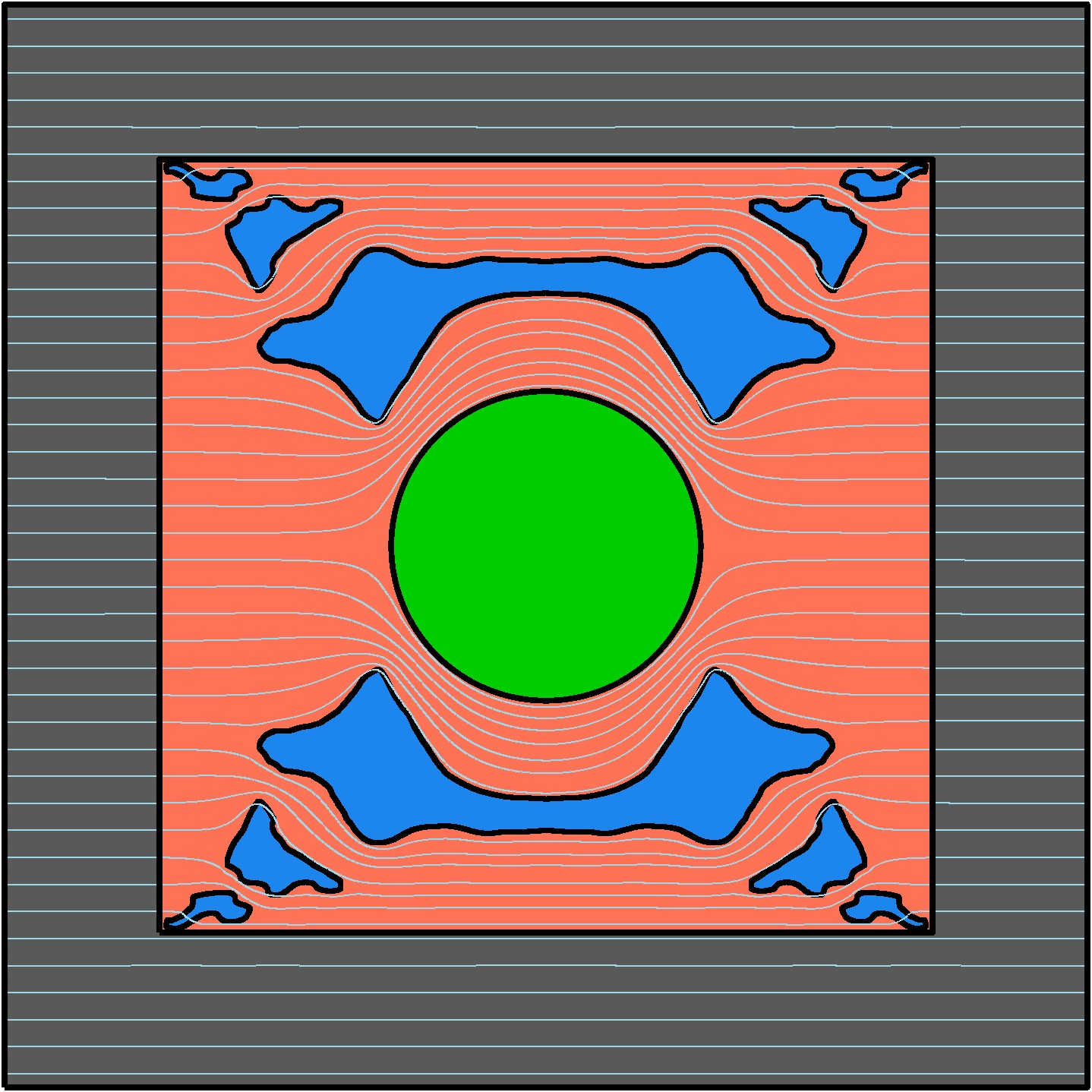}}
        \caption{\centering $J_{\rm cloak}=3.5944\times 10^{-7}$}
       \label{fig:Mchen2015case optTop l}
    \end{subfigure}\\
    \hline  

\multicolumn{3}{|c||}{\small \hspace{30mm} Config. III \hspace{1mm}\hfill \scriptsize$(\chi,\rho)=(10^{-4},10^{-2})$} &  \multicolumn{3}{c|}{\small \hspace{28mm} Config. VII \hfill \scriptsize$(\chi,\rho)=(10^{-2},10^{-2})$} \\
 \hline    
   \vspace{0.2cm}
   \begin{subfigure}[t]{0.15\textwidth}{\centering\includegraphics[width=1\textwidth]{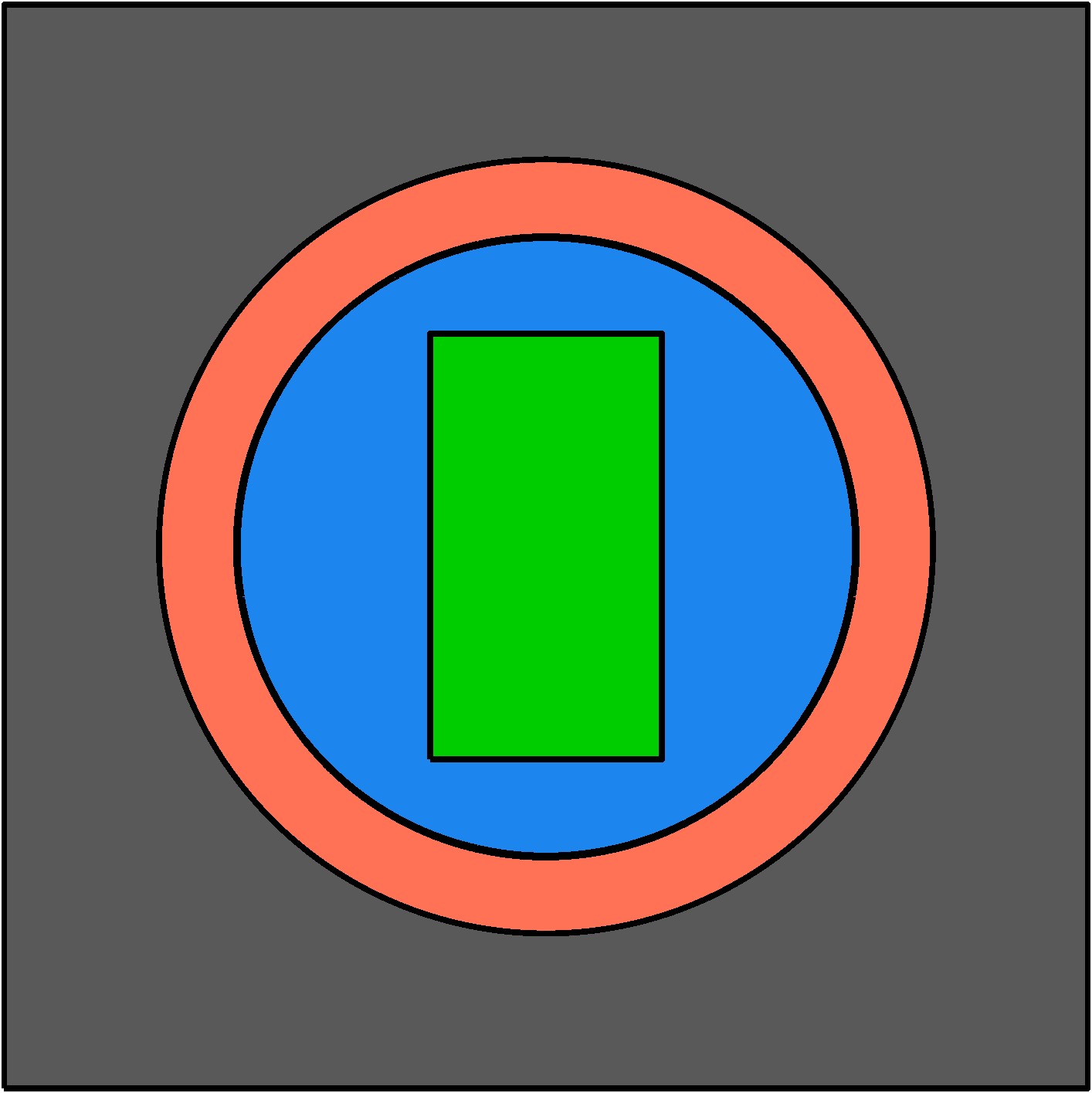}}
        \caption{\centering Initial topology}
       \label{fig:Mchen2015case optTop m}
    \end{subfigure}  & \vspace{0.2cm}
   \begin{subfigure}[t]{0.15\textwidth}{\centering\includegraphics[width=1\textwidth]{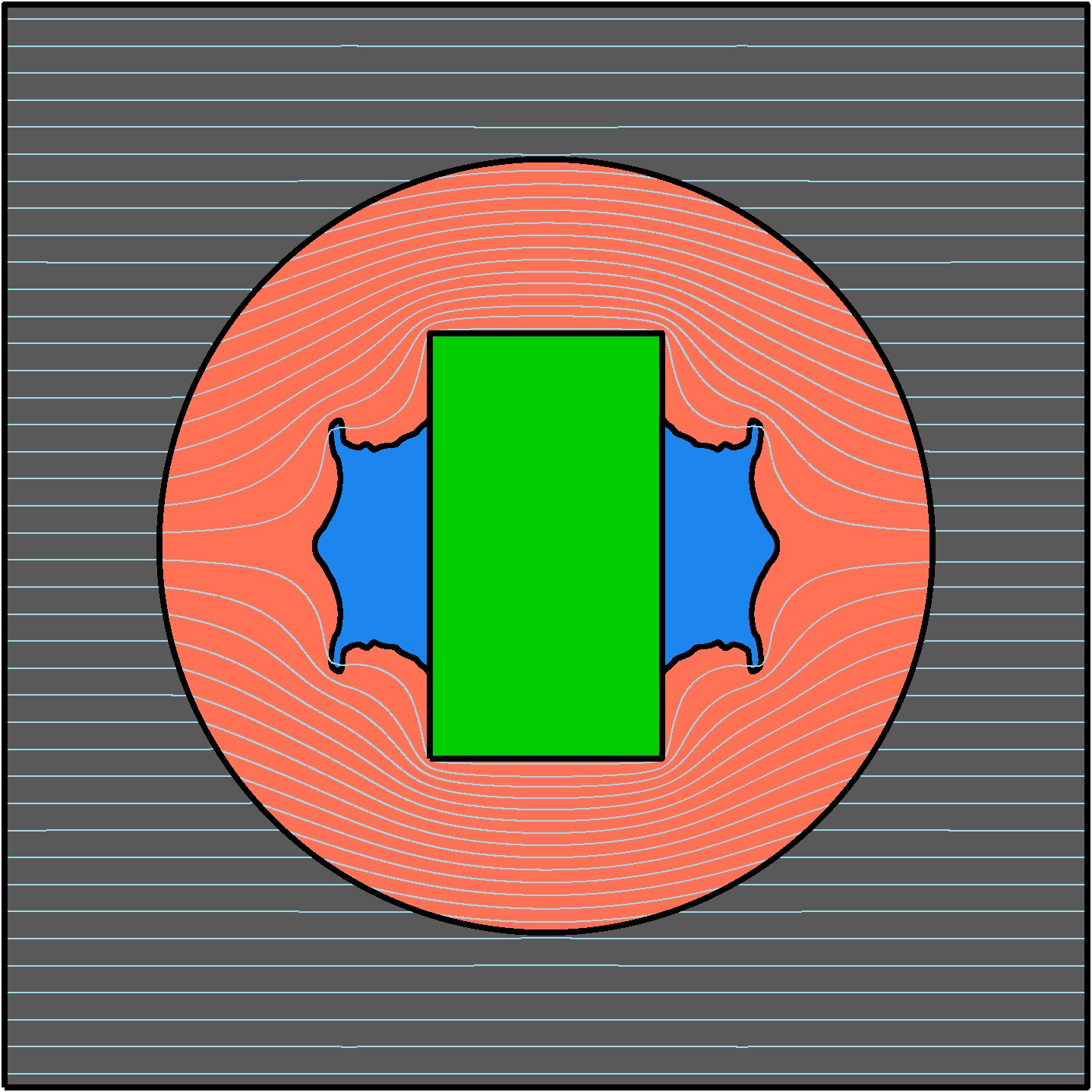}}
        \caption{\centering $J_{\rm cloak}=1.6415\times 10^{-6}$}
       \label{fig:Mchen2015case optTop n}
    \end{subfigure} & \vspace{0.2cm}
    \begin{subfigure}[t]{0.15\textwidth}{\centering\includegraphics[width=1\textwidth]{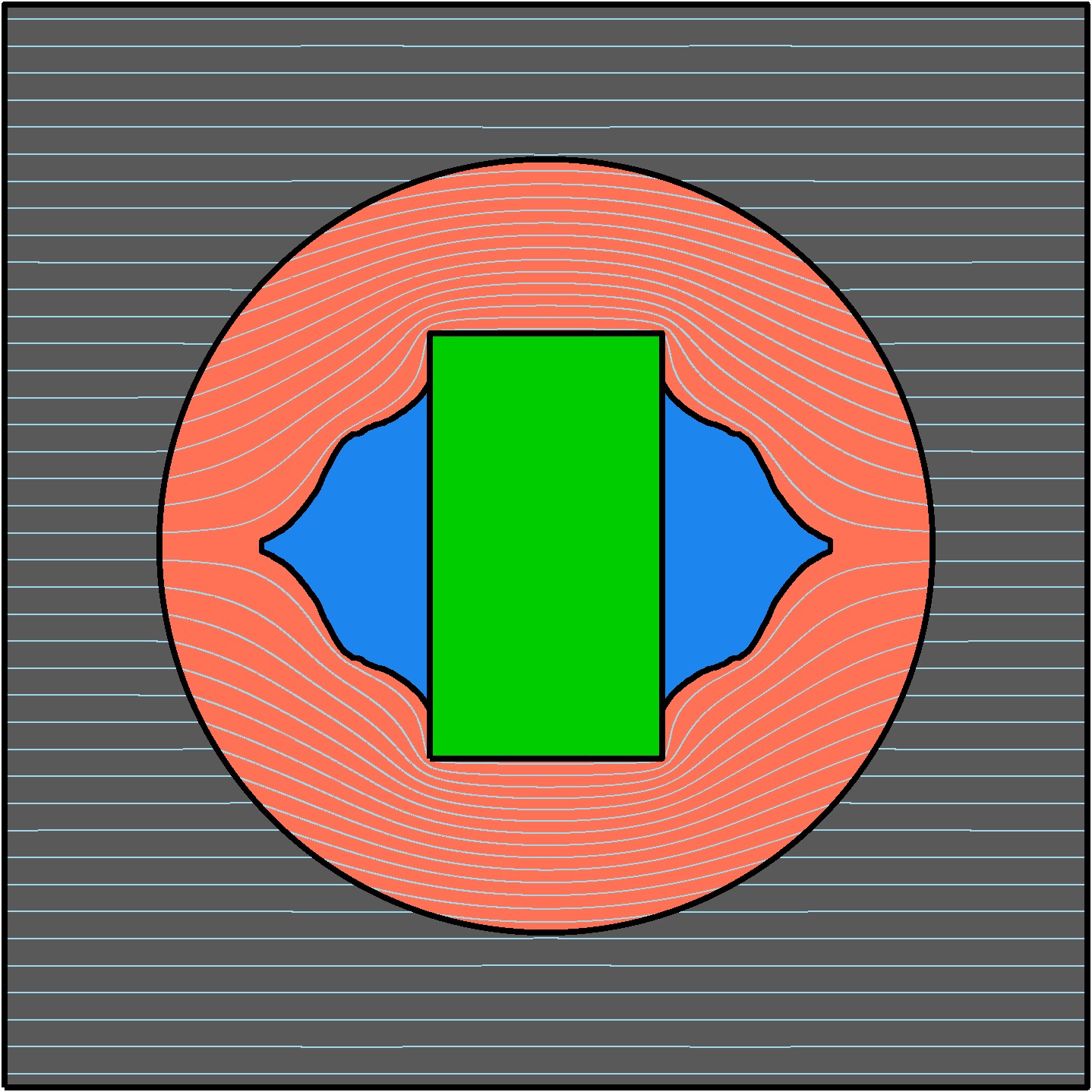}}
        \caption{\centering $J_{\rm cloak}=2.5438\times 10^{-6}$}
       \label{fig:Mchen2015case optTop o}
    \end{subfigure}&

   \vspace{0.2cm}
   \begin{subfigure}[t]{0.15\textwidth}{\centering\includegraphics[width=1\textwidth]{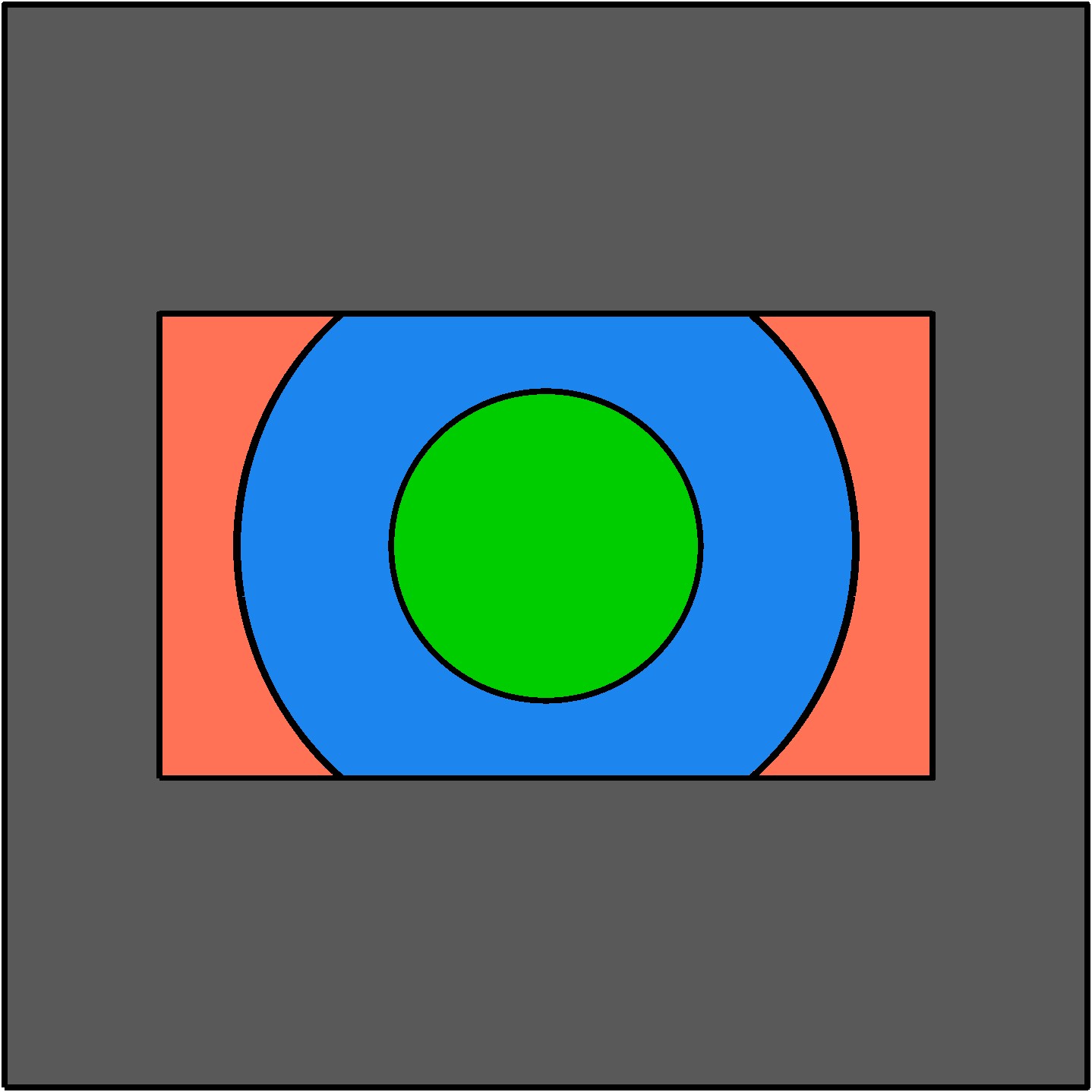}}
        \caption{\centering Initial topology}
       \label{fig:Mchen2015case optTop p}
    \end{subfigure}  & \vspace{0.2cm}
   \begin{subfigure}[t]{0.15\textwidth}{\centering\includegraphics[width=1\textwidth]{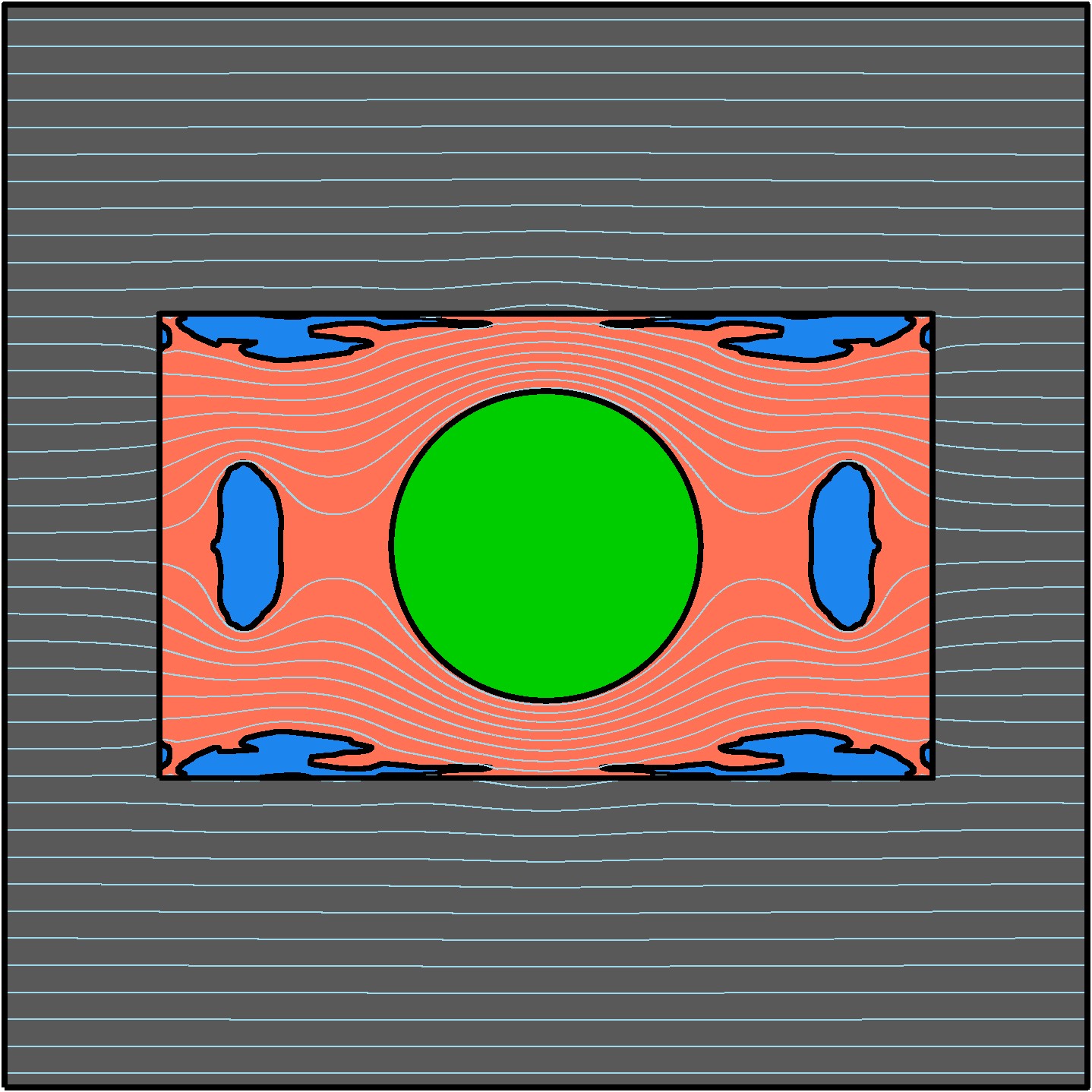}}
        \caption{\centering $J_{\rm cloak}=3.5788\times 10^{-3}$}
       \label{fig:Mchen2015case optTop q}
    \end{subfigure} & \vspace{0.2cm}
    \begin{subfigure}[t]{0.15\textwidth}{\centering\includegraphics[width=1\textwidth]{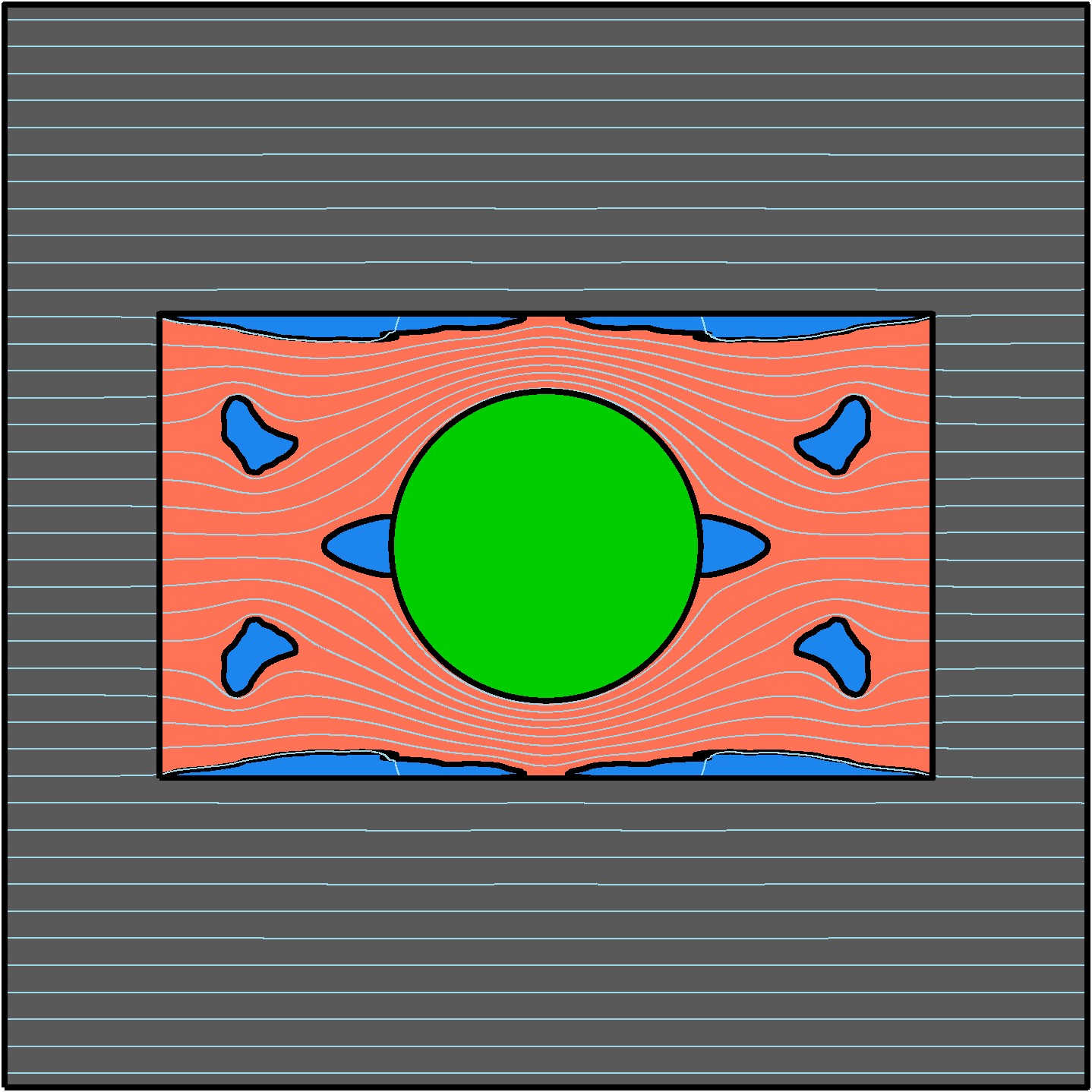}}
        \caption{\centering $J_{\rm cloak}=3.7914\times 10^{-5}$}
       \label{fig:Mchen2015case optTop r}
    \end{subfigure}\\
    \hline  
 
 \multicolumn{3}{|c||}{\small \hspace{30mm} Config. IV \hspace{1mm}\hfill \scriptsize$(\chi,\rho)=(10^{-4},10^{-2})$} &  \multicolumn{3}{c|}{\small \hspace{28mm} Config. VIII\hspace{0mm} \hfill \scriptsize$(\chi,\rho)=(10^{-2},10^{-2})$}
 \\
 \hline   
   \vspace{0.2cm}
   \begin{subfigure}[t]{0.15\textwidth}{\centering\includegraphics[width=1\textwidth]{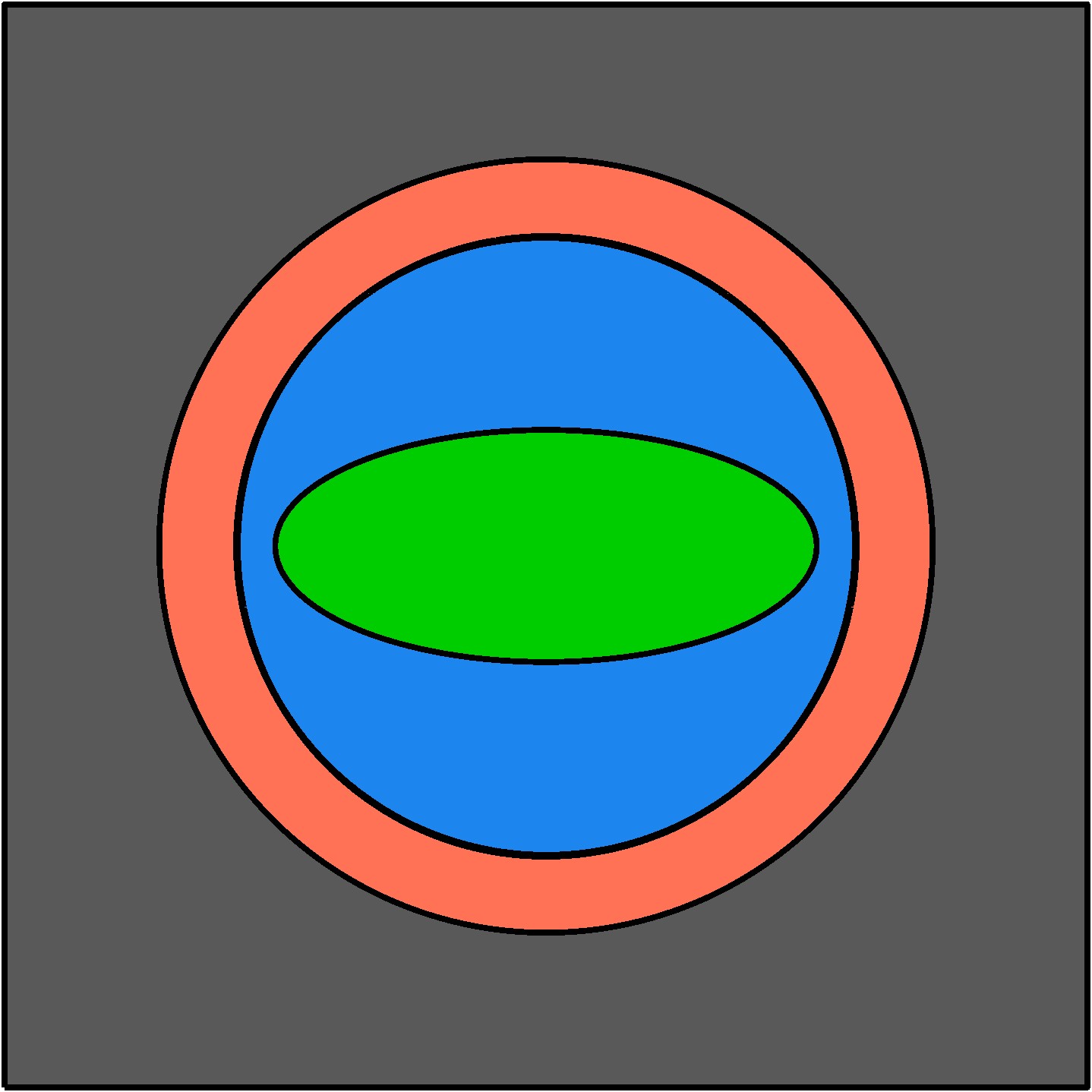}}
        \caption{\centering Initial topology}
       \label{fig:Mchen2015case optTop s}
    \end{subfigure}  & \vspace{0.2cm}
   \begin{subfigure}[t]{0.15\textwidth}{\centering\includegraphics[width=1\textwidth]{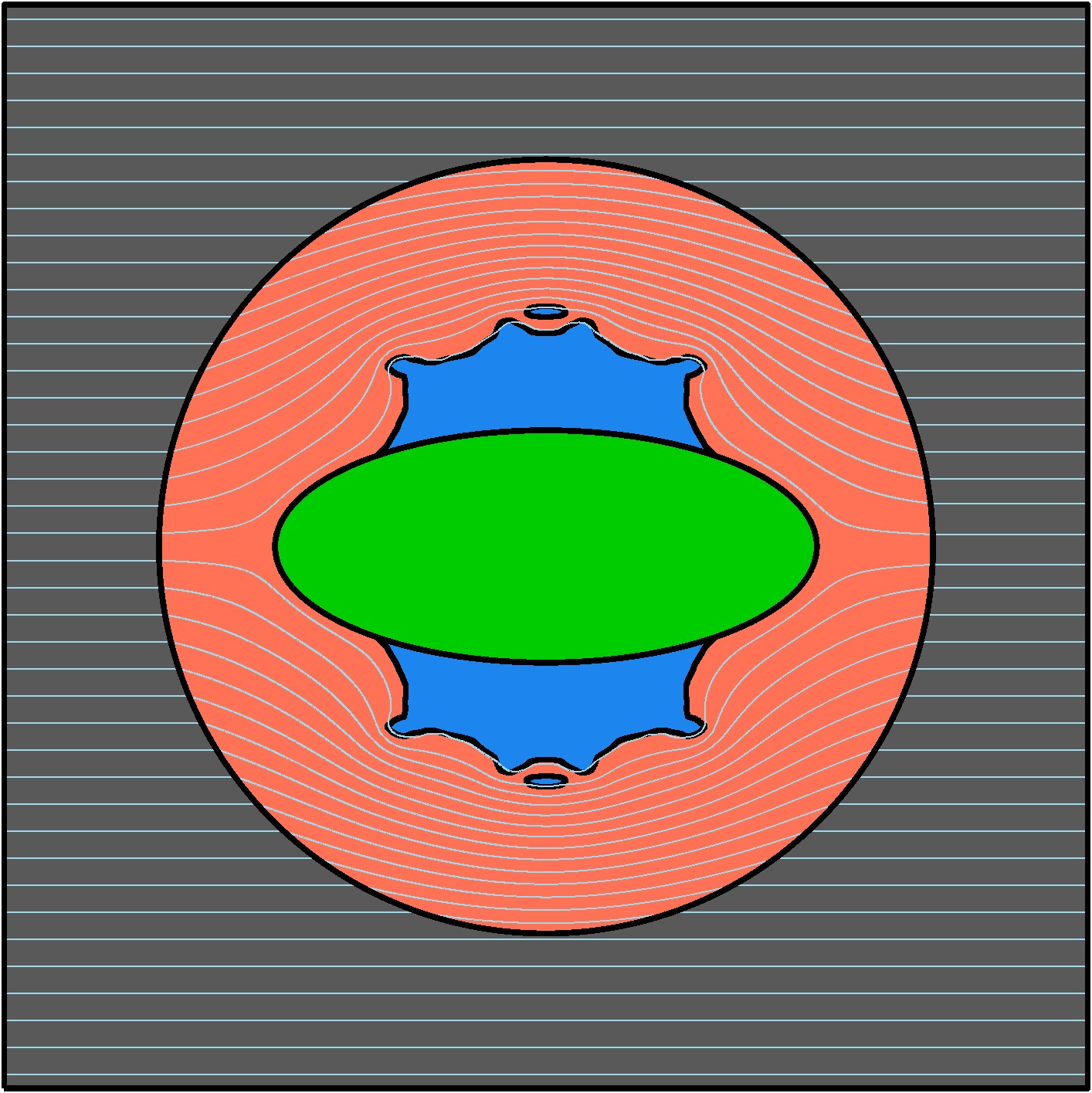}}
        \caption{\centering $J_{\rm cloak}=5.1844\times 10^{-7}$}
       \label{fig:Mchen2015case optTop t}
    \end{subfigure} & \vspace{0.2cm}
    \begin{subfigure}[t]{0.15\textwidth}{\centering\includegraphics[width=1\textwidth]{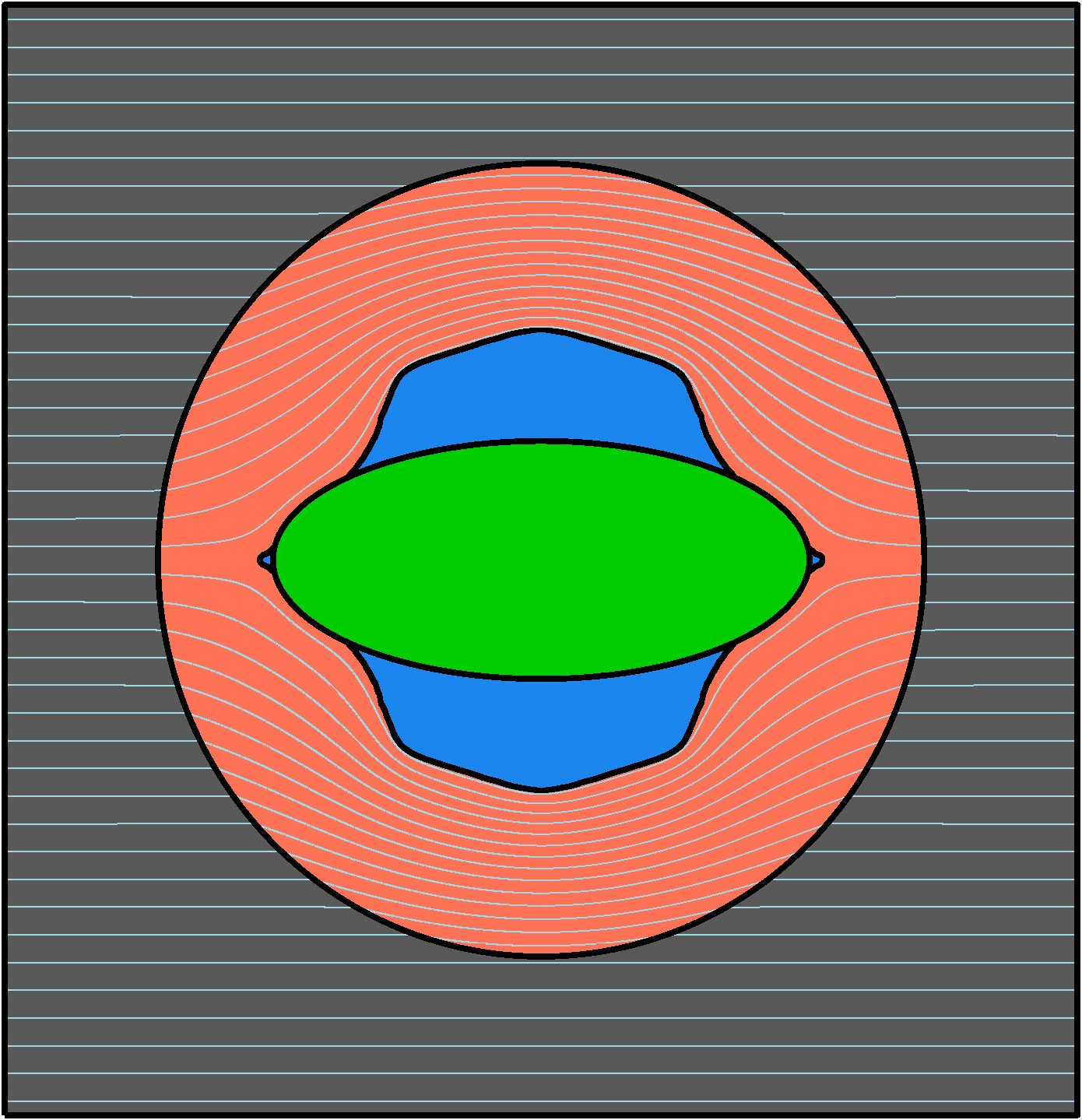}}
        \caption{\centering $J_{\rm cloak}=9.0389\times 10^{-7}$}
       \label{fig:Mchen2015case optTop u}
    \end{subfigure}& 
    
   \vspace{0.2cm}
   \begin{subfigure}[t]{0.15\textwidth}{\centering\includegraphics[width=1\textwidth]{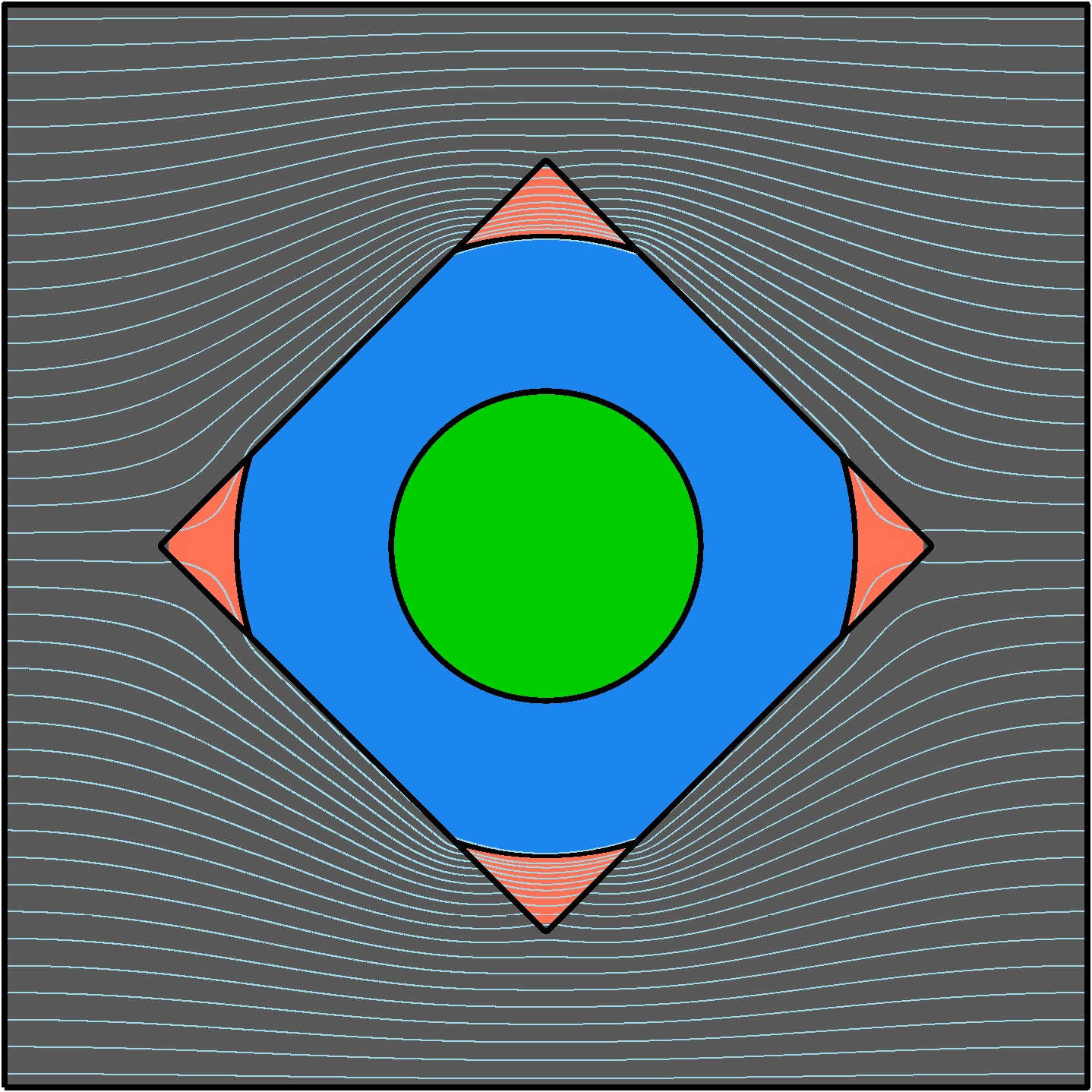}}
        \caption{\centering Initial topology}
       \label{fig:Mchen2015case optTop v}
    \end{subfigure}  & \vspace{0.2cm}
   \begin{subfigure}[t]{0.15\textwidth}{\centering\includegraphics[width=1\textwidth]{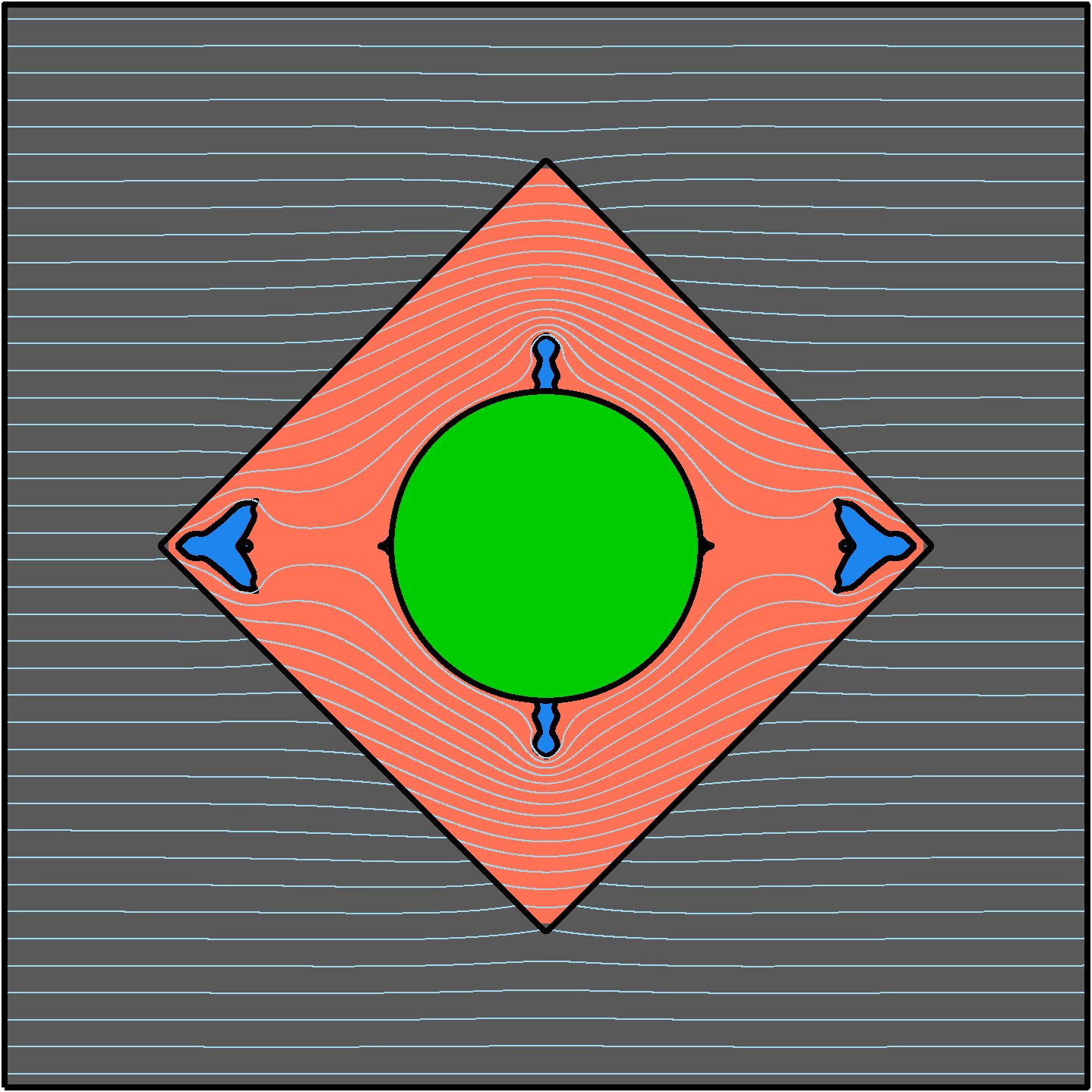}}
        \caption{\centering $J_{\rm cloak}=5.4377\times 10^{-1}$}
       \label{fig:Mchen2015case optTop w}
    \end{subfigure} & \vspace{0.2cm}
    \begin{subfigure}[t]{0.15\textwidth}{\centering\includegraphics[width=1\textwidth]{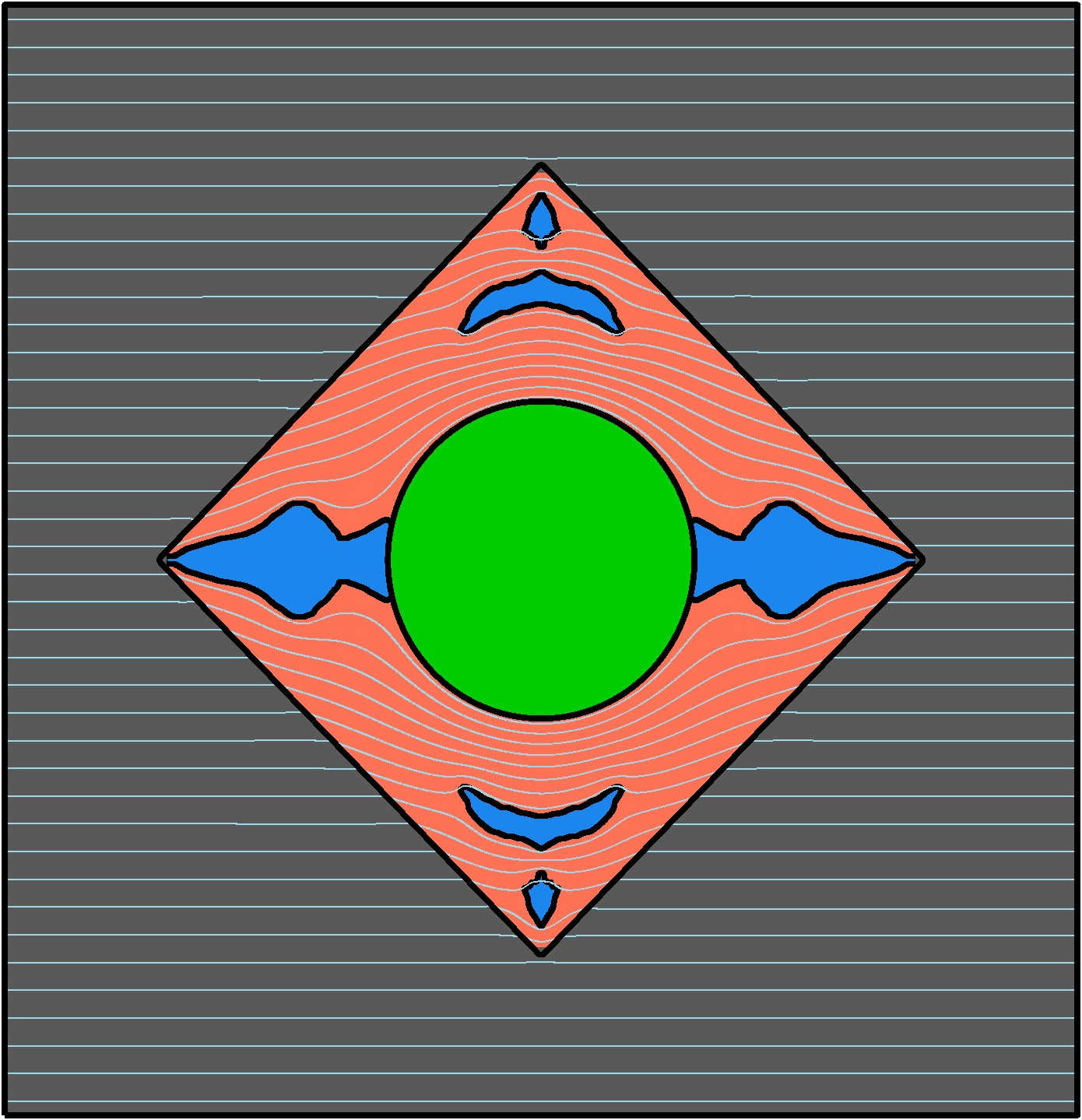}}
        \caption{\centering $J_{\rm cloak}=1.2310\times 10^{-6}$}
       \label{fig:Mchen2015case optTop x}
    \end{subfigure}\\
    \hline  
    \end{tabular}

}
\caption{For the thermal cloak problem, initial topologies, optimized topologies without any regularization, optimized topologies with combined Tikhonov and volume regularizations for different types geometries of a thermal cloak for $N_{\rm var}=1089$ with $\Delta=0.0005$. The proposed method can handle different geometries.}  
    \label{fig:Mchen2015case optTop}
\end{figure}

\par In the next study, we explore several types of geometries of the obstacle and thermal cloak for a given problem. The different configurations under consideration and their dimensions are shown in \fref{fig:Cloak obsctacle schematic}. The plate dimensions are the same as in \fref{fig:Cloak problem schematics}, therefore excluded. For all configurations, we apply symmetry along $x$ and $y$-axis as in the earlier cases except Config. V - inclined ellipsoidal obstacle, for which we remove the symmetry condition due to lack of symmetry in geometry itself. We perform optimization for one value of $N_{\rm var}$, $N_{\rm var}=1089$, and the corresponding results are shown in \fref{fig:Mchen2015case optTop} In terms of the initial topology, the optimized topology (without any regularization), and the optimized topology (with combined Tikhonov and volume regularization). For combined Tikhonov and volume regularization, we present only one case that gives a smoother topology with good accuracy (the corresponding weighing parameters $\chi$ and $\rho$ for each case are also mentioned in \fref{fig:Mchen2015case optTop}). From the results, it can be concluded that the proposed method handles different geometries. In the same vein as the earlier results, regularization can generate smoother geometries. In some cases, the objective function of the regularized problem is better than the unregularized problem. The reason behind it is that the unregularized optimization problem can get stuck in a local minimum without exploring the full scope of the whole design space. After applying regularization, the regularized problem becomes a better-defined optimization problem for those cases. Following it, optimization has the advantage of exploring better designs that were not possible earlier. 

\subsection{Thermal camouflage problem}
\label{sec:Yang2016case_cmflg}
\subsubsection{Problem description}
\label{sec:Yang2016mflg Problem description}
 \begin{figure}[!htbp]
    \centering
    \setlength\figureheight{1\textwidth}
    \setlength\figurewidth{1\textwidth}
     \begin{subfigure}[b]{0.4\textwidth}{\centering\includegraphics[width=0.8\textwidth]{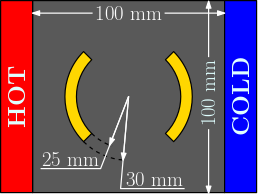}}
        \caption{Two insulator sectors embedded in a base material plate (reference state).}
        \label{fig:camouflage problem schematics a}
    \end{subfigure}\quad
    \begin{subfigure}[b]{0.4\textwidth}{\centering\includegraphics[width=0.8\textwidth]{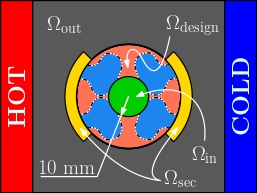}}
             \caption{A conductive object and a metamaterial-based camouflage surrounding it in the plate.}
             \label{fig:camouflage problem schematics b}
    \end{subfigure}
 \caption{Schematic design of (a) A base material (aluminium alloy) plate embedded with two insulator sectors  under constant heat flux applied by high temperature source on the left side and low temperature sink on the right side; (b) A conductive object, a thermal camouflage surrounding the object and two insulator sectors embedded in a base material plate; $\mathrm{\Omega}_{\mathrm{sec}}$ is the region covered by the insulator sectors,  $\mathrm{\Omega}_{\mathrm{in}}$ is the region covered by the object, $\mathrm{\Omega}_{\mathrm{design}}$ is the area of the camouflage where the topology is optimized, $\mathrm{\Omega}_{\mathrm{out}}$ is the outside area of remaining base material, $\mathrm{\Omega}=\mathrm{\Omega}_{\mathrm{in}} \cup \mathrm{\Omega}_{\mathrm{design}}\cup \mathrm{\Omega}_{\mathrm{sec}}\cup \mathrm{\Omega}_{\mathrm{out}}$.}
 \label{fig:camouflage problem schematics}
\end{figure}
\par In this example, we explore the optimization of thermal camouflage. Both thermal cloak and thermal camouflage mimic the thermal signature (in the region of interest) of other reference scenarios. 
Thermal cloaks hide the objects from outside detection; however, they can be detected from inside observation as they have a different heat signature from the outer region. On the other hand, thermal camouflage mimics the heat signature of another object for both inside and outside detection. The schematics of the camouflage problem and the corresponding dimensions are shown in \fref{fig:camouflage problem schematics}. \fref{fig:camouflage problem schematics a} shows the reference state where two insulated sectors are embedded in a square base material-Aluminum alloy (grade 5457) plate ($\kappa_{\rm Al}=177 $~W/mK) with a side length of 100 mm. The insulators' conductivity is taken as $0.0001$ W/mK. \fref{fig:camouflage problem schematics b} shows that the magnesium alloy object (grade AZ91D), with thermal conductivity $\kappa_{\rm Mg}=72.7$~W/mK, is added at the center of two sectors. It also shows a metamaterial-based thermal camouflage covering the area between the sectors and the object. For camouflage, we chose the metamaterial similar to the last example, made of copper and polydimethylsiloxane (PDMS), with thermal conductivity $\kappa_{\rm copper}=398$~W/mK and $\kappa_{\rm PDMS}=0.27$~W/mK. The boundary conditions are identical to those in the previous example.
\subsubsection{Objective function}
\par The objective of thermal camouflage is to reduce the temperature difference with respect to the temperature signature of the reference state in $\mathrm{\Omega}_{\mathrm{in}} \cup \mathrm{\Omega}_{\mathrm{design}} \cup \mathrm{\Omega}_{\mathrm{out}}$. Mathematically, the camouflage function is defined as,
\begin{equation}
    J_{\mathrm{cmflg}}=\dfrac{1}{\widetilde{J}_{\mathrm{cmflg}}} \int_{\mathrm{\Omega}_{\mathrm{in}} \cup \mathrm{\Omega}_{\mathrm{design}} \cup \mathrm{\Omega}_{\mathrm{out}}} \vert T - \overline{T} \vert^2~d\mathrm{\Omega},
    \end{equation}
with $\widetilde{J}_{\mathrm{cmflg}}$ be the normalisation value given as,
\begin{equation}
    \widetilde{J}_{\mathrm{cmflg}}= \int_{\mathrm{\Omega}_{\mathrm{in}} \cup \mathrm{\Omega}_{\mathrm{design}} \cup \mathrm{\Omega}_{\mathrm{out}}} \vert \widetilde{T} - \overline{T} \vert^2~d\mathrm{\Omega},
    \end{equation}
Here, $\overline{T}$ represents the temperature field for the reference case, when $\mathrm{\Omega}_{\mathrm{in}} \cup \mathrm{\Omega}_{\mathrm{design}} \cup \mathrm{\Omega}_{\mathrm{out}}$ is filled with the base material, and $\widetilde{T}$  is the temperature field when $\mathrm{\Omega}_{\mathrm{design}}$ is entirely filled with the insulator.
\par By comparison with \eref{eq:Objective fun definition}, $\Omega_b={\mathrm{\Omega}_{\mathrm{in}} \cup \mathrm{\Omega}_{\mathrm{design}} \cup \mathrm{\Omega}_{\mathrm{out}}}$, $J_b=\dfrac{1}{\widetilde{J}_{\mathrm{cmflg}}}~\vert ~T~-~\overline{T}~\vert^2$, and the surface term is absent.
\subsubsection{Results and discussion}
\par In this example, we consider $N_{\rm var}=1089$ (with $p=2$, $q=1$ and a mesh of 17655 DOF) and $\Delta=0.001$. We also exploit the LSF reinitialization performed after every 10 iterations or 300 function evaluations. All other parameters are taken the same as the last example. 
\par \fref{fig:Yang2016case optTop TVSmth} shows the initial topologies, the optimized topologies without any regularization ($J_{\rm total}=J_{\rm cmflg}$), and the optimized topologies with combined Tikhonov and volume regularization ($J_{\rm total}=J_{\rm cmflg} + \chi J_{\rm Tknv} + \rho J_{\rm vol}$). For combined regularization, we consider four sets of parameters $\chi$ and $\rho$, denoted as set-A ($\chi=1$, $\rho=10^{-2}$), set-B ($\chi=1$, $\rho=10^{-1}$), set-C ($\chi=1$, $\rho=1$) and set-D ($\chi=1$, $\rho=0$). It can be seen that the optimal topologies provide the copper channels to allow the flux to flow similarly to the reference case. The widths of the channels are dependent on numerical accuracy, design freedom, and regularization parameters. As evident from the figure, the optimized geometries without any regularization are impractical considering their too complex features. However, the alternative designs with regularization have better practical topologies. Also, one point to note is that their $J_{\rm cmflg}$-values are in the same range as the unregularized cases, sometimes even smaller. The difference among the optimized topologies from set A to D is very negligible. As discussed in the last example, the parameters $\chi$ and $\rho$ values are difficult to predict apriori and have to be decided based on the trial and error method according to the design requirements.
\par
\begin{figure}
\centering
\scalebox{0.9}{
\begin{tabular}[c]{|m{1em}|m{11.4em} |m{11.4em}|m{11.4em}|}
\hline		
   & \begin{center}
       {\small A base material plate embedded with two insulator sectors (ref. case)} 
   \end{center} & \begin{center}
       {\small A plate embedded with two insulator sectors, a conductive object (at the center), and an optimized thermal camouflage (without regularization)  surrounding the object.}
   \end{center} & \begin{center}
       {\small A plate embedded with two insulator sectors, a conductive object (at the center), and an optimized thermal camouflage (with regularization) surrounding the object.}
   \end{center}\\
\hline 
  \vspace{0.05cm}  
  \rotatebox{90}{\centering Flux flow}  & \vspace{0.2cm} 
    \begin{subfigure}[t]{0.23\textwidth}\begin{center}{
\includegraphics[width=1\textwidth]{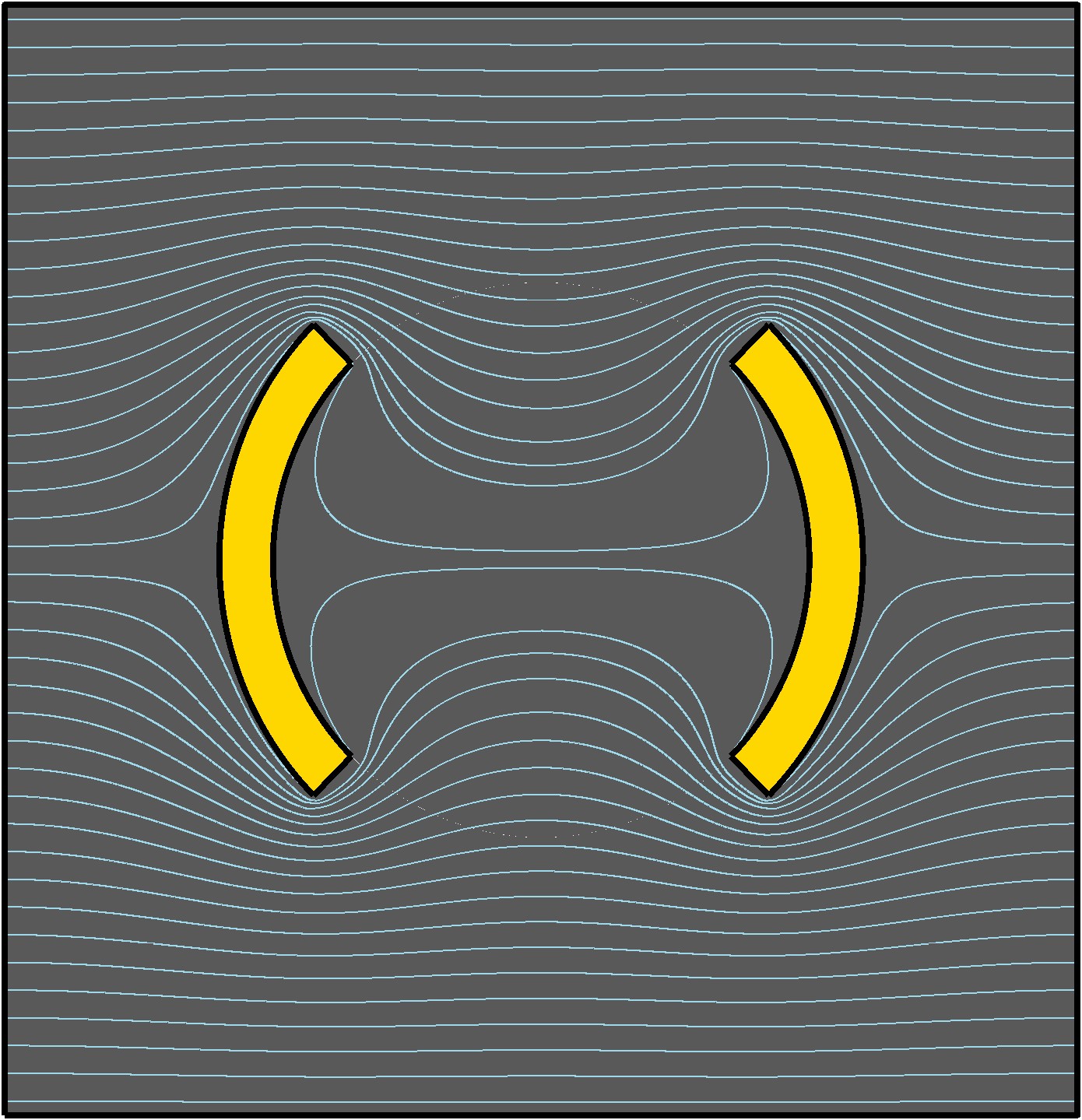}}
    \end{center}
    \end{subfigure} & \vspace{0.2cm} 
    \begin{subfigure}[t]{0.23\textwidth}\begin{center}{
\includegraphics[width=1\textwidth]{Figures_Yang2016case/Laplace_HT_LevelSetTop_Yang2016case_objT_4_ref_p00k00h00_DSNref_p00k55h00_sample1_inA_fluxPlot.jpg}}
    \end{center}
    \end{subfigure} &\vspace{0.2cm} 
    \begin{subfigure}[t]{0.23\textwidth}\begin{center}{
\includegraphics[width=1\textwidth]{Figures_Yang2016case/Laplace_HT_LevelSetTop_TVSmth_Yang2016case_objT_4_ref_p00k00h00_DSNref_p00k55h00_sample1_inA_para3_fluxPlot.jpg}}
    \end{center}
    \end{subfigure} \\  
    \rotatebox{90}{\centering Temp. $T$}  &
      \begin{subfigure}[t]{0.23\textwidth}\begin{center}{
\includegraphics[width=1\textwidth]{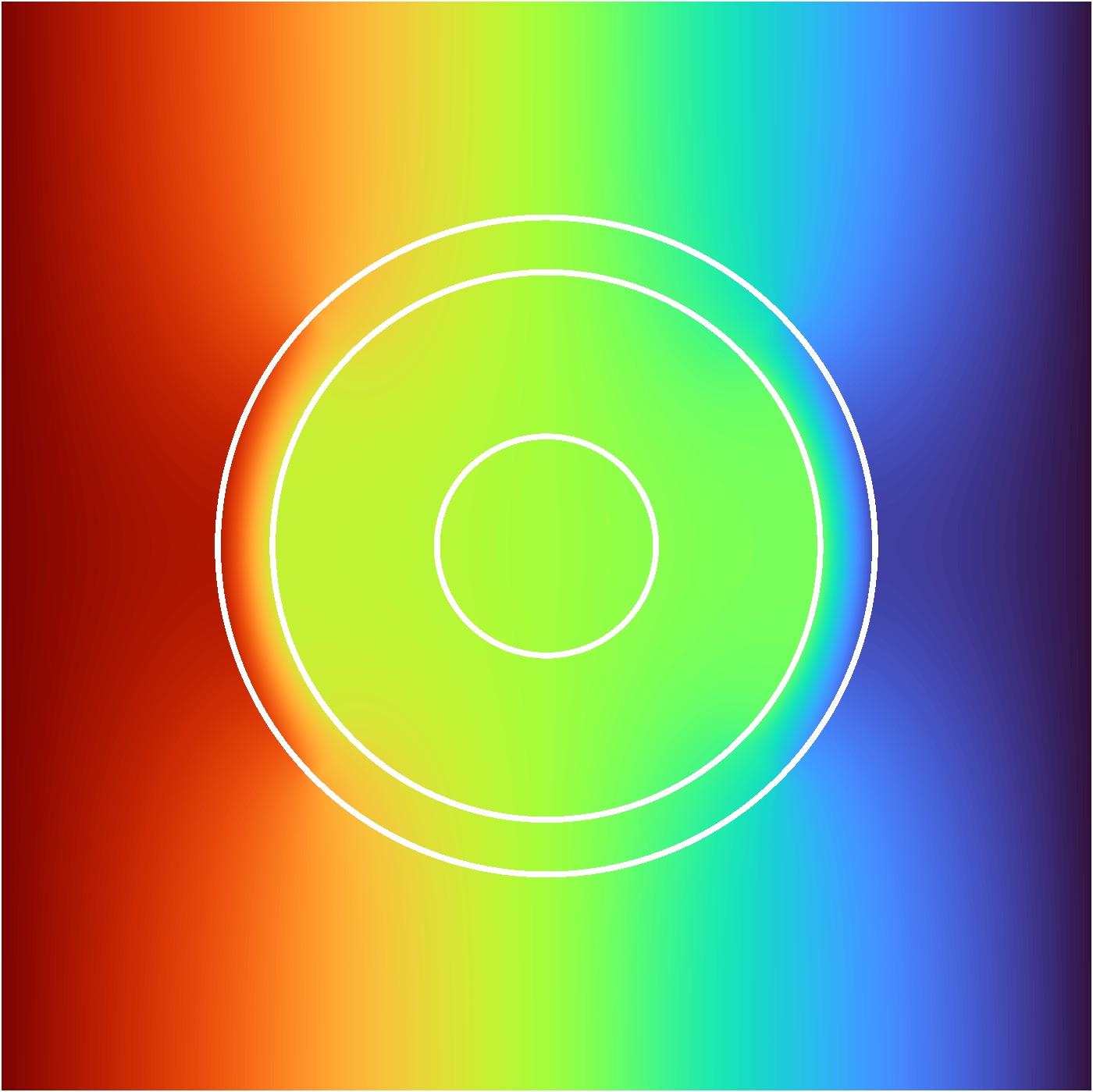}}
    \end{center}
    \end{subfigure}
    \begin{subfigure}[t]{0.07\textwidth}\begin{center}{
\includegraphics[width=1\textwidth]{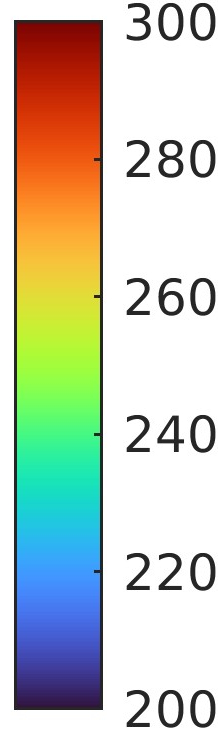}}
    \end{center}
    \end{subfigure} &
    \begin{subfigure}[t]{0.23\textwidth}\begin{center}{
\includegraphics[width=1\textwidth]{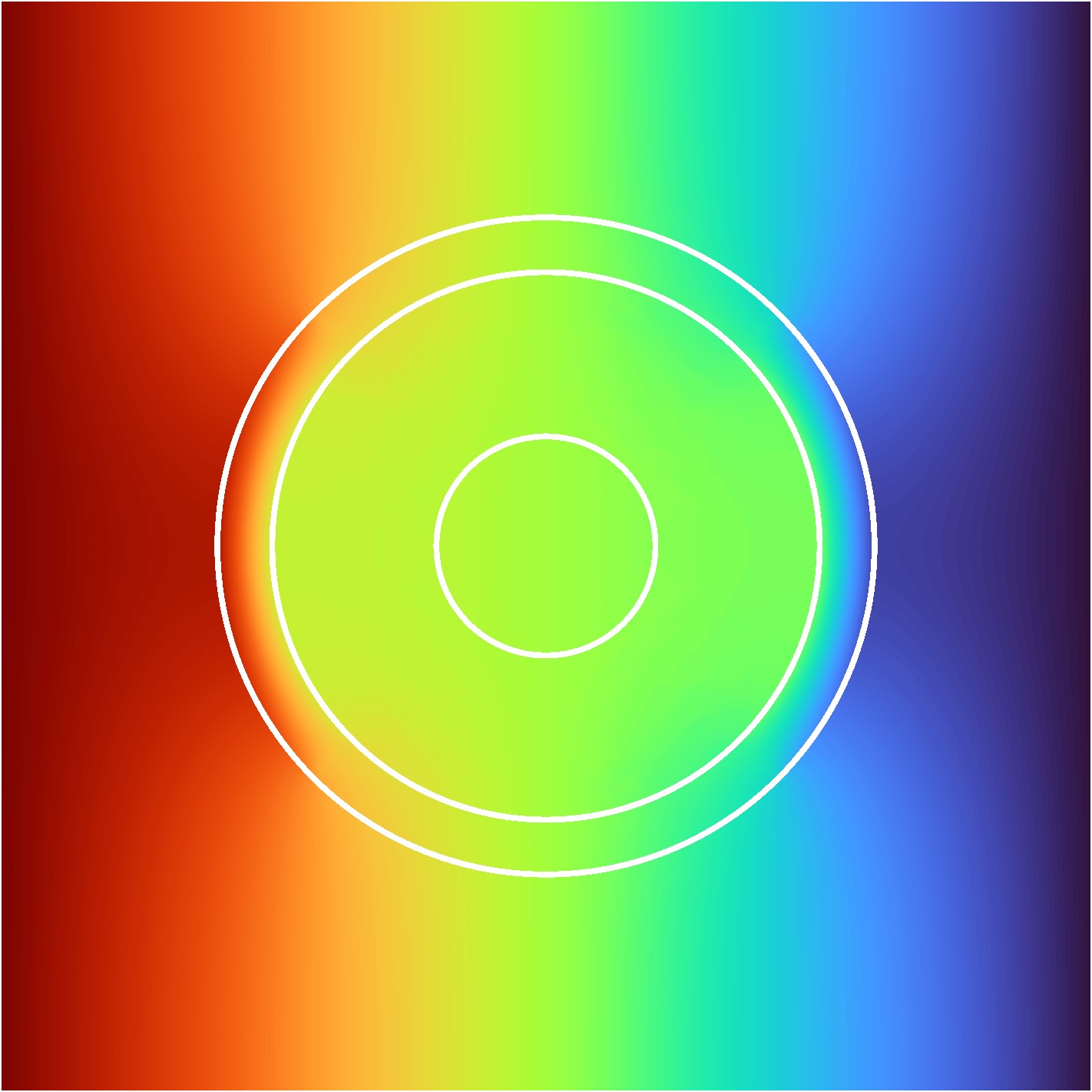}}
    \end{center}
    \end{subfigure} \begin{subfigure}[t]{0.07\textwidth}\begin{center}{
\includegraphics[width=1\textwidth]{Figures_Yang2016case/colorbar1.jpg}}
    \end{center}
    \end{subfigure} &
    \begin{subfigure}[t]{0.23\textwidth}\begin{center}{
\includegraphics[width=1\textwidth]{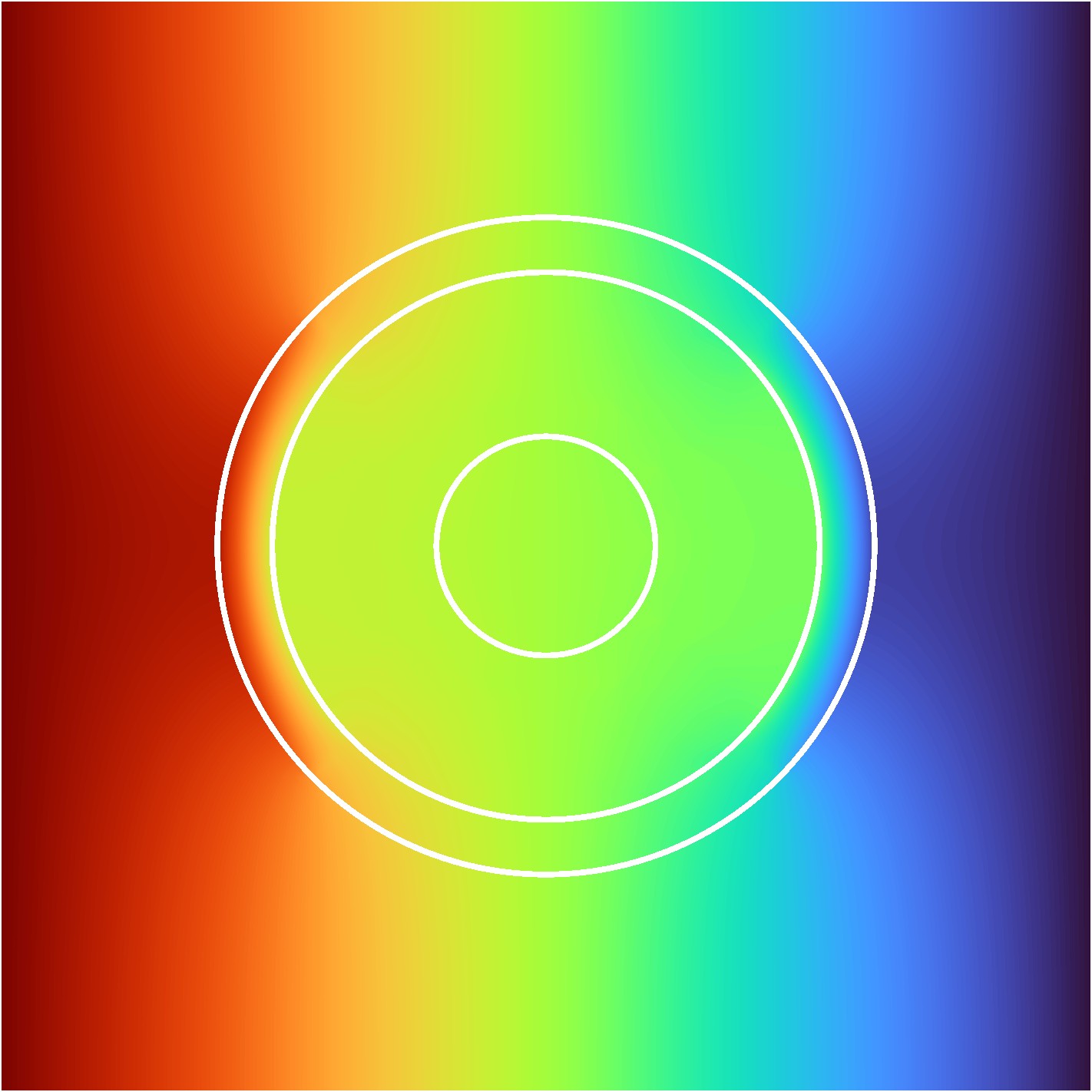}}
    \end{center}
    \end{subfigure} \begin{subfigure}[t]{0.07\textwidth}\begin{center}{
\includegraphics[width=1\textwidth]{Figures_Yang2016case/colorbar1.jpg}}
    \end{center}
    \end{subfigure}\\  
    \rotatebox{90}{\centering Temp. diff.~$T-\overline{T}$}  &
       &
    \begin{subfigure}[t]{0.23\textwidth}\begin{center}{
\includegraphics[width=1\textwidth]{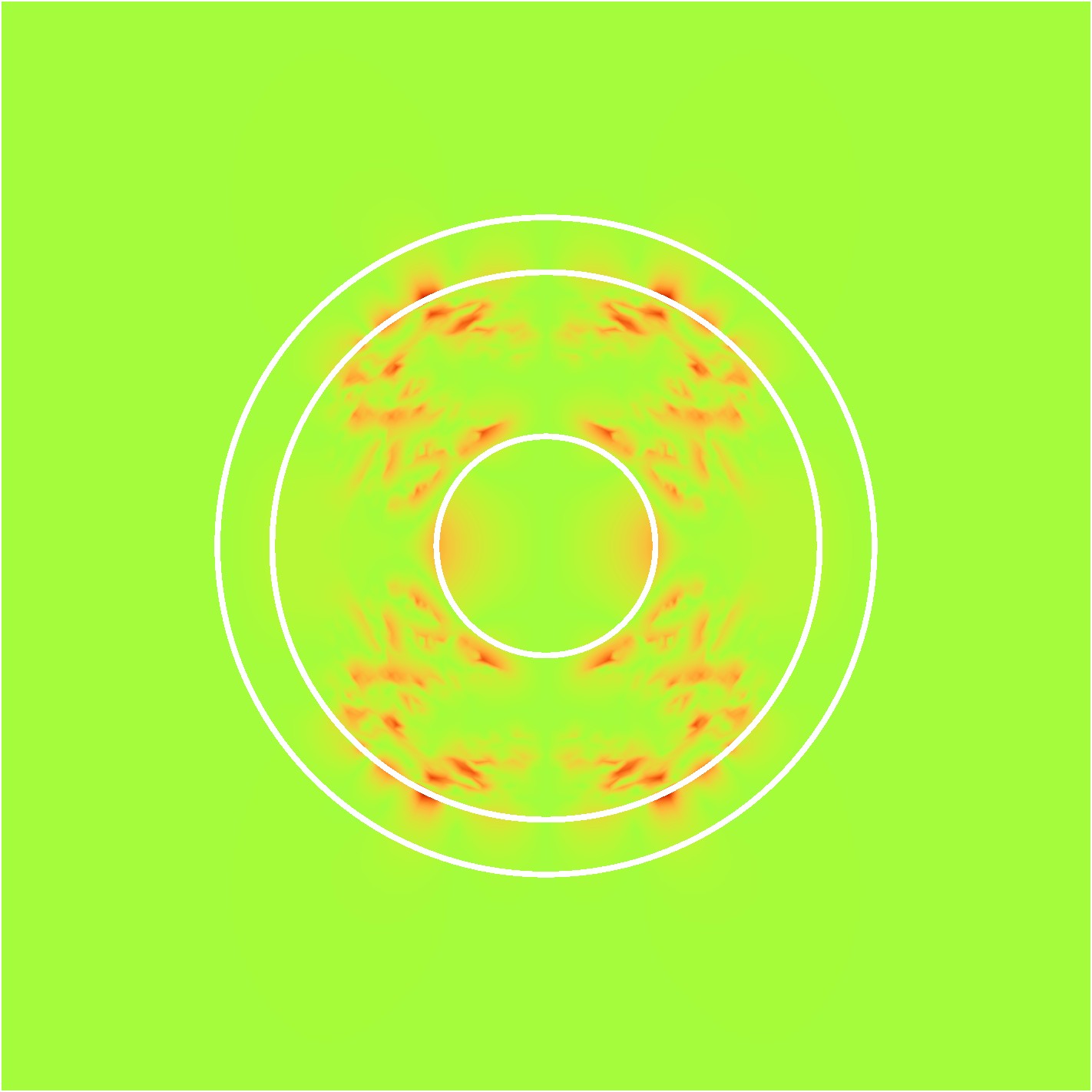}}
    \end{center}
    \end{subfigure} \begin{subfigure}[t]{0.068\textwidth}\begin{center}{
\includegraphics[width=1\textwidth]{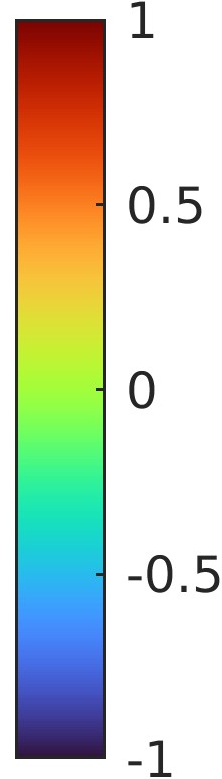}}
    \end{center}
    \end{subfigure} &
    \begin{subfigure}[t]{0.23\textwidth}\begin{center}{
\includegraphics[width=1\textwidth]{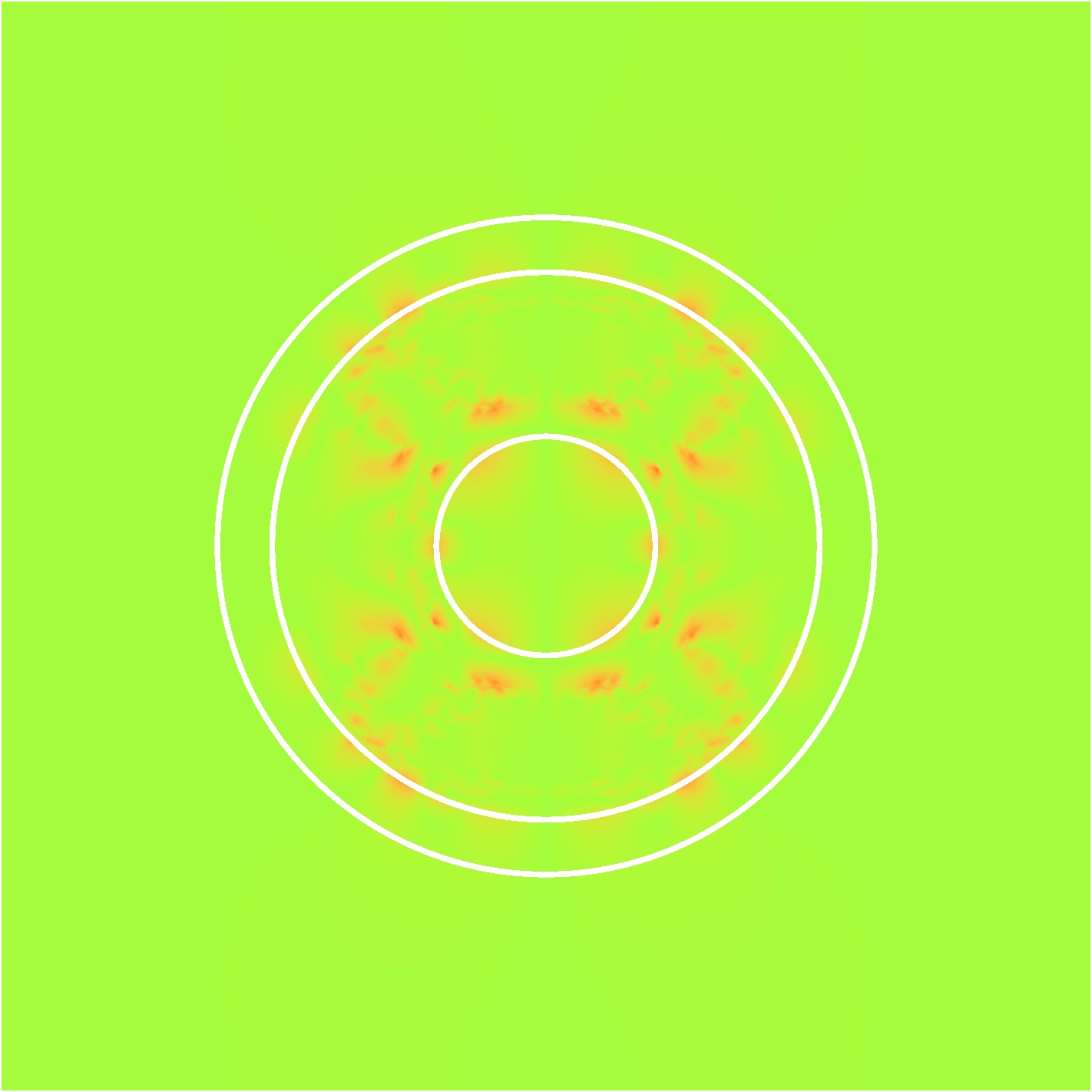}}
    \end{center}
    \end{subfigure} \begin{subfigure}[t]{0.068\textwidth}\begin{center}{
\includegraphics[width=1\textwidth]{Figures_Yang2016case/colorbar2.jpg}}
    \end{center}
    \end{subfigure} \\  
    \hline
    
\end{tabular}

}
\caption{For the thermal camouflage problem, flux flow and temperature distribution for (left column) a base material plate embedded with two insulator sectors (reference case), (middle column) a plate embedded with two insulator sectors, a conductive object (at the center), and an optimized thermal camouflage (without regularization) surrounding the object (sample I), and (right column) a plate embedded with two insulator sectors, a conductive object (at the center), and an optimized thermal camouflage (with regularization) surrounding the object (sample I, set C). The temperature and heat flux distributions are shown. The thermal camouflage reduces the temperature disturbance in $\mathrm{\Omega}_{\mathrm{in}}\cup \mathrm{\Omega}_{\mathrm{design}}\cup\mathrm{\Omega}_{\mathrm{out}}$.}
    \label{fig:Yang2016case_cloak_tempDiff}
\end{figure}

 In \fref{fig:Yang2016case_cloak_tempDiff}, we present the optimized topologies with and without regularizations achieving the desired camouflaging objective for sample I. We compare the flux flow, temperature distribution, and temperature difference with the reference temperature distribution. From the figure, it can be seen that the thermal camouflage produces the temperature signature same as the reference case.

\newcolumntype{U}{>{\centering\arraybackslash}m{5.6em}}
\newcolumntype{V}{>{\centering\arraybackslash}m{5.3em}}
\newcolumntype{W}{>{\centering\arraybackslash}m{0.6em}}
\renewcommand{\arraystretch}{1.5}   
\begin{figure}
\centering
\scalebox{0.9}{
\begin{tabular}[c]{| W | U | U | V | V | V | V |}
\hline		
 & \multirow{2}{5.5em}{\centering Initial topology} & \multirow{2}{5.5em}{\centering W/o any regularization $J_{\rm total}=J_{\rm cmflg}$} & \multicolumn{4}{c|}{Tikhonov +Volume reg. $J_{\rm total}=J_{\rm cmflg} + \chi J_{\rm Tknv} + \rho J_{\rm vol}$ }\\
\cline{4-7}		
  &  & & \centering Set-A $\chi=1 \times10^{0}$ $\rho=1\times10^{-2}$ & \centering Set-B $\chi=1 \times10^{0}$ $\rho=1\times10^{-1}$ & \centering Set-C $\chi=1 \times10^{0}$ $\rho=1\times10^{0}$ & {\centering
     Set-D \newline$\chi=1 \times10^{0}$ $\rho=0\times10^{0}$}\\
\hline 
\vspace{0.2cm}
   \rotatebox{90}{\centering \small Sample I}  &
   \vspace{0.2cm}
   \begin{subfigure}[t]{0.15\textwidth}{\centering\includegraphics[width=1\textwidth]{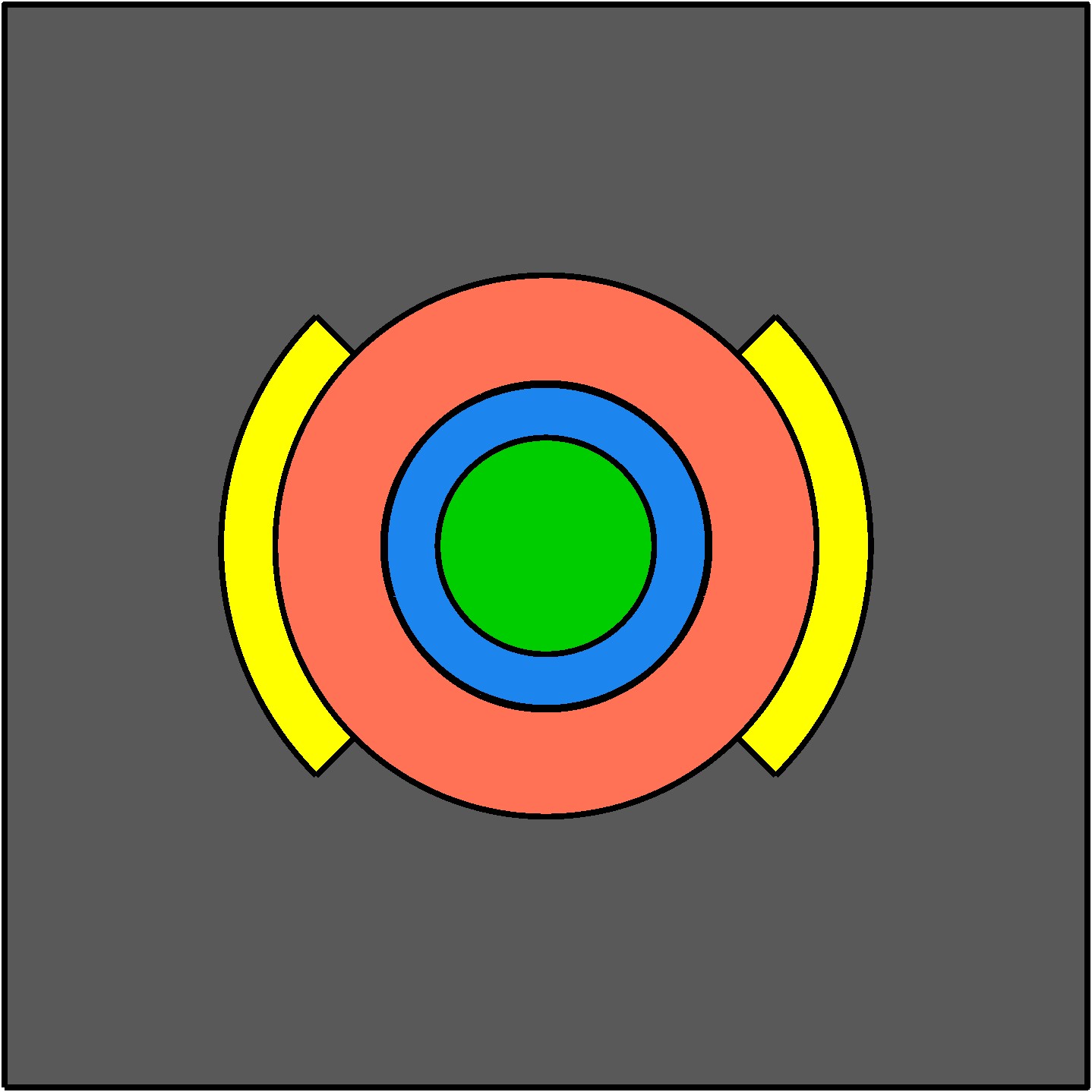}}
        \caption{\centering Initial topology}
       \label{fig:Yang2016case optTop TVSmth a}
    \end{subfigure}  & \vspace{0.2cm}
   \begin{subfigure}[t]{0.15\textwidth}{\centering\includegraphics[width=1\textwidth]{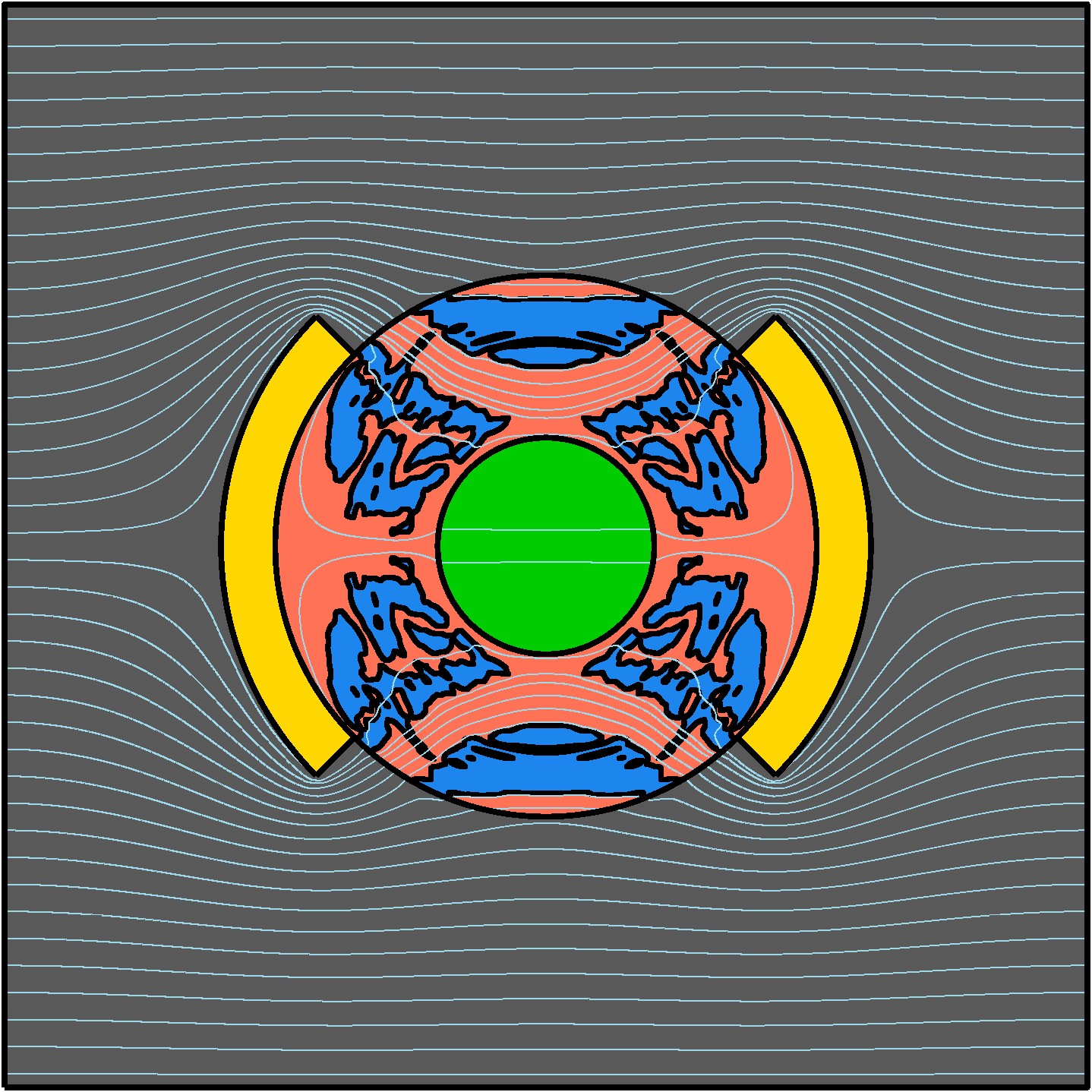}}
        \caption{\centering $J_{\rm cmflg}= 3.4436\times 10^{-3}$}
       \label{fig:Yang2016case optTop TVSmth b}
    \end{subfigure} & \vspace{0.2cm}
    \begin{subfigure}[t]{0.15\textwidth}{\centering\includegraphics[width=1\textwidth]{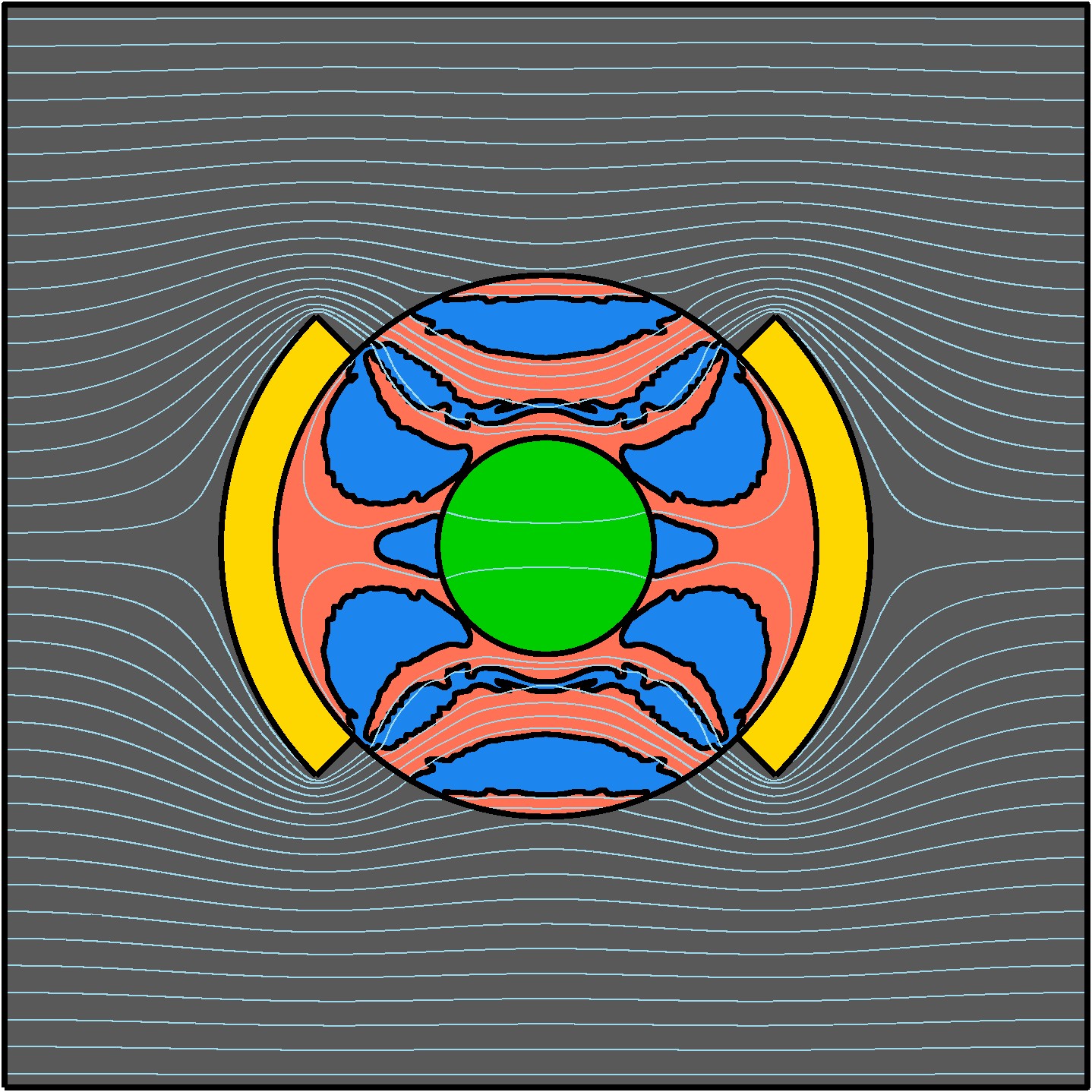}}
        \caption{\centering $J_{\rm cmflg}= 1.3200\times 10^{-3}$}
       \label{fig:Yang2016case optTop TVSmth c}
    \end{subfigure}& \vspace{0.2cm}
    \begin{subfigure}[t]{0.15\textwidth}{\centering\includegraphics[width=1\textwidth]{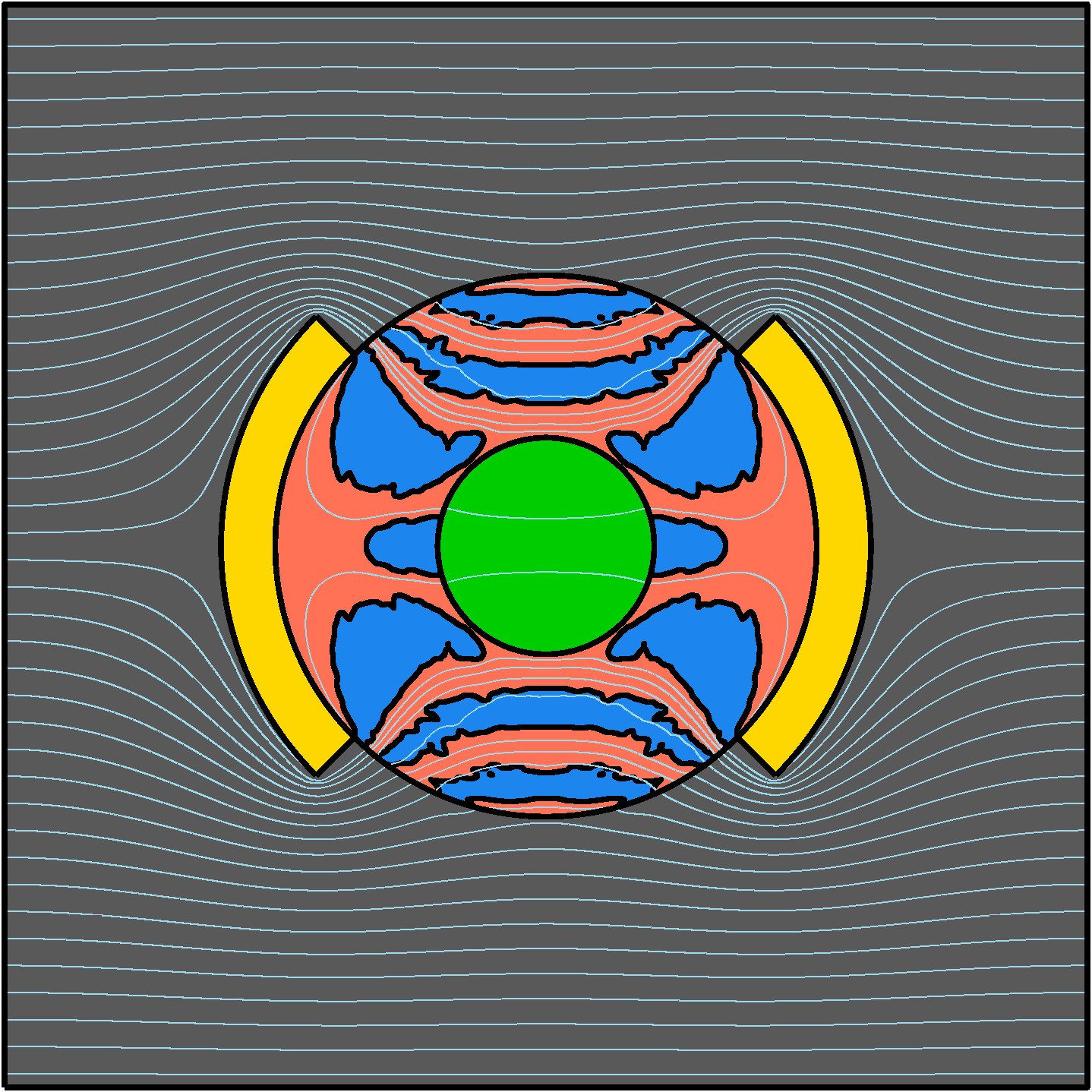}}
        \caption{\centering $J_{\rm cmflg}= 1.4132\times 10^{-3}$}
       \label{fig:Yang2016case optTop TVSmth d}
    \end{subfigure} &\vspace{0.2cm}
    \begin{subfigure}[t]{0.15\textwidth}{\centering\includegraphics[width=1\textwidth]{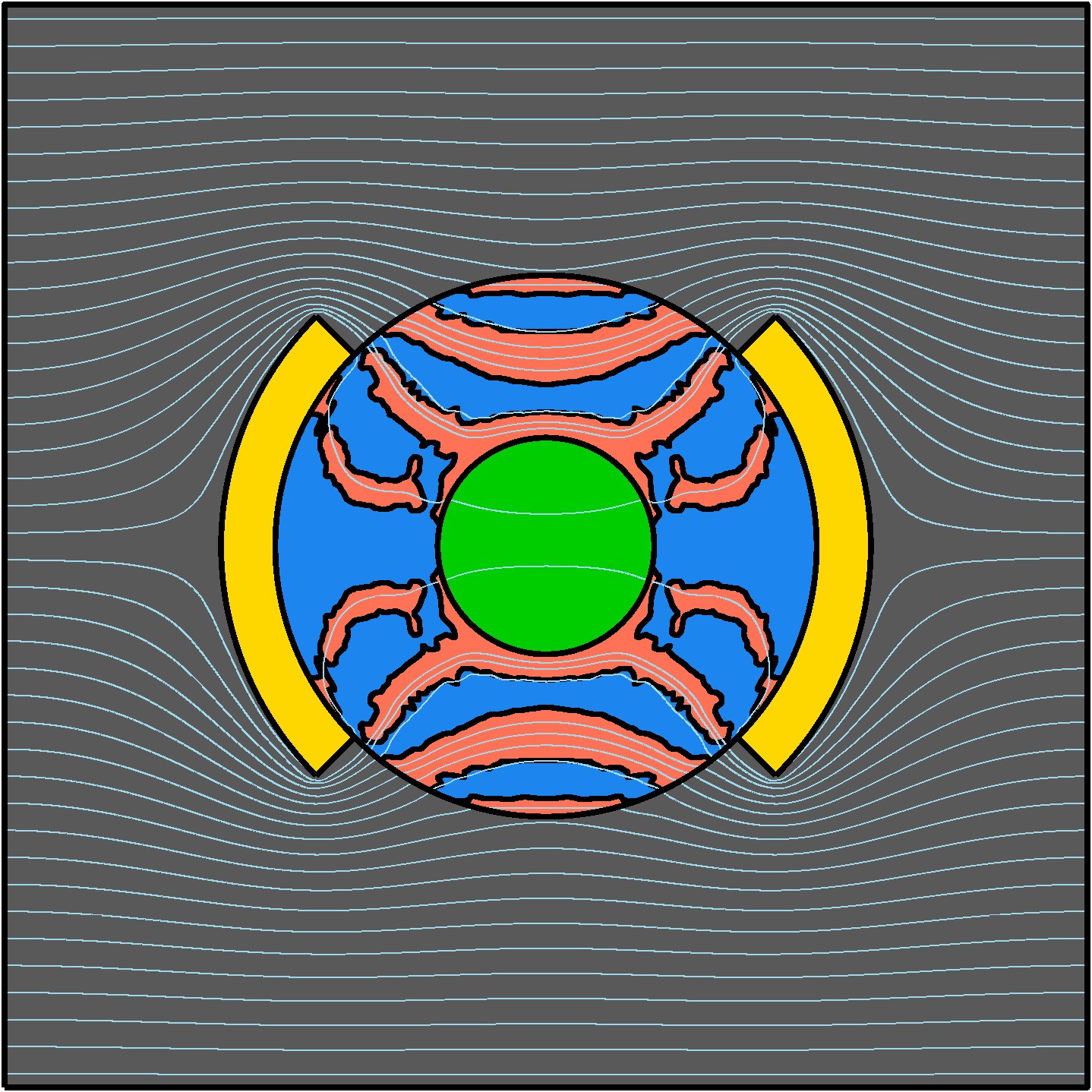}}
        \caption{\centering $J_{\rm cmflg}= 1.6917\times 10^{-3}$}
       \label{fig:Yang2016case optTop TVSmth e}
    \end{subfigure}& \vspace{0.2cm}
    \begin{subfigure}[t]{0.15\textwidth}{\centering\includegraphics[width=1\textwidth]{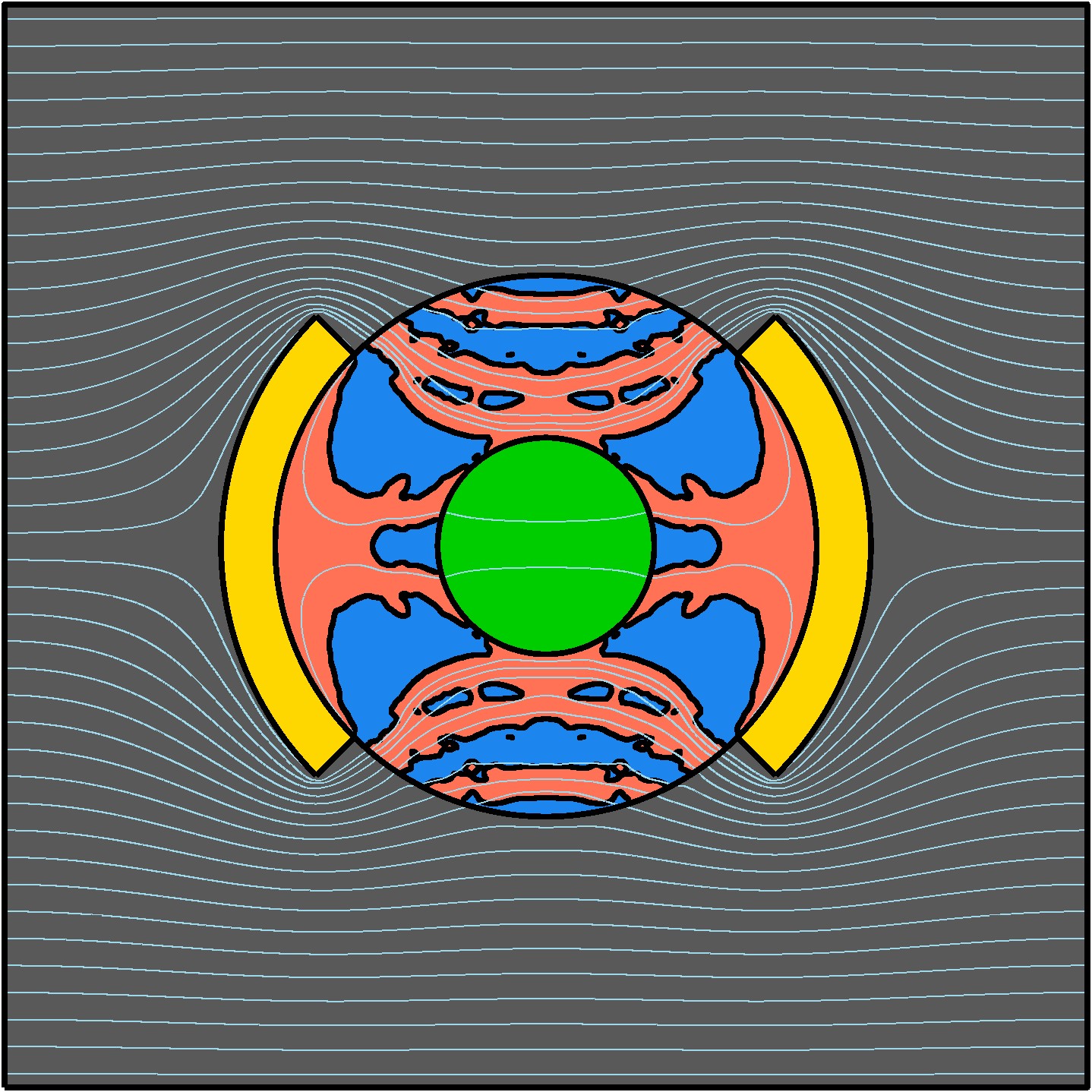}}
        \caption{\centering $J_{\rm cmflg}= 1.3591\times 10^{-3}$}
       \label{fig:Yang2016case optTop TVSmth f}
    \end{subfigure}\\
    \hline
   \rotatebox{90}{\centering \small Sample II}  &
   \vspace{0.2cm}
   \begin{subfigure}[t]{0.15\textwidth}{\centering\includegraphics[width=1\textwidth]{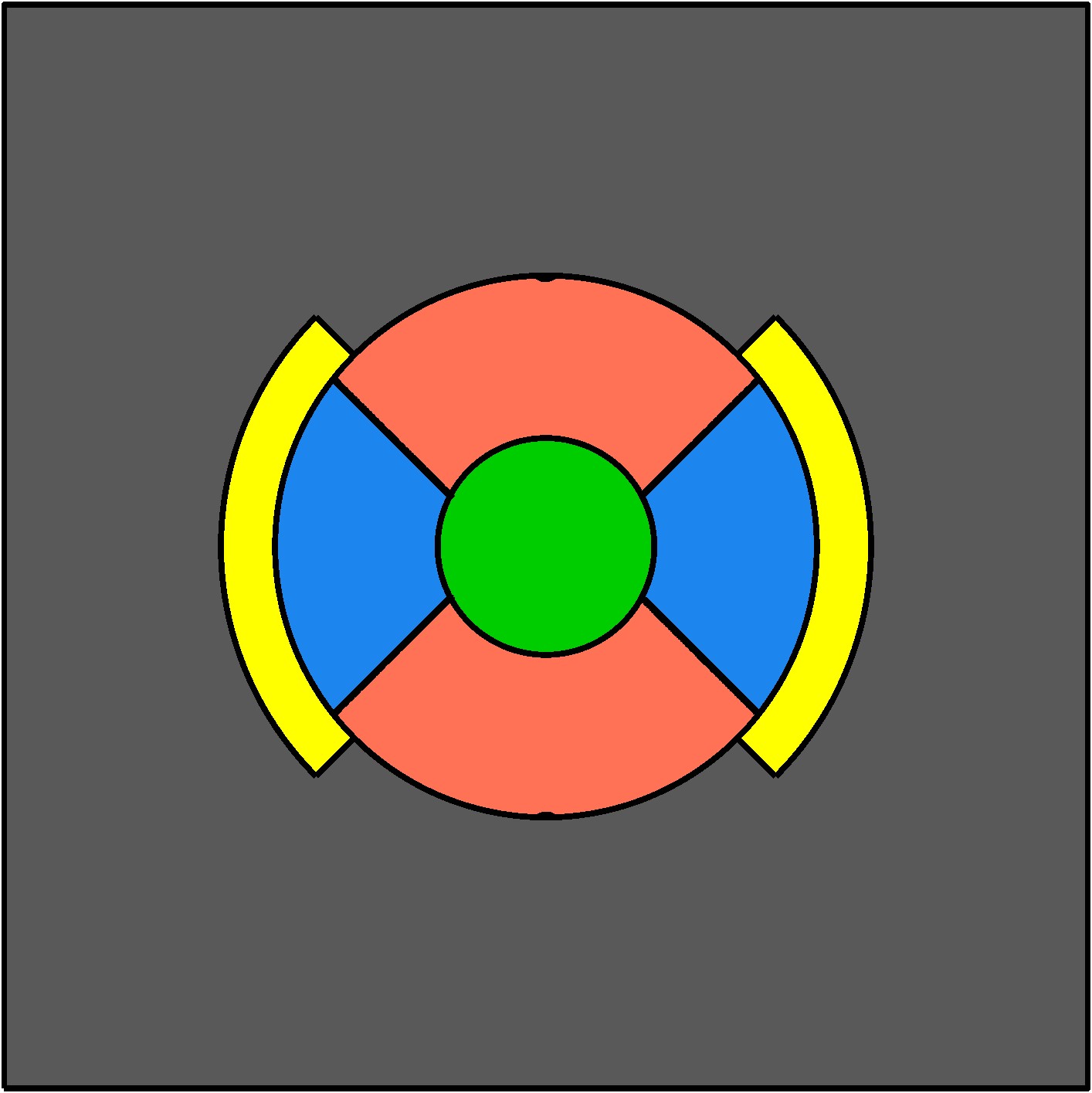}}
       \caption{\centering Initial topology}
       \label{fig:Yang2016case optTop TVSmth g}
    \end{subfigure}  & \vspace{0.2cm}
   \begin{subfigure}[t]{0.15\textwidth}{\centering\includegraphics[width=1\textwidth]{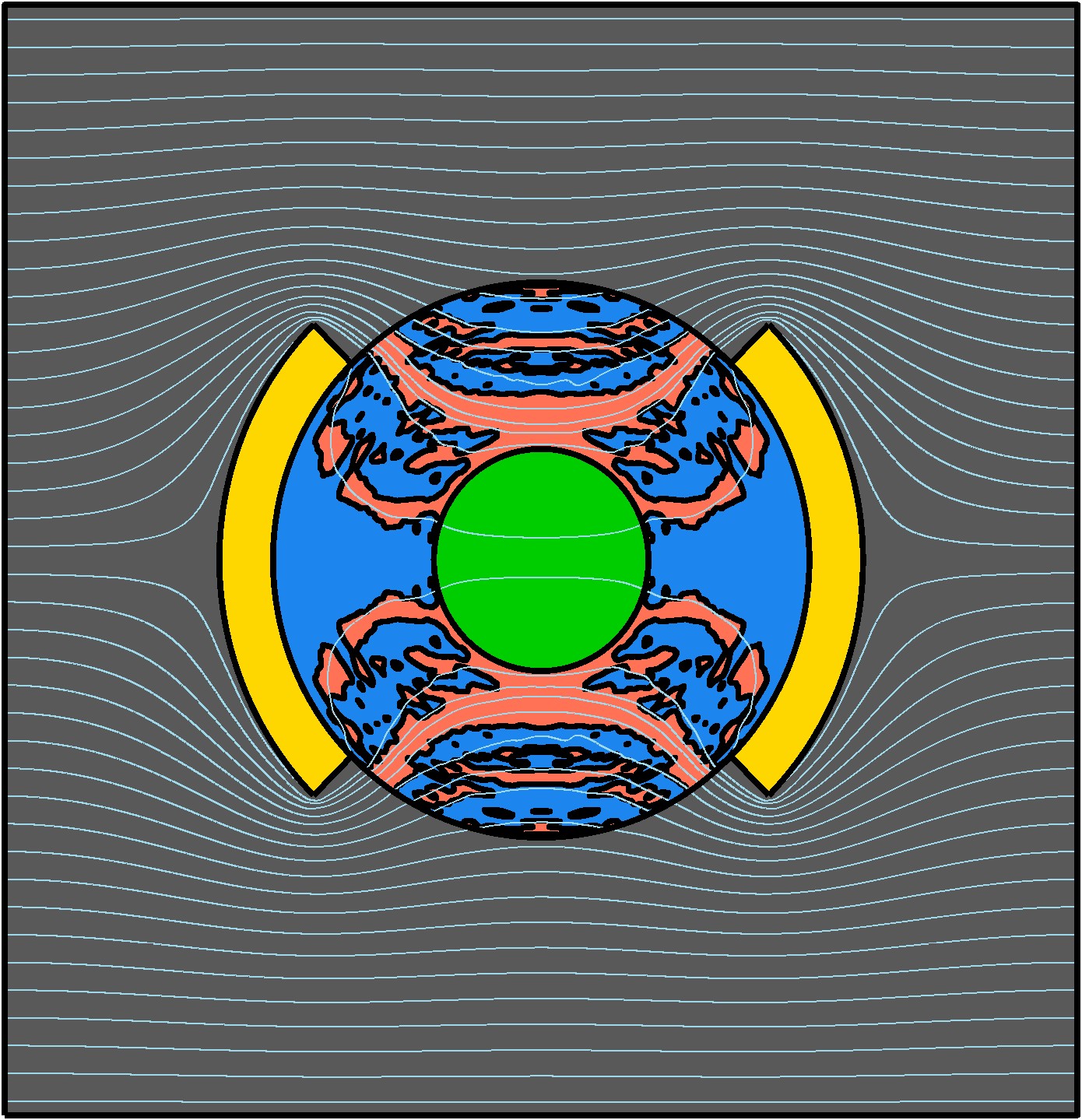}}
        \caption{\centering $J_{\rm cmflg}= 9.1497\times 10^{-4}$}
       \label{fig:Yang2016case optTop TVSmth h}
    \end{subfigure} & \vspace{0.2cm}
    \begin{subfigure}[t]{0.15\textwidth}{\centering\includegraphics[width=1\textwidth]{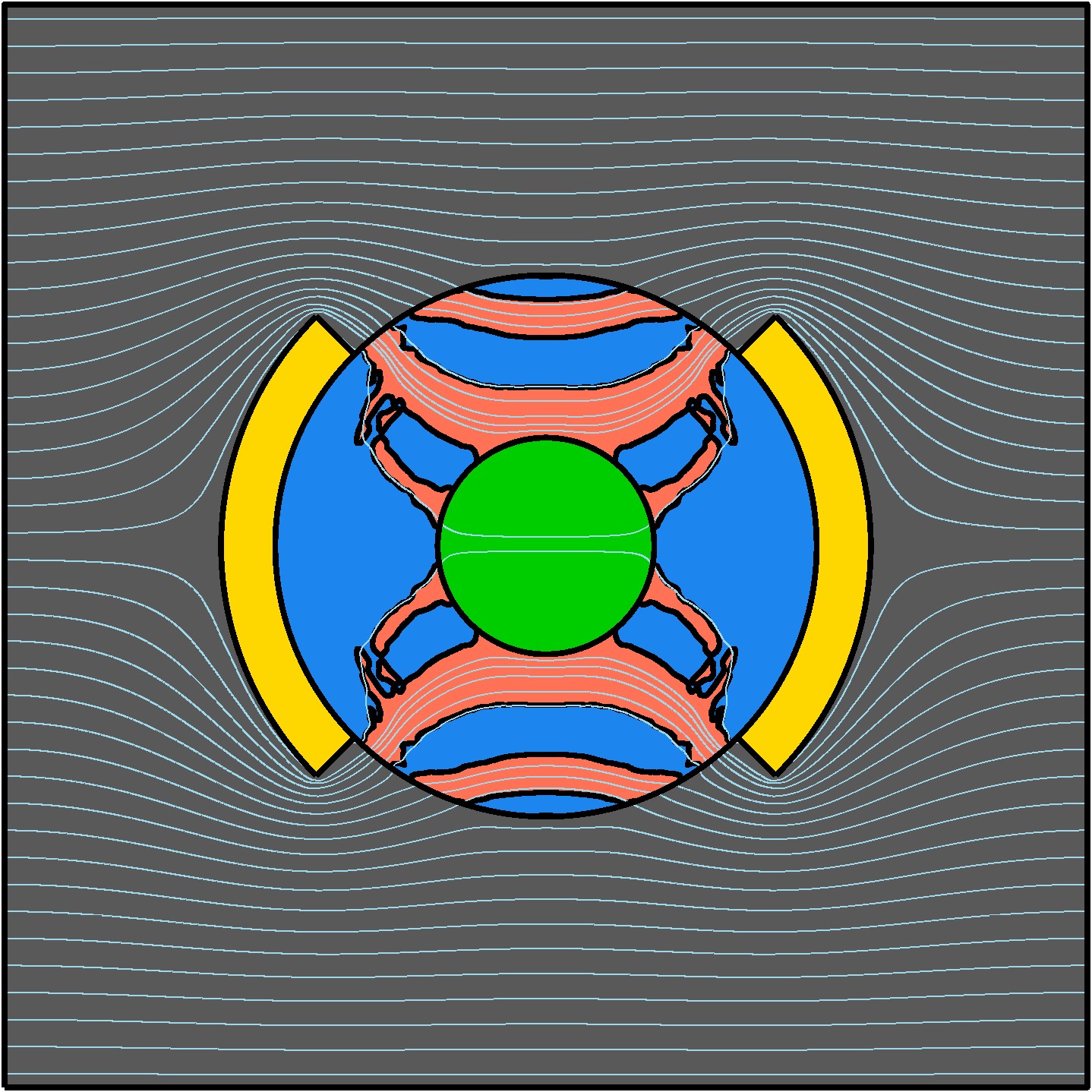}}
       \caption{\centering $J_{\rm cmflg}= 1.3130\times 10^{-3}$}
       \label{fig:Yang2016case optTop TVSmth i}
    \end{subfigure}& \vspace{0.2cm}
    \begin{subfigure}[t]{0.15\textwidth}{\centering\includegraphics[width=1\textwidth]{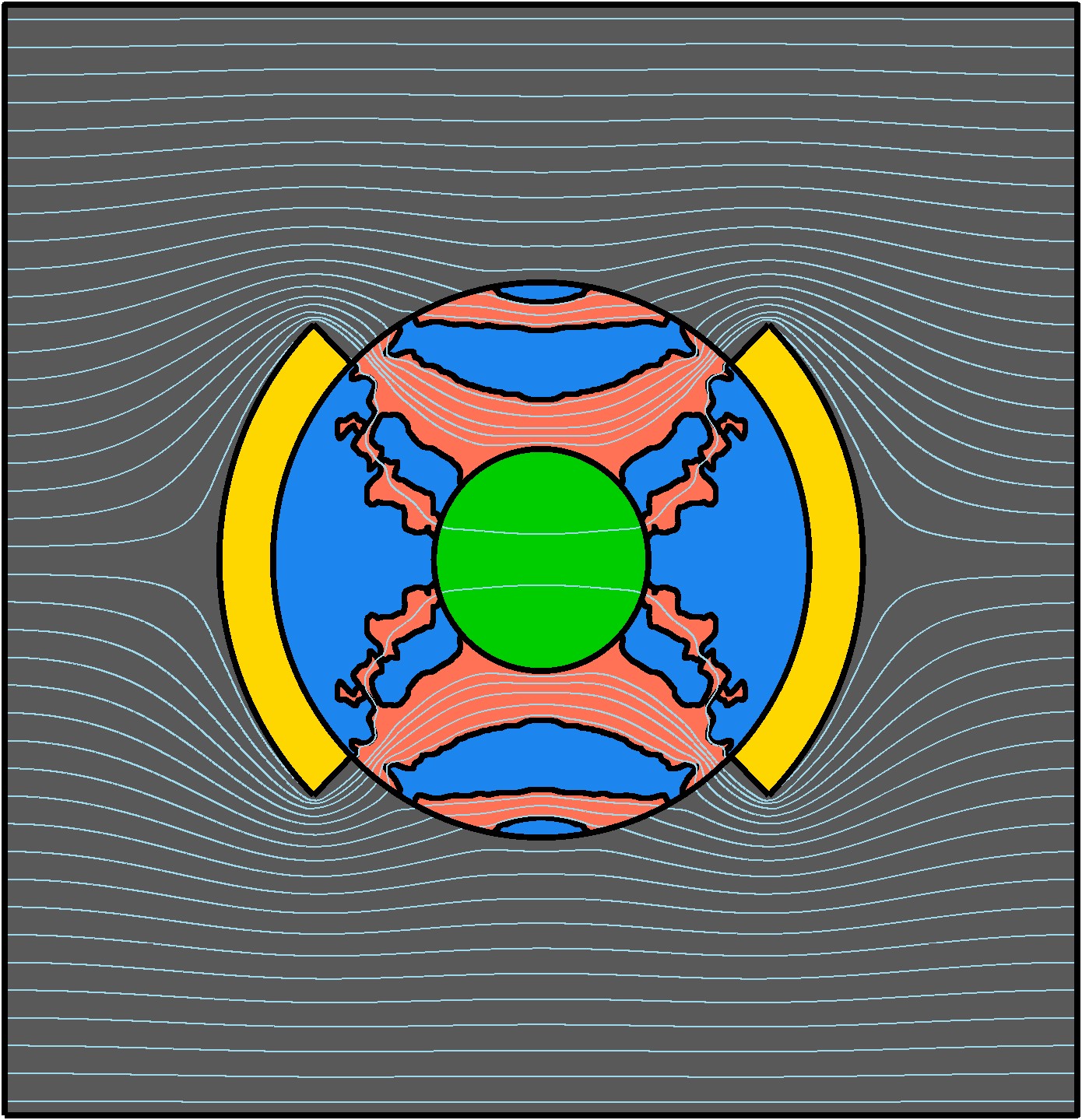}}
        \caption{\centering $J_{\rm cmflg}= 1.2862\times 10^{-2}$}
       \label{fig:Yang2016case optTop TVSmth j}
    \end{subfigure} &\vspace{0.2cm}
    \begin{subfigure}[t]{0.15\textwidth}{\centering\includegraphics[width=1\textwidth]{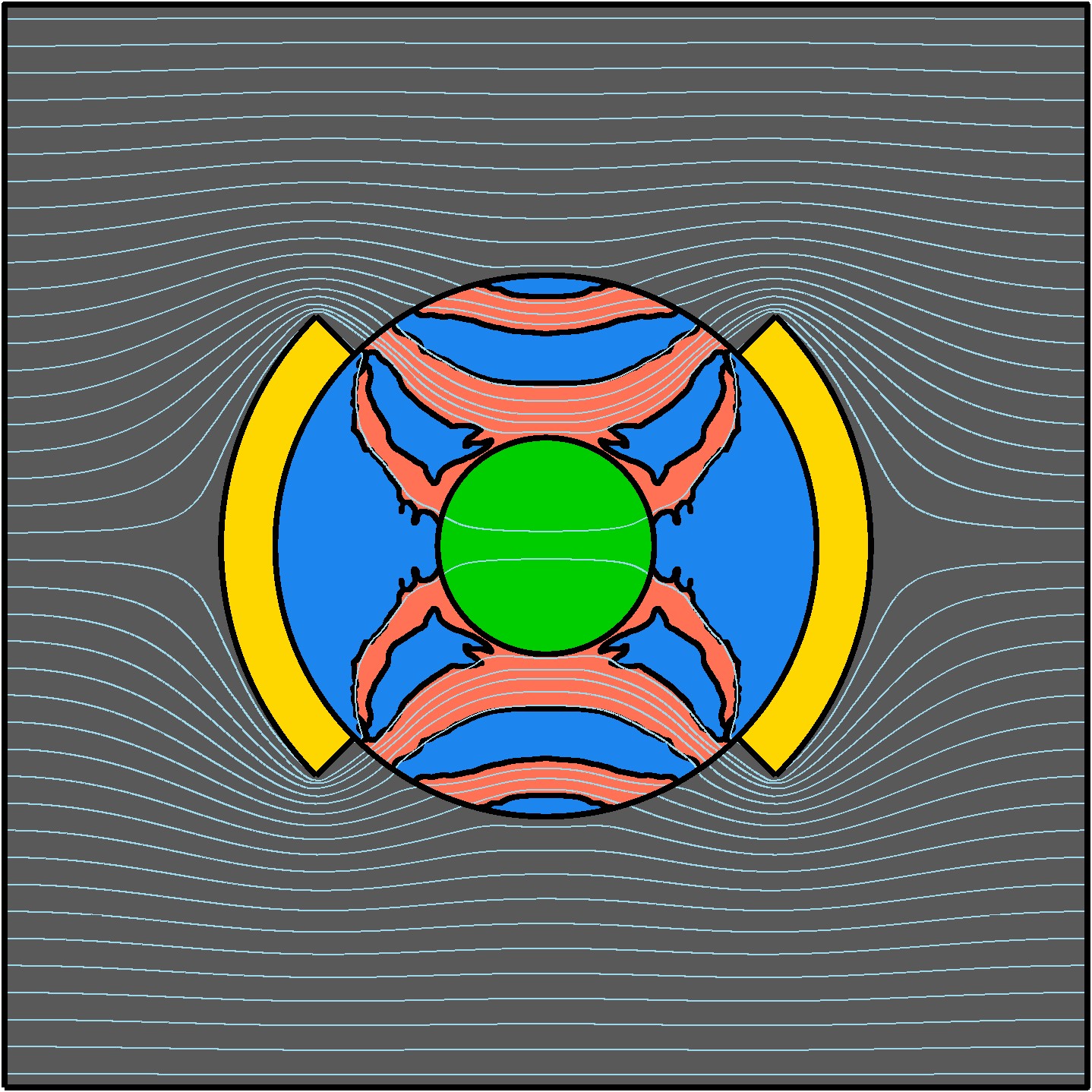}}
        \caption{\centering $J_{\rm cmflg}= 1.4967\times 10^{-3}$}
       \label{fig:Yang2016case optTop TVSmth k}
    \end{subfigure}& \vspace{0.2cm}
    \begin{subfigure}[t]{0.15\textwidth}{\centering\includegraphics[width=1\textwidth]{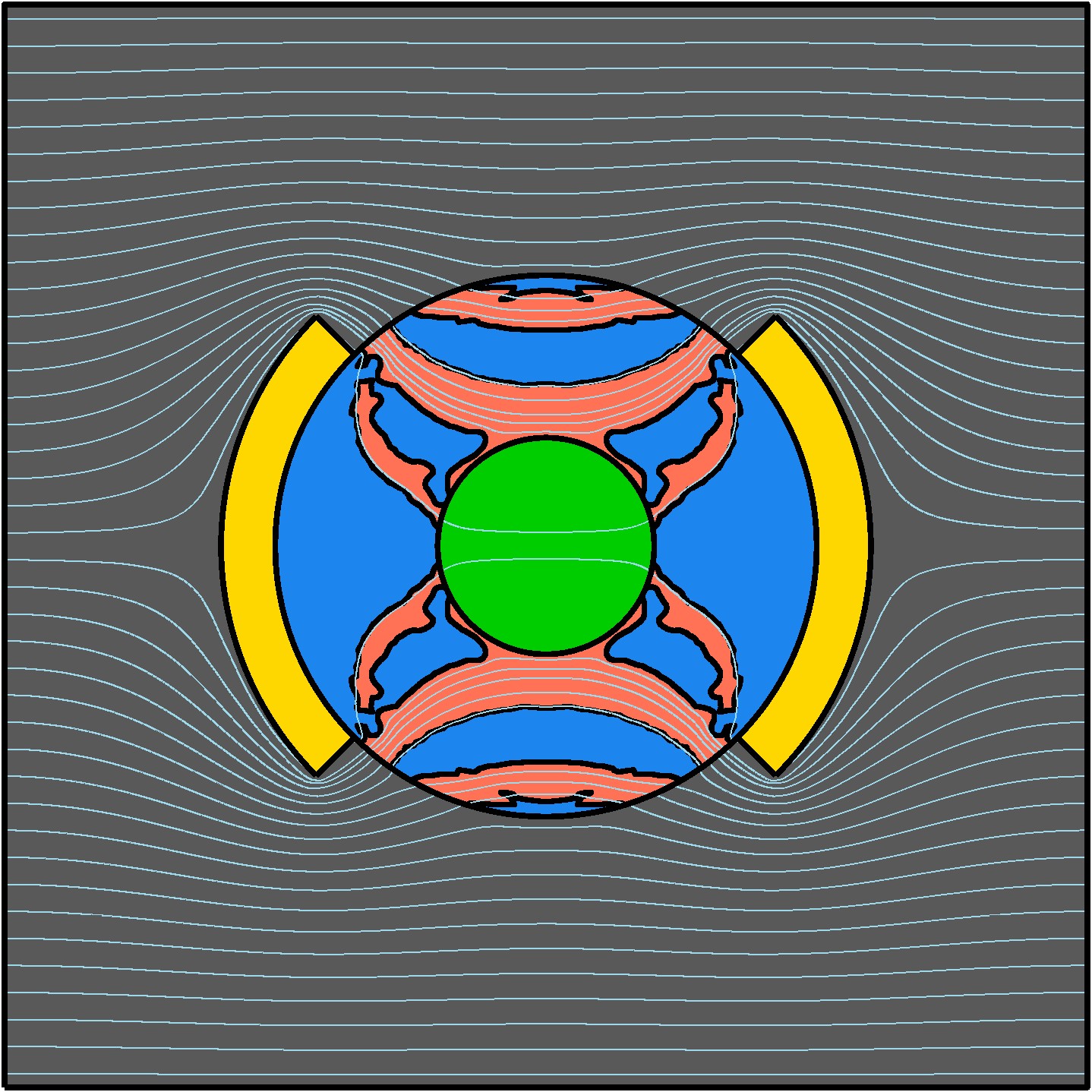}}
        \caption{\centering $J_{\rm cmflg}=9.8891 \times 10^{-4}$}
       \label{fig:Yang2016case optTop TVSmth l}
    \end{subfigure}\\ 
\hline
   \rotatebox{90}{\centering \small Sample III} &
   \vspace{0.2cm}
   \begin{subfigure}[t]{0.15\textwidth}{\centering\includegraphics[width=1\textwidth]{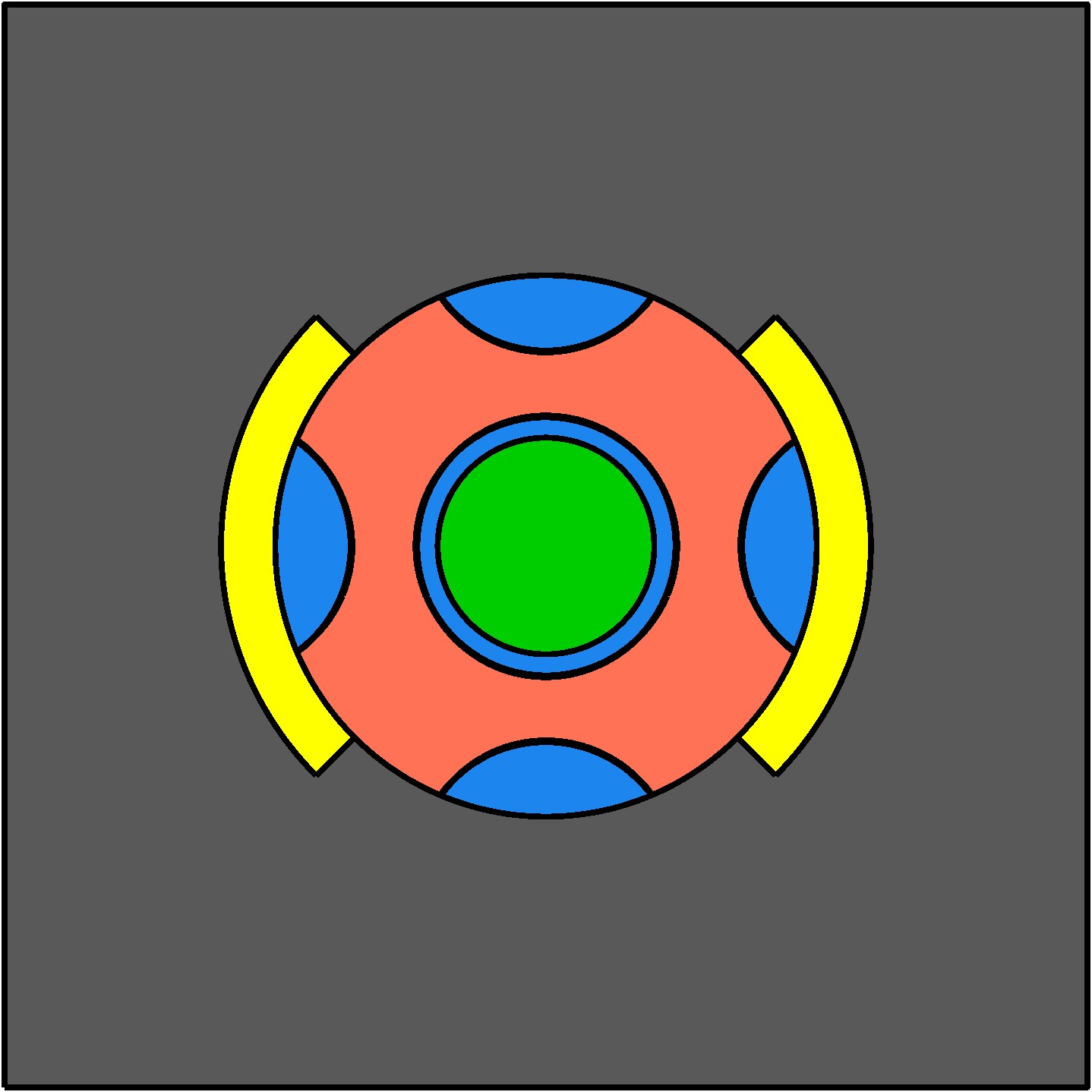}}
        \caption{\centering Initial topology}
       \label{fig:Yang2016case optTop TVSmth m}
    \end{subfigure}   & \vspace{0.2cm}
   \begin{subfigure}[t]{0.15\textwidth}{\centering\includegraphics[width=1\textwidth]{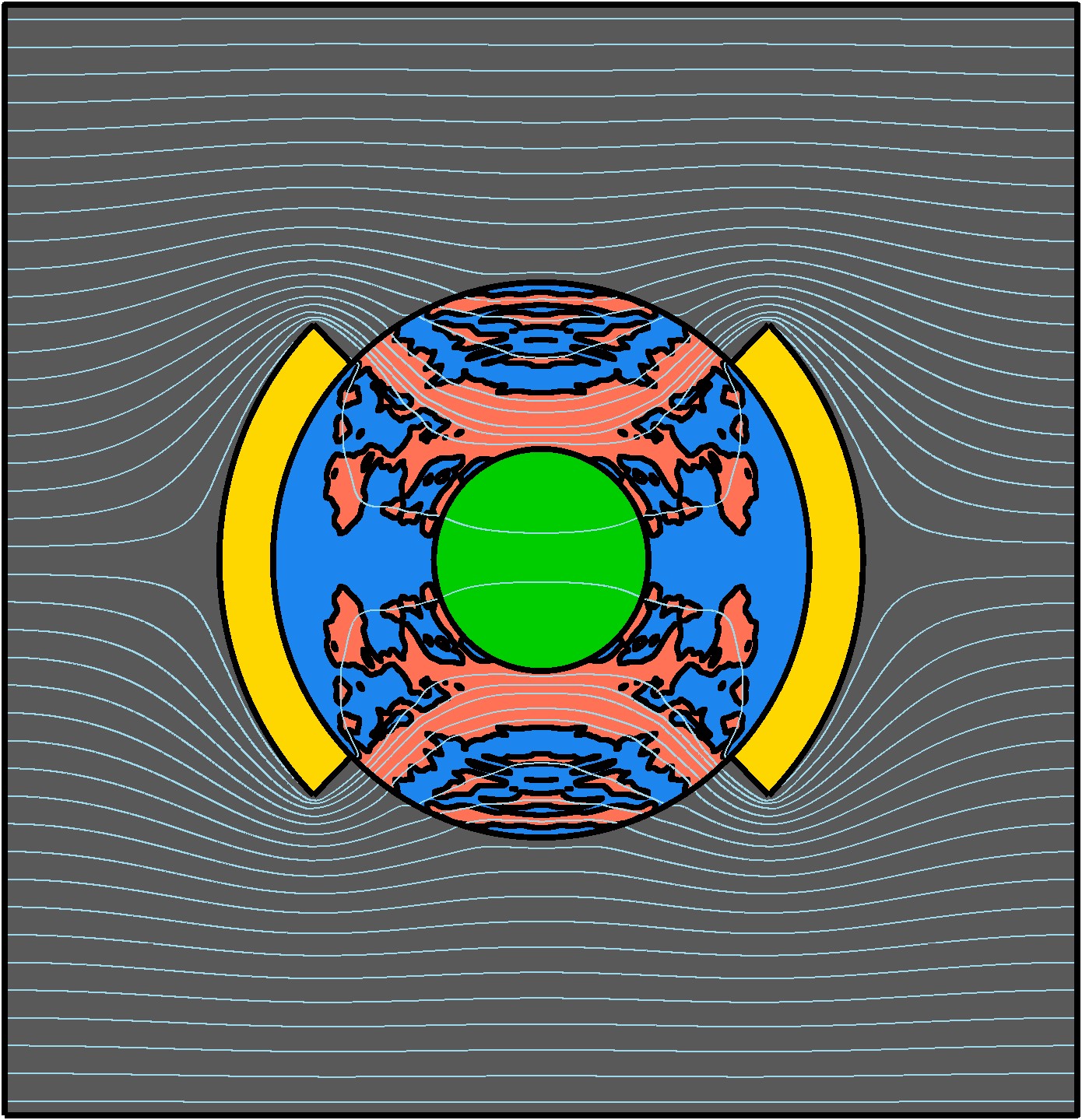}}
        \caption{\centering $J_{\rm cmflg}= 3.6367\times 10^{-3}$}
       \label{fig:Yang2016case optTop TVSmth n}
    \end{subfigure} & \vspace{0.2cm}
    \begin{subfigure}[t]{0.15\textwidth}{\centering\includegraphics[width=1\textwidth]{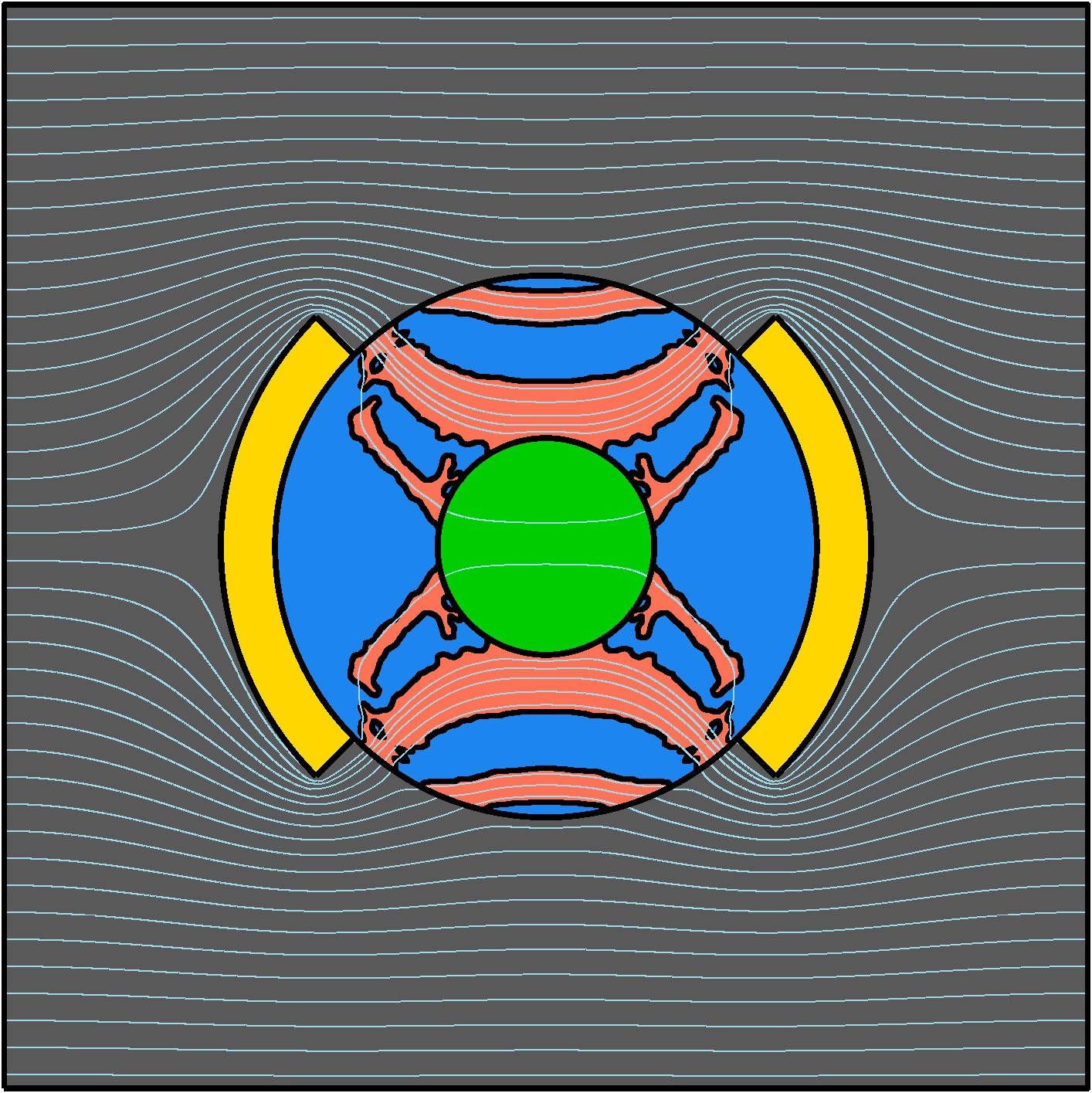}}
        \caption{\centering $J_{\rm cmflg}= 1.6029\times 10^{-3}$}
       \label{fig:Yang2016case optTop TVSmth o}
    \end{subfigure}& \vspace{0.2cm}
    \begin{subfigure}[t]{0.15\textwidth}{\centering\includegraphics[width=1\textwidth]{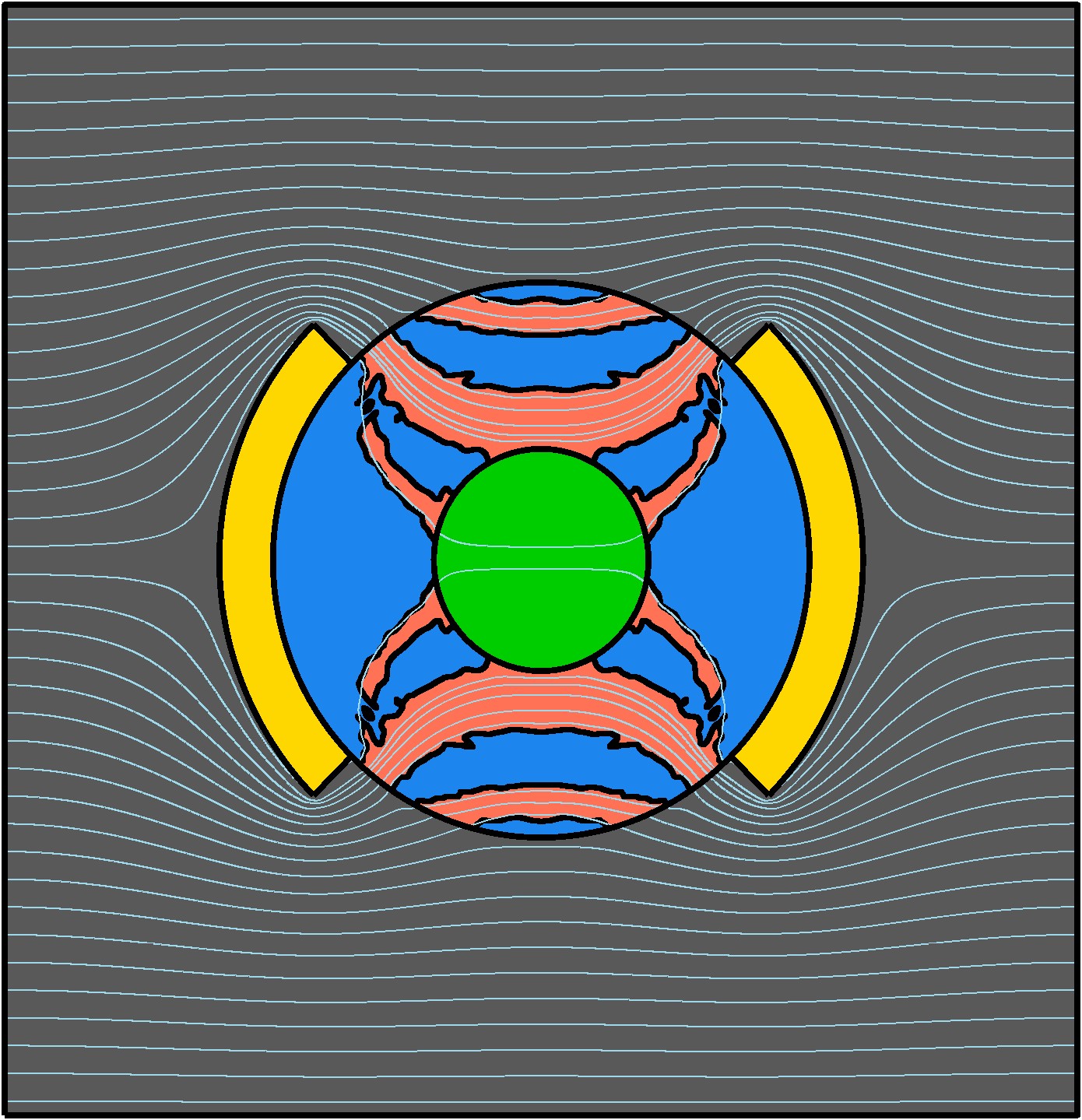}}
        \caption{\centering $J_{\rm cmflg}= 1.0982\times 10^{-3}$}
       \label{fig:Yang2016case optTop TVSmth p}
    \end{subfigure} &\vspace{0.2cm}
    \begin{subfigure}[t]{0.15\textwidth}{\centering\includegraphics[width=1\textwidth]{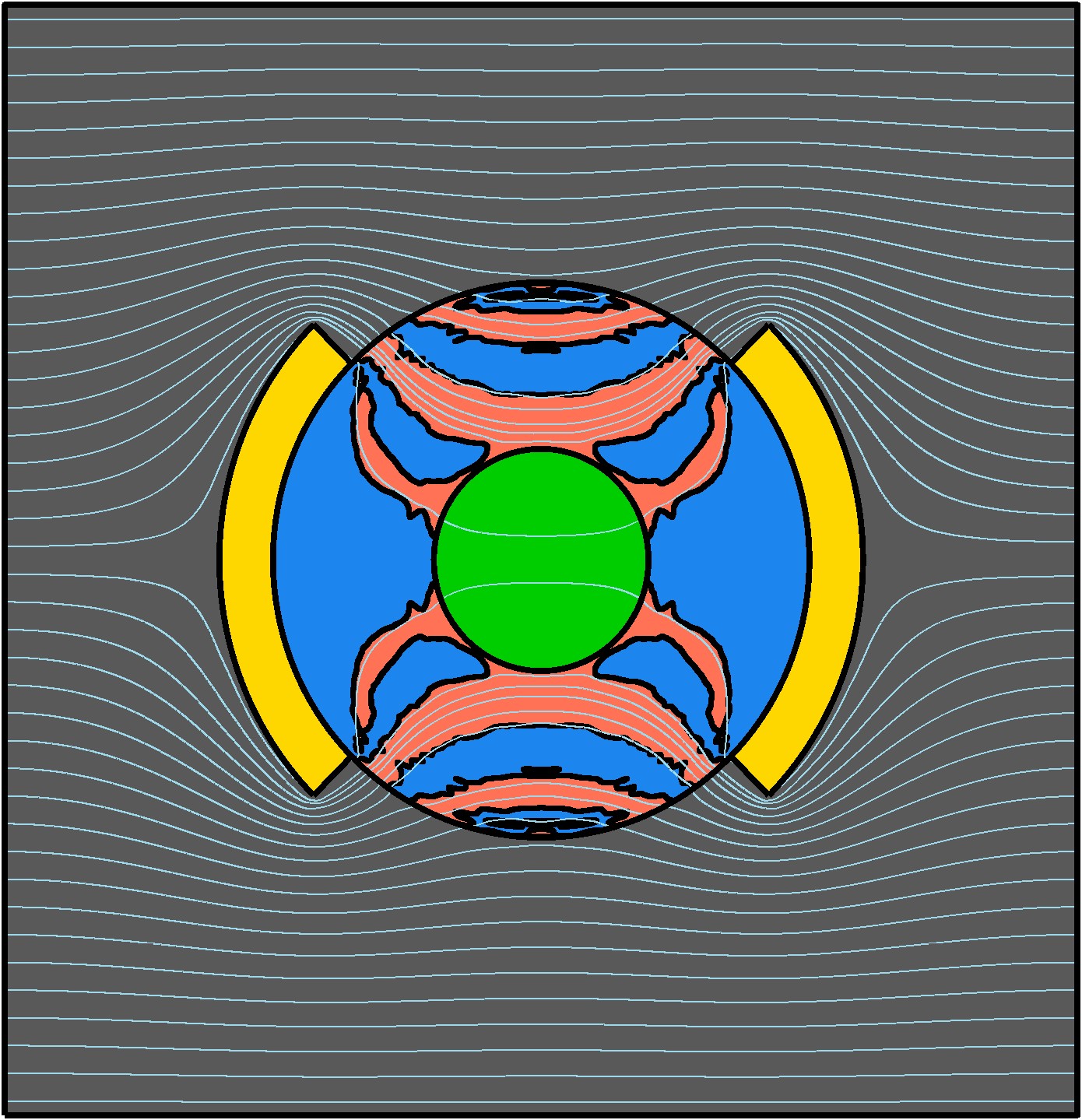}}
        \caption{\centering $J_{\rm cmflg}= 8.6531\times 10^{-4}$}
       \label{fig:Yang2016case optTop TVSmth q}
    \end{subfigure}& \vspace{0.2cm}
    \begin{subfigure}[t]{0.15\textwidth}{\centering\includegraphics[width=1\textwidth]{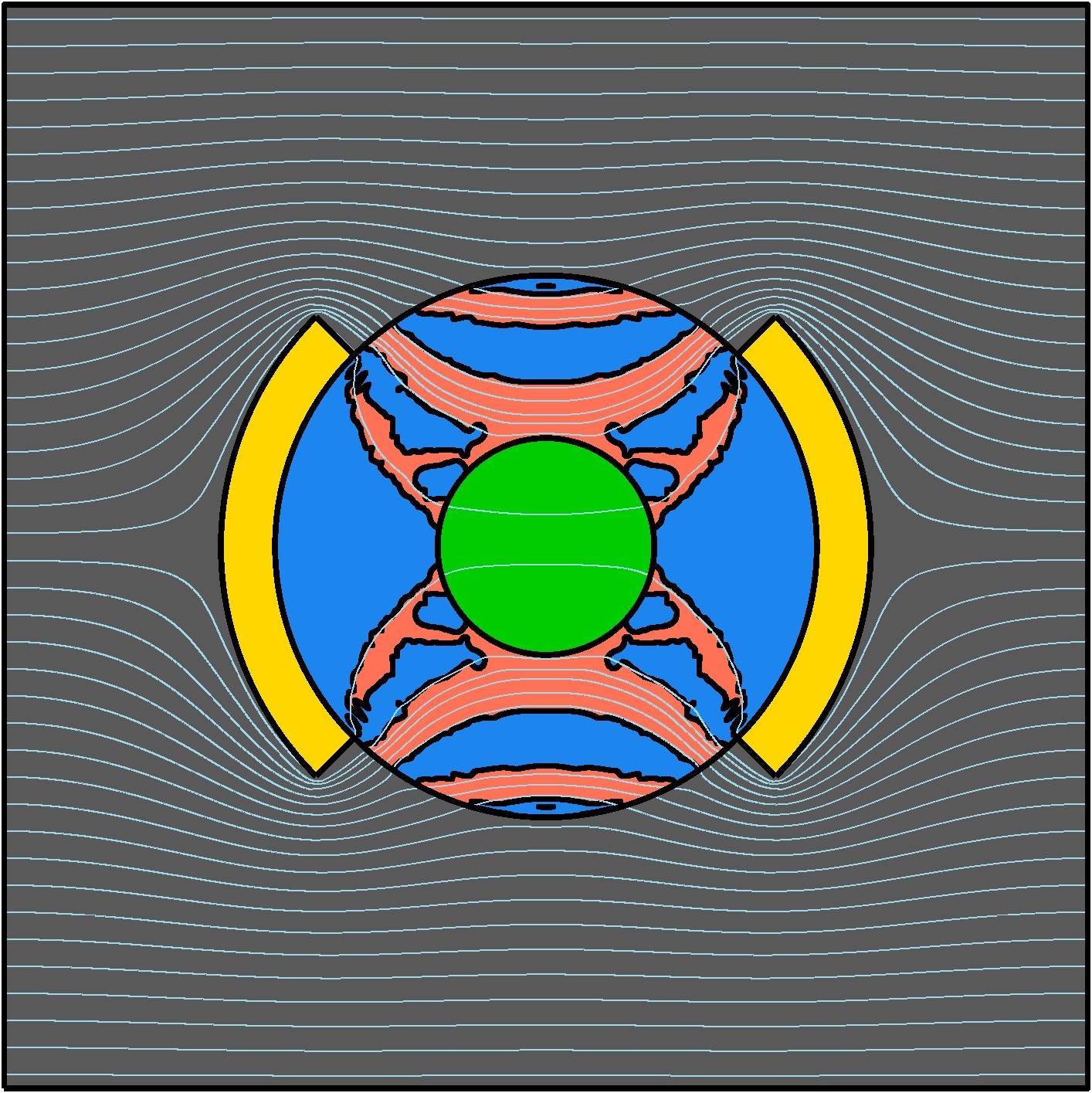}}
       \caption{\centering $J_{\rm cmflg}= 7.4425\times 10^{-4}$}
       \label{fig:Yang2016case optTop TVSmth r}
    \end{subfigure}\\
    \hline
\end{tabular}

}
\caption{For the thermal camouflage problem, initial topologies, optimized topologies without any regularization and with combined Tikhonov and volume regularization for $N_{\rm var}=1089$ with $\Delta=0.001$. Three initial topologies (samples I, II and III) are discretized with the corresponding design basis. Four sets of weighing parameters $\chi$ and $\rho$ are considered.}  
    \label{fig:Yang2016case optTop TVSmth}
\end{figure}

\renewcommand{\arraystretch}{1}   

\section{Conclusions}
\label{sec:Conclusions}
\par In the present article, we explore level set topology optimization method for the design of heat manipulators. The NURBS basis functions are utilized for parameterizations of the geometry, temperature field, and level set function. The thermal boundary value problem is solved using isogeometric analysis. For the optimization problem, a gradient-based advanced mathematical programming technique, Sequential Quadratic Programming (SQP), is used. To calculate the sensitivity, the adjoint method is utilized. Three numerical examples are presented: an annular ring problem, a thermal cloak problem, and a thermal camouflage problem that corroborates the efficiency of the proposed method. 
\par In the annular ring problem, the numerical results match the analytical results, which verifies the accuracy of the proposed method. The results indicate that, 
\begin{itemize}
    \item A smaller support bandwidth of the approximate Heaviside function improves the accuracy of the numerical solution and hence the optimization results. However, a too-small bandwidth affects the numerical sensitivity calculation and produces oscillations in the optimization. This would result in stopping optimization prematurely or making it behave erratically. 
    \item A remedy to oscillations coming from smaller bandwidths is to refine the solution mesh. However, the amount of refinement is also constrained by the associated computational cost. Therefore, a solution mesh and the bandwidth of approximate Heaviside functions are decided on the basis of a compromise among accuracy, stability, and computational cost. However, for the benchmark problem, we developed an empirical lower bound on the required mesh size for a given support bandwidth to give an estimate.  
    \item Considering the uniqueness of the solution in the given design space, the proposed optimization procedure successfully generates the topologies close to the analytical optimal topology irrespective of the initial topologies. The values of the objective function are also within 1\% variation of the optimal value.
\end{itemize}

\par In the thermal cloak problem, we optimized the topology of an annular-shaped thermal cloak.  We also explored two regularizations (Tikhonov regularization and volume regularization) and their combination to generate smoother and more practical optimized topologies. In the end, the efficacy of the current method for other geometries of the thermal cloak was also examined. The results indicate that, 
 \begin{itemize}
    \item The problem is not convex and can have different optimized topologies depending on the initial topology and the number of design variables. Several optimized topologies are close to one known analytical solution-a bilayer cloak with a circular interface. The objective function reaches values of order $10^{-9}-10^{-10}$ for all cases.
    \item By providing regularizations during optimization, smoother and less complex topologies are generated with a slight compromise on the objective function value.
    \item The values of the parameters $\chi$ and $\rho$, the weights of Tikhonov and volume terms, respectively, are very difficult to predict apriori. They are often decided based on a trade-off between the complexity of the topologies and fulfillment of the cloaking objective using the trial and error method.
    \item The proposed method can handle different geometries and shapes with the same effectiveness.
\end{itemize}

\par In the last example, we optimized a thermal camouflage mimicking the temperature signature of a reference scenario. Similar to the thermal cloak problem, we exploited regularizations to generate smoother and more practical designs.  The results are equally satisfactory.
\par The proposed method is generalized to apply to any heat manipulator with varying geometries. Also, keeping in mind that the heat flux manipulation problems are often not unique, the method can find other possible topologies for the already developed heat manipulators based on analytical methods such as transformational thermotics, scattering cancellation method, etc. In addition, many times we deal with a larger number of design variables, and therefore, the gradient-based optimization (with the adjoint method to find sensitivity) works faster than other gradient-free optimization processes. However, the method in its current form lacks the ability to carry out geometry-constrained optimization. Consequently, the method can be extended to impose geometry-constrained optimization in future work. In addition, the optimization problem can also be extended to multiphysics problems such as thermomagnetic, thermoacoustic, thermoelectrics, etc. In another direction, the method can be improved to reduce the computational time by using surrogate models for boundary value problems, acceleration techniques, and NURBS hyper-surfaces~\cite{Montemurro2021topology,Giulio2021nurbs}.

\appendix
\section*{Appendix}

\section{Analytical solutions of the annular ring problem}
\label{sec:Appendix A}
The analytical solutions of the boundary value problem (\eref{eq:BenchAnn BVP}) is given as follows,
\begin{subequations}
\begin{align}
     &T_{a}(r)=c_{a} +d_{a} \log(r),\\
     &T_{b}(r)=c_{b} +d_{b} \log(r),
\end{align}
\end{subequations}
where coefficients $c_a, d_a, c_b, d_b$ are obtained from the following system of algebraic equations, resulted from satisfying boundary conditions (\eref{eq:BenchAnn BVP}) for $r = R_a$, $r = R_b$ and $r = R_L$,
\begin{equation}
   \begin{bmatrix}
           1 & \log(R_a) & 0 & 0\\
           0 & 0 & 1 & \log(R_b) \\
           1 & \log(R_L) & -1 & -\log(R_L) \\
           0 & \kappa_a & 0 & -\kappa_b
         \end{bmatrix} \begin{bmatrix}
           c_a \\
           d_a \\
           c_b \\
           d_b
         \end{bmatrix} = \begin{bmatrix}
           T_a \\
           T_b \\
           0 \\
           0
         \end{bmatrix},
\end{equation} 
which can be written as,
\begin{equation}
   \begin{aligned}
   c_{a}&=-\frac{-\kappa_{b}T_{b} \log(R_{a})-\kappa_{a}T_{a} \log(R_{I})+\kappa_{b}T_{a} \log(R_{I})+\kappa_{a}T_{a}\log(R_{b})}{\kappa_{b}\log(R_{a})+\kappa_{a}\log(R_{I})-\kappa_{b}\log(R_{I})-\kappa_{a}\log(R_{b})},\\
   d_{a}&=\frac{\kappa_{b}(T_{a}-T_{b})}{\kappa_{b}\log(R_{a})+\kappa_{a}\log(R_{I})-\kappa_{b}\log(R_{I})-\kappa_{a}\log(R_{b})},\\
  c_{b}&=-\frac{\kappa_{b}T_{b} \log(R_{a})+\kappa_{a}T_{b} \log(R_{I})-k_{b}T_{b} \log(R_{I})-\kappa_{a}T_{a}\log(R_{b})}{-\kappa_{b}\log(R_{a})-\kappa_{a}\log(R_{I})+\kappa_{b}\log(R_{I})+\kappa_{a}\log(R_{b})},\\
   d_{b}&=\frac{\kappa_{a}(T_{a}-T_{b})}{\kappa_{b}\log(R_{a})+\kappa_{a}\log(R_{I})-\kappa_{b}\log(R_{I})-\kappa_{a}\log(R_{b})}.
 \end{aligned}
\end{equation} 
The adjoint system (\eref{eq:BenchAnn adjoint}) can be reduced to the following two ODEs,
\begin{equation}
\begin{aligned}
P_a^{''}(r) + \dfrac{1}{r} P_a^{'}(r) &= \dfrac{-2}{\kappa_a}(c_a + d_a \log(r)),\\
P_b^{''}(r) + \dfrac{1}{r} P_b^{'}(r) &= \dfrac{-2}{\kappa_b}(c_b + d_b \log(r)),
\end{aligned}    
\end{equation}
whose solution takes the form,
\begin{equation}
\begin{aligned}
P_a(r) &= C_1 \log (r) + D_1 - \dfrac{(c_a - d_a) r^2}{2 \kappa_a} +\dfrac{d_a r^2 \log (r)}{2 \kappa_a}, \\
P_b(r) &= C_2 \log (r) + D_2 - \dfrac{(c_b - d_b) r^2}{2 \kappa_b} +\dfrac{d_b r^2 \log (r)}{2 \kappa_b},
\end{aligned}    
\end{equation}
The coefficients $C_1, D_1, C_2, D_2$ are found by satisfying boundary conditions given in \eref{eq:BenchAnn adjoint}, i.e. from the following system,
\begin{multline}
 \begin{bmatrix}
       1 & \log(R_a) & 0 & 0\\
       0 & 0 & 1 & \log(R_b) \\
       1 & \log(R_L) & -1 & -\log(R_L) \\
       0 & \kappa_a & 0 & -\kappa_b
 \end{bmatrix} 
 \begin{bmatrix}
       C_1 \\
       D_1 \\
       C_2 \\
       D_2
\end{bmatrix} \\
= \begin{bmatrix}
   \dfrac{R_a^2 (c_a - d_a) + R_a^2 d_a \log (R_a)}{2 \kappa_a} \\
   \dfrac{R_b^2 (c_b - d_b) + R_b^2 d_b \log (R_b)}{2 \kappa_b} \\
   \dfrac{R_L^2}{2}\left(\dfrac{c_a - d_a}{\kappa_a} - \dfrac{c_b - d_b}{\kappa_b}\right) + \dfrac{R_L^2\log(R_L)}{2}\left(\dfrac{d_a}{\kappa_a} - \dfrac{d_b}{\kappa_b}\right)\\
   \dfrac{R_L^2}{2} \left(2 c_a - 2 c_b - d_a + d_b\right) + R_L^2 \log(R_L) \left(d_a - d_b\right)
\end{bmatrix}.  
\end{multline}

\bibliographystyle{model1-num-names}
\bibliography{Metamaterials.bib}

\end{document}